\input epsf.sty
\input analyt.sty

\hyphenation{pa-ra-me-tri-zat-ions pa-ra-me-tri-zat-ion dono-ghue}

\baselineskip=12pt
\parindent=18pt
\parskip=0pt plus 1pt
\hfuzz=2pt
\frenchspacing
\tolerance=500

\abovedisplayskip=3mm plus6pt minus 4pt
\belowdisplayskip=3mm plus6pt minus 4pt
\abovedisplayshortskip=0mm plus6pt minus 2pt
\belowdisplayshortskip=2mm plus4pt minus 4pt

\magnification1020
\hsize=15.cm\vsize=20cm

\footline={\hfill}
\rightline{FTUAM 02--28}
\rightline{\timestamp}
\rightline{(hep-ph/0212282)}
\bigskip
\hrule height .3mm
\vskip.6cm
\centerline{{\bigfib Low Energy Pion Physics}}
\hb
\centerline{\medfib -- Completely revised and updated version --}
\medskip
\centerrule{.7cm}
\vskip1cm

\setbox9=\vbox{\hsize65mm {\noindent\fib 
F. J. Yndur\'ain} 
\vskip .1cm
\noindent{\addressfont Departamento de F\'{\i}sica Te\'orica, C-XI,\hb
 Universidad Aut\'onoma de Madrid,\hb
 Canto Blanco,\hb
E-28049, Madrid, Spain.}\hb}
\smallskip
\centerline{\box9}
\vskip1truecm
\setbox0=\vbox{\abstracttype{Abstract} In these notes we present an introductory   
review on various topics about low energy pion physics
 (some kaon  physics is discussed as well). 
Among these, we include the uses of analyticity and unitarity to describe partial
 wave amplitudes (for which we give accurate and economical parametrizations)
 and form factors; 
(forward) dispersion relations; and
the use of the Froissart--Gribov representation to evaluate 
accurately the low energy parameters (scattering lengths and effective ranges)
 for higher ($l\geq1$) waves. 
Finally, we describe some pion physics in QCD and then pass on to
study  the nonlinear sigma model, and the chiral 
perturbation theory approach to low energy pion interactions.\hb
\indent
Most of the results presented here are well known, but we also give a set of 
state of the art,  precise 
determinations of some scattering lengths as well as an independent calculation of 
three of the parameters  $\bar{l}_2$, $\bar{l}_4$, $\bar{l}_6$
 (to one loop), and one $\bar{f}_2$,
  to  two loops,  that appear in chiral perturbation
theory.
The  S waves are  discussed   and compared 
with chiral perturbation theory expectations, and 
the same is done for larger $l$
scattering lengths and effective range parameters.
Also new is the evaluation of some electromagnetic corrections.
}
\centerline{\box0} 
\vskip3truecm
\hrule
\smallskip
{\petit\noindent
Work supported in part by CICYT, Spain.}
\vfill\eject
{Typeset with \physmatex}
\vfill\eject

\noindent {\medfib Contents}
\bigskip
{\petit 
\item{1. }{\fib Introduction}
\itemitem{1.1. }{Foreword\leaderfill 1}
\itemitem{1.2. }{Normalization; kinematics; isospin\leaderfill 2}
\itemitem{}{1.2.1. Conventions\leaderfill 2}
\itemitem{}{1.2.2. Isospin\leaderfill 5}
\itemitem{1.3. }{Field-theoretic, and other models\leaderfill 5}
\smallskip
\item{2. }{\fib Analyticity properties of scattering amplitudes, p.w. 
amplitudes, form factors\hb and correlators. High energy behaviour 
and bounds}
\itemitem{2.1. }{Scattering amplitudes and partial waves\leaderfill 7}
\itemitem{2.2. }{Form factors\leaderfill 9}
\itemitem{2.3. }{Correlators\leaderfill 10}
\itemitem{2.4. }{Scattering amplitudes at high energy: Reggeology\leaderfill 11}
\smallskip
\item{3. }{\fib The effective range formalism for p.w. amplitudes. 
Resonances}
\itemitem{3.1. }{Effective range formalism\leaderfill 17}
\itemitem{3.2. }{Resonances in (nonrelativistic) potential scattering\leaderfill 19}
\smallskip
\item{4. }{\fib The P p.w. amplitude for  $\pi\pi$ scattering in the elementary rho model}
\itemitem{4.1. }{The $\rho$ propagator and the $\pi^+\pi^0$ scattering amplitude\leaderfill 23}
\itemitem{4.2. }{The weak coupling approximation\leaderfill 26}
\itemitem{4.3. }{Low energy scattering\leaderfill 27}
\itemitem{4.4. }{The chiral rho model\leaderfill 27}
\smallskip
\item{5. }{\fib The effective range formalism for p.w. amplitudes; 
resonances (multichannel formalism).\hb Unitarity and form factors; correlators}
\itemitem{5.1. }{General formalism. Eigenphases\leaderfill 29}
\itemitem{5.2. }{The $K$-matrix and the effective range matrix. 
Resonances\leaderfill 31}
\itemitem{5.3. }{Resonance parametrizations in the two-channel case\leaderfill 33}
\itemitem{5.4. }{Reduction to a single channel. Weakly coupled channels\leaderfill 33}
\itemitem{5.5. }{Unitarity for the form factors\leaderfill 35}
\itemitem{5.6. }{Unitarity for correlators\leaderfill 37}
\smallskip
\item{6. }{\fib Extraction and parametrizations of p.w. amplitudes for $\pi\pi$ scattering.\hb
Form factors}
\itemitem{6.1. }{$\pi\pi$ scattering\leaderfill 39}
\itemitem{6.2. }{Form factors and decays\leaderfill 40}
\itemitem{ }{6.2.1. The pion form factor\leaderfill 40}
\itemitem{ }{6.2.2. Form factor of the pion in $\tau$ decay\leaderfill 41}
\itemitem{ }{6.2.3.  $K_{l4}$ decay\leaderfill 42}
\itemitem{ }{6.2.4. The $K\to2\pi$ decays\leaderfill 43}
\itemitem{6.3. }{The P  wave\leaderfill 44}
\itemitem{ }{6.3.1. The P wave in the elastic approximation\leaderfill 44}
\itemitem{ }{6.3.2. The $\rho$ and weakly coupled channels: 
$\omega-\rho$ interference \leaderfill 46}
\itemitem{ }{6.3.3. The P wave for $1\,\gev\leq s^{1/2}\leq 1.42\,\gev$\leaderfill 47}
\itemitem{6.4. }{The S waves\leaderfill 48}
\itemitem{ }{6.4.1. Parametrization of the S wave for  $I=2$\leaderfill 48}
\itemitem{ }{6.4.2. Parametrization of the S wave for  $I=0$\leaderfill 50}
\itemitem{ }{6.4.3. The $I=0$ S wave between $960\,\mev$ and $1420\,\mev$\leaderfill 55}
\itemitem{6.5. }{The D,  F and G waves\leaderfill 57}
\itemitem{ }{6.5.1. Parametrization of the  $I=2$ D wave \leaderfill 57}
\itemitem{ }{6.5.2. Parametrization of the  $I=0$ D wave\leaderfill 58}
\itemitem{ }{6.5.3. The F wave \leaderfill 62}
\itemitem{ }{6.5.4. The G waves \leaderfill 62}
\itemitem{6.6. }{On experimental phase shifts in the range $1.4\,\gev\simeq
s^{1/2}\simeq 2\,\gev$\leaderfill 62}\smallskip
\item{7. }{\fib Analyticity; dispersion relations and the 
Froissart--Gribov representation.\hb
Form factors: the Omn\`es--Muskhelishvili method}
\itemitem{7.1. }{The Omn\`es--Muskhelishvili method\leaderfill 65}
\itemitem{ }{7.1.1. The full Omn\`es--Muskhelishvili problem \leaderfill 65}
\itemitem{ }{7.1.2. The incomplete Omn\`es--Muskhelishvili  problem
\leaderfill 67}
\itemitem{7.2. }{Application to the pion form factors of the
 Omn\`es--Muskhelishvili\hb method\leaderfill 69}
\itemitem{ }{7.2.1. The electromagnetic form factor\leaderfill 69}
\itemitem{ }{7.2.2. The scalar form factor and radius of the pion\leaderfill 72}
\itemitem{ }{7.2.3. The mixed $K\pi$ scalar form factor\leaderfill 75}
\itemitem{7.3. }{Dispersion relations and Roy equations\leaderfill 76}
\itemitem{ }{7.3.1. Fixed $t$ dispersion relations\leaderfill 76}
\itemitem{ }{7.3.2. Forward dispersion relations\leaderfill 77}
\itemitem{ }{7.3.3. The Roy equations\leaderfill 78}
\itemitem{7.4. }{Evaluation of  forward dispersion relations for $\pi\pi$ 
 scattering 
\leaderfill 79}
\itemitem{ }{7.4.1. The Olsson sum rule\leaderfill 80}
\itemitem{ }{7.4.2.   $\pi^0\pi^0$ \leaderfill 81}
\itemitem{ }{7.4.3.   $\pi^0\pi^+$ \leaderfill 82}
\itemitem{7.5.}{The Froissart--Gribov representation and low energy
 P, D, F wave parameters\leaderfill 83}
\itemitem{ }{7.5.1. Generalities\leaderfill 83}
\itemitem{ }{7.5.2. D waves\leaderfill 84}
\itemitem{ }{7.5.3. P and F waves\leaderfill 86}
\itemitem{ }{7.5.4. G waves\leaderfill 87}
\itemitem{7.6.}{Summary and conclusions\leaderfill 87}
\itemitem{ }{7.6.1. The S, P partial waves of Colangelo,
Gasser and Leutwyler\leaderfill 87}
\itemitem{ }{7.6.2. The S wave scattering lengths of Descotes et al., and 
Kami\'nski et al. \leaderfill 89}
\itemitem{ }{7.6.3. Comparison of different calculations. 
Low energy parameters for $\pi\pi$
scattering \leaderfill 90}
\smallskip
\goodbreak
\item{8. }{\fib QCD, PCAC and chiral symmetry for pions and kaons}
\itemitem{8.1. }{The QCD Lagrangian. Global symmetries; conserved currents\leaderfill 93}
\itemitem{8.2. }{Mass terms and invariances; chiral invariance\leaderfill 96} 
\itemitem{8.3. }{Wigner--Weyl and Nambu--Goldston 
realization of symmetries\leaderfill 100}
\itemitem{8.4. }{PCAC, $\pi^+$ decay, the pion propagator and
 light quark mass ratios\leaderfill 101}
\itemitem{}{8.4.1.  The weak axial current and $\pi^+$ decay\leaderfill 101}
\itemitem{}{8.4.2.  The pion propagator; quark mass ratios\leaderfill 102}
\itemitem{8.5. }{Bounds and estimates of light quark masses
 in terms of the pion and kaon masses
\leaderfill 104}
\itemitem{8.6. }{The triangle anomaly; $\pi^0$ decay. 
The gluon anomaly. The $U(1)$ problem\leaderfill 108}
\itemitem{ }{8.6.1. The triangle anomaly and $\pi^0$ decay\leaderfill 108}
\itemitem{ }{8.6.2. The $U(1)$ problem and the gluon anomaly\leaderfill 114}
\smallskip
\item{9. }{\fib Chiral perturbation theory}
\itemitem{9.1. }{Chiral Lagrangians\leaderfill 117}
\itemitem{ }{9.1.1. The $\sigma$ model\leaderfill 117}
\itemitem{ }{9.1.2. Exponential formulation\leaderfill 119}
\itemitem{9.2. }{Connection with PCAC, and a first application\leaderfill 121}
\itemitem{9.3. }{Chiral perturbation theory: general formulation\leaderfill 123}
\itemitem{ }{9.3.1. Gauge extension of chiral invariance\leaderfill 123}
\itemitem{ }{9.3.2. Effective Lagrangians in the chiral limit\leaderfill 124}
\itemitem{ }{9.3.3. Finite pion mass corrections\leaderfill 126}
\itemitem{ }{9.3.4. Renormalized effective theory\leaderfill 127}
\itemitem{ }{9.3.5. The parameters of chiral perturbation theory \leaderfill 128}
\itemitem{9.4. }{Comparison of chiral perturbation theory to one loop 
with experiment\leaderfill 130}
\itemitem{ }{9.4.1. One loop coupling constants, 
and $\pi\pi$ scattering and the electromagnetic\hb
form factor of the pion\leaderfill 130}
\itemitem{ }{9.4.2. The scalar form factor of the pion\leaderfill 133}
\itemitem{ }{9.4.3. Summary of ch.p.t. predictions for $\pi\pi$ scattering\leaderfill 135}
\itemitem{9.5. }{Weak and electromagnetic interactions.\hb The 
accuracy of chiral perturbation theory calculations\leaderfill 136}

\phantom{x\leaderfill 117}

\item{}{\kern-2em{\fib Appendix A: Summary of low energy, $s^{1/2}\leq1.42\,\gev$
 partial waves} \leaderfill 139}
\item{}{\kern-2em{\fib Appendix B: The conformal mapping method}\leaderfill 146}
\item{}{\kern-2em{\fib Appendix C: Sum rules and asymptotic behaviour}\leaderfill 149}
\medskip

\noindent{{\fib Acknowledgements}\leaderfill 153}

\noindent{{\fib References}  \leaderfill 153} 

\noindent{\phantom{x}} 
}
\vfill\eject 

\brochureendcover{}

\pageno=1
\brochureb{\smallsc chapter 1}{\smallsc introduction}{1}

\bookchapter{1. Introduction}
\vskip-0.5truecm
\booksection{1.1. Foreword}
The matter of parametrizations and uses of pion-pion partial waves (p.w.), 
form factors and correlators, in particular in connection with
 resonances, received a great deal of attention in
the late fifties, sixties and  early seventies of last century --until 
QCD emerged as the theory of strong interactions and such studies were 
relegated to a secondary plane. 
In recent times a renewed interest has arisen in this subject and this due to, at least, 
the following  reasons. 
One is the popularity of chiral perturbation theory calculations
 (to which the last part of these notes is
devoted), in  particular of low energy $\pi\pi$ 
parameters: 
scattering lengths and ranges, pion charge radius, etc. 
A second reason is the use of low energy calculations of the pion form factor to get precise
estimates of  the muon magnetic moment or the value of the QED charge on the 
$Z$ particle. And last, but certainly not least, we have the appearance of new 
experimental data on hadronic $\tau$ decay, the pion electromagnetic form factor and on $K_{e4}$ decay. 
The existence of these data allow a much improved determination of low energy pionic observables.

Unfortunately, some of the old lore appears to be lost and indeed many  
modern calculators seem to be unaware of parts of it. 
In the present review we do not present much new knowledge, but mostly intend 
to give an introductory, easily accessible reference to the 
studies of scattering amplitudes, form factors and correlators 
involving pions in the low energy region: our 
aim is, in this respect, mainly pedagogical. 
 
Nevertheless, some very recent results are reported. 
These include parametrizations of the lowest waves (S, P, D, F) in $\pi\pi$ scattering 
which are compatible with analyticity and unitarity and, of course, experimental data, 
depending only on a few (two to four) parameters per wave. 
This is used to evaluate forward dispersion relations and 
the Froissart--Gribov representation of the P, D and higher waves. 
From this there follow very precise determinations of the corresponding 
scattering lengths and effective range parameters. 
Using this, as well as the results on the P wave following from the electromagnetic pion form 
factor and the decay $\tau\to \nu \pi^0\pi^+$, 
we obtain, in particular, 
 a precise
determination (however, only to one loop) of some of  the $\bar{l}$
 parameters in chiral perturbation theory,
as  well as an evaluation  
 of some electromagnetic corrections. 
The scalar radius of the pion is another example.

The plan of this review is as follows. 
In Chapters 1 and 2 we describe briefly the analyticity, 
unitarity and high energy properties of various quantities 
(correlators, form factors, partial waves and scattering amplitudes). 
The elements of the effective range formalism and the characterization of resonances 
are given in Chapter~3. These topics are illustrated in a simple model in Chapter~4,
 while in Chapter~5 we extend the previous analyses (including the requirements of unitarity) 
to the multichannel case.

The core of the review is contained in the last four chapters. In  chapters~6 and 7  
we apply the tools described before to the study of 
partial wave amplitudes and scattering amplitudes 
for $\pi\pi$ scattering and to fit the pion form factors, 
electromagnetic and scalar. 
Here we also implement simple parametrizations of partial wave amplitudes consistent with 
analyticity and unitarity, and fitting experimental data; 
this should be useful to people needing manageable 
representations of  $\pi\pi$ phases, as happens e.g. for 
$J/\psi\to\gamma\pi\pi$ studies. 
Then we discuss (Chapter~7) how the various theoretical requirements (fixed $t$ 
dispersion relations and the Froissart--Gribov representation) may be used 
to check compatibility of the results found with crossing symmetry and 
analyticity for the scattering amplitudes. 
The  Froissart--Gribov representation is also used to get precise determinations of 
low energy parameters for the waves with $l=1$ and higher.
With respect to form factors, the Omn\`es--Muskhelishvili method is 
employed to perform an accurate fit to the 
pion form factor, obtaining in particular precise values of the 
corresponding low energy parameters, and also to give a reliable determination of the 
scalar radius of the pion.
Something which is missing in this review is the Roy equations analysis; 
there are in the literature three recent papers (Ananthanarayan 
et al., 2001, Colangelo, Gasser and Leutwyler, 2001, Descotes et al.,~2002) that 
fill this gap. Some of the
 results we obtain are summarized in \sect~7.6 where, in particular, we present a 
discussion of the results of Descotes et al. (2002) and, 
especially, Ananthanarayan et al.~(2001) and Colangelo, Gasser and Leutwyler~(2001); 
in particular, in view of the criticism of the last 
by Pel\'aez and Yndur\'ain~(2003a), Yndur\'ain~(2003b).

In Chapter~8 we remember that pions are made of quarks, and that we have a theory for 
the interactions of these, QCD. We discuss invariance properties of the QCD 
Lagrangian, in particular chiral invariance that plays a key role for 
the dynamics of pions. 
We use this and PCAC to derive relations between the masses of
 the quarks and the pion and kaon masses, and to
study pion decay. Finally, in Chapter~9 we develop the consistent description of 
pion dynamics based on chiral invariance,  known as chiral perturbation theory. 
In the last sections of this chapter we use the results obtained 
in Chapters~6 and 7 to test the predictions of chiral perturbation theory, 
and show how  to obtain values for the parameters on which it depends.

Before entering into the main body of these notes, it is convenient to clarify what is to be 
understood as ``low energy." 
Above energies $s^{1/2}$ of, say, $1.3\;\sim\;2\,\gev$, perturbative 
QCD (or Regge theory, as the case may be)  is applicable; we will   
be very little concerned with these energies. 
At very low energies, $s^{1/2}\ll\Lambdav_0$, where $\Lambdav_0$ 
is a scale parameter that (depending on the process) may vary 
from $\simeq600\,\mev$ to $4\pi f_\pi\sim 1.1\,\gev$, 
chiral perturbation theory is applicable; this we treat 
in detail in Chapters~8 and 9. Between the two energy scales, analyticity and 
unitarity allow at least an {\sl understanding} of pionic observables. 
This understanding certainly holds until inelastic production begins to become important. 
This means that we are able to cover the energy range of   $s^{1/2}$ 
below 1.42 \gev. 

These notes are primarily about pions. However, in some cases 
kaons and (to a lesser extent) etas are treated as well.

\booksection{1.2. Normalization; kinematics; isospin}
\vskip-0.5truecm
\booksubsection{1.2.1. Conventions}

\noindent
Before entering into specific discussions we will say a few words on 
our normalization conventions.\fnote{We assume here a basic knowledge of $S$ matrix theory, 
in particular of crossing symmetry or partial wave expansions, and of isospin invariance, 
at the level of the first chapters of the texts of 
Martin, Morgan and Shaw~(1976) or Pilkuhn~(1967).} 
If $S$ is the relativistically invariant scattering matrix we 
define the scattering amplitude $F$ for particles $A$, $B$ to give particles $C_i$ 
by
$$\langle C_1,\dots,C_n|S|A,B\rangle=
\ii \delta(P_f-P_i)F(A+B\to C_1+\cdots+C_n).
\equn{(1.2.1)}$$
We take the states to be normalized in a relativistically invariant manner:  
if $p$ is the four-momentum, and $\lambda$ the helicity of a particle, then
$$\langle p,\lambda|p',\lambda'\rangle=2\delta_{\lambda\lambda'}p_0\delta({\bf p}-{\bf p}').
\equn{(1.2.2)}$$
We will seldom consider particles with spin in these notes. 
Particles with spin pose problems of their own; the generalization of 
our discussions to spinning particles is not trivial.

It is the function $F$, defined as in (1.2.1), with the states 
normalized as in (1.2.2), the one which is free of 
kinematical singularities and zeros. 
That is to say, any discontinuity or pole of $F$ is associated with 
dynamical effects. 
If we had used a nonrelativistic normalization 
and defined a corresponding scattering amplitude $T_{\rm NR}$, 
we could write
$$T_{\rm NR}=
\dfrac{1}{\sqrt{2p^0_{A}}}\dfrac{1}{\sqrt{2p^0_{B}}}
\dfrac{1}{\sqrt{2p^0_{{C_1}}}}\dots\dfrac{1}{\sqrt{2p^0_{{C_n}}}}\,F.
\equn{(1.2.3)}$$
Then, no matter which field-theoretic interaction we assumed, 
 $T_{\rm NR}$ would show the branch cuts associated with the factors $1/\sqrt{p^0}$ in (1.2.3).

In what regards form factors care has to be exercised to get 
form factors without kinematic cuts. 
For the simple case of the e.m. (electromagnetic) 
form factor of the pion 
(or any other spinless particle) such form factor is that defined by
$$\langle p_1|J^{\rm e.m.}_\mu(0)|p_2\rangle=
(2\pi)^{-3}(p_1+p_2)_\mu F_\pi(t),\quad t=(p_1-p_2)^2.
\equn{(1.2.4a)}$$
Note that, with this definition, $F_\pi(0)=1$. 
\equn{(1.2.4a)} is valid for spacelike $t\leq0$. 
For timelike $t\geq 4\mu^2$ we write
$$\langle p_1,p_2|J^{\rm e.m.}_\mu(0)|0\rangle
=(2\pi)^{-3}(p_1-p_2)_\mu F_\pi(t),\quad t=(p_1+p_2)^2.
\equn{(1.2.4b)}$$
Both values of $F_\pi$ are particular cases of a single function, $F_\pi(t)$, 
that can be defined for arbitrary, real or complex values 
of the variable $t$ (see below).

\topinsert{
\setbox0=\vbox{\hsize 11.9truecm{\epsfxsize 9.5truecm\epsfbox{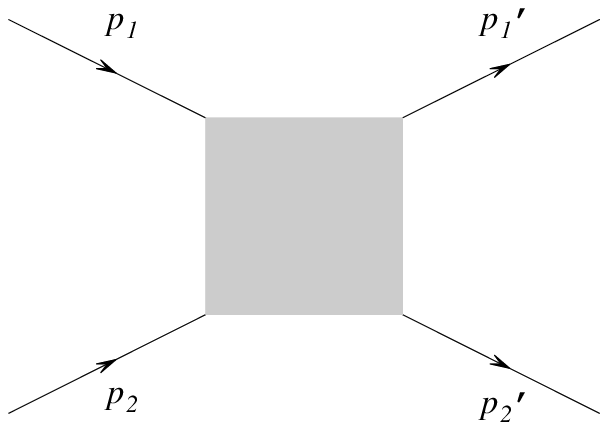}}} 
\setbox6=\vbox{\hsize 7truecm\captiontype\figurasc{Figure 1.1.1. }{ Two 
particle scattering.\hb
\phantom{XX}}\hb
\vskip.1cm} 
\medskip
\centerline{\box0}
\centerline{\box6}
\medskip
}\endinsert

We finish this subsection with a few more definitions. 
Let us consider scattering of two particles, with masses $m_i$ (Fig.~1.1.1):
$$A_1(p_1)+A_2(p_2)\to A'_1(p'_1)+A'_2(p'_2).$$
We define the Mandelstam variables
$$s=(p_1+p_2)^2,\quad u=(p_1-p'_2)^2,\quad t=(p_1-p'_1)^2.$$
They satisfy the equality, when the particles are 
 on their mass shells,
$$s+u+t=\sum_im_i;$$
for pions,
$$s+u+t=4\mu.$$
Here, and throughout these notes, $\mu\equiv138\,\mev$ is the {\sl average} mass of the pions. 
When referring specifically to neutral or charged pion masses 
we will write $m_{\pi^0}$ or $m_{\pi^{\pm}}$. 
The notation $M_\pi$ for $m_{\pi^{\pm}}$ will also be used; see below.

In terms of these variables the modulus of the 
three-momentum, $k$, and the cosine of the 
scattering angle (both in the center of mass) 
are given by
$$k=\dfrac{\sqrt{s-4\mu^2}}{2},\quad \cos\theta=1+\dfrac{2t}{s-4\mu^2}.$$

With our definitions, the two body differential cross 
section in the center of mass is given in terms of $F$ as
$$\left.\dfrac{\dd \sigma}{\dd\omegav}\right|_{\rm c.m.}=\dfrac{\pi^2}{4s}
\dfrac{k'}{k}|F(i\to f)|^2,
\equn{(1.2.5)}$$
with $k$, $k'$ the moduli of the three-momenta of 
initial, final particles. For particles with arbitrary masses $m_i$  
(1.2.5) is still valid but now
$$k=\dfrac{1}{2s^{1/2}}\sqrt{[s-(m_1-m_2)^2][s-(m_1+m_2)^2]}$$ 

The total cross section, also with the same conventions, is
$$\sigma_{\rm tot}(s)=\dfrac{4\pi^2}{\lambda^{1/2}(s,m_1,m_2)}\imag F(s,0);
\equn{(1.2.6)}$$
here we define K\"all\'en's quadratic form
$$\lambda(a,b,c)=a^2+b^2+c^2-2ab-2ac-2bc=
\big[a-(\sqrt{b}-\sqrt{c})^2\big]\,\big[a-(\sqrt{b}+\sqrt{c})^2\big].$$

\booksubsection{1.2.2. Isospin}

\noindent
As we know, there are three kinds of pions: two charged ones, $\pi^\pm$ with a mass
$m_{\pi^\pm}=139.57\,\mev$, and a neutral pion, with mass $m_{\pi^0}=134.98\,\mev$. 
If we neglected electromagnetic interactions, and the 
mass difference between $u$, $d$ quarks, then the interactions of the three pions would be 
identical, and they would have a common mass, that we denote by $\mu$ and take equal to the average: 
$\mu=138\,\mev$.
Unfortunately, and under the influence of Gasser and Leutwyler, 
it has become customary to use the mass of the charged pion as the common pion mass. 
We will (regretfully) follow this convention and will then write $M_\pi=139.57\,\mev$ 
for the common mass of the pions, when identified with the charged pion mass.
When distinguishing individual pion masses we will write, as 
stated above, $m_{\pi^\pm},\,m_{\pi^0}$.

The invariance under rotations of the three pions, called {\sl isospin invariance}, 
is best described by 
introducing a different basis to describe the pions, $|\pi_i\rangle$, $i=1,\,2,\,3$, 
related to the physical pions by
$$|\pi^0\rangle=|\pi_3\rangle,\quad
 |\pi^\pm\rangle=\dfrac{\mp1}{\sqrt{2}}\Big\{|\pi_1\rangle\pm\ii|\pi_2\rangle\Big\}.
$$
Isospin transformations are then just rotations, $|\pi_j\rangle\to\sum_kR_{jk}|\pi_k\rangle$ 
with $R$ a rotation matrix. 
We can then, in the limit of exact isospin invariance, diagonalize the total isospin and its third
component and consider e.g.  scattering amplitudes corresponding to fixed isospin.

The development of the isospin formalism, including explicit expressions, 
may be found in the book of Martin, Morgan and Shaw~(1976); here we only give, for 
ease of reference, the isospin crossing 
matrices:
$$C^{(ts)}=C^{(st)}=\pmatrix{
1/3&1&5/3\cr
1/3&1/2&-5/6\cr
1/3&-1/2&1/6\cr
};
\equn{(1.2.7a)}$$
$$C^{(us)}_{II'}=C^{(su)}_{II'}=(-1)^{I+I'}C^{(st)}_{II'}.
\equn{(1.2.7b)}$$
These matrices act according to 
$$F^{(I_s=I)}=\sum_{I'}C^{(st)}_{II'} F^{(I_t=I')},\quad\hbox{etc.}$$
Note that the order is 0, 1, 2; thus, e.g., 
$$F^{(I_s=2)}=\tfrac{1}{3}F^{(I_t=0)}-\tfrac{1}{2}F^{(I_t=1)}+\tfrac{1}{6}F^{(I_t=2)}.
$$

\booksection{1.3. Field-theoretic, and other models}

\noindent 
It is always convenient to illustrate abstract arguments with model calculations 
in which one can see how the general properties are realized in explicit examples. 
We will take as a very convenient model one in which pions and rho are 
realized as elementary fields, $\phi_a$, $\rho_a$ ($a$ is an 
isospin index). 
The corresponding Lagrangian will be
$${\cal L}=g_{\mu\nu}(\delta_{ab}\partial_\mu-\ii g_\rho\epsilon_{abc}\rho^c_\mu)\phi_b
(\delta_{ad}\partial_\nu-\ii g_\rho\epsilon_{ade}\rho^e_\nu)\phi_d
-\mu^2 \phi_a\phi_a+{\cal L}_\rho.
\equn{(1.3.1)}
$$
Here $\mu$ is the pion mass and ${\cal L}_\rho$ is the pure rho Lagrangian, that need not be specified. 
The mass of the rho particle, $M_\rho$, can be assumed to be introduced by a Higgs-type 
mechanism, with the mass of the associated Higgs particle so large that it will have no 
influence on calculations for energies of the order of $M_\rho$ or lower. 

 The interactions in (1.3.1) induce pion-rho vertices: 
a $\pi\pi\rho$ vertex, which is  
associated with the factor $\ii g_\rho (p_1-p_2)_\mu \epsilon_{abc}$, 
and a seagull one proportional to $\ii g^2_\rho g_{\mu\nu}$.
This model is {\sl not} chiral invariant (see later for a chiral invariant version) 
but it is very simple and thus will be used to illustrate general properties, 
such as analyticity or unitarity, 
independent of the underlying dynamics.

Another explicit model of $\pi\pi$ scattering is that of Veneziano~(1968); see also 
Lovelace~(1968) and  Shapiro~(1969). 
In it the amplitude is written as a combination of beta functions, thus as a sum of poles. 
These are then unitarized (e.g., \`a la Lovelace).

We will not spend any time on this model. While it gives reasonably well the masses 
and widths of the lowest lying resonances, it is a disaster for energies above $\sim 1\,\gev$. 
It predicts elastic heavy resonances and, while its
 high energy behaviour contains the $\rho$ and $P'$ Regge
poles, it 
does not admit a Pomeron.

\bookendchapter
\brochureb{\smallsc chapter 2}{\smallsc analyticity properties of 
scattering
 amplitudes, etc. bounds}{7}
\bookchapter{2. Analyticity properties of 
scattering\hb
 amplitudes, p.w. amplitudes,\hb form factors and correlators.\hb High energy behaviour 
and bounds}
\vskip-0.5truecm
\booksection{2.1. Scattering amplitudes and partial waves}

\noindent
Analyticity of partial waves follows from unitarity and causality. 
In local field theories (such as QCD) 
both properties are, of course, satisfied, but 
locality at least would be violated in a theory of strings; although this 
would occur at energies much higher than the ones in which we are interested here.
In the case of the $\pi\pi$ scattering amplitude, 
$F(s,t)$, one can prove that it is, for  $t$ in the 
Martin--Lehmann 
ellipse,\fnote{With foci at $t=0$ and $t=4\mu^2-s$ (for pions) 
and right extremity at $t=4\mu^2$. This includes the 
physical region, $4\mu^2-s\leq t\leq 0$. For the proof, 
see Martin~(1969) and references there.} 
analytic in the complex $s$ plane with the exception of two cuts: 
a r.h. (right hand) cut, from 
$s=4\mu^2$ to $+\infty$, and a l.h. (left hand) 
cut from $-\infty$ to $-t$.
In addition, if there existed bound states, there would appear poles at the values 
of $s$ or $u$ given by the square of the mass of the bound state.

The p.w. (partial wave) amplitudes, $f_l(s)$, are related to 
$F$ though the  expansion,
$$F(s,t)=\sum_{l=0}^\infty (2l+1)P_l(cos\theta)f_l(s)
\equn{(2.1.1a)}$$
with inverse
$$f_l(s)=\tfrac{1}{2}\int_{-1}^{+1}\dd \cos\theta P_l(cos\theta)F(s,t).
\equn{(2.1.1b)}$$
Here $\theta$ is the scattering angle in the center of mass, and 
$P_l$ are the 
Legendre polynomials. 
We note that the restriction of, say, $s$ to physical values 
produces the {\sl physical} $F(s,t)$ and $f_l(s)$ 
{\sl provided} we take the limit of $s$ real from the upper half plane. 
That is to say, if $s$ is real and physical (and so is $t$), 
the physical values of scattering amplitude and partial waves 
are obtained for
$$F(s,t)=\lim_{\epsilon\to+0}F(s+\ii\epsilon,t);
\quad f_l(s)=\lim_{\epsilon\to+0}f_l(s+\ii\epsilon).$$

For elastic scattering at {\sl physical} $s$ (which, for $\pi\pi$ scattering means $s$ 
real and larger than or equal to $4\mu^2$) and below the opening of the 
first inelastic threshold, that we will denote by $s_0$, one can express the $f_l$ 
in terms of phase shifts:
$$f_l(s)=\dfrac{2s^{1/2}}{\pi k}\sin\delta_l(s)\ee^{\ii\delta_l(s)}
=\dfrac{2s^{1/2}}{\pi k}\dfrac{1}{\cot\delta_l(s)-\ii};\quad
4\mu^2\leq s\leq s_0.
\equn{(2.1.2)}$$

The previous equations are valid assuming that the scattering 
particles are {\sl distinguishable}. 
For $\pi\pi$ scattering, however, the situation is a bit 
complicated. 
One may still write (2.1.1) and (2.1.2) for the processes 
$\pi^0\pi^+\to\pi^0\pi^+$, but not for $\pi^0\pi^0\to\pi^0\pi^0$ or 
$\pi^+\pi^+\to\pi^+\pi^+$. 
The general recipe is the following: if $F^{(I_s)}$ 
is an amplitude with isospin $I_s$ in channel $s$, 
one has to replace (2.1.1) by
$$\eqalign{
F^{(I_s)}(s,t)=&\,2\times\sum_{l={\rm even}}(2l+1)P_l(\cos\theta)f_l^{(I_s)}(s),
\quad I_s=\hbox{even},\cr
F^{(I_s)}(s,t)=&\,2\times\sum_{l={\rm odd}}(2l+1)P_l(\cos\theta)f_l^{(I_s)}(s),
\quad I_s=\hbox{odd,}\cr
f_l^{(I)}(s)=&\,\dfrac{2s^{1/2}}{\pi k}\sin\delta_l^{(I)}(s)\ee^{\ii\delta_l^{(I)}(s)}.\cr 
\cr}
\equn{(2.1.3)}$$
Due to Bose statistics, even waves only exist with isospin $I=0,\,2$ and odd waves must 
necessarily have isospin $I=1$. 
For this reason, we will often omit the 
isospin index for odd waves, 
writing e.g. $f_1$, $f_3$ instead of $f_1^{(1)}$, 
$f_3^{(1)}$.

Another very convenient simplification that we will use here 
is to denote the $\pi\pi$ partial waves by S0, S2, P, D0, D2, F, etc., 
in self-explanatory notation.

When inelastic channels  open (2.1.2) is no more valid, 
but one can still write
$$f_l(s)=\dfrac{2s^{1/2}}{\pi k}
\left[\dfrac{\eta_l\,\ee^{2\ii\delta_l}-1}{2\ii}\right].
\equn{(2.1.4a)}$$
Here $\eta_l$, called the {\sl inelasticity parameter} for wave $l$, 
 is  positive and smaller than or equal to unity. 
The elastic and inelastic cross sections, for a given wave, are given in terms of 
$\delta_l$ and $\eta_l$ by
$$\sigma^{\rm el.}_l=\tfrac{1}{2}\left\{\dfrac{1+\eta_l^2}{2}-\eta\cos2\delta_l\right\},\quad
\sigma^{\rm inel.}_l=\dfrac{1-\eta_l^2}{4};
\equn{(2.1.4b)}$$
$\sigma^{\rm el.}_l,\;\sigma^{\rm inel.}_l$ are defined so that, 
for collision of particles $A$, $B$ (assumed distinguishable), 
$$\sigma_{\rm tot.}=\dfrac{4\pi^2}{\lambda^{1/2}(s,m_A,m_B)}\,
\dfrac{2s^{1/2}}{\pi k}\sum_l(2l+1)
[\sigma^{\rm el.}_l+\sigma^{\rm inel.}_l].$$

 For $\pi\pi$ scattering $s_0=16\mu^2$, but the approximation of 
neglecting inelasticity is valid at the 2\% level or better 
below $s\simeq 1\,\gev$, the precise value depending on 
the particular wave.

The cut structure is more complicated for other processes. 
For example, for $\pi K$ scattering the r.h. cut starts at 
$s=(\mu+m_K)^2$ and the l.h cut also begins at $u=(\mu+m_K)^2$. 
But, since now $s+u+t=2\mu^2+m^2_K$, the l.h. cut in the variable $s$ 
runs from $-\infty$ to $-t+(m_K-\mu)^2$ for the scattering amplitude, and
 from $-\infty$ to $(m_K-\mu)^2$ for the p.w. amplitudes.

For $\bar{K}K\to\pi\pi$ or $\pi\pi\to\bar{K}K$ scattering, 
the  $u$-channel is $\pi K$ scattering. Therefore, the 
r.h. cut starts at the (unphysical) value $s=4\mu^2$, and the l.h. cut at 
$u=(\mu+m_K)^2$, hence the l.h. cut 
runs from $-\infty$ to $-t+(m_K-\mu)^2$ for the scattering amplitude, and
 from $-\infty$ to $(m_K-\mu)^2$ for the p.w. amplitudes.
  Finally, for $\bar{K}K$ scattering, the l.h. cut runs, as for 
$\pi\pi$, from $-\infty$ to 0 for p.w. amplitudes; but there is 
a r.h. cut, both for the amplitude and for p.w.'s 
associated with the unphysical channel $\bar{K}K\to\pi\pi$, and  
starting at $4\mu^2$.

\topinsert{
\setbox0=\vbox{\hsize8.5truecm{\epsfxsize 7truecm\epsfbox{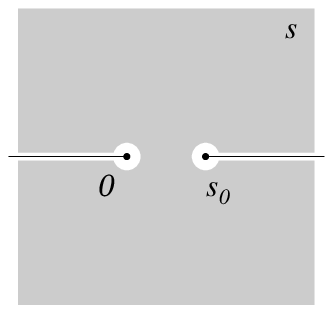}}} 
\setbox6=\vbox{\hsize 4truecm\captiontype\figurasc{Figure 2.1.1. }{\hb The  domain of 
analyticity for $\pi\pi$ partial waves (shadowed region)
 with the cuts of $f_l(s)$ in the complex $s$ plane.\hb
\phantom{XX}}\hb
\vskip.1cm} 
\medskip
\line{
{\box0}\hfil\box6}
\medskip
}\endinsert

Note that from (2.1.1b) it follows that it is $f_l(s)$ 
that satisfies analyticity properties without kinematical zeros or singularities.
These analyticity properties are rather complicated in general; 
in the simple case of $\pi\pi$ scattering 
we have that $f_l(s)$ is analytic in the complex $s$ plane except for two 
cuts, one from $4\mu^2$ to $+\infty$ and another from 
$-\infty$ to 0 (\fig~2.1.1), inherited respectively from the 
r.h. and l.h. cuts of $F(s,t)$.
If there existed bound states (which is not the case 
for $\pi\pi$), there would appear poles at the values 
of $s$ given by the square of the mass of the bound state.

Another property that follows from (2.1.1b) plus the assumption 
(realized in the real world for $\pi\pi$ scattering) that there is no bound state with zero 
energy is the behaviour, as $k\to0$,
$$f_l(s)\simeqsub_{k\to0}\dfrac{4\mu}{\pi}k^{2l}a_l
\equn{(2.1.5)}$$
where $a_l$ is the so-called $l$-th wave {\sl scattering length}. 

Analyticity is seldom of any use without bounds. 
Again, on very general grounds one knows that, for
$t$ physical, 
$|F(s,t)|$ is bounded by 
$C|s|\log^2|s|$ (the Froissart bound). 
Specific behaviours, particularly those that hold in Regge theory, will be discussed below.
For the proof of the Froissart and related bounds we 
require unitarity, causality and the assumption that Green's functions grow, at 
most, like polynomials of the momenta. This last 
assumption holds in renormalizable field theories, 
to all orders in perturbation theory; but may fail 
for nonrenormalizable ones. General discussions of analyticity, 
bounds and expected high energy behaviour of 
scattering amplitudes may be found in Eden, Landshoff, Olive and Polkinghorne~(1966), 
where the analyticity properties of Feynman graphs are also discussed, 
 Martin~(1969), Barger and
Cline~(1969), Sommer~(1970),  Yndur\'ain~(1972), etc. 
For general field theory, cf.~Bogolyubov, Logunov and Todorov~(1975).

\booksection{2.2. Form factors}
Analyticity of form factors, such as the pion or kaon form factors,
 can be proved quite generally using only
causality and  unitarity. 
In particular the electromagnetic pion form factor $F_\pi(t)$ turns out to 
be analytic in the complex $t$ plane cut from $t=4\mu^2$ to 
$+\infty$ (\fig~2.2.1). 
This analyticity, in particular, provides the link between both 
definitions of $F_\pi$, \equs~(1.2.4). 
For timelike, physical $t$, the physical value 
should in fact be defined as
$$F_\pi(t)=\lim_{\epsilon\to+0}F_\pi(t+\ii\epsilon);\quad t\geq 4\mu^2;$$
if we had taken $\lim_{\epsilon\to+0}F_\pi(t-\ii\epsilon)$
 we would have obtained $F^*_\pi(t)$.

Unlike scattering amplitudes, for which bounds hold in {\sl any} 
local field theory, one cannot prove bounds for 
form factors in general. 
However, bounds can be obtained in QCD, where we can even find the high 
momentum behaviour with the Brodsky--Farrar counting rules. 
In particular, for $F_\pi$ one has the Farrar--Jackson~(1979) behaviour
$$F_\pi(t)\simeqsub_{t\to\infty}\;\dfrac{12C_F\pi f^2_\pi\alpha_s(-t)}{-t},
\equn{(2.2.1)}$$
where $f_\pi$ is the pion decay constant, $f_\pi\simeq 93\,\mev$, 
$\alpha_s$ the QCD coupling, and $C_F=4/3$ is a colour factor. 
For scalar form factors (such as the scalar form 
factor of the pion), that we denote by $F_S$, the analyticity is like for 
$F_\pi$, and one can also infer 
a behaviour of the type
$$F_S(t)\simeqsub_{t\to\infty}\;\dfrac{{\rm Const.}}{-t\log^\nu(-t/\hat{t})},
\equn{(2.2.2)}$$
with $\hat{t}$  an unknown scale factor and $\nu$ also unknown although, 
probably, one also has
 $\hat{t}\simeq \lambdav, \quad\nu=1.
$
The corrections to (2.2.1), (2.2.2), however, cannot be calculated. 
These asymptotic behaviours hold, in principle, only on the real axis, but the 
Phragm\'en--Lindel\"of theorem ensures their 
validity in all directions of the complex plane.

\setbox0=\vbox{\hsize8.2truecm{\epsfxsize 7truecm\epsfbox{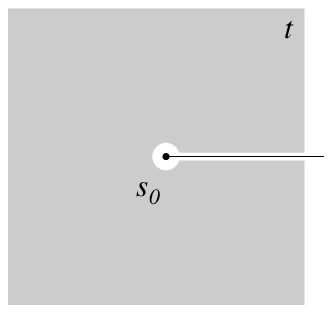}}} 
\setbox6=\vbox{\hsize 4.2truecm\captiontype\figurasc{Figure 2.2.1. }{\hb The  domain of 
analyticity (shadowed region)
 for $F_\pi(t)$, or $\piv(t)$, in the complex $t$ plane.\hb
\phantom{XX}}\hb
\vskip.1cm} 
\medskip
\line{
{\box0}\hfil\box6}
\bigskip

\booksection{2.3. Correlators}
Consider for example the vector current operator $V_\mu(x)=\bar{q}\gamma_\mu q'$
with $q,\,q'$ quark fields. 
We associate to it the correlator
$$\eqalign{
\piv_{\mu\nu}(p)=&\,\ii\int\dd^4x\;\ee^{\ii p\cdot x}
\langle0|{\rm T}V^{\dag}_\mu(x) V_\nu(0)|0\rangle\cr
=&\,
\left(-g_{\mu\nu}t+p_\mu p_\nu\right)\piv_{\rm tr}(t)+
p_\mu p_\nu\piv_S(t),\quad t=p^2.\cr
}
\equn{(2.3.1)}$$

The $\piv(t)$ can be shown, again using only unitarity and causality, to 
be analytic in the complex $t$ plane with a cut from $t_0$ to $+\infty$ 
where $t_0$ is the squared invariant mass of the lightest state with the 
quantum numbers of the current $V_\mu$.  
If $V_\mu$ is the e.m. (electromagnetic) current, that we denote by 
$J_\mu$, and we neglect weak and e.m. interactions, 
then $\piv_S=0$ and $\piv_{\rm tr}$ is analytic except for a cut 
from $t=4\mu^2$ to $+\infty$.

There is no bound with validity for arbitrary field theories for the 
correlators; but, in QCD, we can actually calculate the 
behaviour for large momentum; it is given by 
the parton model result for the $\piv$.

\booksubsection{2.4. Scattering amplitudes at high energy: Reggeology}

\noindent
Although we are here interested only on low and (at times) intermediate energy, it 
is clear that calculations of dispersive type,  like those 
we will discuss in coming Chapters, 
require a model for high energy $\pi\pi$ scattering. 
Regge pole theory provides such a model and, although outside the scope of this notes, 
we will describe here those of its features that are of interest to us. 

Consider the collision of two hadrons, $A+B\to A+B$. 
According to Regge theory, the high energy scattering amplitude, 
at fixed $t$ and large $s$, 
is governed by the exchange of complex, composite objects, known as {\sl Regge poles}, 
related (in some cases) to the resonances that couple to the $t$ channel. 
The same is true for large $t$, dominated by the resonances in the $s$ channel (this 
is the property originally proved, in potential theory, by T.~Regge).
Thus, for isospin 1 in the $t$ channel, high energy scattering
 is dominated by the exchange of a ``Reggeized" 
$\rho$ resonance.
 If no quantum number is exchanged, we say that the corresponding Regge pole is the vacuum, or
Pomeranchuk Regge pole; this name is often shortened to {\sl Pomeron}. 
In a QCD picture, the Pomeron (for example) will be associated with the exchange of 
a gluon ladder between two partons in particles $A$, $B$ (\fig~2.2.2). 
The corresponding formalism was developed by Gribov, Lipatov and
 other physicists in 
the 1970s, and is related to the so-called Altarelli--Parisi, or DGLAP mechanism 
in deep inelastic scattering (see e.g. Barger and Cline~(1969) and Yndur\'ain~(1999) 
and, more recently, Donnachie et al.~(2002), which include 
references to the  original articles).

\midinsert{
\setbox0=\vbox{\hsize 7cm\epsfxsize 5.2cm\epsfbox{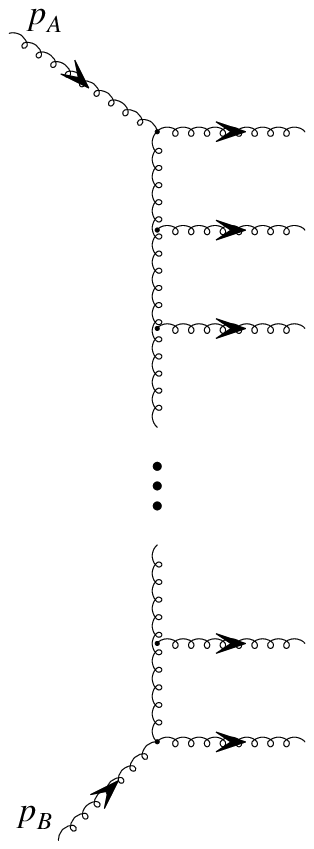}\hfil}
\setbox1=\vbox{\hsize 5cm\captiontype\figurasc{Fig. 2.2.2. }
 {\hb Cut Pomeron ladder exchanged between the partons $p_A$ and $p_B$ in 
hadrons $A$, $B$. The emitted gluons will materialize into a shower of particles.}\hb
\vskip.3cm
\phantom{x}}
\centerline{\box0\hfil\box1}
}\endinsert

One of the useful properties of Regge theory is the so-called {\sl factorization};  
 it can be proved from general properties of Regge theory,\fnote{The proof 
 follows  from extended unitarity
(Gell-Mann, 1962; Gribov and Pomeranchuk, 1962).} or, in QCD, in the DGLAP
formalism,\fnote{Gribov and Lipatov~(1972); Dokshitzer~(1977);  Altarelli and Parisi~(1977).
 See also Kuraev, Lipatov and Fadin~(1976) and Balitskii and Lipatov~(1978).}  
as is intuitively obvious
from
\fig~2.2.2.

Factorization  states that the scattering amplitude $F_{A+B\to A+B}(s,t)$
can be written as a product
$$F_{A+B\to A+B}(s,t)\simeqsub_{{s\to\infty}\atop{t\,{\rm fixed}}}
C_A(t)C_B(t)(s/\hat{s})^{\alpha_R(t)}.
\equn{(2.4.1)}$$ 
$\hat{s}$ is a constant, usually taken to be $1\,\gev^2$ (we will do so here); 
the functions $C_A,\,C_B$ depend on the corresponding particles,  
but the power $(s/\hat{s})^{\alpha_R(t)}$ is universal and
 depends only on the quantum numbers exchanged in 
channel $t$. 
The exponent $\alpha_R(t)$ is called the Regge trajectory associated to the 
quantum numbers in channel $t$ and, for small $t$, 
may be considered linear:
$$\alpha_R(t)\simeqsub_{t\sim0}\alpha_R(0)+\alpha'_Rt.
\equn{(2.4.2)}$$
For the $\rho$ and Pomeron pole, fits to high energy processes give
$$\eqalign{
\alpha_\rho(0)=&\,0.52\pm0.02,\quad\alpha'_\rho(0)=1.01\, {\gev}^{-2},\cr
\alpha_P(0)=&\,1,\quad\alpha'_P=0.11\pm0.03\, {\gev}^{-2},\cr
}
\equn{(2.4.3a)}$$
The 
Regge parameters taken here are essentially those of Rarita et al. (1968);
  for $\alpha_\rho(0)$, 
however, we choose the central value 0.52
  which is more consistent with determinations based on deep inelastic scattering.
There are indications that the Pomeron is not an ordinary Regge pole but 
we will not discuss this here.

For the $\rho$ trajectory, we have enough 
information that we can write a more accurate, quadratic expression that agrees 
on the mean with (2.4.3a) for spacelike $t$ and satisfies the 
Regge constraint $\alpha_\rho(M^2_\rho)=1$:
$$\eqalign{
\alpha_\rho(t)=&\,\alpha_\rho(0)+t\alpha'_\rho(0)+\tfrac{1}{2}t^2\alpha''_\rho(0),
\cr
 \alpha'_\rho(0)=&\,0.90\,{\gev}^{-2},\; \alpha''_\rho(0)=-0.3\,{\gev}^{-4}.\cr
} 
\equn{(2.4.3b)}$$

Let us consider the imaginary part of the spin averaged $\pi N$ or $NN$
 scattering amplitudes (here 
by $NN$ we also understand $\bar{N}N$), which we recall are  
normalized so that
$$\sigma_{\rm tot}(s)=\dfrac{4\pi^2}{\lambda^{1/2}(s,m_A^2,m_B^2)} 
\imag F(s,0).$$
We have, with $f_i$ related to the imaginary part of $C_i$,
$$\eqalign{
\imag F_{NN}(s,t)\simeq &\,f^2_N(t)(s/\hat{s})^{\alpha_R(t)},\cr
\imag F_{\pi N}(s,t)\simeq&\,f_\pi(t)f_N(t)(s/\hat{s})^{\alpha_R(t)},\cr
}
\equn{(2.4.4a)}$$
and therefore, using factorization,
$$
\imag F_{\pi \pi}(s,t)\simeq f^2_\pi(t)(s/\hat{s})^{\alpha_R(t)}.
\equn{(2.4.4b)}$$
The functions $f_i(t)$ 
depend exponentially on $t$ for small $t$ and may be written, approximately, as
$$f_i(t)=\sigma_i\ee^{bt},\quad b=(2.4\pm0.4)\;{\gev}^{-2}.
\equn{(2.4.5)}$$
Consistency 
requires a more complicated form for the residue functions $f_i(t)$; 
below we will give expressions that are sufficiently accurate for 
 the small values of $t$ in which we are interested.

From (2.4.4) we can deduce the relations among the cross sections
$$\dfrac{\sigma_{\pi\pi\to{\rm all}}}{\sigma_{\pi N\to{\rm all}}}=
\dfrac{\sigma_{\pi N\to{\rm all}}}{\sigma_{N N\to{\rm all}}}.
$$
This relation also holds in the naive 
quark model\fnote{Levin and Frankfurter~(1965). 
For a comprehensive review, see Kokkedee~(1969). 
Note, however, that it is not clear why the naive quark model should work, as its mechanism is
very different 
from the orthodox QCD one.}
 in which one considers that scattering of hadrons is given by 
incoherent addition of scattering of their constituent quarks, so we have
$$\sigma_{\pi\pi\to{\rm all}}\;:\;\sigma_{\pi N\to{\rm all}}\;:\;\sigma_{N N\to{\rm all}}=
2\times2\;:\;2\times3\;:\;3\times3.
$$
From any of these relations one can obtain the parameter $\sigma_\pi$
 in (2.4.5) in terms of the 
known $\pi N$ and $NN$ cross sections. 
 
We write explicit formulas for 
$\pi\pi$ scattering, 
taken from 
 the expressions of Rarita et al.~(1968); 
to be precise, as given in
Pel\'aez and Yndur\'ain~(2003). 
 For $I_t=0$ exchange,
$$\eqalign{
\imag F^{(I_t=0)}_{\pi\pi\to\pi\pi}(s,t)&\,\simeqsub_{{s\to\infty}\atop{t\,{\rm fixed}}}
\imag F^{(I_t=0)}_P+\imag F^{(I_t=0)}_{P'};\cr
\imag F^{(I_t=0)}_P=&\,\sigma_\pi(P)\,\dfrac{\left(1+\alpha'_Pt\right)\left(2+\alpha'_Pt\right)}{2}\,
\ee^{bt}(s/\hat{s})^{\alpha_P(0)+\alpha'_P t},\cr 
\imag F^{(I_t=0)}_{P'}=&\,\sigma_\pi(P')\ee^{bt}(s/\hat{s})^{\alpha_\rho(0)+\alpha'_\rho t}.
\cr}
\equn{(2.4.6a)}$$
We have added  the Pomeron and the subleading contribution,
 the so-called $P'$ pole (associated with the $f_2$
resonance) that is  necessary at the lowest energy range.
The slope of the second we have taken as identical to that of the rho. 
As noticed in Pel\'aez and Yndur\'ain~(2003; see also Appendix~C here), this choice gives 
better consistency for crossing sum rules, besides being what one expects in the QCD 
version of Regge theory; the experimental information on 
$\pi N$, $NN$ scattering is not enough to 
fix the slope of the $P'$ with any accuracy.
 
For $I_t=1$, we write 
$$\eqalign{
\imag F^{(I_t=1)}_{\pi\pi\to\pi\pi}(s,t)\simeqsub_{{s\to\infty}\atop{t\,{\rm fixed}}}&\,
\imag F^{(\rho)}(s,t)\cr
=&\,\sigma_\pi(\rho)\dfrac{1+\alpha_\rho(t)}{1+\alpha_\rho(0)}\,
\left[(1+1.48)\ee^{bt}-1.48\right](s/\hat{s})^{\alpha_\rho(0)+\alpha'_\rho t}.\cr
}
\equn{(2.4.6b)}$$
 
From (2.4.5) and the known cross sections for $\pi N$, $NN$ scattering 
we have\fnote{Our Regge parameters $\sigma(i)$ here are slightly smaller than those 
used in Palou and Yndur\'ain~(1974). 
This is because we now add a $P'$ contribution to the 
Pomeron, and a background to the rho piece.}
$$\sigma_\pi(P)=3.0\pm0.30;\quad\sigma_\pi(P')=0.75\pm0.08; \quad \sigma(\rho)=0.84\pm0.10.
\equn{(2.4.6c)}$$

For  isospin $I_t=2$ exchange we cannot fix its parameters 
 from factorization, since $\pi N$ or $NN$ do 
not contain such amplitude. 
It must be due to double rho exchange, so we know that its energy dependence, at $t=0$, 
should be $s^{2\alpha_\rho(0)-1}$, but we know little more. 
We use an empirical formula:
$$\imag F^{(I_t=2)}(s,t)= (0.6\pm0.2)\;\ee^{c_2t}
(s/\hat{s})^{2\alpha_\rho(0)+\alpha'_\rho t-1},
\equn{(2.4.7a)}$$
i.e., a slope like the rho and $P'$.
We have obtained the constant $0.6\pm0.2$ by fitting the difference between the 
experimental $\pi^0\pi^0$ and  $\pi^0\pi^+$ total cross sections at $s^{1/2}=1.42\,\gev$, 
and the Pomeron plus $P'$ values.  
We will, somewhat arbitrarily, take the value $c_2=b$, whose justification is that it 
 produces consistency in crossing sum rules like that in the 
Appendix~C here (see also Pel\'aez and Yndur\'ain,~2003):
$$c_2=b=2.38\pm0.02\;{\gev}^{-2}.
\equn{(2.4.7b)}$$

Another matter is, when one may apply  formulas like (2.4.6).
From the DGLAP version of the Pomeron (for example) 
we expect the following pattern to occur: in the region $|t|\ll s$, $s\gg\lambdav^2$ (with
$\lambdav\sim0.4\,\gev$ the QCD 
parameter) the ladder exchange mechanism will start to dominate the collision $A+B$. 
We then will have the onset of the Regge regime with, at the same time, 
a large increase of inelasticity and a smoothing of the total cross 
section according to the behaviour (2.4.6).  

\topinsert{
\setbox1=\vbox{\hsize12.truecm{\epsfxsize 11.2truecm\epsfbox{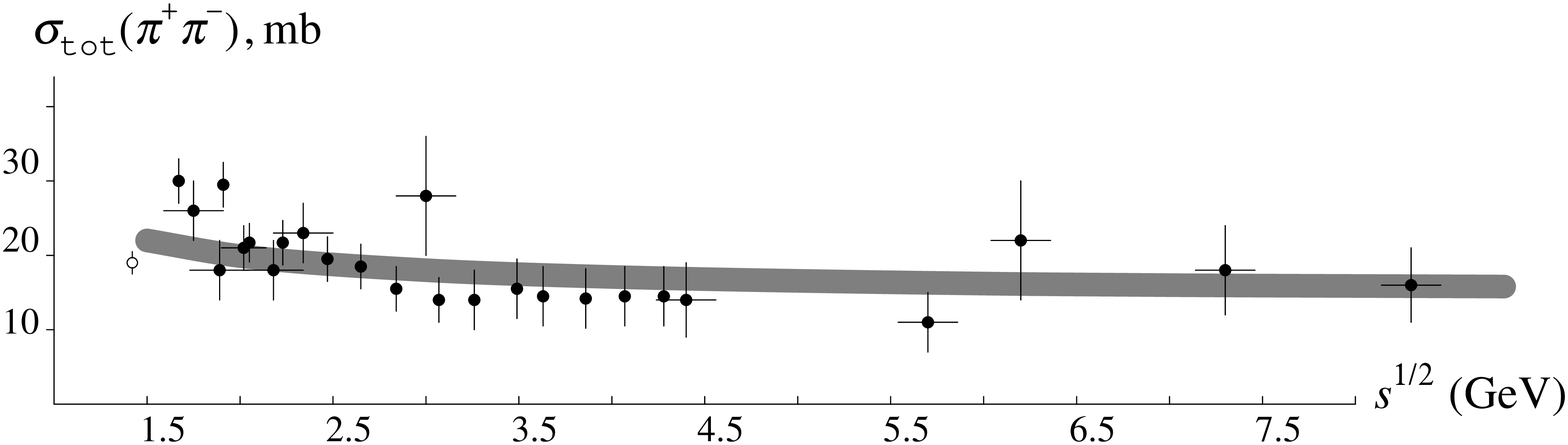}}}
\setbox2=\vbox{\hsize6.3truecm{\epsfxsize 5.7truecm\epsfbox{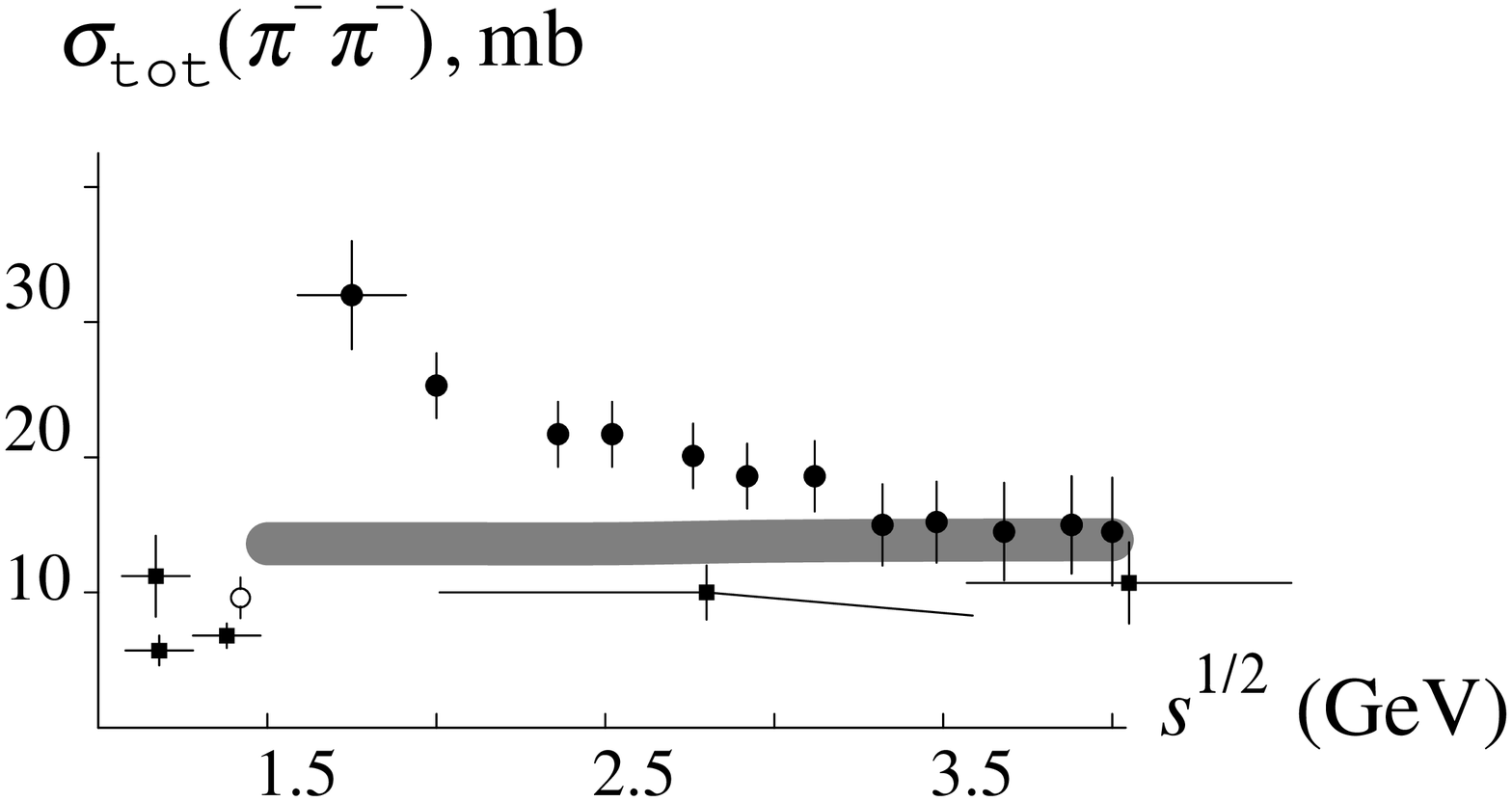}}}
\setbox3=\vbox{\hsize5.7truecm{\epsfxsize 5.0truecm\epsfbox{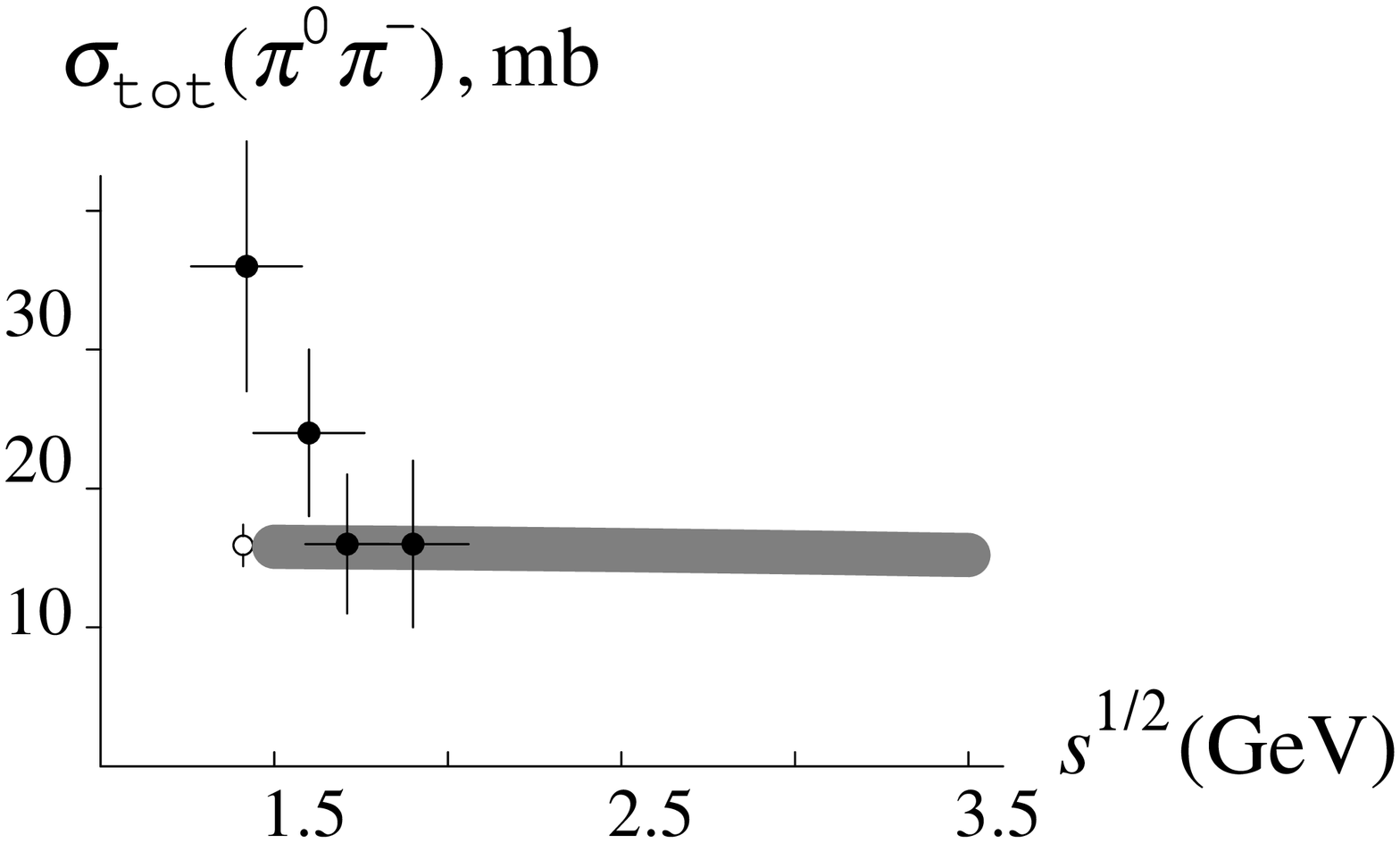}}}
\centerline{\tightboxit{\box1}}
\smallskip
\line{\hfil\tightboxit{\box2}\hfil\tightboxit{\box3}\hfil}
\bigskip
\setbox5=\vbox{\hsize 10truecm\captiontype\figurasc{Figure 2.2.3. }
 {The total cross sections
$\sigma(\pi^-\pi^-)$, $\sigma(\pi^+\pi^-)$ and $\sigma(\pi^0\pi^-)$. 
Black dots and squares: experimental points from  
Robertson, Walker and Davis~(1973), 
Biswas et al.~(1967), Cohen et al.~(1973), Hanlan et al.~(1976), and Abramowicz et al.~(1980). 
Open dots at 1.42 \gev: the 
cross sections that follow from the low energy 
phase shift analyses, see Appendix~A.
Thick gray lines, from 1.42 \gev: Regge formula, with parameters as in (2.4.6,~7) 
(the thickness of
 the line   covers the error in the theoretical value of the 
Regge residue).}}
\centerline{\box5}
\bigskip
}\endinsert

For $\pi N$, $NN$
scattering this occurs as soon as one is
 beyond the region of elastic resonances; in fact  (as can be seen in the cross section summaries in the 
Particle Data Tables) as soon as the kinetic energy in the c.m. is above 
$1\;-\;1.2$ \gev. For $\pi\pi$ 
we thus expect the Regge description to be valid for 
the corresponding energies, that is to say, for  $s^{1/2}\geq 1.4\,\gev$, which is the region 
where we will use it here. Indeed, and as
 we will see, around this energy, it is possible to calculate the 
$\pi\pi$ scattering 
amplitude from 
experiment and indeed it agrees, within a 15\%, with (2.4.6); 
see for example Pel\'aez and Yndur\'ain~(2003). 

It is worth noting that these properties can also be 
verified directly for 
$\pi\pi$ scattering, as has been done by Robertson, Walker and Davis~(1973), 
Biswas et al.~(1967), Cohen et al.~(1973), Hanlan et al.~(1976), and Abramowicz et al.~(1980). 
These authors do not attempt at phase shift reconstruction of 
the amplitude, but measure directly elastic and total $\pi\pi$ cross sections 
at energies between 1.2 and 6 \gev. 
They find a pattern identical to that for $\pi N$, $NN$ or $KN$ 
scattering. In particular, a total cross section 
that flattens out to a value of 15 to 20 mb, precisely as predicted by 
factorization (and in agreement with our numbers here). 
Moreover, the elastic cross section becomes less than a third of the total one 
above 1.7 \gev; see 
for example 
fig.~5 in  Robertson, Walker and Davis~(1973). 
In fact, and as shown in \fig~2.2.3, the experimental $\pi\pi$ cross sections agree very 
well with the prediction obtained from factorization (our equations (2.4.6) here).

As is clear from this minireview, the reliability of the Regge calculation of 
high energy pion-pion scattering cannot go beyond this accuracy of $\sim 15\%$, even 
for small $t$. 
The deviations off simple Regge behaviour are expected to be much larger for large $t$, 
as indeed the counting rules of QCD imply a totally different behaviour for fixed $t/s$. 
This is one of the problems involved in using e.g. the Roy equations that require 
integration up to $-t\sim  0.7\,\gev^2$, $s\sim2\,\gev^2$, where the Regge picture fails completely 
(we expect instead the Brodsky--Farrar behaviour, $\sigma_{{\rm fixed}\,\cos\theta}\sim s^{-7}$). 
However, for forward dispersion relations or the Froissart--Gribov representation 
 we will work only for $t=0$ or $t=4\mu^2$ for 
which  the largest variation, that of $\ee^{bt}$, is still small, 
$b(t=4\mu^2)\simeq0.19$, and we expect no large error 
due to departure off linearity for the exponent in $f_i(t)$ or for the Regge trajectories, 
$\alpha_R(t)$.

\bookendchapter

\brochureb{\smallsc chapter 3}{\smallsc the effective range formalism for p.w.
 amplitudes. 
resonances}{17}
\bookchapter{3. The effective range formalism\hb for p.w.
 amplitudes. \hb
Resonances}
\vskip-0.5truecm
\booksection{3.1. Effective range formalism}

\noindent
We will consider here only the pion-pion case. 
The discontinuity of $f_l(s)$ across the {\sl elastic} 
cut is very easily evaluated. Because all functions (scattering amplitudes, 
form factors and correlators) are real analytic\fnote{A complex function
 $f(z)$ is 
{\sl real analytic} if it satisfies $f^*(z^*)=f(z)$. 
The theorem of Painlev\'e ensures that, if a function 
analytic for $\imag z\neq0$  is real analytic, and is real on a segment
$[a,b]$ of the real axis, it is also 
analytic on the segment. For more information on 
matters of complex analysis, we recommend the texts of Ahlfors~(1953) and Titchmarsh~(1939).} 
we can calculate their discontinuity as
$$\eqalign{
{\rm disc} f(s)=&\,2\ii \imag f(s)=
\lim_{\epsilon\to+0} \left\{f(s+\ii \epsilon)-f(s-\ii \epsilon)\right\}\cr
=&\,
\lim_{\epsilon\to+0} \left\{f(s+\ii \epsilon)-f^*(s+\ii \epsilon)\right\}.\cr
}
\equn{(3.1.1)}$$
For p.w. amplitudes, and for  physical $s$ below the inelastic threshold 
$s_0$,  we have
$$\imag f_l(s)=\dfrac{\pi k}{2s^{1/2}}|f_l(s)|^2,\quad 4\mu^2\leq s\leq s_0.
\equn{(3.1.2)}$$

This suggests how we can form from $f_l$
 a function in which this elastic cut is absent.
This is the function $\phiv_l(s)$ defined for arbitrary complex $s$ by
$$\phiv_l(s)=
\dfrac{\ii k^{2l+1}}{2\sqrt{s}}+\dfrac{k^{2l}}{\pi f_l(s)}.
\equn{(3.1.3a)}$$
We assume
that 
$f_l(s)$ does not vanish for $0\leq s< 4\mu^2$, or for 
$4\mu^2<s\leq s_0$. If $f_l$ vanished below 
threshold, or on the elastic cut, the function $\phiv_l$ would have poles at such zeros; 
the analysis can be generalized quite easily to cope with this, and, for 
$\pi\pi$ scattering, we will 
show explicitly how in the cases of the S waves and the D2 wave. 

We can rewrite (3.1.3a) as
$$\phiv_l(s)=
-2^{-2-2l}(s-4\mu^2)^l\left(\dfrac{4\mu^2}{s}-1\right)^{1/2}+2^{-2l}
\dfrac{(s-4\mu^2)^l}{\pi f_l(s)}.
\equn{(3.1.3b)}$$
In this second form it is obvious that the first term in the r.h. side 
is analytic for all $s$, except for a (kinematic) cut running from 
$-\infty$ to $s=0$ and a cut for $s\geq4\mu^2$. The second term is also analytic 
over the segment $0\leq s< 4\mu^2$, 
and it presents a dynamical cut from $-\infty$ to 0 due to the l.h. cut of $f_l$ 
(\fig~3.1.1).

\midinsert{
\setbox0=\vbox{{\epsfxsize 11.8truecm\epsfbox{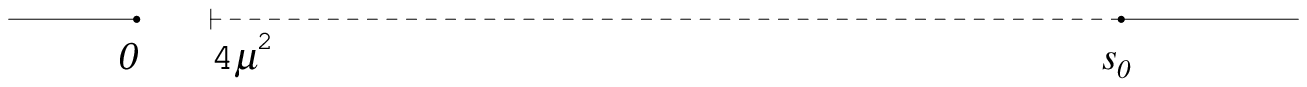}}} 
\setbox6=\vbox{\hsize 10truecm\captiontype\figurasc{Figure 3.1.1. }{The
 cuts in the complex $s$ plane for $\phiv_l(s)$. 
The dotted line shows the absent elastic cut. 
We have taken $s_0=1\,\gev^2$, and the drawing is to scale. \hb
\phantom{XX}}\hb
\vskip.1cm} 
\medskip
\centerline{\tightboxit{\box0}}
\bigskip
\centerline{\box6}
\medskip
}\endinsert

 We next check that $\phiv_l(s)$ is
analytic over the  elastic cut. 
We have, for $4\mu^2<s\leq s_0$,
$$\imag\phiv_l(s)=\dfrac{ k^{2l+1}}{2\sqrt{s}}+k^{2l}
\dfrac{-\imag f_l(s)}{\pi f_l^*(s)f_l(s)}.$$
Using then (3.1.2), the r.h. side is seen to vanish. 
The only point which appears dangerous is the threshold, $s=4\mu^2$, because 
here $f_l$ vanishes for $l\geq 1$; 
but  this zero is exactly compensated by the zero of the factor $k^{2l}$; 
cf.~(2.1.5). 
Therefore it follows that the function $\phiv_l(s)$ is analytic along the elastic cut. 
Its only singularities  are thus (apart from poles due to zeros of $f_l$),
 a r.h. cut from $s=s_0$ to 
$+\infty$; and a l.h. cut, formed by two superimposed cuts, namely, 
the kinematic cut of 
$$2^{-2-2l}(s-4\mu^2)^l\sqrt{\dfrac{4\mu^2}{s}-1}, $$
and the dynamical cut of
$$\dfrac{k^{2l}}{\pi f_l(s)}$$
due to the l.h. cut of $f_l(s)$. 

Eq.~(3.1.3b) defines $\phiv_l(s)$ for all complex $s$; in the particular case where $s$ is 
on the elastic cut, we can use \equn{(2.1.2)} to get
$$\phiv_l(s)=\dfrac{k^{2l+1}}{2\sqrt{s}}\cot \delta_l(s),
\quad 4\mu^2\leq s\leq s_0.
\equn{(3.1.4)}$$
In general, i.e., for any value (complex or real) of $s$, we can solve (3.1.2) and write
$$f_l(s)=\dfrac{2s^{1/2}}{\pi k}\dfrac{1}{2s^{1/2}k^{-2l-1}\phiv_l(s)-\ii}
=\dfrac{k^{2l}}{\pi}\dfrac{1}{\phiv_l(s)-\ii k^{2l+1}/2s^{1/2}} .
\equn{(3.1.5)}$$
 $\phiv_l(s)$ is real on the segment $0\leq s\leq s_0$, but it
 will be {\sl complex} above the 
inelastic threshold, $s_0$, and also for $s\leq0$.

The fact that $\phiv_l(s)$ is analytic across the elastic region is valid not 
only for $\pi\pi$, but also for other p.w. amplitudes; 
for example, for pion-nucleon,  nucleon-nucleon or even nucleon-nucleus. 
This implies that, at low energies ($k\to 0$), one can expand 
$$\phiv_l(s)=\dfrac{1}{4\mu a_l}+R_0k^2+R_1k^4+\cdots.
\equn{(3.1.6)}$$
This is the so-called effective range formalism, widely used in low 
energy nucleon and nuclear physics. 
The quantity $a_l$ is  the scattering length (cf. \equn{(2.1.4)}) and the $R_i$ 
are related to the range of the potential 
(if the scattering is caused by a short-range potential). 
For $\pi\pi$ scattering, the expansion is convergent in the 
disk $|s-4\mu^2|<1$, shown  shaded in \fig~3.1.2.

Besides (3.1.6) we will also use the parameters $b_l$ defined by
$$\dfrac{\pi}{4 \mu k^{2l}}\real f_l(s)\simeqsub_{k\to0}
a_l+b_l k^2+\cdots
\equn{(3.1.7)}$$

\topinsert{
\setbox0=\vbox{{\epsfxsize 11.8truecm\epsfbox{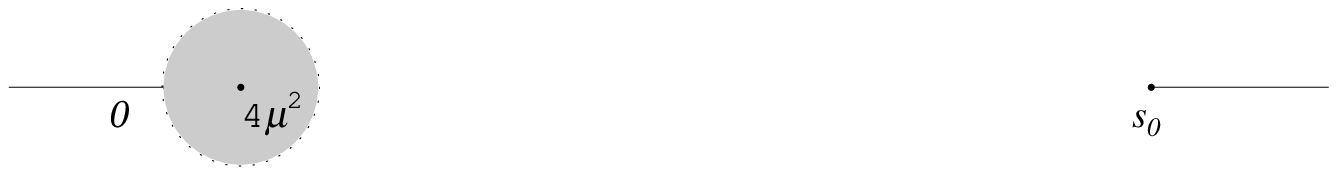}}} 
\setbox6=\vbox{\hsize 10truecm\captiontype\figurasc{Figure 3.1.2. }{The
circle of convergence for the effective range expansion for $\phiv_l(s)$;
 $s_0=1\,\gev^2$. \hb
\phantom{XX}}\hb
\vskip.1cm} 
\medskip
\centerline{\tightboxit{\box0}}
\bigskip
\centerline{\box6}
\medskip
}\endinsert

\booksection{3.2. Resonances in (nonrelativistic) potential scattering}

\noindent
Consider scattering by a  spherical potential, $V(r)$, 
that we assume to be of short range.  
We will simplify the discussion by working  
in the nonrelativistic approximation. 
The nonrelativistic energy $E$ is $E=s^{1/2}-m_1-m_2$ 
with $m_i$ the masses of the particles, and we shall let $m$ be the 
reduced mass. 
To lighten notation, we take mass units so that $2m=1$.

 The $l$-wave Schr\"odinger equation is
$$\dfrac{\dd^2\psi_l(r)}{\dd r^2}+\left[k^2-V(r)-\dfrac{l(l+1)}{r^2}\right]\psi_l(r)=0.
\equn{(3.2.1)}$$
One may express its solutions as
$$\psi_l(r)\simeqsub_{r\to\infty}\dfrac{1}{2\ii}
\left\{\ee^{\ii kr-\ii l\pi/2+\ii \delta_l(E)}-
\ee^{-\ii kr+\ii l\pi/2-\ii \delta_l(E)}\right\}.
\equn{(3.2.2a)}$$
In principle, (3.2.2a) is valid only for 
physical $k\geq0$. However, because (3.2.1) depends explicitly on $k$, we can take the 
solution to be valid for arbitrary, even complex $k$. 

From (3.2.2a) we can find the p.w. amplitudes. First, we rewrite it as
$$\psi_l(r)\simeqsub_{r\to\infty}j^-(k,l)\ee^{\ii kr}+j^+(k,l)\ee^{-\ii kr};
\equn{(3.2.2b)}$$
the $j^\pm$, known as the Jost functions, are identified, 
at large $r$, comparing with (3.2.2a). 
In terms of these we can write the $S$-matrix element,
$$s_l(E)\equiv\ee^{2\ii \delta_l},$$
as
$$s_l(E)=(-1)^{l+1} \dfrac{j^-(k,l)}{j^+(k,l)}.
\equn{(3.2.3)}$$

Now, the exchange of $k\to-k$ does not alter the Schr\"odinger equation, but it 
exchanges the exponentials in (3.2.2a). Therefore, one must have
$$j^-(-k,l)=j^+(k,l).
\equn{(3.2.4)}$$

If we start from the $k$ plane, then the energy plane will be a two-sheeted 
Riemann surface (\fig~3.2.1). We designate {\sl physical} sheet (sheet I) to 
that coming from $\imag k>0$, and {\sl unphysical} sheet (sheet II) 
to that obtained from $\imag k<0$. 
When considering $s_l(E)$ as a function of $E$ it then follows that 
we have two determinations. 
Since obviously $k^{\rm II}=-k^{\rm I}$, we find 
$$s_l^{\rm II}(E)=\left[s_l^{\rm I}(E)\right]^{-1}.
\equn{(3.2.5)}$$
The physical value is  
$$s_l(E)=\lim_{\epsilon\to+0}s_l^{\rm I}(E+\ii\epsilon)=
\lim_{\epsilon\to+0}s_l^{\rm II}(E-\ii\epsilon).$$

\topinsert{
\setbox0=\vbox{\hsize13.truecm{\epsfxsize 11.5truecm\epsfbox{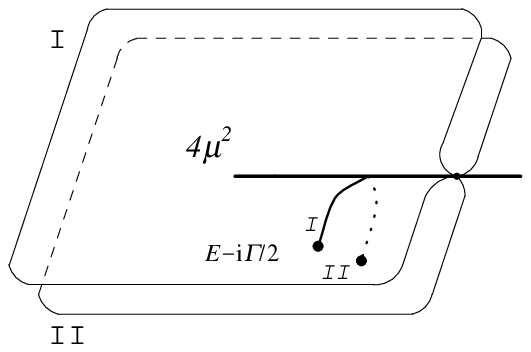}}} 
\setbox6=\vbox{\hsize 10truecm\captiontype\figurasc{Figure 3.2.1. }{The
Riemann sheet for p.w. amplitudes. \hb
\phantom{XX}}\hb
\vskip.1cm} 
\medskip
\centerline{{\box0}}
\vskip-1truecm
\centerline{\box6}
\medskip
}\endinsert

We shall now look for singularities of $s_l(E)$. For 
physical $E>0$ we cannot have poles because $|s_l(E)|=1$. 
For $E=-E_B<0$, a pole of $s_l(E)$ means a zero of $j^+$. 
If the pole occurs in the first sheet, this 
means that the corresponding value of the momentum will be
$k_B=\ii|k|=\ii\sqrt{E_B}$ and hence (3.2.2b) becomes
$$\psi_l(r)\simeqsub_{r\to\infty} j^-(k_B,l)\ee^{-|k|r},$$
i.e., the wave function of a bound state with binding energy $E_B$. 
We thus conclude that poles of the 
$S$-matrix in the physical sheet 
for energies below threshold correspond to bound states.\fnote{We
 will not be interested in poles in the 
unphysical sheet with
 negative energies, known as {\sl antibound} states. 
More details on the subject of this section may 
be found in the treatises of Omn\`es and Froissart~(1963) and Goldberger and Watson~(1964).}

We next investigate the meaning of poles in the lower half-plane in the 
unphysical sheet, that is to say, poles located at $E^{\rm II}=E_R=E_0-\ii \gammav/2$ 
with $E_0,\,\gammav>0$. If there is a pole of $s_l^{\rm II}(E)$ for 
$E=E_R-\ii \gammav/2$, then (3.2.5) implies that 
the physical $S$-matrix element has a zero in the same location:
$$s_l^{\rm I}(E_R-\ii\gammav/2)=0$$
(\fig~3.2.1).
The corresponding wave function is not as easily obtained as for the bound state case; 
a detailed discussion may be found in Godberger and Watson (1966) or Galindo and 
Pascual (1978), but 
an essentially correct result may be obtained by replacing, in the standard 
time dependent wave function for stationary states 
$$\psiv=\ee^{-\ii Et}\psi({\bf r}),$$
$E$ by the complex value $E_R-\ii\gammav/2$. So  we get
$$\psiv=\ee^{-\gammav t/2}\ee^{-\ii E_Rt}\psi({\bf r}):$$
 the probability to find the state decreases with time as 
$|\psiv|^2=\ee^{-\gammav t}$, which can be interpreted as 
a metastable state that decays with a lifetime $\tau=1/\gammav$; that is to say, 
a {\sl
resonance}. 
$\gammav$ is called the {\sl width} of the resonance, and is 
equal to the indetermination in energy of the metastable state.

Let us consider now the corresponding physical phase shift. 
The pole and zero of $s^{\rm II}_l,\,s^{\rm I}_l$ 
imply corresponding zeros of the Jost functions. 
We will assume that $\gammav$ is very small; then, 
 in the neighbourhood of $E_R$ we can write
$$s^{\rm I}_l(E)\simeqsub_{E\sim E_R}
\dfrac{E-E_R-\ii\gammav/2}{E-E_R+\ii\gammav/2}.
\equn{(3.2.6a)}$$
For the phase shift this implies
$$\cot\delta_l(E)\simeqsub_{E\sim E_R}\dfrac{E_R-E}{\gammav/2}.
\equn{(3.2.6b)}$$
This means that at $E_R$ the phase shift goes, {\sl growing}, through $\pi/2$ 
and that it varies rapidly.

We can write the corresponding formulas for the p.w. amplitudes, 
now for the relativistic case. 
We profit from the analyticity of the effective range 
function over the elastic cut to conclude from (3.2.6b) 
and the proportionality between $\cot\delta_l$ and 
$\phiv_l$ that,  for 
$s=M_R^2$ (where $M_R$ is the invariant mass corresponding to 
the energy $E_R$), $\phiv_l(s)$ must have a zero: 
$$\phiv_l(s)\simeqsub_{s\simeq M_R^2}\dfrac{M_R^2-s}{\gamma}.
\equn{(3.2.7a)}$$ 
So we may write the p.w. amplitude in its neighbourhood as
$$f_l(s)\simeqsub_{s\simeq M_R^2}\dfrac{1}{\pi}
\,\dfrac{k^{2l}\gamma}{M_R^2-s-\ii k^{2l+1}\gamma/2s^{1/2}}.
\equn{(3.2.7b)}$$
The residue of $\phiv_l$, $\gamma$, can be related to the 
width of the resonance:
$$\gammav= [k(M_R^2)]^{2l+1}\gamma/2M^2_R.
\equn{(3.2.7c)}$$
\equn{(3.2.7)} is the (relativistic) {\sl Breit--Wigner} formula for the 
p.w. scattering amplitude near a resonance. 
Note however that Eqs.~(3.2.7) are only valid in the vicinity of the resonance; 
away from it, the ratio $\phiv_l(s)/(M_R^2-s)$ will not 
be a constant, so in general we will have to admit  
a dependence of $\gamma$ (and $\gammav$) on $s$.

Let us consider another characterization of a resonance. 
Returning to  nonrelativistic scattering, one can prove 
that the time delay that the interaction causes in 
the scattering of two particles in angular momentum 
$l$ and with energy $E$ is 
$$\Delta t=2\dfrac{\dd\delta_l(E)}{\dd E}.$$
We can say that the particles resonate when this time delay is maximum. 
In the vicinity of a zero of the effective range, we can use 
(3.2.7) to show that $\Delta t(s)$ is maximum for 
$s=M_R^2$ and then the time delay equals $1/\gammav$. 

We have therefore three definitions of an {\sl elastic} resonance: a pole of 
the scattering amplitude in the unphysical Riemann sheet; a zero of the 
effective range function; or a maximum of the quantity
$$\dd\delta_l(s)/\dd s.$$
These three definitions agree, to order $\gamma^2$, when $\gamma$ is small, and neglecting 
variations of $\gamma$; but
a precise description of broad resonances requires 
discussion of these variations. 
In these notes, however, we will only give the value of $s^{1/2}$ at which the phase crosses $\pi/2$. 
Since we will also give explicit parametrizations, to find e.g. the 
location of the poles should not be a difficult matter for the interested reader.

Unstable 
 elementary particles may also be considered a special case of 
resonances; thus, for example, one may treat the
$Z$ particle as a  fermion-antifermion resonance. 
We  discuss this  for a simple model in \sect~4.2.

\bookendchapter
\brochureb{\smallsc chapter 4}{\smallsc the p p.w. amplitude for  $\pi\pi$
 scattering in the elementary rho model}{23}
\bookchapter{4. The P p.w. amplitude for  $\pi\pi$ scattering in the elementary rho model}
\vskip-0.5truecm
\booksection{4.1. The $\rho$ propagator and the $\pi^+\pi^0$ scattering amplitude}

\noindent
Before continuing with general properties of pion interactions, 
it is convenient to illustrate what we have already seen with 
a simple, explicit model.
In the present chapter we do precisely this; 
specifically, we consider the elementary rho model 
and take $\pi^0$, $\pi^+$ interactions to be given by 
the Lagrangian given in (1.3.1).
We start by calculating the $\rho^+$ propagator in dimensional regularization,
 to lowest order and neglecting
the  rho self-interactions. We therefore consider only the diagrams  in \fig~(4.1.1). 
The corresponding vacuum polarization function is then
$$\eqalign{\piv^{(\rho)}_{\mu\nu}(q)=&\,
\ii^2g^2_\rho\int\dd^D\hat{p}\,(2p+q)_\mu(2p+q)_\nu\dfrac{\ii}{p^2-\mu^2}
\dfrac{\ii}{(p+q)^2-\mu^2}\cr
+&\,2\ii g^2_\rho g_{\mu\nu}\int\dd^D\hat{p}\,\dfrac{\ii}{p^2-\mu^2}; \cr}
$$ 
 we have defined
$$\dd^D\hat{p}\equiv\dfrac{\dd^D p}{(2\pi)^D}\nu^{4-D}_0,
$$
and $\nu_0$ is an arbitrary mass parameter. After standard manipulations, 
and with $D=4-\epsilon$, we find
$$\eqalign{
\piv^{(\rho)}_{\mu\nu}(q)=&\,(-q^2g_{\mu\nu}+q_\mu q_\nu)\dfrac{\ii g^2_\rho}{16\pi^2}
\Bigg\{\tfrac{1}{3}\left(\dfrac{2}{\epsilon}-\gammae+\log4\pi-\log \nu^2_0\right)\cr
-&\,\int_0^1\dd x(1-2x)^2\log(\mu^2-x(1-x)q^2)\Bigg\}.\cr
}
\equn{(4.1.1a)}$$
Here $\gammae\simeq0.5772$ is Euler's constant.

\midinsert{
\medskip
\setbox0=\vbox{\hsize13.0truecm{\epsfxsize 11.8truecm\epsfbox{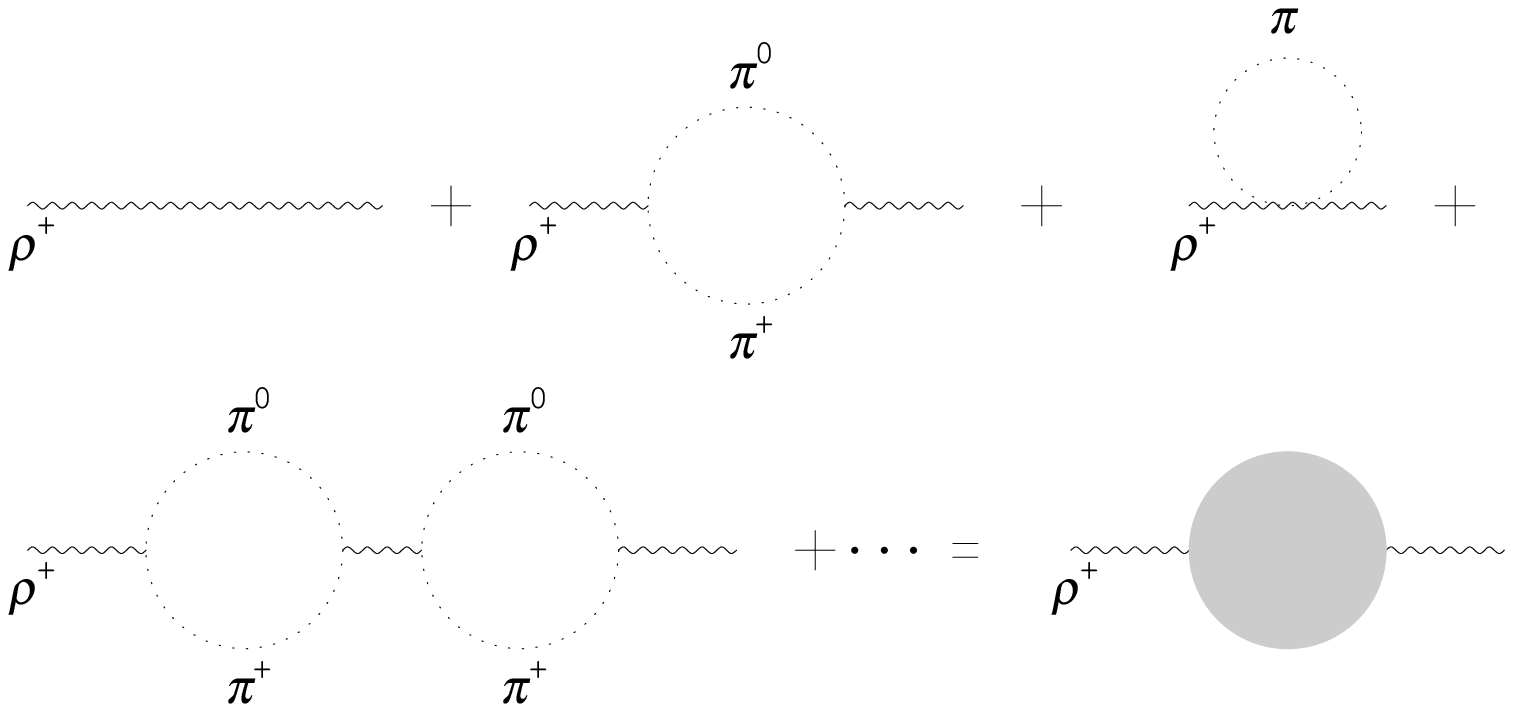}}} 
\setbox6=\vbox{\hsize 10truecm\captiontype\figurasc{Figure 4.1.1. }{The
sum of one loop corrections to the rho propagator. \hb
\phantom{XX}}\hb
\vskip.1cm} 
\medskip
\centerline{{\box0}}
\vskip-1.truecm
\centerline{\box6}
\medskip
}\endinsert

We then calculate the dressed rho propagator. For this, we first rewrite 
(4.1.1a) as
$$\piv^{(\rho)}_{\mu\nu}(q)=(-q^2g_{\mu\nu}+q_\mu q_\nu)\ii\piv_D(s),\quad s=q^2.
\equn{(4.1.1b)}$$
The dressed propagator is then,
$$D_{\mu\nu}^{(\rho;0)}=\dfrac{-\ii g_{\mu\nu}}{s-M_0^2}+
\dfrac{-\ii g_{\mu\nu}}{s-M_0^2}(-\ii s\piv_D)\dfrac{-\ii}{s-M_0^2}+\cdots+
\hbox{gauge terms.}.$$
The gauge terms are terms proportional to $q_\mu q_\nu$. Summing this we find
$$D_{\mu\nu}^{(\rho;0)}=\dfrac{-\ii g_{\mu\nu}}{s-M_0^2+s\piv_D}+\hbox{gauge terms}.$$

This is still unrenormalized, and $M_0$ is the unrenormalized rho mass. 
We renormalize in the \msbar\ scheme, with scale parameter the (renormalized) rho mass, 
$\nu^2_0=M^2$. 
Thus,
$$\piv_{\rm ren.}(s)=
-\dfrac{g^2_\rho}{16\pi^2}\int_0^1\dd x(1-2x)^2\log\dfrac{\mu^2-x(1-x)s}{\bar{M}^2}
\equn{(4.1.2a)}$$
and the renormalized, dressed rho propagator is
$$D_{\mu\nu}^{(\rho)}=
\dfrac{-\ii g_{\mu\nu}}{s-\bar{M}^2+s\piv_{\rm ren.}(s)\vphantom{M^{M^M}}}+\hbox{gauge terms};
\quad \bar{M}=\bar{M}(M^2).
\equn{(4.1.2b)}$$
For $s$ real and larger than $4\mu^2$ we can split $\piv_{\rm ren.}$ into a real and 
an imaginary part as follows:
$$\piv_{\rm ren.}(s)=
-\dfrac{g^2_\rho}{16\pi^2}\int_0^1\dd x(1-2x)^2\log
\left|\dfrac{\mu^2-x(1-x)s}{\bar{M}^2\vphantom{M^{M^M}}}\right|
+\ii \dfrac{g^2_\rho}{16\pi^2}\dfrac{8k^3}{3s^{3/2}},\quad s\geq 4\mu^2.
\equn{(4.1.3)}$$

We next evaluate the scattering amplitude, with the fully dressed propagator. 
We have to calculate the amplitudes $
F^{(s)}$ and  $F^{(u)}$   associated with diagrams 
(s), (u) in \fig~4.1.2, so that the scattering amplitude is
 $F=
F^{(s)}+F^{(u)}$. 
For the first we find,
$$
F^{(s)}=-16\dfrac{g^2_\rho}{16\pi^2}\,
\dfrac{k^2}{s-\bar{M}^2+s\piv_{\rm ren.}\vphantom{M^{M^M}}}\cos\theta,
\equn{(4.1.4)}$$
where $\theta$ is the scattering angle in the c.m.

Projecting $F^{(s)}$ onto the P wave we get
$$
f^{(s)}_1(s)=\tfrac{16}{3}\dfrac{g^2_\rho}{16\pi^2}\,
\dfrac{k^2}{\bar{M}^2-s-s\piv_{\rm ren.}(s)\vphantom{M^{M^M}}}.
\equn{(4.1.5)}$$

We have to add the contribution of diagram (u); note that, in this model, there is no 
contribution from the $t$ channel, at leading order, because you cannot make a $\rho$ 
with two $\pi^0$s.  
We have,
$$F^{(u)}=4\dfrac{g^2_\rho}{16\pi^2}\left(\dfrac{3s-4\mu^2}{2}-2k^2\cos\theta\right)
\dfrac{1}{u-\bar{M}^2+u\piv_{\rm ren.}(u)\vphantom{M^{M^M}}}
\equn{(4.1.6)}$$
and we recall that $u=4\mu^2-s-t=-2k^2(3+\cos\theta)$. 
Projecting into the P wave we find
$$\eqalign{
&\,f^{(u)}_1(s)=\tfrac{1}{2}\dfrac{4g^2_\rho}{16\pi^2}\cr
\times&\,\int_{-1}^{+1}\dd\cos\theta\,\cos\theta
\left(\dfrac{3s-4\mu^2}{2}-2k^2\cos\theta\right)
\dfrac{1}{u-\bar{M}^2+u\piv_{\rm ren.}(u)\vphantom{M^{M^M}}}.\cr
}
\equn{(4.1.7a)}$$

\topinsert{
\medskip
\setbox1=\vbox{\epsfxsize 10.5truecm\epsfbox{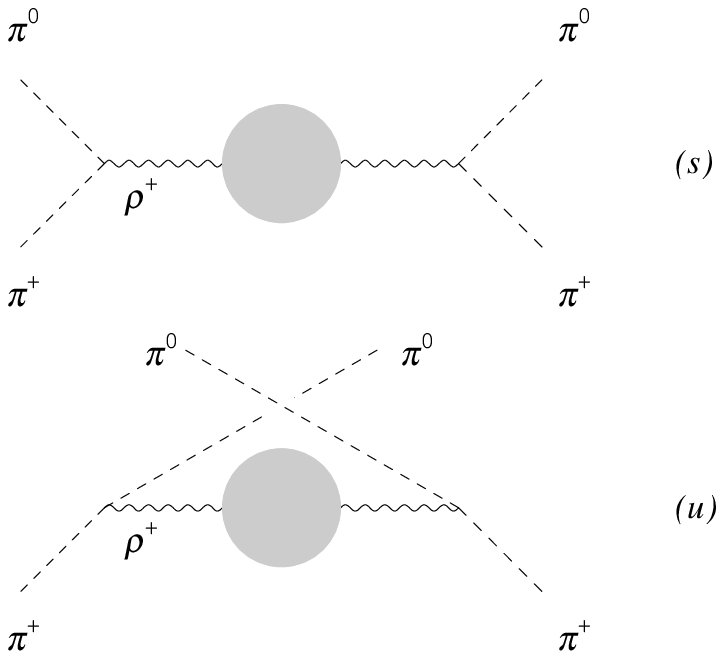}}
\centerline{\kern16em\box1} 
\setbox6=\vbox{\hsize 10truecm\captiontype\figurasc{Figure 4.1.2. }{Diagrams 
for $\pi^0\pi^+$ scattering mediated by the $\rho$.}} 
\bigskip
\centerline{\box6}
\bigskip
}\endinsert
\bigskip

The complete partial wave amplitude is
$$f_1(s)=
f^{(s)}_1(s)+f_1^{(u)}(s).
\equn{(4.1.7b)}$$

This p.w. amplitude does not satisfy 
unitarity. 
This is a general fact; perturbation theory only verifies {\sl perturbative}
 unitarity, that is to say, 
unitarity up to corrections of higher orders. 
We will see in this example how this works; this will allow us to 
see explicitly how in this model the rho behaves as a resonance.

First of all we check that $f_1$ verifies the expected analyticity properties. 
$
f^{(s)}_1(s)$ has a right hand cut, due to that of $\piv_{\rm ren.}(s)$ 
which is the only piece in (4.1.5) which is nonanalytic. 
From \equn{(4.1.3)} we see that it extends from 
$4\mu^2$ to $+\infty$. 
The l.h. cut of $f_1$ comes from the l.h. cut of $F^{(u)}_1(s,t)$. 
From (4.1.7), the only discontinuity occurs when $\piv_{\rm ren.}(u)$  
is discontinuous, which happens when $u\geq 4\mu^2$. 
In terms of $s,\,\cos\theta$ this condition becomes
$$u-4\mu^2=-\tfrac{1}{2}\left[s+4\mu^2
+(s-4\mu^2)\cos\theta\right]\geq0.$$
Therefore, $F^{(u)}(s,t)$ has a discontinuity for $s$ in the range from $-\infty$ to 
$$s_\theta=\dfrac{4\mu^2(\cos\theta-1)}{1+\cos\theta}.$$
Because in (4.1.7) we integrate for $\cos\theta$ between $-1$ and $+1$, it follows 
that the cut of $f^{(u)}_1(s)$, and hence of $f_1(s)$, runs from 
$-\infty$ to 0, as was to be expected on general grounds.

\booksection{4.2. The weak coupling approximation}
Let us now make further approximations. 
If we calculate the rho decay width in our model 
to lowest order  
we obtain, after a simple calculation,
$$\gammav(\rho^+\to\pi^+\pi^0)=\dfrac{g^2_\rho k^3_\rho}{6\pi M^2},
\quad k_\rho=\dfrac{\sqrt{M^2-4\mu^2}}{2}.$$
Putting numbers for the rho mass and width it follows that
$$\dfrac{g^2_\rho}{16\pi^2}\simeq 0.23$$
so the approximation of considering this quantity to be small 
is not too bad.

For $s$ physical, and in particular for $s\sim M^2$, the 
piece $s\piv_{\rm ren.}(s)$ in the expression 
for $
f^{(s)}_1(s)$, although of nominal order 
$g^2_\rho$ (see (4.1.2)) cannot be neglected; else, $
f^{(s)}_1(s)$ 
would be infinite around $s=M^2$. 
However, $u\piv_{\rm ren.}(u)$ can be neglected to a first approximation in $g^2_\rho$ in
 the expression (4.1.7). 
If we do this, $f_1^{(u)}$ can be easily integrated explicitly  and becomes
$$f_1^{(u)}(s)\simeq\dfrac{4g^2_\rho}{16\pi^2}\dfrac{2s-4\mu^2+\bar{M}^2}{k^2}
\left\{1-\dfrac{2k^2+\bar{M}^2}{4k^2}\log\left(1+\dfrac{4k^2}{\bar{M}^2}\right)\right\}.
\equn{(4.2.1)}$$
In this approximation the l.h. cut only runs up to $s=4\mu^2-M^2$: 
the discontinuity across the piece $[4\mu^2-M^2,0]$ is of 
order $(g^2/16\pi^2)^2$, and can be neglected (within the model) in a first approximation. 
With the value found for $g_\rho$, we expect this to be valid to some 6\%.

If we consider the region near $s=\bar{M}^2$, then $|f_1^{(s)}(s)|\sim 1$ 
while $f_1^{(u)}(s)$ is of order $g^2/16\pi^2$. 
We can further approximate $f_1$ by neglecting the whole of $f_1^{(u)}(s)$, 
 and thus write
$$f_1(s)\simeqsub_{s\sim M^2}
f^{(s)}_1(s)=
\tfrac{16}{3}\dfrac{g^2_\rho}{16\pi^2}\dfrac{k^2}{\bar{M}^2-s-s\piv_{\rm
ren.}(s)\vphantom{M^{M^M}}}.
\equn{(4.2.2)}$$
It is important to notice that 
this approximation 
is only valid when $f_1^{(s)}(s)$ is of order unity; otherwise, both $s$ and $u$ channel 
pieces are of comparable order of magnitude.
Another interesting point is that 
this approximation is unitary and indeed it is very similar to the
Breit--Wigner approximation.  To see this more clearly, we define the 
(resonance) mass of the rho as the solution of the 
 equation
$$\bar{M}^2=M^2_\rho\left\{1+\dfrac{g^2_\rho}{16\pi^2}\int_0^1\dd
x(1-2x)^2\log\left|\dfrac{\mu^2-x(1-x)\bar{M}^2}{\bar{M}^2\vphantom{M^{M^M}}}\right|\right\}.
\equn{(4.2.3)}$$
From (4.1.3) and (4.1.5) it follows that, in the present approximation,
we can identify, for physical $s$,
$$\cot\delta_1(s)=\ii+\dfrac{6\pi}{k^3g^2_\rho}
\left[\bar{M}^2-s-s\piv_{\rm ren}(s)\right]=
\dfrac{6\pi s^{1/2}}{k^3g^2_\rho}
\left(\bar{M}^2-s-s\real\piv_{\rm ren}(s)\right),$$
which, because of (4.2.3), vanishes at $s=M^2_\rho$.

In this model, the effective range function is
$$\phiv_1(s)=\dfrac{s-4\mu^2}{4}\sqrt{\dfrac{4\mu^2}{s}-1}
+\dfrac{3\pi}{g^2_\rho}\left[\bar{M}^2-s-s\piv_{\rm ren}(s)\right].$$

\booksection{4.3. Low energy scattering}
We now calculate the low energy scattering in the elementary rho model; 
to be precise, we will evaluate the scattering length, $a_1$. 
For this calculation a number of approximations can be made. 
Although, in the real world, $\gammav_\rho$ and $\mu$ 
are very similar, it is believed that they 
have a different origin. 
$\mu^2$ is supposed to 
be proportional to the sum of $u$ and $d$ quark masses, whereas $\gammav_\rho$ 
is related to the QCD parameter $\lambdav$.
We will thus  make a calculation neglecting the $u$ channel contribution 
and evaluating $\real\piv$ in leading order in $\log M^2_\rho/\mu^2$. 
In this approximation we have (cf. Eqs.~(2.1.4), (4.1.5)) 
$$a_1^{(s)}=\dfrac{g^2_\rho}{12\pi\mu M^2_\rho}\,
\dfrac{1}{1-(4\mu^2/M^2_\rho)\left[1+\dfrac{g^2_\rho}{48\pi^2}\log\dfrac{M^2_\rho}{\mu^2}\right]}
\simeq36\times10^{-3}\,\mu^{-3}.
\equn{(4.3.1)}$$

The experimental value is 
$$a_1=(39.1\pm1.4)\times10^{-3}\,\mu^{-3}.$$
We see that, for such a crude model, the 
agreement with experiment is quite good; in fact, as we will see in Chapter~8, 
comparable to what one 
gets with sophisticated calculations. 
On the other hand, of course, the model is only valid for the P 
wave; 
for example, it gives zero (to order $g^2_\rho/16\pi^2$) for $\pi^0\pi^0$ 
scattering, although the interaction here is very strong.

\booksection{4.4. The chiral rho model}
The model we have developed for $\rho$ mediated pion interactions is not compatible 
with chiral symmetry. 
A model compatible with this has been developed by Gasser and Leutwyler;\fnote{In fact, 
the chiral rho model is much older; see e.g. Coleman, Wess and Zumino~(1969) or Weinberg~(1968b).} 
in it the $\rho$ is   coupled through the field strengths, 
$F^{(a)}_{\mu\nu}$, with $a$ an isospin index, to the 
pions. The model is rather complicated and  can be found in the paper of these authors 
(Gasser and Leutwyler,~1984; see also Ecker et al.~1989 where it is further developed). 
This coupling produces a nonrenormalizable interaction (as opposed to the 
previous rho model, which was renormalizable) so 
only tree level calculations are, in principle, allowed with it. 

In fact, it is possible to make loop calculations with this model, but 
to get finite results we will have to add extra interactions (and extra 
coupling constants) every time we go to a higher order in the number of 
loops taken into account; the model soon loses 
its predictive power and, in this respect, it is inferior to the nonchiral model we have studied 
before. 
Moreover, it cannot satisfy rigorous unitarity (that requires 
an infinite number of loops), although Dyson resumed versions 
of it are available in the literature (Guerrero and Pich,~1997).  
Its main interest lies in providing an {\sl explicit} realization 
for chiral perturbation theory calculations, and a way to extrapolate these to the resonance region. 

We will not give the details of such calculations here, that 
the interested reader may find  in  
the literature quoted.

\bookendchapter
\brochureb{\smallsc chapter 5}{\smallsc the effective range formalism. 
resonances (multichannel), etc.}{29}
\bookchapter{5. The effective range formalism\hb for p.w. amplitudes;
resonances \hb
(multichannel formalism).\hb
Unitarity and form factors;\hb correlators}
\vskip-0.5truecm
\booksection{5.1. General formalism. Eigenphases}

\noindent
The extension of the developments of the previous section to the case where we have 
several channels open is very simple, provided 
these channels are all two-particle channels. 
To a good approximation this is the case for pion-pion scattering up to energies 
of about $s^{1/2}\simeq 1.3\,\gev$.

In the general case, we label the various two-body channels 
by  letters $a,b,\dots$, each with values $1,2,\dots,n$ (for $n$ channels). 
So, we have the p.w. amplitudes\fnote{We put in this Chapter
 the angular momentum variable $l$ as 
an index or superindex, according to convenience. 
So we write $f_l$ or $f^{(l)}$, $\delta^{(l)}$ or $\delta_l$.}
$f^{(l)}_{ab}(s)$ that describe scattering of particles\fnote{Note the reversed order; 
this is because the S matrix elements are usually defined by 
$$\langle P_1(a),P_2(a)|S|P_1(b),P_2(b)\rangle.$$}  $P_1(b)+P_2(b)\to
P_1(a)+P_2(a)$.  

As an example, we may have the channels
$$\matrix{\pi^+\pi^-,\quad &a=1\cr
\pi^0\pi^0,\quad &a=2\cr
K^+K^-,\quad &a=3\cr
K^0\bar{K}^0,\quad &a=4.\cr}$$
This would be simplified to two uncoupled two-channel problems 
(for isospin 0 and 1)  
if assuming isospin invariance.

We  define the (modulus of the) three-momentum, in channel $a$, as 
$k_a$. 
Then, the unitarity condition may be written  as
$$\imag f^{(l)}_{ab}(s)=\dfrac{\pi}{2s^{1/2}}\sum_ck_cf^{(l)}_{ac}(s)f^{(l)}_{bc}(s)^*,
\equn{(5.1.1)}$$
and we have used time-reversal invariance which implies that
$$f^{(l)}_{ab}=f^{(l)}_{ba}.$$ 
If we had only one channel, or if there were only diagonal interactions 
($f^{(l)}_{ab}=f^{(l)}_a\delta_{ab}$), (5.1.1) would tell us that
one can write
$$f^{(l)}_a=\dfrac{2s^{1/2}}{\pi k_a}\sin\delta^{(l)}_a\ee^{\ii \delta^{(l)}_a},
$$ 
i.e., \equn{(2.1.2)}.

To treat the general case it is convenient to use a matrix formalism. 
Denoting the matrices by boldface letters, we define 
$$
{\bf f}_l=\left(f^{(l)}_{ab}\right),\quad {\bf k}=(k_a\delta_{ab}).$$
We will also define the multichannel $S$-matrix elements,
$$s^{(l)}_{ab}(s)=\dfrac{2s^{1/2} }{\pi k_a}\delta_{ab}+2\ii f^{(l)}_{ab}(s)
\equn{(5.1.2a)}$$
or, in matrix notation,
$${\bf s}_l=(2s^{1/2}/\pi){\bf k}^{-1}+
{\bf f}_l.
\equn{(5.1.2b)}$$
If we had uncoupled channels, (2.1.2) would tell us immediately that
$$s^{(l)}_a=\dfrac{2s^{1/2}}{\pi k}\ee^{2\ii \delta_a^{(l)}}.$$
To see what the unitarity relations imply in the multichannel case,
 it is convenient to form the matrix $\bf u$
with
$$u_{ab}\equiv k_a^{1/2}s^{(l)}_{ab}k_b^{1/2}.$$
After a simple calculation, using (5.1.1) and 
time reversal invariance, we find that
$$\sum_c u^*_{ac}u_{bc}=\dfrac{4s}{\pi^2}\delta_{ab}.$$
Therefore, $(\pi/2s^{1/2}){\bf u}={\bf D}_l$ is a unitary matrix. 
We let ${\bf C}_l$ be the unitary matrix 
that diagonalizes it, and denote by $\widetilde{\bf D}_l$
to the diagonalized matrix, with elements
$(\exp{2\ii\widetilde{\delta}^{(l)}_a})\delta_{ab}$. 
The $\widetilde{\delta}^{(l)}_a(s)$ are called the {\sl eigenphases}, and are the
generalization to the multichannel case of the 
ordinary phase shifts.  We find that we can write:
$${\bf s}_l=
\dfrac{2s^{1/2}}{\pi}\,{\bf k}^{-1/2}{\bf C}_l\widetilde{\bf D}_l
{\bf C}_l^{-1}{\bf k}^{-1/2}.
\equn{(5.1.3)}$$
Note that, because of time reversal invariance, the matrix $\bf C$ may in fact be chosen to be 
{\sl real}. 

Inverting these relations we obtain the general form for the p.w. amplitudes,
$$\eqalign{
{\bf f}_l=&\dfrac{2s^{1/2}}{\pi}
{\bf k}^{-1/2}{\bf C}_l
\widetilde{\bf f}_l{\bf C}_l^{-1}{\bf k}^{-1/2},\cr
\widetilde{\bf f}_l=&\pmatrix{
\sin\widetilde{\delta}^{(l)}_1\ee^{\ii\widetilde{\delta}^{(l)}_1}&0&\dots&0\cr
0&\sin\widetilde{\delta}^{(l)}_2\ee^{\ii\widetilde{\delta}^{(l)}_2}&\dots&0\cr
\vdots&\vdots&\ddots&\vdots\cr
0&0&\dots&\sin\widetilde{\delta}^{(l)}_n\ee^{\ii\widetilde{\delta}^{(l)}_n}\cr}.
\cr}
\equn{(5.1.4)}$$
This is the generalization of (2.1.2) to the quasi-elastic multichannel case.

\booksection{5.2. The $K$-matrix and the effective range matrix. 
Resonances}

\noindent
We define the $K$-matrix, ${\bf K}_l$, 
such that
$$
{\bf f}_l=\left\{{\bf K}_l^{-1}-\dfrac{\ii\pi}{2s^{1/2}}{\bf k}\right\}^{-1}.
\equn{(5.2.1)}$$
In terms of it we can write the matrix ${\bf D}_l$ as
$${\bf D}_l=\dfrac{
1+\ii(\pi/2s^{1/2})\,{\bf k}^{1/2}{\bf K}_l{\bf k}^{1/2}}
{1-\ii(\pi/2s^{1/2})\,{\bf k}^{1/2}{\bf K}_l{\bf k}^{1/2}}.
\equn{(5.2.1)}$$
The unitarity and symmetry of ${\bf D}_l$ in the quasi elastic region
 means that ${\bf K}_l$ 
will be hermitean and symmetric there, hence it will be {\sl real} 
across the two particle cuts: 
$${\bf K}_l={\bf K}_l^{\dag}={\bf K}_l^*.$$

The definition of ${\bf K}_l$ does not take into account the 
behaviour at the thresholds. To do so we define the 
{\sl effective range} matrix, ${\bf \Phi}_l$ by
$${\bf \Phi}_l=\dfrac{1}{\pi}{\bf k}^l{\bf K}_l^{-1}{\bf k}^l.$$
In terms of it we find
$$
{\bf f}_l=\dfrac{1}{\pi}
{\bf k}^l\left({\bf \Phi}_l-\dfrac{\ii}{2s^{1/2}}{\bf k}^{2l+1}\right)^{-1}{\bf k}^l,
\equn{(5.2.2)}$$
an obvious generalization of (3.1.5). 
${\bf \Phi}_l$ is  real and symmetric. It is therefore analytic 
except for the l.h. cut of the $f^{(l)}_{ab}$, and for 
the r.h. cut that occurs when $s$ is above a true inelastic (multiparticle) threshold, 
$s>s_{\rm mult.}$.

Let us now discuss resonances in the multichannel case. 
It is clear that the  eigenstates of the time evolution operator 
will correspond to the eigenphases, as they are eigenstates of the 
$S$-matrix. 
We will therefore identify resonances with a resonant-like behaviour of the 
eigenphases: we will say that we have a resonance at $s=M^2$ 
provided one of the eigenphases crosses $\pi/2$ and varies rapidly there. 
We will assume that resonances are simple, i.e., only one eigenphase resonates 
at a given $s=M^2$, and moreover we suppose that $M$ does not 
coincide with the thresholds. 
The resonance condition, in eigenchannel $r$, is then
$$\widetilde{\delta}^{(l)}_r(s=M^2)=\pi/2;\quad
\left.\dfrac{\dd\widetilde{\delta}^{(l)}_r(s)}{\dd s}\right|_{s=M^2}=\hbox{maximum},
\equn{(5.2.3a)}$$
but
$$\widetilde{\delta}^{(l)}_{i\neq r}(s=M^2)\neq\pi/2.
\equn{(5.2.3b)}$$

Let us see what this implies in terms of ${\bf\Phi}_l$. From (5.1.4), (5.2.2) 
we can write
$$\widetilde{\bf f}_l=2s^{1/2}
\left({\bf C}_l^{-1}{\bf k}^{-l-1/2}{\bf \Phi}_l{\bf k}^{-l-1/2}{\bf C}_l-\ii\right)^{-1}.$$
Because $\widetilde{\bf f}_l$ and $\ii$ are diagonal, so must be
${\bf g}_l\equiv{\bf C}_l^{-1}{\bf k}^{-l-1/2}{\bf \Phi}_l{\bf k}^{-l-1/2}{\bf C}_l$. 
Recalling again (5.1.4), it follows that its elements are such that
$$(2s^{1/2}g^{(l)}_a-\ii)^{-1}=\sin\widetilde{\delta}^{(l)}_a\ee^{\ii\widetilde{\delta}^{(l)}_a},$$
i.e., one can write 
$$2s^{1/2}g^{(l)}_a=\cot\widetilde{\delta}^{(l)}_a.$$
The resonance condition then is equivalent (forgetting for the 
moment the requisite of rapid variation of the derivative of the phase) to 
the condition
$$g^{(l)}_r(s=M^2)=0;\quad g^{(l)}_{a\neq r}(s=M^2)\neq0.$$
Therefore, the quantity $\det({\bf g}_l(s))$ has a simple zero at $s=M^2$. 
Since, for this value of $s$, the determinants of ${\bf k},\,{\bf C}_l$ are finite, 
we have obtained that the condition of resonant behaviour 
(above all thresholds) is that the 
determinant of the  effective 
range matrix,
$$\det({\bf\Phi}_l(s))$$
has a simple zero at $s=M^2$.

We will next incorporate the condition of rapid variation, and calculate the 
{\sl partial widths}, that generalize the quantity $\gammav$ 
of the one-channel case. Near $s=M^2$ we write
$$\cot\widetilde{\delta}^{(l)}_r(s)\simeq \dfrac{M^2-s}{M\gammav}.\equn{(5.2.4)}$$
The condition of rapid variation is that $\gammav$ be small. 
Next, and using (5.1.4), we have
$$f^{(l)}_{ab}\simeqsub_{s\sim M^2}\dfrac{2s^{1/2}}{\pi}
\dfrac{1}{\sqrt{k_ak_b}}\left\{C^{(l)}_{ar}C^{(l)}_{br}\dfrac{M\gammav}{M^2-s-\ii M\gammav}+
\sum_{i\neq r}C^{(l)}_{ai}C^{(l)}_{bi}
\sin\widetilde{\delta}^{(l)}_i\ee^{\ii \widetilde{\delta}^{(l)}_i}\right\}.
\equn{(5.2.5a)}$$
We have profited from the unitarity and reality of ${\bf C}_l$ to write 
${\bf C}_l^{-1}={\bf C}_l^{\rm T}$.

We then define the {\sl partial widths}, $\gammav_a$, and {\sl inelasticity parameters} 
$x_a$ as
$$\gammav_a^{1/2}\equiv C^{(l)}_{ar}\gammav^{(1/2)};\quad
x_a=\gammav_a/\gammav.$$
Since the matrix ${\bf C}_l$ is orthogonal, 
one has $\sum_a\gammav_a=\gammav$. 
In terms of the $\gammav_a$ we can rewrite (5.2.5a) as
$$f^{(l)}_{ab}\simeqsub_{s\sim M^2}\dfrac{2s^{1/2}}{\pi}
\dfrac{1}{\sqrt{k_ak_b}}\left\{\dfrac{M\gammav_a^{1/2}\gammav_b^{1/2}}{M^2-s-\ii M\gammav}+
\sum_{i\neq r}C^{(l)}_{ai}C^{(l)}_{bi}\sin\widetilde{\delta}^{(l)}_i
\ee^{\ii \widetilde{\delta}^{(l)}_i}\right\}.
\equn{(5.2.5b)}$$
Thus we see that in the presence of a resonance all channels show a Breit--Wigner 
behaviour, plus a background due to the reflection of all the nonresonant eigenphases.

If, for a given channel, $x_a\simeq 1$, then we say that, in this channel, the resonance is
{\sl elastic}; if $x_a<1/2$, we say that it is {\sl inelastic}. 
For elastic resonances, and if
 the phase is near $\pi/2$ at the resonance, the parameter $\eta$ of (2.1.4) 
is related to $x$ by
$$\eta=2x-1.
\equn{(5.2.6)}$$

In general, when we have a resonance (even in the presence of multiparticle channels) 
we can write, for a given two-particle channel $a$,
$$\imag f^{(l)}_{aa}(s)\simeqsub_{s\sim M^2}
\dfrac{2s^{1/2}}{\pi k_a}\,\dfrac{M^2\gammav^2}{(M^2-s)^2+M^2\gammav^2}\,\times{\rm BR}
\equn{(5.2.7)}$$
with $\gammav$ the total width, and BR the branching 
ratio into channel $a$, 
${\rm BR}=\gammav_a/\gammav$. 

\booksection{5.3. Resonance parametrizations in the two-channel case}

\noindent
We will now present explicit formulas for parametrizations of resonances in 
the important case where only two channels are open. 
We start by changing a little bit the notation, writing, for obvious reasons, 
$g_l^{(\pm)}$ for the two 
eigenvalues of ${\bf g}_l$. 

We want to present parametrizations that profit from the analyticity of 
${\bf \Phi}_l$ so that they are not only valid on the resonance; thus,  
we will write our formulas in terms of ${\bf \Phi}_l$. 
Actually, we will use as parameters the 
diagonal elements of ${\bf \Phi}_l$,  $\phiv^{(l)}_{11}(s)$, $\phiv^{(l)}_{22}(s)$, and 
its determinant, that, because we have a resonance at $s=M^2$, we may write as  
$\det{\bf \Phi}_l(s)=\gamma(s)(s-M^2)$, with 
$\gamma(s)$ a smooth function (that can in most cases be approximated by a constant).

Next, we express the $g_l^{(\pm)}$ in terms of these parameters. 
We let $\deltav$ and $\tau$ be the determinant and trace of  ${\bf g}_l$. 
We have, on one hand, and in the physical region for both channels,    
$$g_l^{(\pm)}=\dfrac{\tau\mp\sqrt{\tau^2-4\deltav}}{2};\quad
{\bf g}_l=\pmatrix{g_l^{(+)}&0\cr
0&g_l^{(-)}\cr};\quad k_1,\,k_2\geq0 
\equn{(5.3.1a)}$$
and, on the other,
$$\eqalign{\deltav=&\det({\bf g}_l)=(k_1k_2)^{-2l-1}\det{\bf
\Phi}_l(s)=(k_1k_2)^{-2l-1}\gamma(s)(s-M^2),\cr 
\tau=&\trace {\bf g}_l=k_1^{-2l+1}\phiv^{(l)}_{11}+k_2^{-2l-1}\phiv^{(l)}_{22}.
\cr}
\equn{(5.3.1b)}$$
The resonating phase is  $\delta_l^{(+)}$ if $\tau$ is positive 
and  $\delta_l^{(-)}$ if $\tau$ is negative 
because, from (5.3.1), it follows that $\deltav$ 
 vanishes for $s=M^2$.

The mixing matrix ${\bf C}_l$ can also be obtained explicitly. 
One has,
$$\eqalign{{\bf C}_l=\pmatrix{\cos\theta&\sin\theta\cr
-\sin\theta&\cos\theta\cr};\quad
\cos\theta=
\left\{\dfrac{k_1^{-2l-1}\phiv^{(l)}_{11}-g_l^{(-)}}{g_l^{(+)}-g_l^{(-)}}\right\}^{1/2}.\cr}
\equn{(5.3.2)}$$

\booksection{5.4. Reduction to a single channel. Weakly coupled channels}
We will now consider the case in which one has two channels, but we are interested 
chiefly on one of them, that we will denote by channel 1. 
We will further assume that this channel opens before channel 2. 
Below the opening of channel 2, the formulas reduce to those of 
one single channel, so we can write (cf. (3.1.5))
$$f^{(l)}_{11}=\dfrac{1}{\pi}
\dfrac{k_1^{2l}}{\phiv^{(l)}_{\rm el}-\dfrac{\ii}{2s^{1/2}}k_1^{2l+1}}.
\equn{(5.4.1)}$$
 $\phiv^{(l)}_{\rm el}$ may be expressed in terms of ${\bf{\Phi}}^{(l)}$ using 
(5.2.2). 
We  define $\kappa_a=\ii k_a$ and get,
$$\phiv^{(l)}_{\rm el}=\,
\dfrac{\vphantom{\Bigg|}\dfrac{(-1)^l}{2s^{1/2}}\kappa_2^{2l+1}\phiv^{(l)}_{11}+\det {\bf{\Phi}}^{(l)}}
{\vphantom{\Bigg|}\dfrac{(-1)^l}{2s^{1/2}}\kappa_2^{2l+1}+\phiv^{(l)}_{22}}.
\equn{(5.4.2)} 
$$
Before the opening of channel 2, 
and above the l.h. cut, $\phiv^{(l)}_{\rm el}$ 
is, as expected, real and analytic.

It is worth noting that \equn{(5.4.2)} is 
still valid above the opening of channel 2, but $\kappa_2$ will now be 
{\sl imaginary}. 
Because of this some care has to be exercised to identify 
the quantity $\delta^{(l)}_{11}$. 
From (2.1.2), which is valid above threshold for channel 1, but  
below  channel 2 threshold we have, using (3.1.5), 
$$\cot\delta^{(l)}_{11}(s)=\dfrac{2s^{1/2}}{k_1^{2l+1}}\,
\phiv^{(l)}_{\rm el}(s),
\equn{(5.4.3)}$$
with $\phiv^{(l)}_{\rm el}$ given by (5.4.2). 
But, because $\kappa_2$ becomes imaginary above the opening of channel 2, 
it follows that $\cot\delta^{(l)}_{11}(s)$ will 
be complex there. 
This is of course to be expected; a real $\delta^{(l)}_{11}(s)$ 
implies strict elastic unitarity.

We next continue with  two channels, but now assume that they are 
weakly coupled. 
This is made transparent by writing
$$\phiv^{(l)}_{12}\equiv \epsilon_{12},$$ 
and we will work to lowest nontrivial order in $\epsilon_{12}$. 
We can write,
$$f^{(l)}_{11}=\dfrac{1}{\pi}
\dfrac{\vphantom{\Bigg|}\phiv^{(l)}_{22}-\dfrac{\ii}{2s^{1/2}}k_2^{2l+1}}
{\vphantom{\Bigg|}\left(\phiv^{(l)}_{11}-\dfrac{\ii}{2s^{1/2}}k_1^{2l+1}\right)
\left(\phiv^{(l)}_{22}-\dfrac{\ii}{2s^{1/2}}k_2^{2l+1}\right)-\epsilon^2_{12}}.
$$
Expanding to lowest order in the mixing, this becomes 
$$f^{(l)}_{11}=\dfrac{1}{\pi}\dfrac{k_1^{2l+1}}{\phiv^{(l)}_{11}-\dfrac{\ii}{2s^{1/2}}k_1^{2l+1}}
\left\{1+\dfrac{\phiv^{(l)}_{11}}{\phiv^{(l)}_{11}-\dfrac{\ii}{2s^{1/2}}k_1^{2l+1}}
\,\dfrac{\epsilon^2_{12}}{\phiv^{(l)}_{22}-\dfrac{\ii}{2s^{1/2}}k_2^{2l+1}}\right\}
\equn{(5.4.4)}$$
i.e., like an effective one-channel amplitude, 
$$\bar{f}^{(l)}_{11}=
\dfrac{1}{\pi}\dfrac{k_1^{2l+1}}{\phiv^{(l)}_{11}-\dfrac{\ii}{2s^{1/2}}k_1^{2l+1}}
\equn{(5.4.5a)}$$
modulated by the factor
$$G^{(l)}_1=
1+\dfrac{\phiv^{(l)}_{11}}{\phiv^{(l)}_{11}-\dfrac{\ii}{2s^{1/2}}k_1^{2l+1}}
\,\dfrac{\epsilon^2_{12}}{\phiv^{(l)}_{22}-\dfrac{\ii}{2s^{1/2}}k_2^{2l+1}}:
\equn{(5.4.5b)}$$
one has,
$$f^{(l)}_{11}=\bar{f}^{(l)}_{11}G^{(l)}_1.
\equn{(5.4.5c)}$$

In the case in which we have a resonance in each channel, we 
write
$$\phiv^{(l)}_{11}(s)\simeq(M_1^2-s)/\gamma_1,\quad
\phiv^{(l)}_{22}(s)\simeq(M_2^2-s)/\gamma_2.
\equn{(5.4.6)}$$
In this case (5.4.4) becomes
$$\eqalign{f^{(l)}_{11}\simeq&\dfrac{1}{\pi}
\dfrac{k_1^{2l+1}\gamma_1}{M_1^2-s-\ii k_1^{2l+1}\gamma_1/2s^{1/2}}\cr
\times&\left\{1+\dfrac{M_1^2-s}
{k_2^{2l+1}\left(M_1^2-s-\ii k_1^{2l+1}\gamma_1/2s^{1/2}\right)}\,
\dfrac{\epsilon^2_{12}\gamma_2k_2^{2l+1}}{M_2^2-s-\ii k_2^{2l+1}\gamma_2/2s^{1/2}}\right\}.\cr
}
\equn{(5.4.7)}$$
It is noteworthy that, if the resonances are narrow, 
and not too near the thresholds, the 
modulation of the first ($M_1$) by the second is negligible 
(of order $\gamma_2\epsilon^2_{12}$) except on top of the 
second, $s\simeq M^2_2$.

The mixing angle also has a simple expression now:
$$\sin\theta=
\dfrac{(k_1k_2)^{l+1/2}|\epsilon_{12}|^2}
{\left|k_2^{2l+1}\phiv^{(l)}_{11}-k_1^{2l+1}\phiv^{(l)}_{22}\right|}.
\equn{(5.4.8)}$$

We note to finish that the coupling of the channels displaces the resonances. 
Defining them as solutions of the equation
$$\det{\bf\Phi}(\widetilde{M}^2_a)=0,\quad a=1,2,
\equn{(5.4.9a)}$$
we see that e.g. for the first we have
$$\widetilde{M}_1^2=M_1^2+\dfrac{\gamma_1\gamma_2\epsilon^2_{12}}{M_2^2-M_1^2}.
\equn{(5.4.9b)}$$

\booksection{5.5. Unitarity for the form factors}

\noindent
The expression for the form factor of scalar particles $A,\, \bar{A}$ 
(which we consider with electric charge $\pm e$) in
the timelike region is defined, for example, in terms of the process 
$$e^+e^-\to A\bar{A}.$$
The corresponding matrix element may be written, to lowest order
 in the electromagnetic interaction, 
and with the effective photon-hadron interaction 
${\cal L}_{\rm eff}=eJ_\mu(x)A^\mu(x)$, as
$$\eqalign{
\langle A(p_1)\bar{A}(p_2)&|S|e^+(k_1)e^-(k_2)\rangle
=\ii e^2\dfrac{1}{(2\pi)^3}\bar{v}(k_1)\gamma_\mu u(k_2)\dfrac{-\ii}{(p_1+p_2)^2}\cr
\times& 
(2\pi)^4\delta(k_1+k_2-p_1-p_2)\langle A(p_1)\bar{A}(p_2)|J^\mu(0)|0\rangle,\cr
}
$$
and we recall that the form factor is defined  (for spinless particles) as
$$\langle A(p_1)\bar{A}(p_2)|J^\mu(0)|0\rangle=
(2\pi)^{-3}(p_1-p_2)^\mu F(s),\quad s=(p_1+p_2)^2.$$

Let us write the $S$ matrix as
$S=1+\ii{\cal T}$ so that 
$$\langle f|{\cal T}|i\rangle=\delta(p_f-p_i)F(i\to f).$$
 Unitarity of $S$ implies the relation
$${\cal T}-{\cal T}^+=\dfrac{1}{\ii}{\cal T}{\cal T}^+.$$
Taking matrix elements, we get
$$\imag \langle  A_a(p_1)\bar{A}_a(p_2)|{\cal T}|e^+e^-\rangle
=\tfrac{1}{2}\langle  A_a(p_1)\bar{A}_a(p_2)|{\cal T}{\cal T}^+|e^+e^-\rangle,$$
and we have assumed that we have several two particle channels, denoted with the 
index $a$. 
Summing now over intermediate states, we find 
$$\eqalign{
&\imag \langle  A_a(p_1)\bar{A}_a(p_2)|{\cal T}|e^+e^-\rangle\cr
=\sum_b&\,\int\dfrac{\dd^3{\bf q}_1}{2q_{10}}\dfrac{\dd^3{\bf q}_2}{2q_{20}}
\tfrac{1}{2}\langle  A_a(p_1)\bar{A}_a(p_2)|{\cal T}|A_b(q_1)\bar{A}_b(q_2)\rangle
\langle A_b(q_1)\bar{A}_b(q_2)|{\cal T}^+|e^+e^-\rangle.\cr
}
$$
In terms of the form factors and scattering amplitudes, therefore,
$$\eqalign{
&\imag (p_1-p_2)^\mu F_a(s)\cr
=&\,
\tfrac{1}{2}\sum_b \int\dfrac{\dd^3{\bf q}_1}{2q_{10}}\,\dfrac{\dd^3{\bf q}_2}{2q_{20}}
(q_1-q_2)^\mu F_b^*(s_q)
\langle  A_a(p_1)\bar{A}_a(p_2)|{\cal T}|A_b(q_1)\bar{A}_b(q_2)\rangle\cr
=&\,\tfrac{1}{2}\sum_b \int\dfrac{\dd^3{\bf q}_1}{2q_{10}}\,\dfrac{\dd^3{\bf q}_2}{2q_{20}}
\delta(q_1+q_2- p_1-p_2)(q_1-q_2)^\mu F_b^*(s_q)F_{ab}(q_1,q_2\to p_1,p_2),\cr
}$$
$s_q=(q_1+q_2)^2$.
In the c.m., $(p_1-p_2)^0=0$, $(p_1-p_2)^i=2k_i$ 
with $\bf k$ the c.m. three-momentum. Considering the spacelike
 part of above equation (the timelike part
is trivial) we find, after simple manipulations,\fnote{We hope there will be no confusion 
between the form factors, $F_a$, and scattering amplitudes, 
$F_{ab}=F_{ab}(s,t).$}
$$\imag  F_a(s)=
\tfrac{1}{2}\dfrac{1}{2{\bf k}^2}\sum_b  F_b^*(s_q)
\int\dfrac{\dd^3{\bf q}_1}{4q_{10}^2}\,({\bf q}_1{\bf p}_1)\delta(2p_{10}-q_{10}) F_{ab}.
$$
Writing
$$F_{ab}=\sum_l(2l+1)P_l(\cos\theta)f^{(l)}_{ab},$$
$\cos\theta=({\bf q}_1{\bf p}_1)/k_a k_b$, $k_a\equiv|{\bf p}_1|$, 
 $k_b\equiv|{\bf q}_1|$ we finally obtain the expression of unitarity in terms of form factors and 
p.w. amplitudes:
$$\imag  F_a(s)=
\dfrac{3\pi}{8s^{1/2}}\sum_b\dfrac{k^2_b}{k_a} F_b^*(s)f^{(1)}_{ab}(s).
\equn{(5.5.1)}$$

One can diagonalize this. With  the formulas for the $f^{(1)}_{ab}$ 
in terms of the eigenphase shifts, $\widetilde{\delta}^{(l)}_a$, 
and the diagonal p.w. amplitudes, 
we find (matrix notation)
$$\imag {\bf C}^{-1}{\bf k}^{3/2}{\bf F}=
\tfrac{3}{8}\widetilde{\bf f}^{(1)}{\bf C}^{-1}{\bf k}^{3/2}{\bf F}^*.$$ 
It follows that the combination
$$\sum_b C_{ba}k^{3/2}_b F_b$$ 
has a phase equal to $\widetilde{\delta}^{(1)}_a$. 
For the one channel case this proves the equality of the 
phases of form factor and p.w. amplitudes 
with the appropriate quantum numbers; 
for the electromagnetic form factor,  the P wave and for the scalar one, the S0 wave.

\booksection{5.6. Unitarity for correlators}

\noindent
We will for definiteness consider a correlator\fnote{We 
write, generally, 
$\langle A\dots B\rangle_0\equiv \langle0|A\dots B|0\rangle$.} of vector currents (not 
necessarily conserved), $V_\mu$:
$$\piv_{\mu\nu}(q)=\ii\int\dd^4x\,\ee^{\ii q\cdot x}
\langle{\rm T}V_\mu(x)V^{\dag}_\nu(0)\rangle_0
\equiv(-q^2g_{\mu\nu}+q_\mu q_\nu)\piv_{\rm tr}(q^2)+q_\mu q_\nu \piv_S(q^2),
\equn{(5.6.1)}$$
and we have split it into a transverse component ($\piv_{\rm tr}$) and a scalar one ($ \piv_S$). 
If the current was conserved, $\partial\cdot V=0$, then $\piv_S=0$.

The imaginary part of the correlator is given by the expression

$$\eqalign{
I_{\mu\nu}(q)=&\,\imag \piv_{\mu\nu}(q)=
\tfrac{1}{2}\int\dd^4x\,\ee^{\ii q\cdot x}\langle[V_\mu(x),V^{\dag}_\nu(0)]\rangle_0,
\quad q^2\geq0;\cr
I_{\mu\nu}(q)=&\,0,\quad q^2\leq0.\cr
}
\equn{(5.6.2a)}$$
Inserting a complete sum of states, $\sum_\gammav|\gammav\rangle\langle\gammav|$, this becomes
$$I_{\mu\nu}(q)=\tfrac{1}{2}\int\dd^4x\,\ee^{\ii q\cdot x}
\sum_\gammav\langle0|V_\mu(x)|\gammav\rangle
\langle\gammav|V^{\dag}_\nu(0)|0\rangle.$$
Writing also
$$\langle0|V_\mu(x)|\gammav\rangle=\ee^{-\ii p_\gammav\cdot x}
\langle0|V_\mu(0)|\gammav\rangle$$
we get the result
$$I_{\mu\nu}(q)=\tfrac{1}{2}(2\pi)^4
\sum_\gammav\delta(q-p_\gammav)\langle0|V_\mu(0)|\gammav\rangle
\langle0|V_\nu(0)|\gammav\rangle^*.
\equn{(5.6.2b)}$$
(Of the two terms in the commutator only the first gives a nonzero result, 
because necessarily the momentum of $\gammav$, $p_\gammav$, has to be timelike). 
In particular, (5.6.2b) implies that $I_{\mu\nu}$ is positive definite, i.e., for any $p$ 
(even complex),
${p^\mu}^* I_{\mu\nu}p^\nu\geq0$. 
If we write
$$I_{\mu\nu}=(-q^2g_{\mu\nu}+q_\mu q_\nu)\imag\piv_{\rm tr}(q^2)+q_\mu q_\nu \imag\piv_S(q^2)
\equn{(5.6.3a)}$$
then
$$\imag \piv_{\rm tr}\geq0,\quad \imag\piv_S\geq0.
\equn{(5.6.3b)}$$

We will consider two important cases of intermediate states: 
when $|\gammav\rangle$ is a single particle state of mass $m$,  any spin, and when 
it is the state of two spinless particles. 
In the first case,
$$\sum_\gammav\to\sum_\lambda\int\dfrac{\dd^3p}{2p_0}|p,\lambda\rangle\langle p,\lambda|$$
and $\lambda$ is the third component of the spin. 
Then, and working in the c.m. reference system where $q_0=\sqrt{q^2}\equiv s^{1/2}$, ${\bf q}=0$,
$$\eqalign{
I_{\mu\nu}(q)=&\,\tfrac{1}{2}(2\pi)^4\sum_\lambda\int\dfrac{\dd^3p}{2p_0}\,\delta(q-p)
\langle0|V_\mu(0)|p,\lambda\rangle\langle0|V_\nu(0)|p,\lambda\rangle^*\cr
=&\,\dfrac{(2\pi)^4}{4s^{1/2}}\delta(s^{1/2}-m)
\sum_\lambda\dfrac{\sqrt{2}\, F_\mu(q,\lambda)}{(2\pi)^{3/2}}
\dfrac{\sqrt{2}\, F^*_\nu(q,\lambda)}{(2\pi)^{3/2}};
\cr}
$$
 we have defined
$$\langle0|V_\mu(0)|p,\lambda\rangle=\dfrac{\sqrt{2}}{(2\pi)^{3/2}}F_\mu(p,\lambda).
\equn{(5.6.4a)}$$
If the particle is a pion $\pi^-$ and $V_\mu$ is the weak axial current, 
$V_\mu=\bar{u}\gamma_\mu\gamma_5 d$, then $ F_\mu(q,\lambda)$ is related to 
the {\sl pion decay constant}, $f_\pi$:
$$ F_\mu(q,\lambda)=f_\pi p_\mu,\quad f_\pi\simeq 93\,\mev,
\equn{(5.6.4b)}$$
see Chapter~8. 
In general we have
$$I_{\mu\nu}(q)=2\pi\delta(s-m^2)\sum_\lambda F_\mu(q,\lambda)F_\nu(q,\lambda)^*.
\equn{(5.6.5)}$$

For the case of a two-particle intermediate state, with spinless particles,
$$\sum_\gammav\to \int\dfrac{\dd^3 p_1}{2p_{10}}\dfrac{\dd^3 p_2}{2p_{20}}\,
|p_1p_2\rangle\langle p_1p_2|$$
and then
$$\eqalign{
I_{\mu\nu}(q)=&\,\tfrac{1}{2}(2\pi)^4\int\dfrac{\dd^3 p_1}{2p_{10}}\dfrac{\dd^3 p_2}{2p_{20}}\,
\delta(p-p_1-p_2)\langle0|V_\mu(0)|p_1p_2\rangle\langle0|V_\nu(0)|p_1p_2\rangle^*\cr
=&\,\dfrac{(2\pi)^4}{2 s^{1/2}}\int\dd^3k\,
\delta(s-(p_1+p_2)^2)\langle0|V_\mu(0)|p_1p_2\rangle\langle0|V_\nu(0)|p_1p_2\rangle^*;
\cr}
$$
 ${\bf k}={\bf p}_1=-{\bf p}_2$. 
If we assume that the current is conserved, we can express the expectation value of the current 
in terms of a form factor,\fnote{If the current is not conserved we will have terms 
proportional to $p_1+p_2$ in (5.6.6).}
$$\langle0|V_\mu(0)|p_1,p_2\rangle=\dfrac{1}{(2\pi)^3}(p_1-p_2)_\mu F(s),
\equn{(5.6.6)}$$
hence
$$I_{\mu\nu}(q)=\dfrac{|F(s)|^2}{2(2\pi)^2s^{1/2}}\int\dd^3k\,
\delta(s-(p_1+p_2)^2)(p_1-p_2)_\mu(p_1-p_2)_\nu.
$$
The integral is easiest calculated in the c.m. reference system. 
Here $(p_1-p_2)_\mu=2k_\mu$, and we have defined $k_0|_{\rm c.m.}=0$. If $\mu$ is the mass of the particles in
the  intermediate state (assumed equal, as they have to be if the current is conserved), then 
$(p_1+p_2)^2=2(\mu^2+{\bf k}^2)$. 
In spherical coordinates,
$$I_{\mu\nu}(q)=\dfrac{\sqrt{s-4\mu^2)}}{4(2\pi)^2s^{1/2}}|F(s)|^2\int\dd\omegav_{\bf k}k_\mu k_\nu.$$
The angular integral, returning to an arbitrary reference system is
$$\int\dd\omegav_{\bf k}k_\mu k_\nu=\dfrac{4\pi}{3s}
\left(\dfrac{s}{4}-\mu^2\right)(-g_{\mu\nu}s+q_\mu q_\nu),$$
so we get the final expression
$$\imag \piv_{\rm tr}(s)=\dfrac{1}{6\pi}\left(\dfrac{s/4-\mu^2}{s}\right)^{3/2}|F(s)|^2.
\equn{(5.6.7)}$$

\bookendchapter
\brochureb{\smallsc chapter 6}{\smallsc p.w. amplitudes for $\pi\pi$ scattering. 
form factors}{39}
\bookchapter{6. Extraction and parametrizations\hb of p.w. amplitudes
 for $\pi\pi$
scattering.\hb Form factors}
\vskip-0.5truecm
\booksection{6.1. $\pi\pi$ scattering }

\noindent
There is of course no possibility to arrange collisions of {\sl real} pions. 
One can get information on some phase shifts, at a few energies, 
from processes such as  kaon decays, or from 
the pion electromagnetic or weak form factors (about which more later). 
But a lot of, unfortunately not very precise, information comes from 
peripheral pion production, that we now 
briefly discuss.

What one does in these types of experiments
 is to collide pions with protons and produce two pions and either 
a nucleon, $N$, or a resonance $\Delta$:
$$\pi p\to\pi\pi N;\quad \pi p\to \pi\pi\Delta.$$
One selects events where the momentum $p_\pi$ transferred by the incoming 
pion to the proton
 is small and thus one can assume that the process is mediated 
by  exchange of a virtual pion (\fig~6.1.1). 
The process $\pi p\to \pi\pi\Delta$ is in principle more difficult to analyze than 
$\pi p\to\pi\pi N$; but the last presents a zero for $p_\pi\sim0$, thus suppressing it in the more 
interesting region: 
both processes are, in consequence, equally well (or equally poorly) suited for extracting 
$\pi\pi$ scattering data. 
We then expect that the scattering amplitude for the full process 
will factorize into the $\pi\pi$ scattering amplitude, with one pion off-shell, 
$F(s,t;p_\pi^2)$, and the matrix element $\langle H|\phi_\pi|p_\pi\rangle$.
Here $H=N,\;\Delta$ 
and $\phi_\pi$ is the pion field operator.

It is clear that the method presents a number of drawbacks. 
First of all, a model is necessary for the dependence on $p_\pi$ 
of $F(s,t;p_\pi^2)$ and $\langle H|\phi_\pi|p_\pi\rangle$. 
Indeed, a model is required for  $\langle H|\phi_\pi|p\rangle$ itself. 
Secondly, in factorizing the full processes one is neglecting final state interactions between 
the pions and the $N$ or $\Delta$. These are presumably small, but only rather crude 
models exist for them.
 
\midinsert{
\setbox1=\vbox{\hsize9truecm{\epsfxsize 7.3truecm\epsfbox{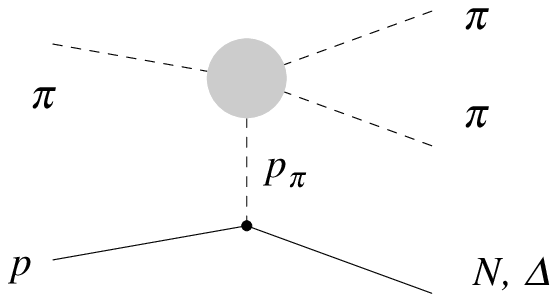}}}
 \setbox6=\vbox{\hsize 4.5truecm\captiontype\figurasc{Figure 6.1.1. }{\hb 
Diagrams 
for $\pi p\to\pi\pi N,\,\Delta$.\hb
\phantom{X}\hb}} 
\line{\box1\hfil\box6}
\medskip
}\endinsert

Another very important problem is that, as soon as inelastic
 channels become important for $\pi\pi$ scattering, which
occurs for $s^{1/2}\sim1\,\gev$ for the S wave and for $s^{1/2}\gsim 1.4\,\gev$ for 
P, D waves, the analysis becomes impossibly complicated: 
the errors grow very fast.\fnote{In fact, 
it can be proved (Atkinson, Mahoux and Yndur\'ain, 1973) that,
 even if one only has two channels, 
say, $\pi\pi$ and $\bar{K}K$,
there is no unique solution (at fixed energy) unless one also measured $\pi\pi\to\bar{K}K$, and
 $\bar{K}K\to\bar{K}K$ as well. 
That inelastic channels are important for $s^{1/2}\gsim1.4\,\gev$ 
is clear by looking at the branching ratios of resonances with higher mass.}
Indeed, above $s^{1/2}\sim 1.5\,\gev$
 it is impossible to disentangle the interesting processes
from 
a number of other ones and, as a consequence, there are 
no any reliable  data. 
We will discuss this more in \sect~6.6.

As a consequence of all these difficulties, it happens that the sets
 of phase shifts one extracts from 
data present unknown biases and, 
in particular,  are dependent on the models used to perform the fits. 
This is very clear in the several sets of solutions presented by Protopopescu et al. (1973), 
Estabrooks and Martin~(1974), and 
in the large errors of the analysis of Hyams et al. (1973) or Grayer et al.~(1974). 
We could have tried to quantify this 
by introducing systematic errors (for example, the difference between 
various determinations). This we do in some cases; 
in others  
we  
simply admit that a \chidof\ of up to $\sim2\,\sigma$, with only statistical errors, may 
be  acceptable.

A help out of these difficulties is to use supplementary information from 
processes like 
$$e^+e^-\to\pi^+\pi^-,\quad \tau^+\to\bar{\nu}_\tau\pi^+\pi^0,\quad K\to l{\bar{\nu}_l}\pi\pi,
\quad K\to 2\pi.$$
We will discuss them later later, but note already that this only provides information on 
the S, P waves at low energy ($s\lsim 1\,\gev^2$). 
Another possibility is to supplement the experimental information with theory; 
in \sects~6.3 to 5 of this chapter we take into account the analyticity
 properties of p.w. amplitudes to write
economical and accurate 
parametrizations of these; the implementation of other constraints, such 
as dispersion relations, is left for next chapter. 

\booksection{6.2. Form factors and decays}
\vskip-0.5truecm
\booksubsection{6.2.1. The pion form factor}

\noindent 
The process $e^+e^-\to\pi^+\pi^-$ (\fig~6.2.1) can, at low energy $t^{1/2}\lsim1\,\gev$, 
be related to the pion form factor.
We  write 
$$\dfrac{\sigma^{(0)}(e^+e^-\to{\rm hadrons})}{\sigma^{(0)}(e^+e^-\to\mu^+\mu^-)}
=12\pi\imag \Piv(t),
$$
where $\Piv$ is hadronic part of the photon polarization function and 
the superindices (0) mean that we evaluate the so tagged quantities to lowest order in 
electromagnetic interactions. 
At low energy this is dominated by the $2\pi$ state and we have
$$\imag\Piv=\imag \piv_{2\pi}(t)=\dfrac{1}{48\pi}\left(1-\dfrac{4\mu^2}{t}\right)^{3/2} 
|F_\pi(t)|^2.\equn{(6.2.1)}$$

\midinsert{
\setbox1=\vbox{\hsize8.9truecm{\epsfxsize 7.6truecm\epsfbox{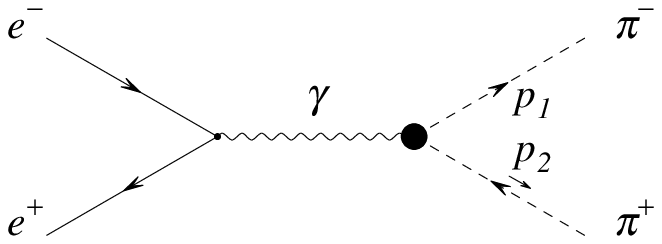}}}
 \setbox6=\vbox{\hsize 3.truecm\captiontype\figurasc{Figure 6.2.1. }{\hb
Diagram 
for $e^+e^-\to\pi^-\pi^+$.\hb
\phantom{X}\vskip0.5truecm
\phantom{X}}} 
\line{\box1\hfil\box6}
\medskip
}\endinsert

The evaluation of the pion form factor is slightly complicated by the phenomenon 
of $\omega-\rho$ interference. This can be solved by considering only the 
isospin $I=1$ component, and adding later the $\omega\to2\pi$ 
and interference 
separately; 
that is to say,  in a first approximation we neglect the breaking of isospin 
invariance. We will also neglect for now  electromagnetic corrections. 
In this approximation the properties of $F_\pi(t)$ are the following: 
\item{(i) }{$F_\pi(t)$ is an analytic function of $t$, with a cut 
from $4\mu^2$ to infinity.}
\item{(ii) }{On the cut, the phase of $F_\pi(t)$ is, because of unitarity, identical to 
that of the P wave, $I=1$, $\pi\pi$ scattering, $\delta_1(t)$, and 
this equality 
holds until the opening of the inelastic threshold at $t=s_0$. 
This we showed in \sect~5.4, and the property is known as the
 Fermi--Watson final state interaction 
theorem.}
\item{(iii) }{For large $t$, $F_\pi(t)\sim 1/t$. This follows from 
perturbative QCD.}
\item{(iv) }{$F(0)=1$.}
 
The inelastic threshold occurs, rigorously speaking, at $t=16\mu^2$. 
However, it is an experimental fact that inelasticity is negligible 
until the quasi-two~body channels $\omega\pi,\,a_1\pi\,\dots$ are open. 
In practice we will take
$$s_0\simeq 1\;\gev^2,$$
and fix the best value for $s_0$  empirically. 
It will be $s_0=1.05^2\,\gev^2$, and it so happens that, if we keep close to 
 this value, the dependence of the results of our analysis on  $s_0$ is very slight. 

\booksubsection{6.2.2. Form factor of the pion in $\tau$ decay}

\noindent 
Besides the process $e^+e^-\to\pi^+\pi^-$ one can get  data on the vector pion 
form factor  from the decay
 $\tau^+\to\bar{\nu}_\tau \pi^+\pi^0$ (\fig~6.2.2) 
For this 
 we have to assume isospin invariance, to write the form factor $v_1$ 
for $\tau$ decay 
 in terms of $F_\pi$:
$$v_1=\tfrac{1}{12}\left(1-\dfrac{4\mu^2}{t}\right)^{3/2}|F_\pi(t)|^2,
\equn{(6.2.2a)}$$
where, in terms of the weak vector current $V_\mu=\bar{u}\gamma_\mu d$, 
and in the exact isospin approximation,
$$\eqalign{
\Piv^V_{\mu\nu}=&\,\left(-p^2g_{\mu\nu}+p_\mu p_\nu\right)\Piv^V(t)=
\ii\int\dd^4x\,\ee^{\ii p\cdot x}\langle0|{\rm T}V^+_\mu(x)V_\nu(0)|0\rangle;\cr
&v_1=2\pi\imag \Piv^V.\cr
}
\equn{(6.2.2b)}$$
\equn{(6.2.2)} may be verified inserting a complete set of states in the 
expression for $\imag \piv^V$, and 
assuming it to be saturated by the states $|\pi^+\pi^0\rangle$; 
cf.~\equn{(5.6.7)}.

\setbox1=\vbox{\hsize7truecm\hfil {\epsfxsize 6.truecm\epsfbox{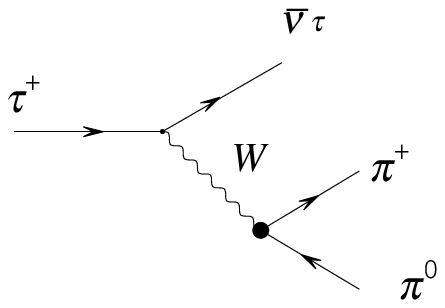}}\hfil}
 \setbox6=\vbox{\hsize 5truecm\captiontype\figurasc{Figure 6.2.2 }{\hb 
Diagram 
for $\tau\to\bar{\nu}_\tau\pi^0\pi^+$.\hb
\phantom{X}\vskip0.5truecm
\phantom{X}}} 
\line{\box1\hfil\box6}
\smallskip

We next make a few remarks concerning  the matter of isospin breaking, due to 
electromagnetic interactions or the  mass difference between $u,\;d$
 quarks, that would spoil the equality (6.2.2a). 
It is not easy to estimate this. A large part of the breaking, the 
$\omega\to2\pi$ contribution and $\omega-\rho$ mixing, may 
 be taken into account explicitly (for the form factor in $\pi^+\pi^-$) 
with the Gounnaris--Sakurai~(1968) method, but this does not exhaust the effects. 
\eqs~(6.2.2) were obtained neglecting 
the mass difference $m_u-m_d$ 
and electromagnetic corrections, 
in particular the $\pi^0 - \pi^+$ mass difference. 
We can take the last partially into 
account by distinguishing between the pion masses in the 
phase space factor in (6.2.2a). 
To do so, we write now (6.2.2b) as
$$\Piv^V_{\mu\nu}=
\ii\int\dd^4x\,\ee^{\ii p\cdot x}\langle0|{\rm T}V^+_\mu(x)V_\nu(0)|0\rangle=
\left(-p^2g_{\mu\nu}+p_\mu p_\nu\right)\Piv^V(t)+p_\mu p_\nu \piv^{S};\quad
v_1\equiv2\pi\imag \Piv^V.
\equn{(6.2.3a)}$$  
We find 
$$v_1=\tfrac{1}{12}
\left\{\left[1-\dfrac{(m_{\pi^+}-m_{\pi^0})^2}{t}\right]
\left[1-\dfrac{(m_{\pi^+}+m_{\pi^0})^2}{t}\right] \right\}^{3/2}|F_\pi(t)|^2.
\equn{(6.2.3b)}$$
To compare with the experimentally measured quantity, 
which involves all of $\imag \piv^V_{\mu\nu}$, we have to 
neglect the scalar component $\piv^S$. This is reasonable, as it is proportional to $(m_d-m_u)^2$, 
and thus likely very small. 
This matter of isospin breaking one thus treats in successive steps. 
First, we neglect isospin breaking. Then we 
take it into account by 
admitting different masses and widths for the resonances $\rho^0,\,\rho^+$, 
 including $\omega-\rho$ mixing, and taking into account the
difference in phase space, etc. 
Before doing so, however, we must develop the 
necessary mathematical tools, which we will do in next chapter.

\booksubsection{6.2.3. $K_{l4}$ decay}

\noindent
We now consider the so-called $K_{l4}$ decay ($K_{l4}$ 
stands for {\sl leptonic four body} decay), 
$$K\to l\bar{\nu}_l\pi^+\pi^-,$$
with $l$ an electron or a $\mu^-$. 
The effective lagrangian for the decay is
$${\cal L}_{\rm int,eff}=\dfrac{G_F\sin\theta}{\sqrt{2}}\,
\bar{l}\gamma_\mu(1-\gamma_5)\nu_l\,\bar{s}\gamma^\mu(1-\gamma_5)u,
$$
where $G_F$ is Fermi's constant, $\theta$ the Cabibbo angle, 
and $s$, $u$ the field operators for the 
corresponding quarks. 
The decay amplitude is then
$$F(K\to l\bar{\nu}_l\pi^+\pi^-)=\dfrac{G_F\sin\theta}{\sqrt{2}(2\pi)^2}
\bar{v}_l\gamma_\mu(1-\gamma_5)u_{\nu_l}F^\mu(s);
\equn{(6.2.4a)}$$
 the form factor $F^\mu$ is
$$F^\mu(s)=
\langle \pi^+(p_+)\pi^-(p_-)|\bar{s}(0)\gamma^\mu\gamma_5u(0)|K\rangle,
\quad s=(p_++p_-)^2.
\equn{(6.2.4b)}$$
If we expand $F^\mu$ into a scalar ($F_S$) and a vector piece, $F_P$ ($P$ for P wave),
$$F^\mu=(p_+^\mu+p_-^\mu)F_S+(p_+^\mu-p_-^\mu)F_P,
\equn{(6.2.5)}$$
then one can, with an argument like that of \sect~5.5, show that
$${\rm arg}\,F_S(s)=\delta_0^{(0)}(s),\quad
{\rm arg}\,F_P(s)=\delta_1(s).
\equn{(6.2.6)}$$
It follows that, by measuring the differential decay rate
$$\dfrac{\dd\Gammav(K\to l\bar{\nu}_l\pi^+\pi^-)}{\dd s\dd \Omegav_{\bf p}},$$
with ${\bf p}={\bf p}_+|_{\rm c.m.}$, 
we can separate 
the contributions from $|F_S|^2$, $|F_P|^2$ 
and the interference piece,
 $F_S^*F_P=|F_S|\,|F_P|\cos[\delta_0^{(0)}(s)-\delta_1(s)]$,
and thus get 
 the difference of phases $\delta_0^{(0)}(s)-\delta_1(s)$.
This provides very important information on  low energy 
$\pi\pi$ scattering, particularly since, in this process, both pions 
are on their mass shell.

\booksubsection{6.2.4. The $K\to2\pi$ decays}

\noindent
If we denote by $H_{W}$ to the weak interaction hamiltonian, 
we will consider the matrix elements 
related to the decays $K^+\to\pi^+\pi^0$, 
$K_S\to\pi^+\pi^-$ and $K_S\to\pi^0\pi^0$:
$$\eqalign{
\langle\pi^+\pi^0|H_W|K^+\rangle=&\,-\langle2,1|H_W|K^+\rangle,\cr
\langle\pi^+\pi^-|H_W|K_S\rangle=&\,-\sqrt{\tfrac{1}{3}}\,\langle2,0|H_W|K_S\rangle-
\sqrt{\tfrac{2}{3}}\,\langle0,0|H_W|K_S\rangle,\cr
\langle\pi^0\pi^0|H_W|K_S\rangle=&\,\sqrt{\tfrac{2}{3}}\,\langle2,0|H_W|K_S\rangle-
\sqrt{\tfrac{1}{3}}\,\langle0,0|H_W|K_S\rangle.\cr
\cr}
\equn{(6.2.7)}$$
Here the labels in the $\langle I,I_3|$ refer to isospin and third component thereof. 
From the Fermi--Watson 
final state interaction theorem, it follows that 
$$\langle  I,I_3|H_W|K\rangle=\big|\langle  I,I_3|H_W|K\rangle\big|\,\ee^{\ii
\delta_0^{(I)}(m^2_K)},
\equn{(6.2.8)}$$
and, therefore, measuring the three decays provides a determination of the difference 
of phase shifts
$$\delta_0^{(0)}(m^2_K)-\delta_0^{(2)}(m^2_K).
\equn{(6.2.9)}$$

A precise analysis requires considerations of isospin violation, especially by 
electromagnetic interactions,\fnote{Belavin and Navodetsky~(1968);
 Nachtmann and de~Rafael~(1969); 
Cirigliano, Donoghue and 
Golo\-wich~(2000).} 
that shift the phase by some $4\degrees$. 
The old experimental determinations gave (see, e.g. Pascual and Yndur\'ain,~1974) 
$$\delta_0^{(0)}(m^2_K)-\delta_0^{(2)}(m^2_K)=58.0\pm4.6\degrees,
\equn{(6.2.10a)}$$
while more modern determinations of the kaon decays (Aloisio et al.,~2002; Gatti,~2003)
 have led to the numbers
$$\eqalign{
\delta_0^{(0)}(m^2_K)-\delta_0^{(2)}(m^2_K)=&\,48.5\pm2.6;\cr
\delta_0^{(0)}(m^2_K)-\delta_0^{(2)}(m^2_K)=&\,47.8\pm2.8.\cr
} 
\equn{(6.2.10b)}$$

\booksection{6.3. The P  wave}

\noindent
We present in this and the following two  sections of the present chapter 
parametrizations of the S, P, D and F waves in $\pi\pi$ scattering 
that follow from the theoretical requirements
 we have discussed in previous chapters, and which agree
with  {\sl experimental} data (we will also say a few words on G waves). 
To check that the scattering amplitude that one obtains in this way 
is consistent with dispersion relations or the 
Froissart--Gribov representation 
will be done in the following chapter. 
When neglecting isospin violations we will take the 
Gasser--Leutwyler convention of approximating the pion mass by 
$M_\pi=m_{\pi^\pm}$.

\booksubsection{6.3.1. The P  wave  in the elastic approximation}
We will consider first the P wave for $\pi\pi$ scattering, because it is 
obtained at low energy with a method different from those used for the other waves. 
We start thus considering the region of energies 
 where the inelasticity is below the  2\% level; say,  
$s_0\leq 1.1\,\gev^2$. 
We will neglect for the moment isospin invariance violations due 
to e.m. interactions or the $u\,-\,d$ quark mass difference. 
This implies, in particular, neglecting the $\omega$ and $\phi$ interference  
effects.
 
We may use the analyticity properties of $\phiv_1(s)$ to write a 
simple parametrization of $\phiv_1(s)$, hence of $\delta_1(s)$. 
An effective range expansion is not enough, as it only converges in the region 
$|s-4M^2_\pi|<0$ (\fig~3.1.2). To take fully advantage
 of the analyticity domain, shown in \fig~2.1.1, 
the simplest procedure is to make a conformal mapping of the 
cut plane into the unit disk 
(\fig~6.3.1) by means of the transformation\fnote{This 
type of parametrization 
presents a number of advantages with respect to less efficient ones 
used in the literature. The  gain 
obtained by taking into account the correct 
analyticity properties is enormous; see the Appendix~B here for a
 discussion and an explicit example, and 
 Pi\u{s}ut~(1970) for other examples and applications to $\pi\pi$ scattering.
 Moreover, the physical meaning of, say, (6.3.3) is very clear: 
$B_0$ gives the normalization, and 
$B_1$ 
is related to the average intensity of the l.h. cut and the inelastic cut.}
$$w=
\dfrac{\sqrt{s}-\sqrt{s_0-s}}{\sqrt{s}+\sqrt{s_0-s}}.
\equn{(6.3.1)}$$

One can then expand $\phiv_1(s)$ in powers of $w$, and, reexpressing 
$w$ in terms of $s$, the expansion will be convergent over all the cut $s$-plane. 
Actually, and because we know that the P wave resonates at $s=M^2_\rho$, 
it is more convenient to expand not  $\phiv_1(s)$ 
itself, but ${\psi}(s)$ given by
$$\phiv_1(s)=(s-M^2_\rho){\psi}(s)/4;
\equn{(6.3.2a)}$$
so we write
$${\psi}(s)=\left\{B_0+B_1w+
\cdots\right\}.
\equn{(6.3.2b)}$$

\topinsert{
\setbox0=\vbox{{\epsfxsize 9.2truecm\epsfbox{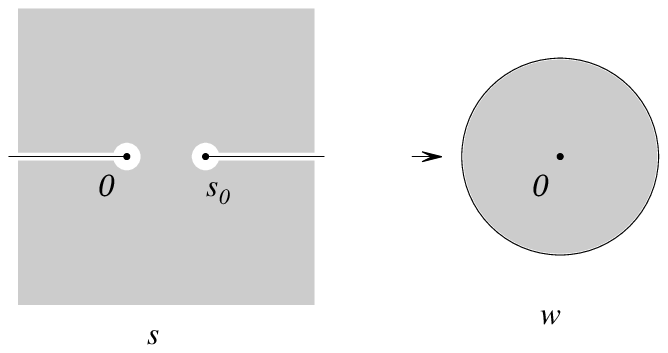}}} 
\setbox6=\vbox{\hsize 6truecm\captiontype\figurasc{Figure 6.3.1. }{ The  
mapping $s\to w$.\hb
\phantom{XX}}\hb
\vskip.1cm} 
\medskip
\centerline{\box0}
\centerline{\box6}
\medskip
}\endinsert

In terms of  $\phiv_1(s)$ we find the expression for the phase shift, 
 keeping two terms in the expansion,
$$\cot\delta_1(s)=\dfrac{s^{1/2}}{2k^3}
(M^2_\rho-s)\left[B_0+B_1\dfrac{\sqrt{s}-\sqrt{s_0-s}}{\sqrt{s}+\sqrt{s_0-s}}
\right];
\equn{(6.3.3)}$$
$M_\rho,\,B_0,\,B_1$ are free parameters to be fitted to experiment.
In terms of $\phiv_1,\,\psi$ we have, for the rho width,
$$\gammav_\rho=\dfrac{2k^3_\rho}{M^2_\rho{\psi}(M^2_\rho)},
\quad
k_\rho=\tfrac{1}{2}\sqrt{M^2_\rho-4M^2_\pi},
\equn{(6.3.4a)}$$
and the scattering length, $a_1$, is
$$a_1=\dfrac{1}{4M_\pi\phiv_1(4M_{\pi}^2)}=\dfrac{1}{M_\pi\psi(4M^2_\pi)}.
\equn{(6.3.4b)}$$

The values $B_0=\hbox{const.}$, $B_{i\geq 1}=0$ would 
correspond to a perfect Breit--Wigner. 
Actually, it is known that the $\rho$ deviates from a pure Breit--Wigner 
and for a precision parametrization 
 two terms, $B_0$ and $B_1$, have to be kept in (6.3.3). 
Note that the parametrization holds not only on the physical region 
$4M_{\pi}^2\leq s\leq s_0$, but on the unphysical region $0\leq s\leq 4M_{\pi}^2$ and 
also over the 
whole region of the complex $s$ plane with $\imag s\neq 0$.
The parametrization given now is the one that has less biases, in the sense 
that no model has been used: we have imposed only the highly safe requirements of 
analyticity and unitarity, depending only on 
causality and conservation of probability.

\topinsert{
\setbox0=\vbox{\hsize11truecm{\hfil\epsfxsize10truecm\epsfbox{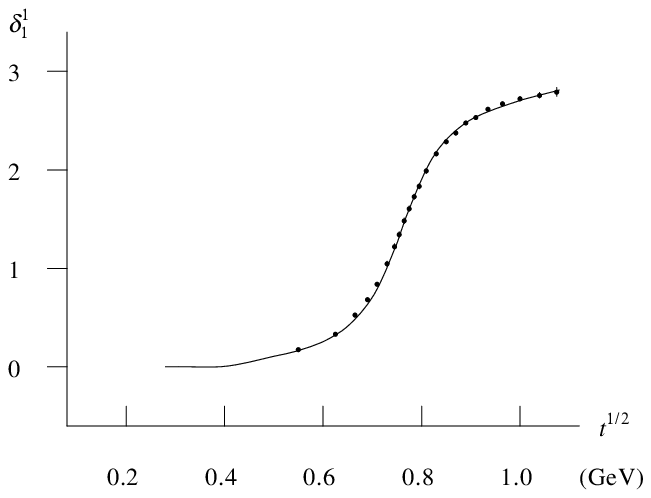}\hfil}} 
\setbox6=\vbox{\hsize 10truecm\captiontype\figurasc{Figure 6.3.2 }{The  
phase shifts of solution 1 from Protopopescu et al.~(1973) (the dots, 
with errors of  the size of the dots) compared with the prediction with 
the parameters (6.3.5a), described by the solid line. 
We emphasize that this solid line is {\sl not} a fit 
to the data of Protopopescu et al., but is obtained 
from the pion form factor.}
} 
\medskip
\centerline{\tightboxit{\box0}}
\bigskip
\centerline{\box6}
\bigskip
}\endinsert 

The best values for our parameters are actually obtained from fits to the 
pion form factor, that we will discuss in \sect~7.2. 
Including systematic experimental errors in the fits, and fitting also the 
value $a_1=(38\pm3)\times10^{-3}\,M_{\pi}^{-3}$ for the scattering length we have,
$$B_0=1.071\pm0.007,\quad B_1=0.18\pm0.05;\quad M_\rho=773.5\pm0.85\,\mev.
\equn{(6.3.5a)}$$
The corresponding values for the width of the $\rho$ and
 for the scattering length and effective range parameter are
$$\eqalign{
a_1=&\,(38.6\pm1.2)\times10^{-3}M_{\pi}^{-3},
\quad b_1=(4.47\pm0.29)\times10^{-3}M_{\pi}^{-5};\cr
\gammav_\rho=&\,145.5\pm1.1\,\mev.\cr
}
\equn{(6.3.5b)}$$
Although the values of the experimental $\pi\pi$ 
phase shifts were {\sl not} included in the fit, the 
phase shifts that (6.3.5a) implies are en very good agreement with them, 
as shown in \fig~6.3.2.

Eqs.~(6.3.5) above were evaluated with an average of information on the 
two channels that contain the $I=1$ P wave, $\pi^+\pi^-$ (dominated by the 
$\rho^0$) and $\pi^0\pi^+$, dominated by the $\rho^+$. 
The values for a pure $\rho^0$ ($\pi^+\pi^-$) are slightly different; we find
$$\eqalign{B_0=&\,1.065\pm0.007,\quad B_1=0.17\pm0.05,\quad M_{\rho^0}=773.1\pm0.6,\cr 
\gammav_{\rho^0}=&147.4\pm1.0\,\mev,
}
\equn{(6.3.5c)}$$
and $a_1,\,b_1$ do not change appreciably. 
However, this last feature occurs only because 
the fit was made including the constraint  
$a_1=(38\pm3)\times10^{-3}\,M_{\pi}^{-3}$; 
see \sect~9.5 for more on this.
Eqs.~(6.3.5)  provide an estimate of the importance of  isospin breaking.

\booksubsection{6.3.2. The $\rho$ and weakly coupled inelastic channels:
  $\omega-\rho$ interference}

\noindent
Because of the different masses of the $u,\,d$ quarks, isospin invariance is broken 
and there is a nonzero probability of transition 
between $\pi^+\pi^-$ in isospin 1 and isospin 0
states: hence, a small --but nonzero-- mixing of the $\rho$ and $\omega$ resonances.

To study this phenomenon a popular approximation is that of Gounnaris and 
Sakurai (1968). A  consistent 
treatment requires a two-channel analysis. 
We denote by channel 1 to the P wave isospin 1  $\pi^+\pi^-$  state, and channel 2 will 
be a P wave isospin zero $3\pi$ state. 
To be fully rigorous, we would have to set up a three-body formalism for the last; 
but we will simply take this into account replacing the two body by 
three body phase space for the $\omega$. Using now Eqs.~(5.4.4) to (5.4.7) we write

$$\eqalign{f^{(1)}_{11}=&\dfrac{1}{\pi}
\dfrac{k_1^{3}\gamma_1}{M^2_\rho-s-\ii k_1^{3}\gamma_\rho(s)/2s^{1/2}}\cr
\times&\left\{1+\dfrac{M^2_\rho-s}
{k_\omega^{3}(s)\left(M^2_\rho-s-\ii k_1^{2l+1}\gamma_\rho(s)/2s^{1/2}\right)}\,
\dfrac{\epsilon^2_{12}\gamma_\omega k_2^{3}}{M^2_\omega-s-\ii k_\omega^{3}(s)\gamma_\omega/2s^{1/2}}\right\}.\cr
}
\equn{(6.3.7)}$$
Here we still have
$$k_1=\tfrac{1}{2}\sqrt{s-4M_{\pi}^2}
\equn{(6.3.8a)}$$
but for $k_\omega(s)$ 
we have to take the value following from three-body phase space. 
Because the interference effect is only important 
near $s=M^2_\omega$,  a reasonable approximation for it is 
to take $k_\omega$ constant: this is the model of Gounnaris and Sakurai~(1968). 
The model is completed if we
take a constant width for the $\omega$, justified in view of its narrowness, but a full effective range
formula for the 
$\rho$:
$$\gamma_\omega=\gammav_\omega^2/2f_\omega(M^2_\omega),
\quad \gamma_\rho(s)=1/\bar{\phiv}(s),
\equn{(6.3.8b)}$$
with $\bar{\phiv}(s)$ given by a parametrization like (6.3.3). 
The effect of this modulation is a shoulder above the $\rho$ that may be seen in e.g. the 
pion form factor (cf.~\fig~7.2.1).

\booksubsection{6.3.3. The P wave for $1\gev\leq s^{1/2}\leq 1.42\gev$}
In the range $1\,\gev\leq s^{1/2}\leq 1.3\,\gev$ one is sufficiently far away from thresholds to neglect 
their influence (the coupling to $\bar{K}K$ is negligible) and, 
moreover, the inelasticity is reported  small: 
according to Protopopescu et al.~(1973) and Hyams et al.~(1973), below the 7\% level. 
A purely empirical parametrization that agrees with the data in this references 
 up to 
1.2 \gev, within errors, 
is given by a modulated $\rho$ tail,
$$\eqalign{
\delta_1(s)=&\,{\rm arc}\,\cot\dfrac{\eta\,(M_\rho-s)}{M_\rho\gammav_\rho}-
\epsilon\,\left(1-\dfrac{4m_K^2}{s}\right)^{3/2},\cr
 \eta=&\,0.75\pm0.10,\quad\epsilon=0.08\pm0.02.\cr
}
\equn{(6.3.9)}$$
and the second term takes into account the effects of the inelasticity.

For larger $s^{1/2}$, (6.3.9) is incompatible with the properties of the P wave as measured in 
the analysis of Hyams et al.~(1973), or in 
$e^+e^-$ annihilations, where a highly inelastic resonance,that we here denote by $\rho'$,
 occurs around 1450~\mev.
An alternate parametrization for the imaginary part of the p.w. 
amplitude  that takes this into account is obtained by adding to the imaginary part produced by 
(6.3.9) the inelastic piece
$$\eqalign{
\imag f_{1;{\rm inel}}(s)=&\,\dfrac{2s^{1/2}}{\pi k}\dfrac{{\rm BR}\times
M^2_{\rho'}\gammav^2[k/k(M^2_{\rho'}]^6}{(s-M_{\rho'}^2)^2+M_{\rho'}^2\gammav^2[k/k(M^2_{\rho'}]^6};\cr
 M_{\rho'}=&\,1.45\,\mev, \quad\gammav=310\,\mev,\quad{\rm BR}\simeq0.15.\cr
}
\equn{(6.3.10)}$$
The value of BR could vary by 50\%.
For more details, see   Appendix~A.

\goodbreak

\booksection{6.4. The S waves}
\vskip-0.5truecm
\booksubsection{6.4.1. Parametrization of the S wave for  $I=2$}

\noindent
We consider two sets of experimental data.
 The first, that we will denote by
``Hoogland~A", 
 corresponds to solution A in the paper by 
Hoogland et al.~(1977), who use the reaction $\pi^+ p\to\pi^+\pi^+n$; and 
the second set, denoted by ``Losty," corresponds 
 to that from the work of Losty
et al.~(1974), who analyze instead  $\pi^- p\to\pi^-\pi^-\Delta$. 
We will not consider the so-called solution B in the paper of Hoogland et al.~(1977); 
while it produces results similar to the other two, its errors 
are clearly underestimated. 
We will also not include in the fit the data of Cohen et al.~(1973); 
 it may be biased at low energy because it is obtained from scattering on 
neutons bound in deuterium. 
Neverteless, our fits go nicely over these experimental points.
 The 
result of Losty et al and Hoogland et al. (also those 
of Cohen et al.) represent a substantial improvement over 
previous ones; since they produce two like charge pions, only isospin 2 
contributes, and one gets rid of the large isospin zero S wave and P wave contamination. 
However, they still present the problem that 
one does not have scattering of real pions.

For isospin 2, there is no low energy resonance, but $f_0^{(2)}(s)$
 presents the feature that a zero 
is expected (and, indeed, confirmed by the fits) 
in the region $0<s<4M_{\pi}^2$.  If we neglected this and wrote
$$\eqalign{\cot\delta_0^{(2)}(s)=&\,\dfrac{2s^{1/2}}{2k}\,\dfrac{B_0+B_1w(s)}{4};\cr
w=&\,
\dfrac{\sqrt{s}-\sqrt{s_0-s}}{\sqrt{s}+\sqrt{s_0-s}},\quad s_0=(1.450\;\gev)^2,\cr}
$$
then we could fit the data
with the parameters
$$B_0=-1.87,\quad B_1=5.56.$$
We have a not too bad $\chidof=13.8/(14-2)$ but the expansion has poor convergence 
properties as, in most of the region, $|B_1w|$ is rather larger than $|B_0|$. 
The corresponding value of the scattering length would be $a_0^{(2)}=-0.16\,M_{\pi}^{-1}$, 
way too large (that a naive fit gives a scattering length of this order 
has been known for a long time; see Prokup et al., 1974). 
Clearly, we have to take the zero of the partial wave into account.

\topinsert{
\setbox0=\vbox{\hsize12.truecm{\epsfxsize 11.2truecm\epsfbox{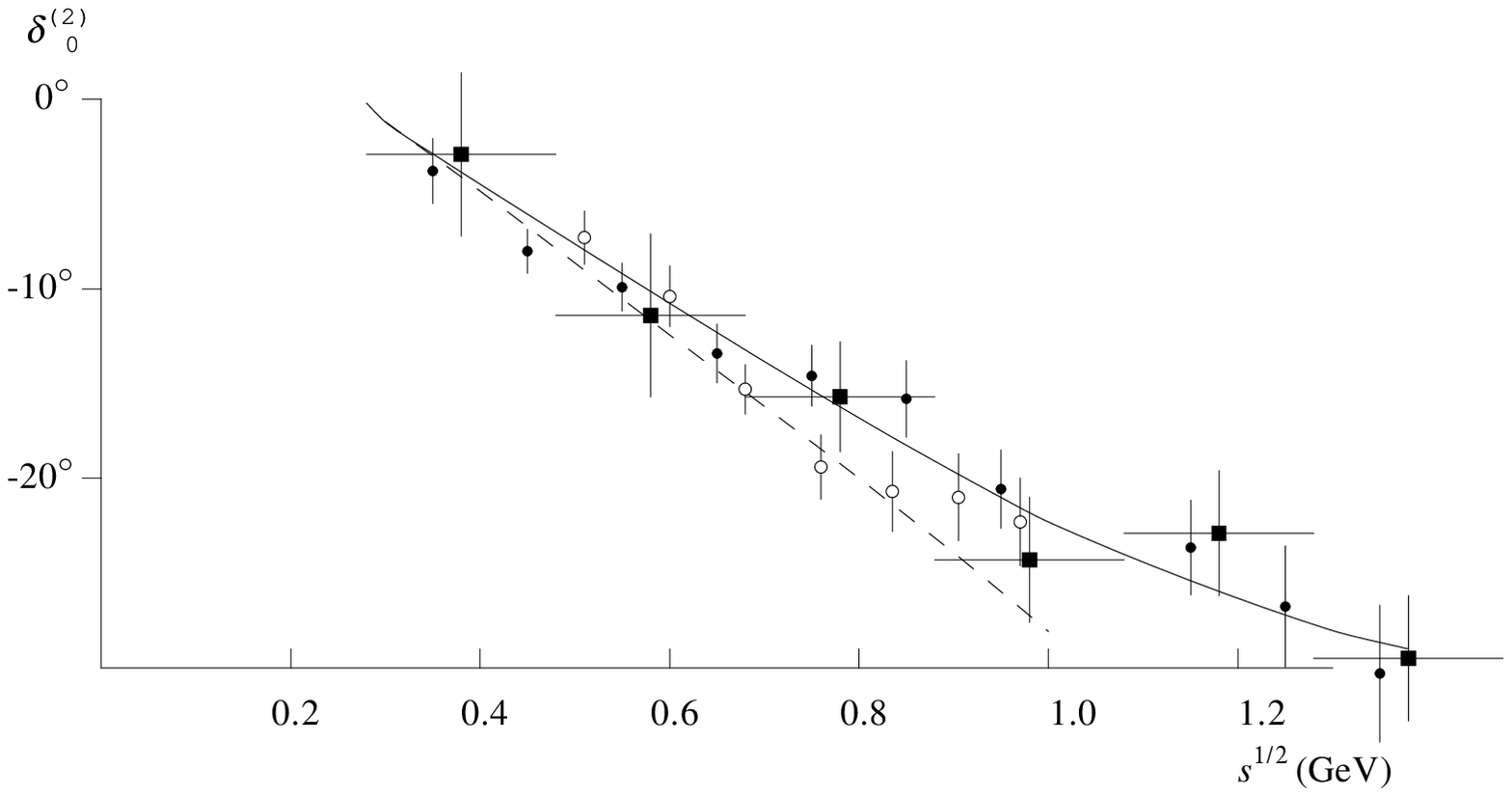}}} 
\setbox6=\vbox{\hsize 10truecm\captiontype\figurasc{Figure 6.4.1. }{
The  
$I=2$, $S$-wave phase shifts corresponding to (6.4.1), 
with experimental points from Losty et al.~(1974) (open circles), Hoogland et al.~(1977),
 solution A 
(black dots) and Cohen et al.~(1973) (crosses). 
The dashed line is the S2 phase of Colangelo, Gasser and Leutwyler~(2001).  
}\hb} 
\centerline{\tightboxit{\box0}}
\bigskip
\centerline{\box6}
\medskip
}\endinsert

The zero of $f_0^{(2)}(s)$ is related to the so-called Adler zeros (see Chapter~9) 
and, to lowest order in chiral perturbation theory,  
occurs at $s=2z_2^2$ with $z_2=M_\pi$. In view 
of this, in a first fit  
we extract the zero (leaving its value as a {\sl free} 
parameter) and write
$$\cot\delta_0^{(2)}(s)=\dfrac{s^{1/2}}{2k}\,\dfrac{M_{\pi}^2}{s-2z_2^2}\Big\{B_0+B_1w(s)\Big\}.
\equn{(6.4.1a)}$$
The quality of the fit improves substantially: we get $\chidof=8.0/(14-3)$ 
and a second order term such that $|B_1w|<|B_0|$.
 The parameters are now
$$B_0=-116\pm5.6,\quad B_1=-127\pm9,\qquad z_2=145\pm21\,\mev.
\equn{(6.4.1b)}$$
$w(s)$ is as for the D2 wave, \equn~{(6.4.1b)}.

In the fit (6.4.1) we have not considered experimental data above 0.97 \gev.
The result for the scattering length,
$$a_0^{(2)}=(-0.061\pm0.023)\,M_{\pi}^{-1}.
\equn{(6.4.1c)}$$
 is compatible (within $\sim1\,\sigma$), as we will see, with the values
suggested by  chiral perturbation theory;\fnote{The 
discrepancy in the central value is likely due to 
a systematic bias of the experimental data. 
In fact, if we also include in the fit the data of the Cern-Munich collaboration, in the 
version of Estabrooks and Martin~(1974), the central value of $a_0^{(2)}$ becomes 
$0.46\,M_{\pi}^{-1}$.} and this agreement is 
satisfactory also in another respect: 
 the value for $z_2$ which the fit returns, $z_2=145\pm21\,\mev$, comprises the 
value expected from second order chiral perturbation theory that gives $z_2=131\,\mev$.

One can improve on this fit by using  
forward dispersion relations, in the form of the Olsson sum rule, see \subsect~7.4.3. 
We moreover fix $z_2=M_\pi$ and fit all experimental data, up to $s^{1/2}=1350\,\mev$. 
One finds  slightly different parameters:
$$\eqalign{
\cot\delta_0^{(2)}(s)=&\,\dfrac{s^{1/2}}{2k}\,\dfrac{M_{\pi}^2}{s-2z_2^2}\,
\left\{B_0+B_1\dfrac{\sqrt{s}-\sqrt{s_0-s}}{\sqrt{s}+\sqrt{s_0-s}}\right\};\cr
s_0^{1/2}=1.45\;\gev;&\quad\chi^2/{\rm d.o.f.}=17.2/(19-2).\cr
 B_0=&\,-118\pm2.5,\quad B_1=-105\pm2.5,\quad z_2=139.57\;\mev\;\hbox{[fixed]}.\cr
}
\equn{(6.4.2a)}$$
The value of the scattering length which this implies, 
$$a_0^{(2)}=(-0.0422\pm0.0022)\,M_{\pi}^{-1},
$$
is  between what is obtained with the help of Roy equations by different groups:
$$\eqalign{
a_0^{(2)}=&\,(-0.0444\pm0.0010)\,M_{\pi}^{-1}\quad\hbox{(Colangelo, Gasser ad Leutwyler, 2001)},\cr
a_0^{(2)}=&\,(-0.0382\pm0.0038)\,M_{\pi}^{-1}\quad\hbox{(Descotes et al., 2002)},\cr
a_0^{(2)}=&\,(-0.0343\pm0.0036)\,M_{\pi}^{-1}\quad\hbox{(Kami\'nski, Le\'sniak and
Loiseau~(2003))}.\cr
}
$$ 

We have, for a precision representation, to include the 
inelasticity as determined in the 
experiments of Cohen et el.~(1973) and Losty et al.~(1974). 
We will take a purely empirical fit,
$$\eta_0^{(2)}(s)=1-c(1-M^2_{\rm eff}/s)^{3/2},\quad 
c=0.28\pm0.12.
\equn{(6.4.2b)}$$
The fit is good, $\chidof=4.2/(5-1)$, and this formula is supposed to hold 
for $s^{1/2}\geq0.96\,\gev$.

We will take (6.4.2) to be valid up to $1.42$ \gev.

\booksubsection{6.4.2. Parametrization of the S wave for  $I=0$}

\noindent
The S wave with isospin zero is by far the most difficult to parametrize.
 Here we have a very broad 
enhancement, variously denoted as $\epsilon,\,\sigma,\,
f_0$, around
$s^{1/2}\equiv M_\sigma\sim800\,\mev$; we will use the name $\sigma$. 
We will not discuss here whether
 this enhancement is a {\sl bona fides} 
resonance; we merely remark that in all experimental
 phase shift analyses $\delta_0^{(0)}(s)$ crosses 90\degrees\ 
somewhere between 600 and 900 \mev. (This is {\sl not} enough to class 
the object as a resonance. For example, the derivative $\dd \delta_0^{(0)}(s)/\dd s$ is  
more a minimum than a maximum 
 at $M_\sigma$).

There is also a possible resonance, which used to be called
$S^*$ and is now denoted by $f_0(980)$, and another resonance (which was
called
$\epsilon'$ in the seventies), labeled 
as $f_0(1370)$ in the Particle Data Tables, with a mass
around $1.37\,\gev$. 
Moreover,we expect  a zero of $f_0^{(0)}(s)$ (Adler zero), hence a pole of 
the effective range function 
$\phiv_0^{(0)}(s)$, for $s=z_0^2$ with $z_0^2$ in the region $0< s<4M_{\pi}^2$. 
In fact, chiral perturbation theory suggests that this zero is located at 
$z_0^2=
\tfrac{1}{2}M_{\pi}^2$ but,  as we will discuss in 
\subsect~9.3.5, one cannot trust the accuracy of this prediction, unlike what happened for the 
$I=2$ zero, $z_2$.
 
We can distinguish two energy regions: below $s^{1/2}_0=2m_K$ we are under 
the $\bar{K}K$ threshold. 
Between $s^{1/2}_0$ and $s^{1/2}\sim1.2$ there is a strong coupling between the 
 $\bar{K}K$ and $\pi\pi$ channels and the analysis becomes very unstable, because there 
is little information on the process $\pi\pi\to\bar{K}K$ and even less on 
$\bar{K}K\to\bar{K}K$. 
We will not treat this case here in any detail; the interested reader may find details and references in 
Yndur\'ain~(1975), Aguilar-Ben\'{\i}tez et al.~(1978). 
We will merely present, in the next subsection, an empirical fit in the 
region of energies around and above 1 \gev, and 
 we will now concentrate our efforts in the low energy region.

Below the $\bar{K}K$ threshold we can write a one-channel formula:
$$\cot\delta_0^{(0)}(s)=\dfrac{2s^{1/2}}{k}\,\phiv_0^{(0)}(s).
\equn{(6.4.4)}$$
To parametrize $\phiv_0^{(0)}$ we have, as stated, a difficult situation, from the theoretical 
as well as  from the
experimental point of view.  From the first, 
 and because of the strong coupling
 of the $\bar{K}K$ channel above $s=4m^2_K$, it is
essential to take into account the presence of the associated  cut. 
Moreover, and to reproduce correctly the low energy region data, 
the Adler zero cannot be neglected:
 we must necessarily use a complicated parametrization.

On the experimental side the situation is still a bit confused,  although it has cleared up
substantially in the last years. The experimental information we have on this S0 wave is of
three kinds:  from phase shift analysis in collisions $\pi p\to\pi\pi N,\Delta$; 
from the decay $K_{l4}$; and  from the decay $K_{2\pi}$ 
(\subsects~6.2.3, 6.2.4). 
The last gives the value of the combination 
$\delta_0^{(0)}-\delta_0^{(2)}$ at $s^{1/2}=m_K$; 
the decay $K_{l4}$ gives $\delta_0^{(0)}-\delta_1$ at low energies, $s^{1/2}\lsim380\;\mev$. 
If using the more recent $K_{2\pi}$ information
 (Aloisio et al.,~2002) together with  the $I=2$ phase obtained in the previous subsection,  
this gives the $I=0$ phase
$$\delta_0^{(0)}(m^2_K)=43.3\degrees\pm2.3\degrees.
\equn{(6.4.6)}$$
The change is substantial from the previous experimental values 
that implied (Pascual and
Yndur\'ain,~1974)
 $$\delta_0^{(0)}(m^2_K)=51\degrees\pm8\degrees.
$$

From the various phase shift analyses one concludes that there is {\sl not} a unique solution 
if fitting only $\pi\pi$ data; 
one can get an idea of the uncertainties in old analyses by having a look at 
\fig~3.3.6 in the book by Martin, Morgan and Shaw~(1975) or realizing that 
 the values of the scattering length $a_0^{(0)}$ that the various experimental fits 
(Protopopescu et al., 1973; Hyams et al., 1973; Grayer et al.~1974) gave  varied in the range
$$0.1\leq a_0^{(0)}\leq 0.9\;M_{\pi}^{-1}.
$$
Today one can improve substantially on this thanks to the appearance of $K_{l4}$ decay data 
and to  use of consistency conditions, 
but, as we will see, the situation is not as 
satisfactory as for the P wave.

We will here  consider here the following set of data  to be fitted. 
First of all we take the low energy data from $K_{l4}$ decay 
 (Rosselet et al.,~1977; 
Pislak et al.,~2001).\fnote{As a technical point, we mention that we have 
increased by 50\% the error in  the point at highest energy, $s^{1/2}=381.4\,\mev$, 
from the  $K_{e4}$ compilation of Pislak et al.~(2001), whose 
status is dubious; the experimental value represents an average
 over a long energy range that extends 
to the edge of phase space.} 
Then  
we impose the  value of $\delta_0^{(0)}(m_K^2)$ in (6.4.6).

The main virtue of these data is that they 
refer to pions on their mass shell; but, unfortunately,
 this is not sufficient to stabilize the fit at high
energy,
$s^{1/2}\gsim0.8\,\gev$.  For this we have to add further data: 
$$\eqalign{
\delta_0^{(0)}(0.870^2\,\gev^2)=&\,91\pm9\degrees;\quad
\delta_0^{(0)}(0.910^2\,\gev^2)=\,99\pm6\degrees;\cr
\delta_0^{(0)}(0.935^2\,\gev^2)=&109\pm8\degrees;\quad 
\delta_0^{(0)}(0.965^2\,\gev^2)=134\pm14\degrees.\cr
}
\equn{(6.4.7a)}$$
These points are taken from solution 1 of Protopopescu et al.~(1973) 
(both with and without modified moments), with the error 
increased by the difference between this and solution~3 data in the same reference. 
These data points have the rare virtue of agreeing, within errors, 
with the results of  other experimental analyses. 
Their inclusion is essential; if we omit them, the fits would 
produce results at total variance with experimental information above $s^{1/2}=0.5\,\gev$.
We will also include in the fit the data, at similar energies,
of Grayer et al.~(1974):
$$\eqalign{
\delta_0^{(0)}(0.912^2\,\gev^2)=&\,103\pm8\degrees;\quad
\delta_0^{(0)}(0.929^2\,\gev^2)=112.5\pm13\degrees;\cr
\delta_0^{(0)}(0.952^2\,\gev^2)=&\,126\pm16\degrees;\quad 
\delta_0^{(0)}(0.970^2\,\gev^2)=141\pm18\degrees.\cr
}
\equn{(6.4.7b)}$$
The central values are obtained averaging the three solutions given by 
Grayer et al., and the error is calculated adding quadratically
 the statistical error of the highest point, the 
statistical error of the lowest point (for each energy) and the difference 
between the central value and the farthest point. 
Moreover, we add three points between 0.8 and 0.9 \gev\ obtained 
averaging the $s$-channel solution 
of Estabrooks and Martin~(1974), which consistently 
provides the less biased data (see for example for the D0 wave, fig.~6.4.1), 
and solution 1 of Protopopescu et al.~(1973), which represent two extremes. 
The error is obtained adding the difference between these two 
in quadrature to the largest statistical error. 
In this way we obtain the numbers,
$$\eqalign{
\delta_0^{(0)}(0.810^2\,\gev^2)=&\,88\pm6\degrees;\quad
\delta_0^{(0)}(0.830^2\,\gev^2)=92\pm7\degrees;\cr
\delta_0^{(0)}(0.850^2\,\gev^2)=&\,94\pm6\degrees. 
\cr
}
\equn{(6.4.7c)}$$

We will not add points at lower energies ($0.5\,\gev\leq s^{1/2}\leq0.8\,\gev$); 
the difference among the values found for the phases in different experiments 
is such that no meaningful value could be given for the errors.

For the theoretical formulas we consider two basic possibilities. 
We  impose the Adler zero at $s=\tfrac{1}{2}M_{\pi}^2$ (no attempt is made to vary this),
 and a resonance with mass
$M_\sigma$, a free parameter.
 Then we map the $s$ plane cut along the left hand
cut ($s\leq0$) and  the $\bar{K}K$ cut, writing
$$\cot\delta_0^{(0)}(s)=\dfrac{s^{1/2}}{2k}\,
\dfrac{M_{\pi}^2}{s-\tfrac{1}{2}M_{\pi}^2}\,\dfrac{M^2_\sigma-s}{M^2_\sigma}\psi(s),
\equn{(6.4.8)}$$
and
$$\psi(s)=\big[B_0+B_1w(s)+B_2w(s)^2\big];\quad
w(s)=\dfrac{\sqrt{s}-\sqrt{s_0-s}}{\sqrt{s}+\sqrt{s_0-s}},\quad
s_0=4m^2_K$$ (we have taken $m_K=(0.496\,{\gev})$). 
The complicated structure of this wave requires two or three parameters $B_0$, $B_1$ and 
$B_2$ (besides $M_\sigma$) for an acceptable fit.

This parametrization does not represent fully the 
coupling of the $\bar{K}K$ channel and, indeed, the corresponding phase shift deviates 
somewhat from 
experiment at the upper energy range ($s^{1/2}>0.96\,\gev$; see \fig~6.4.3). 
We can, alternatively, try to use the reduction to one channel of the 
two channel formulas (5.4.1,2) and write
$$\cot\delta_0^{(0)}(s)=\dfrac{s^{1/2}}{2k}\psi_{\rm el}(s),\quad
\psi_{\rm el}(s)=\dfrac{\dfrac{\kappa_2}{2s^{1/2}}\,\phiv_{11}(s)+\det {\bf \Phi}\vphantom{\Big|}}
{\dfrac{\kappa_2}{2s^{1/2}}+\phiv_{22}(s)\vphantom{\Big|}}
\equn{(6.4.9a)}$$
where $\kappa_2=\tfrac{1}{2}\sqrt{4m_K^2-s}$. We take a linear 
approximation for the $\phiv_{ii}$, and a constant for $\phiv_{12}$, requiring a zero of 
$\det{\bf \Phi}$ at $s=M_{f_0}^2$, and we allow $M_{f_0}$ to vary between 1 and 1.4 \gev. 
So we write,
$$\eqalign{
\phiv_{11}(s)=&\,\alpha_1+\beta_1s,\quad
\phiv_{22}(s)=\alpha_2+\beta_2s,\cr
\det{\bf \Phi}=&\,(\alpha_1+\beta_1s)(\alpha_2+\beta_2s)-
(\alpha_1+\beta_1M^2_{f_0})(\alpha_2+\beta_2M^2_{f_0}).\cr
}
\equn{(6.4.9b)}$$
This represents correctly the $\bar{K}K$ cut, but does not 
allow for the Adler zero or produce a dynamical 
left hand cut.
Therefore we expect reliability of (6.4.9) near $4m^2_K$, but 
poor description near threshold, which is indeed the case. We do not try to combine the two 
parametrizations as this would lead to a hopeless tangle due to the 
large number of parameters and also to the appearance of 
left hand cut of  $\bar{K}K$, that the $\phiv_{ij}$
inherit (but which must cancel for $\psi_{\rm el}$).

\topinsert{
\setbox0=\vbox{\hsize12.truecm\hfil{\epsfxsize 11.truecm\epsfbox{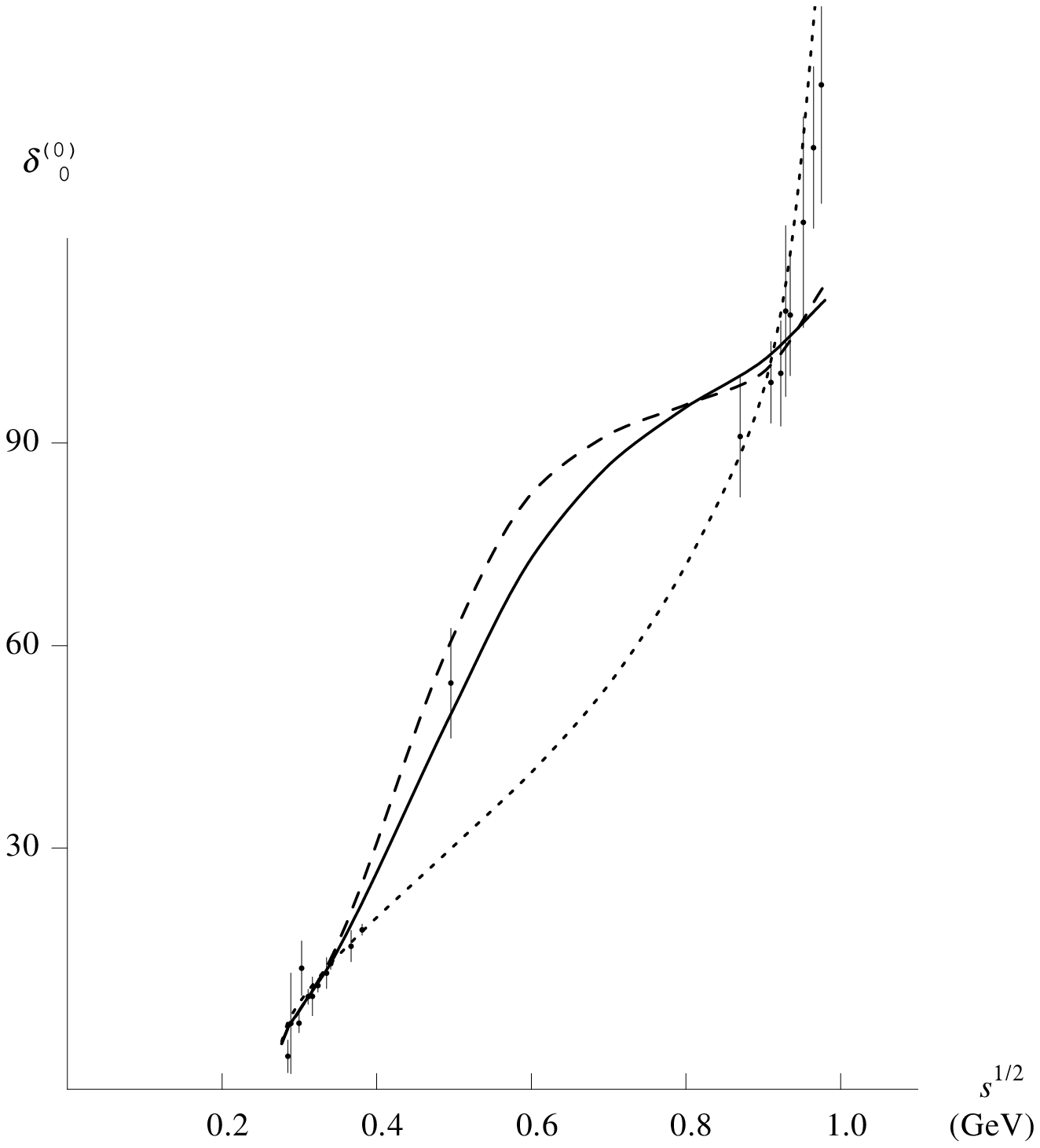}}\hfil} 
\setbox6=\vbox{\hsize 11truecm\captiontype\figurasc{Figure 6.4.2. }{
The  
$I=0$, $S$-wave phase shifts corresponding to (6.4.10) (dashed line) and 
what one would get with the old value 
$\delta_0^{(0)}(M^2_K)=49\degrees\pm5\degrees$, continuous line. 
Also shown are the points, at low energies, from the $K_{l4}$ experiments, 
the point from $K_{2\pi}$ decay (the more recent value),
 and the high energy data of Protopopescu et al. and
Grayer et al. included in the fits.  }\hb} 
\centerline{\tightboxit{\box0}}
\bigskip
\centerline{\box6}
\medskip
}\endinsert

Let us now turn to the results of the fits. 
First of all, we note that the inclusion of the value of the 
phase at $s=m^2_K$ is essential; for example, if we 
had not included it we would have found the following minimum:\fnote{That using only 
phase shifts data there are two alternate 
possibilities for the intermediate energy S0 wave  
 was recognized already  by, e.g.,  Estabrooks and Martin~(1974); 
see also the textbook of Martin, Morgan and Shaw~(1975) for a discussion.}
$$\eqalign{ 
{B}_0=&\,46.87\pm0.68,\quad {B}_1=92.72\pm1.47,\quad{B}_2=60.59\pm3.24,\cr
M_\sigma=&\,874\pm30\,\mev;\cr
a_0^{(0)}=&\,(0.274\pm0.024)\,M_{\pi}^{-1};\quad \delta_0^{(0)}(m_K^2)=30\degrees.\cr    }
\equn{(6.4.10)}$$
The $\chidof=20.0/(20-4)$ is reasonable; the value of
$\delta_0^{(0)}(m_K^2)$ is not.

If we impose $\delta_0^{(0)}(m_K^2)$ as given in (6.4.6) 
we find quite  different results. With only two $B_i$s, we have 
what we will call the {\sl B2 Solution},
$$\eqalign{
\cot\delta_0^{(0)}(s)=&\,\dfrac{s^{1/2}}{2k}\,\dfrac{M_{\pi}^2}{s-\tfrac{1}{2}M_{\pi}^2}\,
\dfrac{M^2_\sigma-s}{M^2_\sigma}\,
\left\{B_0+B_1\dfrac{\sqrt{s}-\sqrt{s_0-s}}{\sqrt{s}+\sqrt{s_0-s}}\right\};\cr
{B}_0=&\,21.04,\quad {B}_1=6.62,\quad
M_\sigma=782\pm24\,\mev;\quad \dfrac{\chi^2}{\rm d.o.f.}=\dfrac{15.7}{19-3}.\cr
\quad 
a_0^{(0)}=&\,(0.230\pm0.010)\times M_{\pi}^{-1};\quad\delta_0^{(0)}(M_K)=
41.0\degrees\pm2.1\degrees;
\cr    }
\equn{(6.4.11a)}$$
this fit we take to be valid for $s^{1/2}\leq0.96\,\gev$. 
The errors of the $B_i$ are strongly correlated; uncorrelated errors are obtained if 
replacing the $B_i$ by the 
parameters $x,\,y$ with
$$B_0=y-x;\quad B_1=6.62-2.59 x;\quad y=21.04\pm0.75,\quad x=0\pm 2.4.
\equn{(6.4.11b)}$$

\midinsert{
\setbox0=\vbox{\hfil{\epsfxsize 11.9truecm\epsfbox{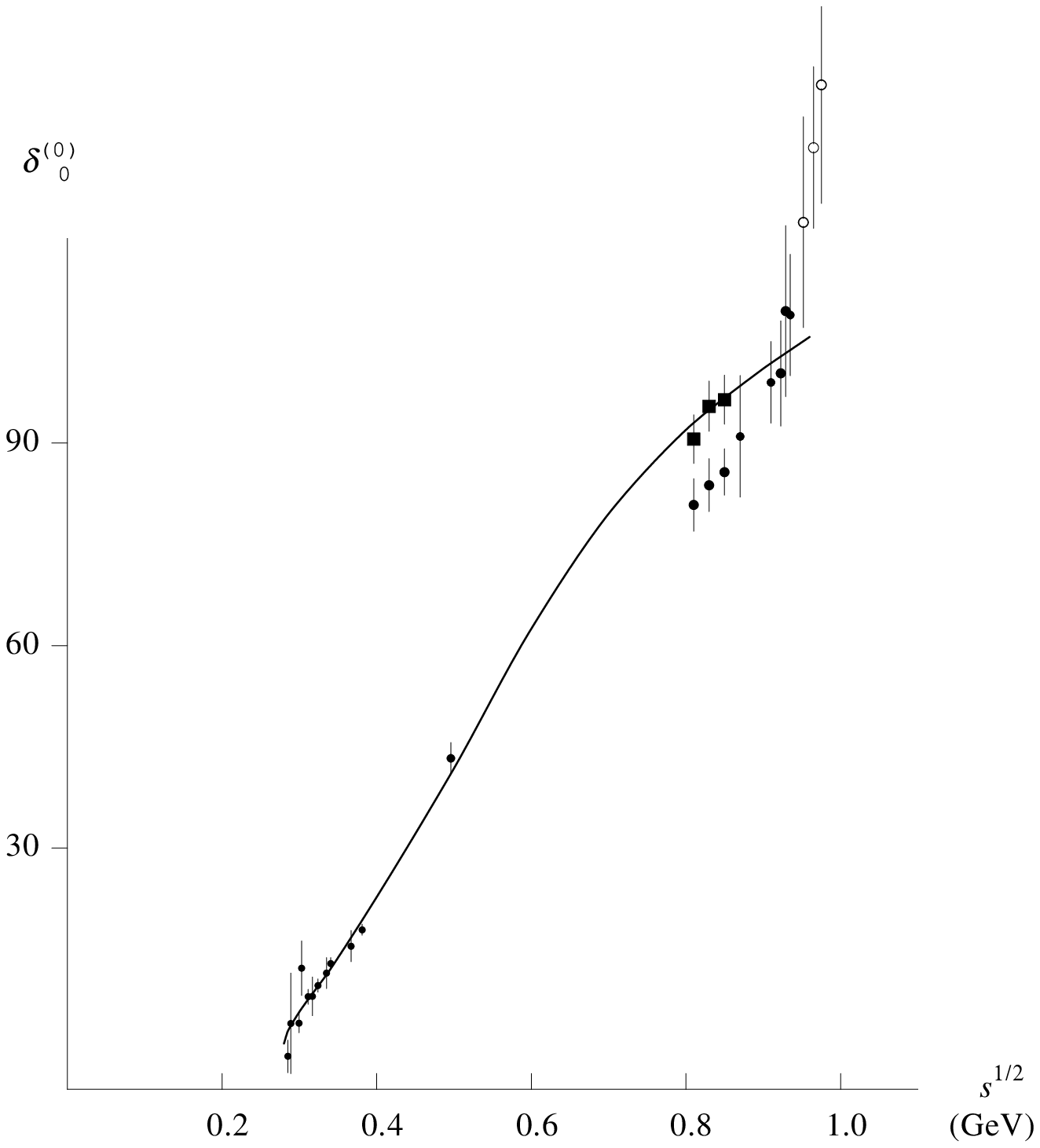}}\hfil} 
\setbox6=\vbox{\hsize 10truecm\captiontype\figurasc{Figure 6.4.3. }{
The  
$I=0$, $S$-wave phase shifts corresponding to Eq.~(6.4.11) (continuous line). 
Also shown are the points from  $K_{l4}$ and $K_{2\pi}$ decays,
 and the high energy data of Protopopescu et al. (black dots),
Grayer et al. (open circles), and the $s$-channel solution 
of  Estabrooks and Martin (black squares) 
 included in
the fits.}\hb} 
\centerline{\tightboxit{\box0}}
\bigskip
\centerline{\box6}
\medskip
}\endinsert

If we 
 allow for an extra parameter a new minimum appears:
$$\eqalign{
\cot\delta_0^{(0)}(s)=&\,\dfrac{s^{1/2}}{2k}\,\dfrac{M_{\pi}^2}{s-\tfrac{1}{2}M_{\pi}^2}\,
\dfrac{M^2_\sigma-s}{M^2_\sigma}\cr
\times&
\left\{B_0+B_1\dfrac{\sqrt{s}-\sqrt{s_0-s}}{\sqrt{s}+\sqrt{s_0-s}}+
B_2\left[\dfrac{\sqrt{s}-\sqrt{s_0-s}}{\sqrt{s}+\sqrt{s_0-s}}\right]^2\right\};\cr
s_0^{1/2}=2M_K;&\quad\chi^2/{\rm d.o.f.}=11.1/(19-4).\cr
\quad M_\sigma=806\pm21,&\,\; B_0=21.91\pm0.62,\; B_1=20.29\pm1.55, \;
B_2=22.53\pm3.48;\cr a_0^{(0)}=&\,(0.226\pm0.015)\;M_{\pi}^{-1}.\cr
\cr
}
\equn{(6.4.12)}$$
This, that we may denote by {\sl B3 Solution},
 is something between (6.4.11) and the solution of
Colangelo, Gasser and Leutwyler~(2001), which indeed is comprised inside the errors of (6.4.12).

We next say a few words on results using (6.4.9). 
The quality of the fit is substantially lower than all the fits given in (6.4.11,12). 
Although we expect  (6.4.9) to reproduce better the high energy range, 
the lack of correct left-hand cut structure 
clearly disrupts the lower range. 
Thus, for the fit not imposing $\delta_0^{(0)}(m_K^2)$ we find a \chidof\ of 45/(15-4), 
certainly excessive; so we stick to (6.4.11).

It is difficult to give reasons
 to prefer any of the two sets of parameters (6.4.11,12).  
Both give essentially identical 
results for the Olsson and Froissart--Gribov sum rules, 
but  (6.4.11) looks more appealing in that 
the convergence properties of the conformal expansion 
are clearly superior to those of (6.4.12). 
For these reasons  
we will only give results using (6.4.11).

The value of the scattering length that (6.4.11) gives compares well with the recent 
value of Descotes et al.~(2002) who impose also the Roy 
equations, and less so with the solution of  Colangelo, Gasser and Leutwyler,~2001:
$$\eqalign{
a_0^{(0)}=&\,(0.230\pm0.010)\times M_{\pi}^{-1},\quad\hbox{[Eq.~(6.4.11)]}\cr
a_0^{(0)}=&\,(0.228\pm0.012)\times M_{\pi}^{-1},\quad\hbox{[Descotes et al.,~2002]}\cr
a_0^{(0)}=&\,(0.220\pm0.005)\times M_{\pi}^{-1},\quad\hbox{[Colangelo, Gasser and
Leutwyler,~2001]}\cr
a_0^{(0)}=&\,(0.224\pm0.013)\times M_{\pi}^{-1},\quad\hbox{[Kami\'nski, Le\'sniak and
Loiseau,~2003].}\cr }
\equn{(6.4.13)}$$

\booksubsection{6.4.3. The $I=0$ S wave between $960\,\mev$ and $1420\,\mev$}

\noindent
As we have already commented, the description of 
pion-pion scattering above the $\bar{K}K$ threshold requires a full two-channel 
formalism. To determine the three independent components of the 
effective range matrix $\bf \Phi$, $\phiv_{11}$, 
$\phiv_{22}$ and $\phiv_{12}$, one requires 
measurement of  three cross sections. 
Failing this, one gets an indeterminate set, which is reflected 
very clearly in the wide variations of the effective range matrix parameters in the 
energy-dependent fits of Protopopescu et al.~(1973) and 
Hyams et al.~(1973), Grayer et al.~(1974).
 
The raw data themselves are also incompatible; 
Protopopescu et al. find a phase shift that flattens above $s^{1/2}\simeq 1.04\,\gev$, 
while that of Hyams et al. or Grayer et al. continues to grow. 
This incompatibility is less marked if we choose the 
solution with modified higher moments by 
Protopopescu et al. (Table~XIII there). 
The inelasticities are more compatible among the various determinations, although the errors
 of Protopopescu et al. appear to be underestimated. 

\topinsert{
\setbox0=\vbox{\hsize10.truecm{\epsfxsize 8.truecm\epsfbox{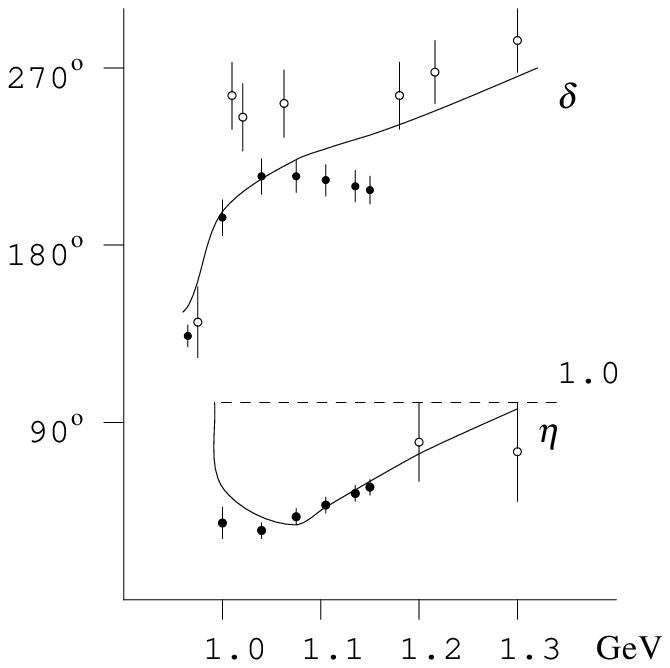}}} 
\setbox6=\vbox{\hsize 10truecm\captiontype\figurasc{Figure 6.4.4. }{
Fits to the  
$I=0$, $S$-wave phase shift and inelasticity from 960 to 1300 \mev. 
Data  from solution~1 of Protopopescu et al.~(1973) (black dots) and 
  Grayer et al.~(1974) (open circles).
}\hb} 
\centerline{\tightboxit{\box0}}
\bigskip
\centerline{\box6}
\medskip
}\endinsert

In spite of this it is possible to give a reasonable semi-phenomenological 
fit to $\delta_0^{(0)}$ and $\eta_0^{(0)}$, defined 
as in (2.1.4). 
We write
$$\cot\delta_0^{(0)}(s)=c_0\,\dfrac{(s-M^2_\sigma)(M^2_{f}-s)}{M^2_{f} s^{1/2}}\,
\dfrac{|k_2|}{k^2_2},\quad k_2=\dfrac{\sqrt{s-4m^2_K}}{2}
\equn{(6.4.14a)}$$
and
$$\eta_0^{(0)}=1-\left(c_1\dfrac{k_2}{s^{1/2}}+c_2\dfrac{k_2^2}{s}\right)
\,\dfrac{M'^2-s}{s}.
\equn{(6.4.14b)}$$
In the first, $c_0$ and $M_\sigma$ are free parameters and we fix $M_{f}=1320\,\mev$. 
In (6.4.14b),  the free parameters are  $c_1,\,c_2$ and we adjust $M'$ 
to get the inelasticity agreeing with the central value given by 
Hyams et al.~(1973) on the $f_0(1370)$. 
We choose to fit the data points of solution~1 of Protopopescu et al. above 
 $\bar{K}K$ threshold, plus two values  at 1.2 and 1.3 \gev\ of Hyams et al. 
for 
the  inelasticity. 
For the phase shift, more conflictive as there is clear  
incompatibility between the two sets of experiments, 
we include the seven values of Protopopescu et al. for $s^{1/2}\geq965\,\mev$, and another
seven  points of Grayer et al.~(1974), in the same range. 
The errors of these data have been evaluated as for (6.4.7).
We find,
$$\eqalign{
c_0=&\,1.36\pm0.05,\quad M_\sigma=802\pm11\,\mev;\quad \chi^2/{\rm dof}=36.2/(14-2)
\cr
c_1=&\,6.7\pm0.17,\quad c_2=-17.6\pm0.8;\quad\chi^2/{\rm dof}=7.7/(8-2).\cr
}
\equn{(6.4.14c)}$$
The errors for $c_0$, $M_\sigma$ correspond to {\sl three} standard deviations, 
since  we have a $\chidof\simeq3$.
The fit (6.4.14c) presents the nice feature that the value of $M_\sigma$ coincides, 
{\sl grosso modo}, with what we found below $\bar{K}K$ threshold. 
The qualitative features of the fits may be seen in \fig~6.4.4, where the incompatibility of the  
data of both sets of experiments is apparent.

\booksection{6.5. The D,  F and G waves}
\vskip-0.5truecm
\booksubsection{6.5.1. Parametrization of the $I=2$ D wave}

\noindent
For isospin equal 2, there are no resonances in the D wave (or, indeed, in any other wave), 
at least at low energies. 
This is an experimental fact that can be understood theoretically by recalling that 
one cannot have $I=2$ with a quark-antiquark state.
 
We would only expect important inelasticity when the channels 
 $\pi\pi\to\rho\rho$ $\pi\pi\to\rho\pi\pi$ open up, 
so we will take
$$w(s)=
\dfrac{\sqrt{s}-\sqrt{s_0-s}}{\sqrt{s}+\sqrt{s_0-s}},\quad
s_0=1.45^2\,\gev^2\sim4M^2_\rho.
\equn{(6.5.1)}$$  
 But life is complicated: 
a pole term is necessary to get an acceptable fit down to low energy
since we expect $\delta_2^{(2)}$ to change sign near threshold.
The experimental measurements (Losty et al.,~1974; Hoogland et al.,~1977) give 
negative and small values for the phase above some $500\,\mev$, while we will 
see that chiral perturbation calculations (\sect~9.4) 
and the Froissart--Gribov representation\fnote{An interesting feature of the Froissart--Gribov 
calculation is that the structure of $\delta_2^{(2)}$, in 
particular the zero near threshold, was 
in fact {\sl predicted} from it (Palou and Yndur\'ain,~1974).}  (\sect~7.5) 
indicate a positive scattering length, $a_2^{(2)}\simeq(2.2\pm0.2)\times10^{-4}\,M_{\pi}^{-5}$.

If we want a parametrization that 
applies down to threshold, we must incorporate this  
zero of the phase shift. So 
we write
$$\cot\delta_2^{(2)}(s)=
\dfrac{s^{1/2}}{2k^5}\,\left[B_0+B_1 w(s)\right]\,\dfrac{{M_\pi}^4 s}{4({M_\pi}^2+\deltav^2)-s}
\equn{(6.5.2a)}$$
with $\deltav$ a free parameter and
$$w(s)=\dfrac{\sqrt{s}-\sqrt{s_0-s}}{\sqrt{s}+\sqrt{s_0-s}},\quad
 s_0=1450\,\mev.$$  
 Moreover, we impose  the 
value for the scattering length 
that follows from the Froissart--Gribov representation, 
$a_2^{(2)}=(2.22\pm0.33)\times10^{-4}$, in units of $M_\pi$ (see below, \sect~7.5). 

We first perform a fit up to $s^{1/2}$; we do not include in it the 
data of Cohen et al.~(1973). 
Since these are obtained from scattering off bound neutrons (in deuterium) 
they are more liable to systematic errors at low energy and, in fact, 
if we included them the resulting effective range parameter $b_2^{(2)}$ 
would be far from its expected value (see below).
We only include two parameters $B_0$, $B_1$ ({\sl Solution B2}\/);  
we get a mediocre fit, $\chidof=53/(16-3)$, and the values of the parameters
 are\fnote{This fit, due to
Pel\'aez and Yndur\'ain~(2003) corrects an error of the previous version of the 
present paper.}
$$B_0=(2.30\pm0.17)\times10^3,\quad B_1=-267\pm750,\quad \deltav=103\pm11\,\mev;
\quad s^{1/2}\lsim1.2\;\gev.
\equn{(6.5.2b)}$$
Doubtlessly  the incompatibilities between the 
 experimental data (which is obvious from a look 
at \fig~6.5.1), 
probably related to those for the  S2 wave, 
preclude a better fit.

  The fit returns a good value for the scattering length, and also for the effective range parameter,
$b_2^{(2)}$: 
$$a_2^{(2)}=(2.20\pm0.16)\times10^{-4}\,{M_\pi}^{-5};\quad
b_2^{(2)}=(-5.75\pm1.26)\times10^{-4}\,{M_\pi}^{-7},
\equn{(6.5.2c)}$$
to be compared with what we will get from the Froissart--Gribov representation
(\sect~7.5),
$$a_2^{(2)}=(2.22\pm0.33)\times10^{-4}\,{M_\pi}^{-5};\quad
b_2^{(2)}=(-3.34\pm0.24)\times10^{-4}\,{M_\pi}^{-7}.
\equn{(6.5.2c)}$$ 
For once, the value of $a_2^{(2)}$ is more accurate than the value following from the 
Froissart--Gribov calculation;
 the value of $b_2^{(2)}$differs by less than $2\,\sigma$ from the expected one.

To get the parameters for the region above 1.0 \gev, we 
take simply a quadratic fit. 
We find, 
$$\eqalign{
\delta_2^{(2)}(s)=&\,(-0.051\pm0.003)+
a\left(\dfrac{s}{1\,{\gev}^2}-1\right)+b\left(\dfrac{s}{1\,{\gev}^2}-1\right)^2;\cr
a=&\,-0.081\pm0.033,\quad b=0.042\pm0.005;
\quad s\geq1.0\;\gev.\cr
}
\equn{(6.5.2d)}$$
We can add inelasticity to the D2 wave by assuming that it is 
something between zero and  what one has for the S2 wave  
(\subsect~6.5.1). 
So we would write,
$$\eqalign{\eta_2^{(2)}(s)=&\,1-c(1-M^2_{\rm eff}/s)^{3/2},\quad M_{\rm eff}=0.96\,\gev,\quad
c=0.12\pm0.12;
\cr
s^{1/2}\geq&\,0.96\,\gev.\cr
}
\equn{(6.5.2e)}$$
We should add that it is possible to get a reasonable fit to data at all energies, 
with a formula like (6.5.2a),
but we require 
 {\sl four} parameters $B_i$ ({\sl Solution B4}\/). 
Including also the data of Cohen et al.~(1973) one gets,
$$\eqalign{
B_0=&\,(1.94\pm0.14)\times10^3,\quad B_1=(10.15\pm1.3)\times10^3,\quad
B_2=(18.68\pm2.4)\times10^3,\cr
B_3=&\,(-31.04\pm5.5)\times10^3;\qquad\deltav=218\pm22\,\mev.\cr
}
\equn{(6.5.3a)}$$
The errors here correspond to $3\,\sigma$. 
One has $\chidof=57/(25-5)$ and the fit returns the values 
of the low energy parameters
$$a_2^{(2)}=(2.04\pm0.5)\times10^{-4}\,M_{\pi}^{-5},\quad
b_2^{(2)}=(1.6\pm0.3)\times10^{-4}\,M_{\pi}^{-7}.
\equn{(6.5.3b)}$$
The large values of the parameters $B-i$, and the incompatibility of the three data sets, 
makes one suspect that the corresponding minimum is  spureous.

\topinsert{
\setbox0=\vbox{{\hfil\epsfxsize 11.5truecm\epsfbox{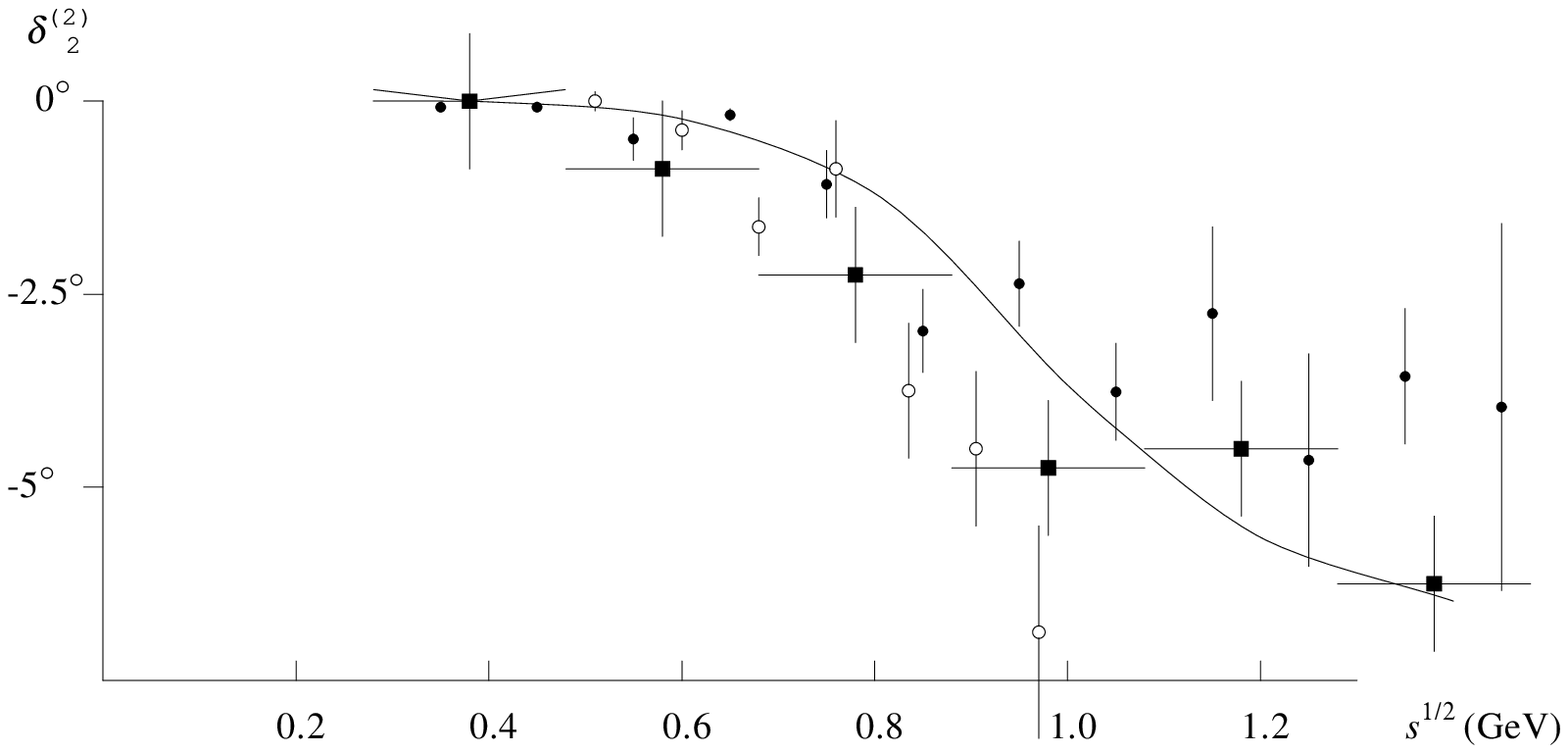}\hfil}} 
\setbox6=\vbox{\hsize 10truecm\captiontype\figurasc{Figure 6.5.1. }{
Fits to the  
$I=2$, $D$-wave phase shift. 
Also shown are the data points of Losty et al.,~1974 (open circles),
 from solution~A of Hoogland et al.,~1977 (black dots) and from Cohen et al.~(1973) (crosses).
}\hb} 
\centerline{\tightboxit{\box0}}
\bigskip
\centerline{\box6}
\medskip
}\endinsert

\booksubsection{6.5.2. Parametrization of the $I=0$ D wave}

\noindent
The D wave with isospin 0 in $\pi\pi$ scattering presents two resonances 
below $1.7\,\gev$: the $f_2(1270)$ and the $f_2(1525)$, 
that we will denote respectively by $f_2$, $f'_2$. 
Experimentally, 
$\gammav_{f_2}=185\pm4\,\gev$ and $\gammav_{f'_2}=76\pm10\,\gev.$ 
The first, $f_2$,  
couples mostly to $\pi\pi$, with small  couplings to 
$KK$ ($4.6\pm0.5\,\%$), $4\pi$ ($10\pm3\,\%$) and 
$\eta\eta$. The second 
couples mostly to $2K$, with a small coupling to $\eta\eta$ and $2\pi$, 
respectively $10\pm3\,\%$ and $0.8\pm0.2\,\%$. 
This means that the channels $\pi\pi$ and $KK$ are 
essentially decoupled: they only connect indirectly, 
so it is not very profitable to set up a multiple channel calculation.
To a 15\% accuracy we may neglect inelasticity up to $s_0=1.42^2\gev^2$. 
The formulas are like those for the P wave; 
we will discuss them presently. 

There are not many experimental data on the D wave which, at accessible energies, 
 is  small. 
So, the compilation of  $\delta_2^{(0)}$ 
phase shifts of Protopopescu et al. (1973) covers 
only the range $810\leq s^{1/2}\leq 1150\,\mev$. 
In view of this, it is impossible to get accurately the D wave scattering lengths, 
or indeed any other low energy parameter, from this information.
We give here a parametrization whose use lies in that 
it represents with reasonable accuracy the data, something that will be useful later on. 
We write
$$\cot\delta_2^{(0)}(s)=\dfrac{s^{1/2}}{2k^5}\,(M^2_{f_2}-s)\,M^2_\pi\,{\psi}(s),\quad
{\psi}(s)=B_0+B_1w(s)+\cdots\,,
\equn{(6.5.4a)}$$
and 
$$w=\dfrac{\sqrt{s}-\sqrt{s_0-s}}{\sqrt{s}+\sqrt{s_0-s}},\quad s_0^{1/2}=1.43\;\gev.$$

\topinsert{
\setbox0=\vbox{{\hfil\epsfxsize 11.truecm\epsfbox{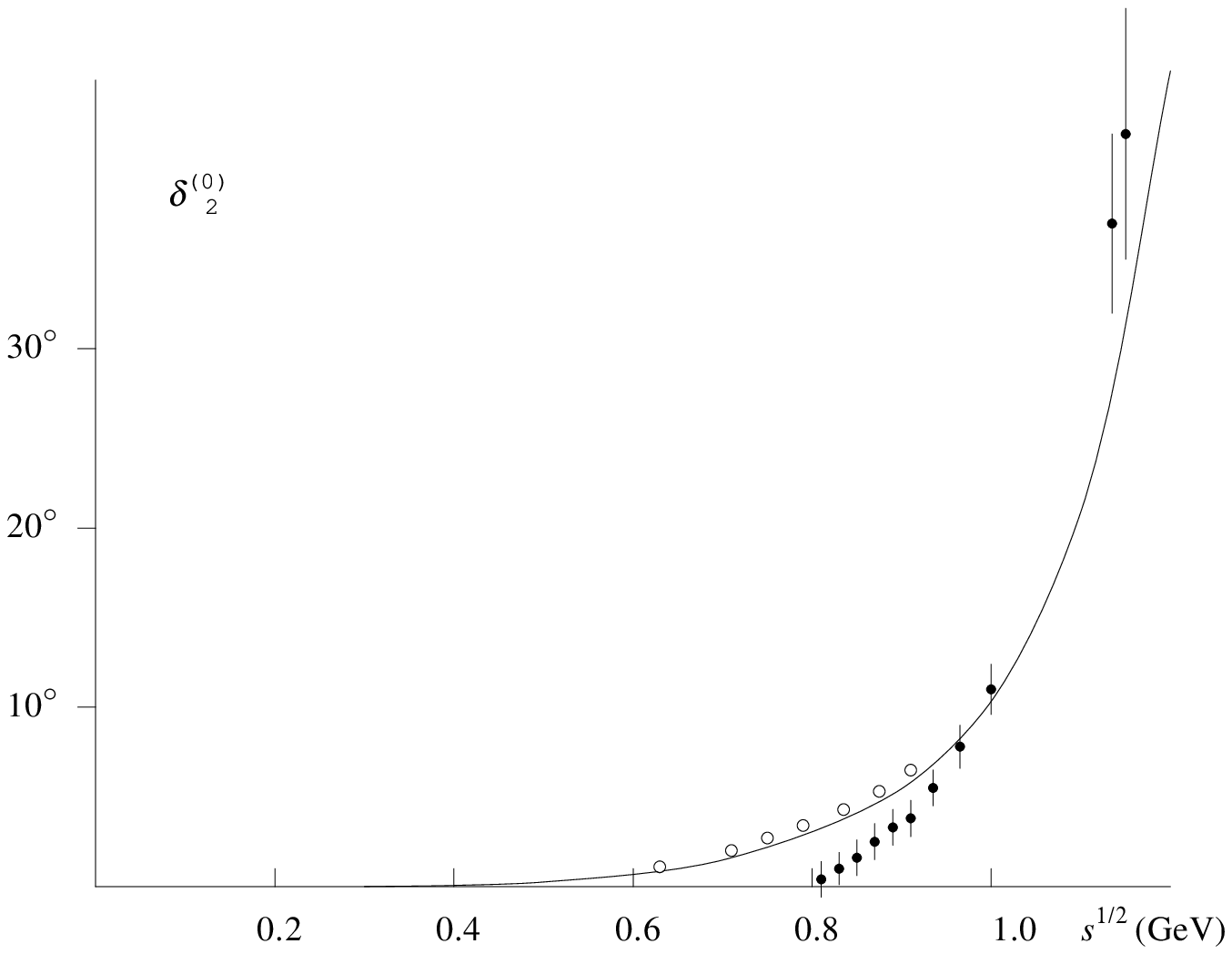}\hfil}} 
\setbox6=\vbox{\hsize 10truecm\captiontype\figurasc{Figure 6.5.2. }{
Fits to the  
$I=0$, $D$-wave phase shift. 
Also shown are the data points from solution~1 of Protopopescu et al.~1973 (black dots) and 
some data  of  Estabrooks and Martin~1974 (open circles).
}\hb} 
\centerline{\tightboxit{\box0}}
\bigskip
\centerline{\box6}
\medskip
}\endinsert

We take the data of Protopopescu et al. (1973) and consider the so-called  ``solution 1",
with the two possibilities given in Table~VI and 
Table~XIII (with modified higher moments). 
These data cover the range 
mentioned before, $s^{1/2}=0.810\,\gev$ to $1.150\,\gev$.
The problem with these data points is that they 
are contaminated, for $s\gsim1.1\,\gev^2$, by the 
bias of the S wave with $I=0$ in the same region, whose values there are
quite incompatible with
those of other experiments 
(see \subsect~6.5.3). 
For this reason we perform two fits: either including or excluding the data points for
$s^{1/2}\geq1.075\,\gev$. 
In both cases we present results only for the version with 
modified higher moments (Table~XIII in Protopopescu et al., ~1973) 
as they are the ones that show better compatibility with 
other experiments. 
We also impose the fit to the width of the $f_2$ resonance, 
with the condition $\gammav_{f_2}=185\pm10\,\mev$. 
We find,
$$\eqalign{
\dfrac{\chi^2}{\rm d.o.f.}=&\,\dfrac{46.9}{14-2},\quad B_0=20.16,\quad B_1=19.48,\quad
\hbox{[All points]};\cr
\gammav_{f_2}&\,=213\,\mev,\quad a_2^{(0)}=17\times10^{-4}\,M_{\pi}^{-5}\cr
}
$$
and
$$\eqalign{
\dfrac{\chi^2}{\rm d.o.f.}=&\,\dfrac{20.5}{10-2},\quad B_0=23.95,\quad B_1=18.91,\quad
\hbox{[Points for $s^{1/2}<1.075\,\gev$]};\cr
\gammav_{f_2}&\,=187\,\mev,\quad a_2^{(0)}=11\times10^{-4}\,M_{\pi}^{-5}.\cr
}
$$
The drastic decrease of the \chidof\ when eliminating the higher
 energy points signals clearly their biased
character.\fnote{We remark again that the \chidof\ is less poor than it looks at first sight, 
as it only takes into account statistical errors, while systematic
 ones are certainly is large as  these.} 
However, the parameters of the fits are reasonably stable, 
no doubt because we have imposed the 
correct width of the resonance $f_2$. 
We can therefore take as our best result an average of 
the two determinations, with half their 
difference as an estimated error:
$$B_0=22.1\pm1.9,\quad B_1=19.2\pm0.3
\equn{(6.5.4b)}$$
and this corresponds to
$$\gammav_{f_2}=200\pm13\,\mev,\quad a_2^{(0)}=(14\pm3)\times10^{-4}\,M_{\pi}^{-5},
$$
 reasonably close to their experimental values, the 
second as deduced from the Froissart--Gribov 
representation (cf.~\sect~7.6):
$$a_2^{(0)}=(18.1\pm0.4)\times 10^{-4}\,M_{\pi}^{-5},\quad 
b_2^{(0)}=(-3.60\pm0.25)\times 10^{-4}\,M_{\pi}^{-7}$$
(we also give the effective range parameter from the same source).

An alternate possibility is to include the scattering length, as deduced from 
the Froissart--Gribov representation, in the fit, which completely  
 stabilizes the results.  Moreover, we take into account the inelasticity 
iteratively. We write
$$\cot\delta_2^{(0)}(s)=\dfrac{s^{1/2}}{2k^5}\,(M^2_{f_2}-s)\,{{M_\pi}^2}\,{\psi}(s),\quad
{\psi}(s)=B_0+B_1w(s)+\cdots\,
\equn{(6.5.5a)}$$
and
$$w(s)=\dfrac{\sqrt{s}-\sqrt{s_0-s}}{\sqrt{s}+\sqrt{s_0-s}},\quad
 s_0=1430\,\mev;\quad
M_{f_2}=1275.4\,\mev.$$ 
We find, fitting also the points of Estabrooks and Martin~(1974),
$$\eqalign{
\dfrac{\chi^2}{\rm d.o.f.}=&\,74/(21-2),\quad B_0=22.4\pm0.1,\quad B_1=23.3\pm3.0;\cr
}
$$
The very poor \chidof\ is obviously due to the strong 
 bias of the data of Protopopescu et al.,~1973, clearly seen in \fig~6.5.2. 

Above values of the $B_i$ 
would give $\gammav_{f_2}=196\pm6\,\mev$. 
We will then take this solution to be valid up to $\bar{K}K$ threshold; on it, 
we  join 
the solution to a new one, for which we impose the $f_2$ width; we get 
$$B_0=22.5\pm0.1,\quad B_1=28.5\pm3.6.$$
Therefore, we have
$$B_0=\cases{22.4\pm0.1,\quad s< 4M_K^2,\cr22.5\pm0.1,\quad s>4M_K^2\cr};\qquad
B_1=\cases{23.3\pm3.0,\quad s< 4M_K^2,\cr28.5\pm3.6,\quad s>4M_K^2.\cr}
\equn{(6.5.5b)}$$ 
We then take into account the inelasticity by writing
$$\eta_2^{(0)}(s)=\cases{1,\qquad  s< 4M_K^2,\;
\cr
1-2\times\epsilon_f\,\dfrac{k_2(s)}{k_2(M^2_{f_2})},\quad
 \epsilon_f=0.131\pm0.015;\quad s> 4M_K^2.\cr} 
\equn{(6.5.5c)}$$
$ k_2=\sqrt{s/4-M^2_K}.$
We have fixed the coefficient $\epsilon_f$ fitting the inelasticities
 of Protopopescu et al.,~1973,  
and the experimental inelasticity of the $f_2$; the overall $\chidof$ of this 
fit is $\sim1.8$.
The fit returns the values
$$\eqalign{
a_2^{(0)}=&\,(18.4\pm7.6)\times10^{-4}\,\times M_{\pi}^{-5},\quad
b_2^{(0)}=(-7.9^{+4.1}_{-11.0})\times10^{-4}\,\times M_{\pi}^{-7};\cr
\gammav_{f_2}=&\,185\pm5\,\mev.\cr
}
$$
One could try to improve the fit by adding an extra term, $B_2w^2$, 
and requiring also the value of $b_2^{(0)}$ to agree with the 
chiral perturbation theory  value (or with that obtained from the 
Froissart--Gribov representation, see below). 
In fact, we prefer to keep the larger errors given above; the values
 of the inelasticities and low energy parameters are compatible at the $1.5\;\sigma$ level with 
experimental information, and we feel that the improvement obtained by diminishing the
errors would be 
made at the cost of reliability.

\booksubsection{6.5.3. The F wave}

\noindent
The experimental situation for the F wave is somewhat confused. 
According to Protopopescu et al.~(1973) it starts negative
 (but compatible with zero at the $2\,\sigma$ level) and
 becomes positive around $s^{1/2}=1\,\gev$. 
Hyams et al.~(1973) and Grayer et al.~(1974) report a positive $\delta_3(s)$ 
when it differs from zero (above $s^{1/2}=1\,\gev$). 
In both cases the inelasticity is negligible up to, at least, $s^{1/2}=1.5\,\gev$.

The corresponding scattering length may be calculated with the help of the Froissart--Gribov 
representation and one finds (\sect~7.6)
$$a_3=(6.00\pm0.07)\times10^{-5}\,M_{\pi}^{-7}.
$$
It could in principle be
 possible that $\delta_3(s)$ changes sign {\sl twice}, once near
threshold  and once near $s^{1/2}=1\,\gev$.    
However,  we disregard 
this possibility and write, simply,
$$\cot\delta_3(s)=\dfrac{s^{1/2}}{2k^7}\,\Big\{B_0+B_1w(s)\big\}\,M_{\pi}^6,\quad
w(s)=
\dfrac{\sqrt{s}-\sqrt{s_0-s}}{\sqrt{s}+\sqrt{s_0-s}},
\equn{(6.5.6a)}$$
with $s_0^{1/2}=1.5\,\gev$, and {\sl impose} (6.5.5). 

It is to be understood
 that this parametrization provides only an empirical representation of the available data, 
and that it may not be reliable except at very low energies, 
where it is dominated by the scattering length, and for  
$s^{1/2}\sim 1\,\gev$. 
We find 
$$\dfrac{\chi^2}{\rm d.o.f.}=\dfrac{5.7}{7-2},\quad B_0=(1.07\pm0.03)\times 10^5,\quad
B_1=(1.35\pm0.03)\times 10^5.
\equn{(6.5.6b)}$$
For $s^{1/2}\geq1.1\,\gev$, however, the effect of the $\rho_3(1690)$ resonance 
should be included; see Pel\'aez and Yndur\'ain~(2003) 
and Appendix~A.8 here for an explicit expression valid for $\imag f_3$.

\booksubsection{6.5.4. The G waves}

\noindent
The experimental information on the G waves is very scarce. 
For the wave G2, we have six nonzero values for $\delta_4^{(2)}$, 
two from Cohen et al.~(1973) and four from Losty et al.~(1974); 
they are somewhat incompatible. 
We then fit the data separately, with a scattering length formula;
we write
$$\cot\delta_4^{(2)}(s)=\dfrac{s^{1/2}M^8_\pi}{2k^9}\,B.$$
If we fit he data of Cohen et al.~(1973) we find $B=(-9.5\pm2.7)\times10^{6}$, 
while from Losty et al.~(1974) we get  $B=(-0.6\pm0.1)\times10^{6}$. 
Fitting both sets together we find $B=-7.8\times10^{6}$, and a very poor  $\chidof=30/(6-1)$. 
Enlarging the resulting  error so
to cover $6\,\sigma$ we  obtain our best result, 
$$\cot\delta_4^{(2)}(s)=\dfrac{s^{1/2}M^8_\pi}{2k^9}\,B,\quad B=(-7.8\pm3.3)\times 10^6.
\equn{(6.5.7)}$$

This formula can only be considered as 
an empirical fit, valid for a limited range, $0.9\leq s^{1/2}\leq1.5\,\gev$. 
In fact, from the Froissart--Gribov representation it follows that the G2 scattering length is
{\sl positive},
$$a_4^{(2)}=(4.5\pm0.2)\times10^{-6}\,M_{\pi}^{-9}$$
while (6.5.7) would give a negative value.

For the G0 wave, the situation is worse: there are no experimental data on the phase shift. 
All we know is the existence of a very inelastic resonance with mass  around 2 \gev. 
An effective value for the imaginary part of the corresponding partial wave may be found 
in Appendix~A.9. 
At low energy we can use a scattering length approximation with 
$$a_4^{(0)}=(8.0\pm0.2)\times10^{-6}\,M_{\pi}^{-9}.$$

\booksection{6.6. On experimental phase shifts in the range $1.4\,\gev\simeq
s^{1/2}\simeq 2\,\gev$}

\noindent 
As we mentioned in \sect~6.1, we expect that, as soon as the 
center of mass kinetic energy  in a reaction, $E_{\rm kin}$, 
increases beyond 1 \gev, inelastic processes  become more and more
 important with increasing
energy,  so much so that, 
for $E_{\rm kin}\gsim1.2\,\gev$, they should dominate elastic ones. 
This is easily understandable in the QCD, ladder version of the Regge picture, as discussed in
\sect~2.4; and indeed, it is verified experimentally in
 the hadronic processes  
$\pi N$, $KN$ and $NN,\,\bar{N}N$ where, for   $E_{\rm kin}>1.2\,\gev$, the 
elastic
 cross section is smaller than the inelastic
one and, for   $E_{\rm kin}>1.5\,\gev$, 
the elastic  cross sections are a third or less 
than the total cross sections.  There is no reason to imagine
that
$\pi\pi$ scattering would follow a different pattern. 
In fact,  
the experimental results on $\pi\pi$ cross sections 
at high energies (like e.g., those of Robertson, Walker and
Davis,~1973)  have checked unambiguously all these features. 

 In this case in which inelastic cross sections are large, and again as mentioned in
\sect~6.1,
 it can be proved theoretically that 
there is not a unique solution to the phase shift analysis: 
some sets of $\eta$s and $\delta$s may fit the data; but so would others. 

In spite of this, the Cern-Munich experiments\fnote{Hyams et al.,~(1973);
 Grayer et al.~(1974).} have produced a
set of phase shifts and  inelasticities which go up to $s^{1/2}\simeq2\,\gev$, 
which have been used in several theoretical analyses. 
Unfortunately, these phase shifts are likely to diverge 
more and more from reality 
as   
$s^{1/2}=E_{\rm kin}+2M_\pi$ beomes larger and larger than 
(say) $1.5\,\gev$. 
This is suggested, besides the theoretical reasons
  mentioned in \sect~6.1, because the Cern-Munich phase shifts and inelasticities
clearly contradict a number of
physical properties related to their 
(lack of) inelasticity: we will here mention a few.

First of all, the inelasticities ($1-\eta_l^{(I)}$) for all the waves in the Cern-Munich 
results remain small in the range $1.6\,\gev$ 
to 2 \gev. However,  as we have remarked above, one would 
expect dominant inelastic cross sections there. 
For a given wave, equality of elastic 
and inelastic cross sections occurs for 
$\eta_l^{(I)}=\cos2\delta_l^{(I)}$ (cf.~(2.1.4b)), and the condition 
 to have the inelastic cross section much
larger than the elastic one is  $\eta_l^{(I)}\ll\cos2\delta_l^{(I)}$.\fnote{For the 
P and D0 waves, this suggests that one should have $\delta_1>\pi$, 
$\delta_2^{(0)}>\pi$ near 2 \gev. 
As we show below, there is extra evidence for the first inequality (violated 
by the Cern--Munich phase) from a different source.} 
This inequality is not satisfied by any of  the Cern--Munich phases 
and elasticity parameters, except for the F wave 
 on 
the $\rho_3(1690)$ resonance.
  The Cern--Munich elastic cross sections are larger or comparable to the
inelastic ones up to 
$s^{1/2}=2\,\gev$ and, what is worse, 
 their inelastic cross sections, alone of all 
hadronic cross sections, {\sl decrease} 
when the kinetic energy grows from  1 \gev\ to 1.7 \gev; 
 for  e.g., $\pi^+\pi^-$ scattering this is clearly shown in Fig.~7 in the paper 
of Hyams et al.~(1973).  

Secondly, the combination of $\delta$ and $\eta$ for both P and S0 
waves at an energy around $1.8\,\gev$ is incompatible with what QCD implies for the 
electromagnetic and scalar form factors of the pion. 
In fact, as we will show in \sect~7.2, the Brodsky--Farrar counting rules for these form 
factors imply that their phases $\delta(t)$ behave like
$$\delta(t)\simeqsub_{t\to\infty}
\pi\left(1+\dfrac{\log s/\hat{t}}{\log\log s/\hat{t}}\right),\quad s\gg
\lambdav^2;\quad
\hat{t}\sim\lambdav^2 
\equn{(6.6.1)}$$
($\lambdav$ is the QCD parameter). 
One may take (6.6.1) to hold for $s\gsim3\,\gev^2$ ($s^{1/2}>1.6\,\gev$).
If one had negligible inelasticity for these waves 
somewhere in the region $1.6\,\gev\lsim s^{1/2}\lsim 2\,\gev$, as the 
Cern--Munich data seem to imply, form factors and 
partial waves should have the same phase at such energies, and thus 
the same behaviour (6.6.1) should hold 
for $\delta_1(s)$, $\delta_0^{(0)}(s)$. 
But the phases the Cern--Munich experiment gives  clearly contradict (6.6.1) 
around $1.8$ \gev. For example, the Cern--Munich phase 
$\delta_1$ stays consistently {\sl below}
$\pi$, while (6.6.1) implies that it should be {\sl above}. 
We have already seen evidence on this coming from 
inelasticity inequlalities above.

Thirdly, and as happens for  $\pi N$, $KN$,  $NN$ and even $\gamma N$, $\gamma \gamma$,
 scattering, 
we would expect a levelling off of the total cross section for 
$E_{\rm kin}>1.3\,\gev$. 
However, the Cern--Munich total cross section 
decreases roughly like $1/s$ up to 2 \gev. 
This is because P, D0 phases are $\sim\pi$, and $\delta_3\sim0$, 
in the range $E_{\rm kin}>1.3\,\gev$, while 
the corresponding parameters $\eta_l^{(I)}$ are near unity there: the 
Cern--Munich scattering amplitude is almost exclusively S waves 
for $s^{1/2}>1.55\,\gev$. 

Fourthly, 
we would expect large isospin  S2 and D2 waves 
as we approach the $2\rho$ thresold. 
However, these waves are essentially ignored in the Cern--Munich analysis.
Thus, besides the general problem for the cross sections 
we have individual problems 
for each of the S0, S2, P and D2 phases.

Finally,   both the  Regge picture 
and the experimental cross sections for all hadronic processes 
indicate that the number of waves that contribute 
effectively to the imaginary part (say) of the scattering amplitudes 
grows with the kinetic energy as $E_{\rm kin}/\lambdav$ for $E_{\rm kin}$ upwards of 1 \gev.
We thus expect 2 to 3 waves (for fixed isospin) 
at  $E_{\rm kin}\sim1\,\gev$, and almost double this, 3 to 5 
waves at $E_{\rm kin}\sim1.7\,\gev$. 
In fact, for $\pi\pi$ scattering  at this energy,  the 
contribution of the  F wave is as large as that of the P wave, 
the D0 wave is as large as the S0 wave, and the contribution of the
 D2 wave is as large or larger
than that of the S2 wave: the partial wave series with only two waves per isospin channel 
is not convergent. 
The approximations  that neglect all higher waves  at 
such energies have another reason for being irrealistic.

All these arguments indicate that, in particular, the inelasticities of the Cern--Munich 
phase shifts are much underestimated beyond $\sim1.5\,\gev$. 
It is possible that the results presented by the Cern--Munich group 
fit the {\sl elastic} $\pi\pi$ cross section, but, because they 
feature insufficient inelasticity, they certainly
 misrepresent the {\sl total} cross section.
They must therefore lead to a distorted imaginary part of the
 $\pi\pi$  scattering amplitude. 
It is thus not surprising that Pennington~(1975), Ananthanarayan et
al.~(2001) and Colangelo, Gasser  and Leutwyler (2001), who 
fix their Regge parameters by balancing them above $2\,\gev$ with 
phase shifts below $2\,\gev$, get incorrect values for the first.

We would like to emphasize that what has been said should not be taken as implying 
criticism of the 
 Cern--Munich experiment which, for $s^{1/2}\lsim1.4\,\gev$, 
 produced what are probably the best determinations of phase shifts and inelasticities. 
Above $1.4$ \gev, they did what they could: it is for 
theorists to realize that this was not enough to produce acceptable phase shifts 
and inelasticities at these higher energies.

\bookendchapter

\brochureb{\smallsc chapter 7}{\smallsc  analyticity. dispersion relations.
form factors}{65}
\bookchapter{7. Analiticity: dispersion relations 
and the Froissart--Gribov representation.\hb
Form factors: the Omn\`es--Muskhelishvili\hb method}
\vskip-0.5truecm

\booksection{7.1. The Omn\`es--Muskhelishvili method}

\noindent
In the analysis of the pion form factors we have the following situation: we
 have information on the phase of a quantity, $F$, and, 
in some cases,  know 
{\sl experimentally} its 
modulus. 
We would like to translate this into a general parametrization of the quantity. 
This last problem was first solved by Muskhelishvili~(1958) and later applied to the 
physical case by Omn\`es~(1958). 
We turn to this method.

\booksubsection{7.1.1. The full Omn\`es--Muskhelishvili problem}

\noindent
We want to find the most general representation for a function, $F(t)$, 
of which we know that it is analytic in the complex $t$ plane, cut from $t=4\mu^2$ to $\infty$, 
assuming that we know its phase on the cut, 
$${\rm arg}\;F(t)=\delta(t),\quad 4\mu^2\leq t.
\equn{(7.1.1)}$$
This is the so-called (full) Omn\`es--Muskhelishvili problem. 
Note that we do {\sl not} take the principal value of the argument here, except for $s$ near
threshold;  
we have to assume $\delta$ to be continuous, so it could go above $2\pi$ 
at high energy.

First of all, it is clear that, unless we have further information on $F$, the solution to 
this equation is highly nonunique. 
For, if $F_0(t)$ is a solution to (7.1.1), then any
$$\ee^{at}F_0(t),\quad{\rm or}\quad \ee^{a\ee^{bt}}F_0(t),\;\dots$$
would also be a solution. 
Fortunately, in the physically interesting cases we have information on the growth 
of $F(t)$ at large $t$ that precludes such functions. 
For example, and as already discussed, the Brodsky--Farrar counting rules imply that, 
for meson form fators,
$$F(t)\lsimsub{t\to\infty}\dfrac{{\rm Const.}}{-t\log^\nu (-t)}.
\equn{(7.1.2)}$$

We will restrict our analysis to the case where $\delta(t)$ is H\"older continuous 
(Muskhelishvili,~1958). 
We will also, in this subsection, assume
that the phase has a finite, positive limit  as $t\to\infty$:
$$\delta(t)\rightarrowsub_{t\to\infty}\delta(\infty),\quad\delta(\infty)>0. 
\equn{(7.1.3)}$$
In fact, (7.1.2) implies $\delta(\infty)=\pi$, so it 
 will turn out that (7.1.3) is really  the condition that is relevant for 
physical applications; so we assume it. 

To solve our problem the first step is to form the auxiliary function
$$J(t)=\exp\dfrac{t}{\pi}\int_{4\mu^2}^\infty\dd s\,
\dfrac{\delta(s)}{s(s-t-\ii0)}.
\equn{(7.1.4)}$$ 
We will assume that $F(0)=1$; otherwise, we would consider the function $F(t)/F(0)$. 
From (7.1.4) two properties of $J$  are immediately obvious: $J(0)=1$ and 
$J$ has no zero in the complex plane (the last because, due to  the continuity of 
$\delta$, the integral in the exponent is finite).

It is easy to verify that the function $J$ has the same analyticity properties and 
 the same phase as $F$. 
For example, using the relation $1/(x\pm\ii0)=\pepe (1/x)\mp\ii\pi\delta(x)$, we have
$$\dfrac{t}{\pi}\int_{4\mu^2}^\infty\dd s\,
\dfrac{\delta(s)}{s(s-t-\ii0)}=\dfrac{t}{\pi}\pepe\int_{4\mu^2}^\infty\dd s\,
\dfrac{\delta(s)}{s(s-t)}+\ii\delta(t).
\equn{(7.1.5a)}$$
At large $t$, the real part of the integral above dominates over its imaginary part and we 
find
$$\dfrac{t}{\pi}\int_{4\mu^2}^\infty\dd s\,
\dfrac{\delta(s)}{s(s-t-\ii0)}\simeqsub_{t\to\infty}-\dfrac{\delta(\infty)}{\pi}\log|t|.
\equn{(7.1.5b)}$$
In view of this last relation we obtain the behaviour,
$$J(t)\simeqsub_{t\to\infty}|t|^{-\delta(\infty)/\pi}.
\equn{(7.1.6)}$$ 

Next step is to form the function $G(t)$, defined by
$$F(t)=G(t)J(t).
$$
Because $J$ never vanishes, $G(t)$ is, at least, analytic in the same domain as $F(t)$. 
Moreover, since $J$ and $F$ have the same phase on the cut, it follows that 
$G(t)$ is real on 
the cut.  
According to the theorem of Painlev\'e, this implies that $G$ is also analytic on the cut, hence $G(t)$ is
analytic in the whole $t$ plane, i.e., it is an entire function.

It is now that the growth condition (7.1.2)   enters. 
The only entire functions that do not grow  exponentially (or faster) in some direction are the 
polynomials. Hence, (7.1.2) implies that $G(t)=P_N(t)$, 
where $P_N(t)$ is a polynomial of degree $N$: we have found the general representation
$$F(t)=P_N(t)J(t).
\equn{(7.1.7)}$$

Now, which polynomials are allowed depends on the value of $\delta(\infty)$. 
We will simplify the discussion by assuming that  $\delta(\infty)=n\pi$, 
$n$ an integer; the interested reader may find information on 
other situations in the text of  Muskhelishvili~(1958). 
On comparing (7.1.6) and (7.1.2) it follows 
immediately that $N=n-1$. 
Thus, in the case (that will turn out to be the more interesting one for us here) in which $n=1$, 
the function $J$ is actually the most general solution to the problem:
$$F(t)=J(t)\quad[\delta(\infty)=\pi].
\equn{(7.1.8)}$$

\booksubsection{7.1.2. The incomplete Omn\`es-Muskhelishvili problem}

\noindent
In the physically relevant cases we do not know $\delta(t)$ for all $t$, but only up 
to a certain $s_0$, typically the energy squared at which inelastic channels 
start becoming important. 
 We will make the calculations for the form factor $F$
of  spinless particles with mass 
$\mu$, so the results will be directly applicable to pions.

The idea for the treatment of this case is to extend $\delta(t)$
 to the full $t$ range, in an appropriate manner, so as to reduce the problem to the previous one. 
Let us call $\delta_{\rm eff}(t)$ to this extension, so that
$\delta_{\rm eff}(t)=\delta(t)$ for $t\leq s_0$. 
We then form 
$$J_{\rm eff}(t)=\exp\dfrac{t}{\pi}\int_{4\mu^2}^\infty\dd s\,
\dfrac{\delta_{\rm eff}(s)}{s(s-t-\ii0)}
\equn{(7.1.9a)}$$
and define $G$ by 
$$F(t)=G(t)J_{\rm eff}(t).
\equn{(7.1.9b)}$$
Because now $\delta_{\rm eff}(t)$ equals $\delta(t)$ only for 
$4\mu^2\leq t\leq s_0$, $G(t)$ will not be analytic on the whole $t$ plane, but will 
retain a cut from $t=s_0$ to $\infty$. $G$ will be an unknown function, that will 
have to be obtained from a model or fitted to experiment. 
Because of this, we have interest to have it as smooth as possible, so that a few 
terms will represent it. 
Since discontinuities of $\delta_{\rm eff}(t)$ will generate infinities of $J_{\rm eff}(t)$, 
and  of $G(t)$, we must choose a smooth continuation of $\delta(t)$ above 
$t=s_0$. Moreover, if we do not want to have a $G$ growing without limit  for 
large $t$, we have to construct a $J_{\rm eff}(t)$ that decreases at infinity like 
$F(t)$. 
These conditions are fulfilled if we simply define
$$\delta_{\rm eff}(t)=\cases{\quad
\delta(t),\quad\phantom{-\pi\dfrac{t_0}{t}}\quad t\leq s_0;\cr
\pi+\left[\delta(s_0)-\pi\right]\dfrac{s_0}{t},\quad t\geq s_0.\cr
}
\equn{(7.1.10)}$$
In this case the piece from $s_0$ to $\infty$ in the integral in \equn{(7.1.9a)} 
can be performed explicitely and we get
$$\eqalign{
F(t)&\,=G(t)J_{\rm eff}(t)
=G(t)\ee^{1-\delta(s_0)/\pi}
\left(1-\dfrac{t}{s_0}\right)^{[1-\delta(s_0)/\pi]s_0/t}
\left(1-\dfrac{t}{s_0}\right)^{-1}\cr
\times&
\exp\left\{\dfrac{t}{\pi}\int_{4\mu^2}^{s_0} \dd s\;
\dfrac{\delta(s)}{s(s-t-\ii0)}\right\}.\cr
}
\equn{(7.1.11)}$$
If we knew that $F(t)$ behaves exactly as $1/t$, for $t\to\infty$, 
it would follow that its phase has to tend to $\pi$ at infinity. 
More generally, if one has, as in (7.1.2),
$$F(t)\simeqsub_{t\to\infty}\dfrac{{\rm Const.}}{-t\log^\nu (-t)},$$
then this implies
$$\delta(t) \simeqsub_{t\to\infty}\pi\left\{1+\nu\dfrac{\log\log t}{\log t}\right\}.$$ 
Therefore, $\delta_{\rm eff}$ in (7.1.10) may be then considered as a linear interpolation 
(in $t^{-1}$) for
$\delta(t)$ between 
$s_0$ and infinity, and $G(t)$ may be interpreted as giving the 
correction to this.

It only remains to write a general parametrization of $G(t)$ compatible with 
its known properties. 
To do so, we map the cut $t$ plane into the unit disk in the variable $z$ 
(\fig~7.1.1),
$$z=\dfrac{\tfrac{1}{2}\sqrt{s_0}-\sqrt{s_0-t}}{\tfrac{1}{2}\sqrt{s_0}+\sqrt{s_0-t}}.
\equn{(7.1.12)}$$
The most general $G$ is analytic inside this disk, so we can write 
a Taylor expansion for it, which conveniently we set in the form
$$G(t)=1+A_0+c_1z(t)+c_2 z(t)^2+c_3 z(t)^3+\cdots
\equn{(7.1.13a)}$$
This expansion that will be uniformly convergent for all $t$ inside the cut plane. 
We can implement the condition $G(0)=1$, necessary to ensure 
$F(0)=1$ order by order, by putting 
$$A_0=-\left[c_1z_0+c_2 z_0^2+c_3 z_0^3+\cdots\right],\quad
z_0\equiv z(t=0)=-1/3.
\equn{(7.1.13b)}$$
We remark in passing that since, inside the unit circle, one has $|z|\leq 1$, it follows that 
to every finite order in the expansion (7.1.13a), $G(t)$ is bounded in 
the $t$ plane. Hence $F(t)$ and $J_{\rm eff}(t)$ have the same 
asymptotic behaviour, as desired.
 
\topinsert{
\setbox0=\vbox{\hsize11.truecm{\epsfxsize 10truecm\epsfbox{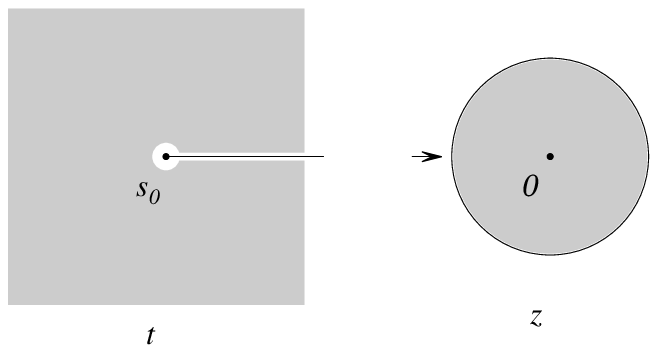}}} 
\setbox6=\vbox{\hsize 8truecm\captiontype\figurasc{Figure 7.1.1 }{
The  
mapping $t\to z$.}\hb
\vskip.1cm} 
\medskip
\centerline{{\box0}}
\centerline{\box6}
\smallskip
}\endinsert

We end this section with a simple example that shows clearly the desirability of 
expanding a function which, like $G$, is regular at the frontier of the 
domain of analyticity (which happens because we were careful to extrapolate $\delta$ without introducing 
singularities and keeping the correct asymptotic behaviour). 
Consider the three series
$$\eqalign{
\dfrac{1}{1-z}=&\,\sum_{n=0}^\infty z^n\quad ({\rm A});\cr
\log(1-z)=&\,\sum_{n=1}^\infty\dfrac{1}{n} z^n\quad ({\rm B});\cr
\int_0^z\dd t\,\dfrac{\log(1-t)}{t}=&\,\sum_{n=1}^\infty\dfrac{1}{n^2} z^n\quad ({\rm C}).\cr
}
$$
The first has a pole, the second a logarithmic singularity and the third is regular 
at the edge of the convergence disk. 
The first series diverges there, the second is conditionally convergent at all points except at 
$z=1$, and the third is convergent even at the edge of the disk. 
This pattern is general.

\booksection{7.2. Application 
to the pion form factors of the Omn\`es-Muskhelishvili method}
\vskip-0.5truecm
\booksubsection{7.2.1. The electromagnetic form factor}

\noindent
The application to 
the pion form factors 
of the formalism presented in the previous section is straightforward as, indeed, it was
tailored for precisely this case. We start with the electromagnetic form factor.
The function $\delta(t)$ is now the  P wave phase in $\pi\pi$ scattering, 
that we have denoted by $\delta_1(t)$. 
If we consider $\pi^+\pi^-$ scattering, then we have experimental information on 
$|F(t)|$ from $e^+e^-\to\pi^+\pi^-$ and, at $t<0$, we can use 
data on $F(t)$ from $\pi e^-$ scattering. If we take $\pi^+\pi^0$, then 
the information, at positive $t$, comes from the decay
$\tau^+\to\bar{\nu}_\tau\pi^+\pi^0$.  We may parametrize $\delta_1(t)$ as in (6.3.3); 
as for $G(t)$, we take two terms in (7.1.13) and write
$$G(t)=1+
c_1\left[\dfrac{\tfrac{1}{2}\sqrt{s_0}-\sqrt{s_0-t}}{\tfrac{1}{2}\sqrt{s_0}+\sqrt{s_0-t}}
+\tfrac{1}{3}\right]+
c_2\Bigg[\left(\dfrac{\tfrac{1}{2}\sqrt{s_0}-\sqrt{s_0-t}}{\tfrac{1}{2}\sqrt{s_0}+\sqrt{s_0-t}}\right)^2
-\tfrac{1}{9}\Bigg],
\equn{(7.2.1)}$$
$c_1,\,c_2$ free parameters. 
We remark that, although there are only 
 two free parameters, this is because we have imposed the condition
$G(0)=1$; 
the expansion (7.2.1) gives correctly the first three terms.

The quality of the fits, with only five free parameters
 ($B_0,\,B_1,\,M_\rho;\,c_0,\,c_1$) is remarkable, 
as can be seen in the accompanying figures 7.2.1,\/2; the $\chi^2$ is, including systematic and 
statistical errors, 
$\chidof=213/204$ (the $\omega - \rho$ interference 
effect was treated with the Gounnaris--Sakurai method).

\topinsert{
\setbox0=\vbox{\hsize14.truecm\hfil{\epsfxsize 11.9truecm\epsfbox{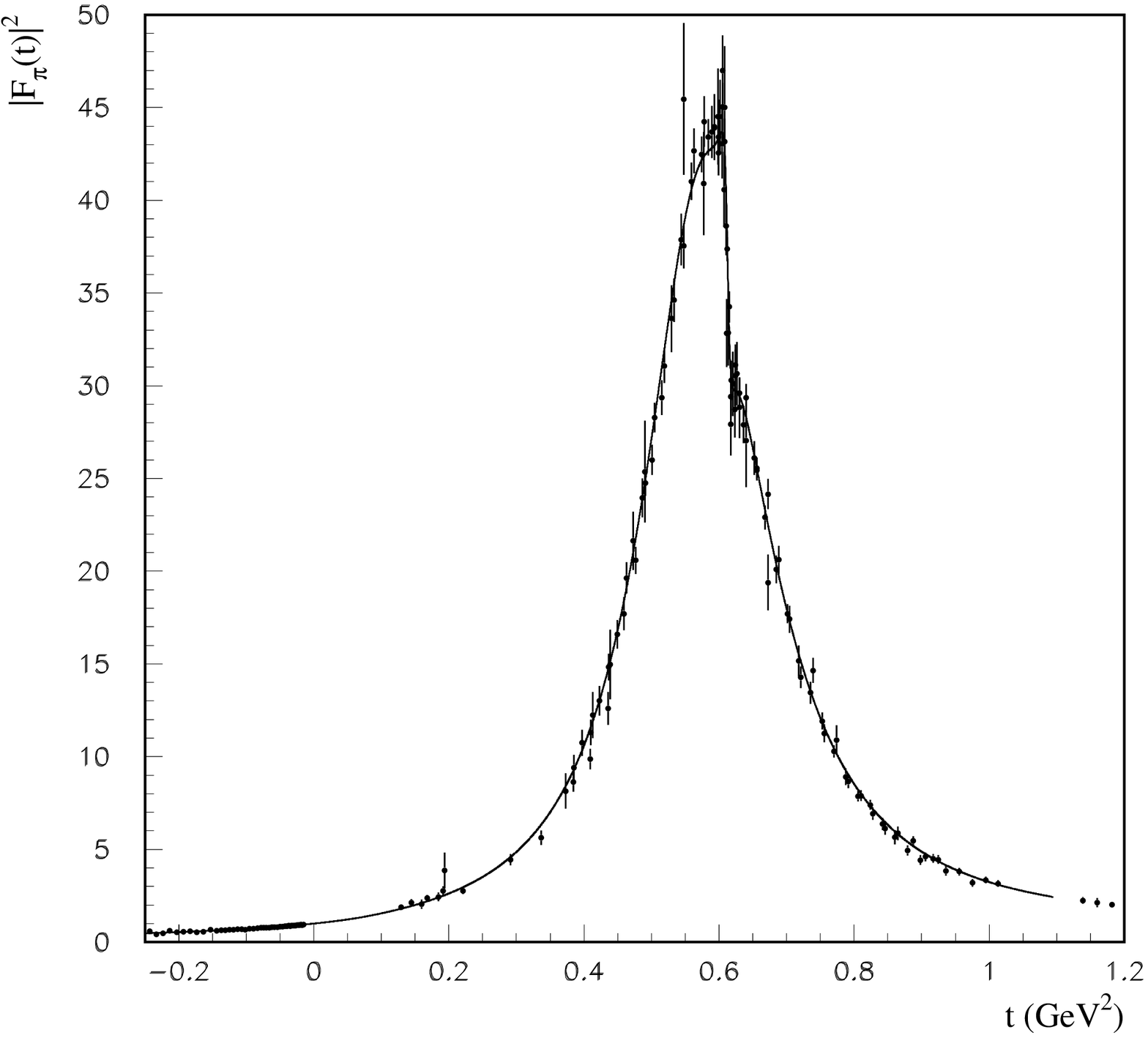}}\hfil} 
\setbox6=\vbox{\hsize 10truecm\captiontype\figurasc{Figure 7.2.1. }{Plot of 
the fit to $|F_\pi(t)|^2$, timelike and spacelike data. The 
theoretical curve actually drawn is that obtained by fitting also $\tau$ data, but 
the curve obtained fitting only $e^+e^-$ and $\pi e$ data 
 could not be distinguished from that drawn if we plotted it.}\hb
\vskip.1cm} 
\medskip
\centerline{\box0}
\centerline{\box6}
\medskip
}\endinsert

Because we are interested not only on (relatively) rough 
estimates, but aim at pinning down fine details of isospin breaking as well, 
we will spend some time presenting the results. 
These results have been obtained in the course of the work reported by 
de~Troc\'oniz and Yndur\'ain~(2002), but not all of them have been published.

We consider the following types of fits. 
Firstly, we may take into account $\pi^+\pi^-$ form factor data (in the spacelike 
as well as the timelike regions) and data from  $\tau$ decay into $\nu\pi^+\pi^0$. 
Isospin breaking is incorporated by using the correct phase space for each case,
 and allowing for different masses and
widths for $\rho^0$, $\rho^+$; but 
 the function $G(t)$, whose cut only starts at $t\sim1\,\gev^2$, 
is assumed isospin independent. This produces the best results for 
the hadronic contributions to the $g-2$ of the muon and for the mass and width of the $\rho$: 
$$\eqalign{
M_{\rho^0}=&\;772.6\pm0.5\;\mev,\quad \gammav_{\rho^0}=147.4\pm0.8\,\mev;\cr
M_{\rho^+}=&\;773.8\pm0.6\;\mev,\quad \gammav_{\rho^+}=147.3\pm0.9\,\mev.\cr
}
\equn{(7.2.2)}$$

\topinsert{
\setbox9=\vbox{
\setbox0=\vbox{\hsize16.truecm\line{\hfil{\epsfxsize 6.truecm\epsfbox{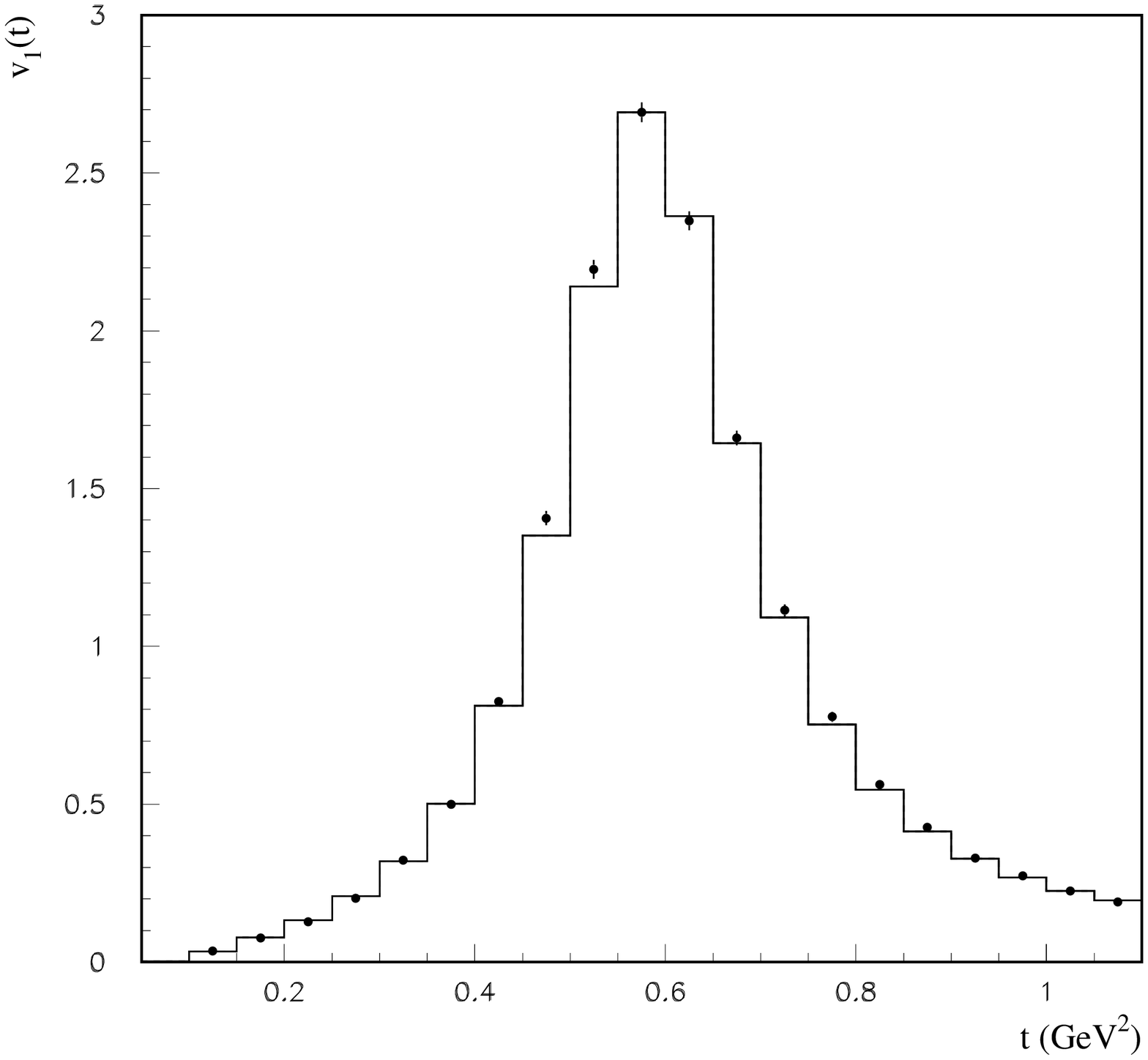}}\hfil
{\epsfxsize 6.truecm\epsfbox{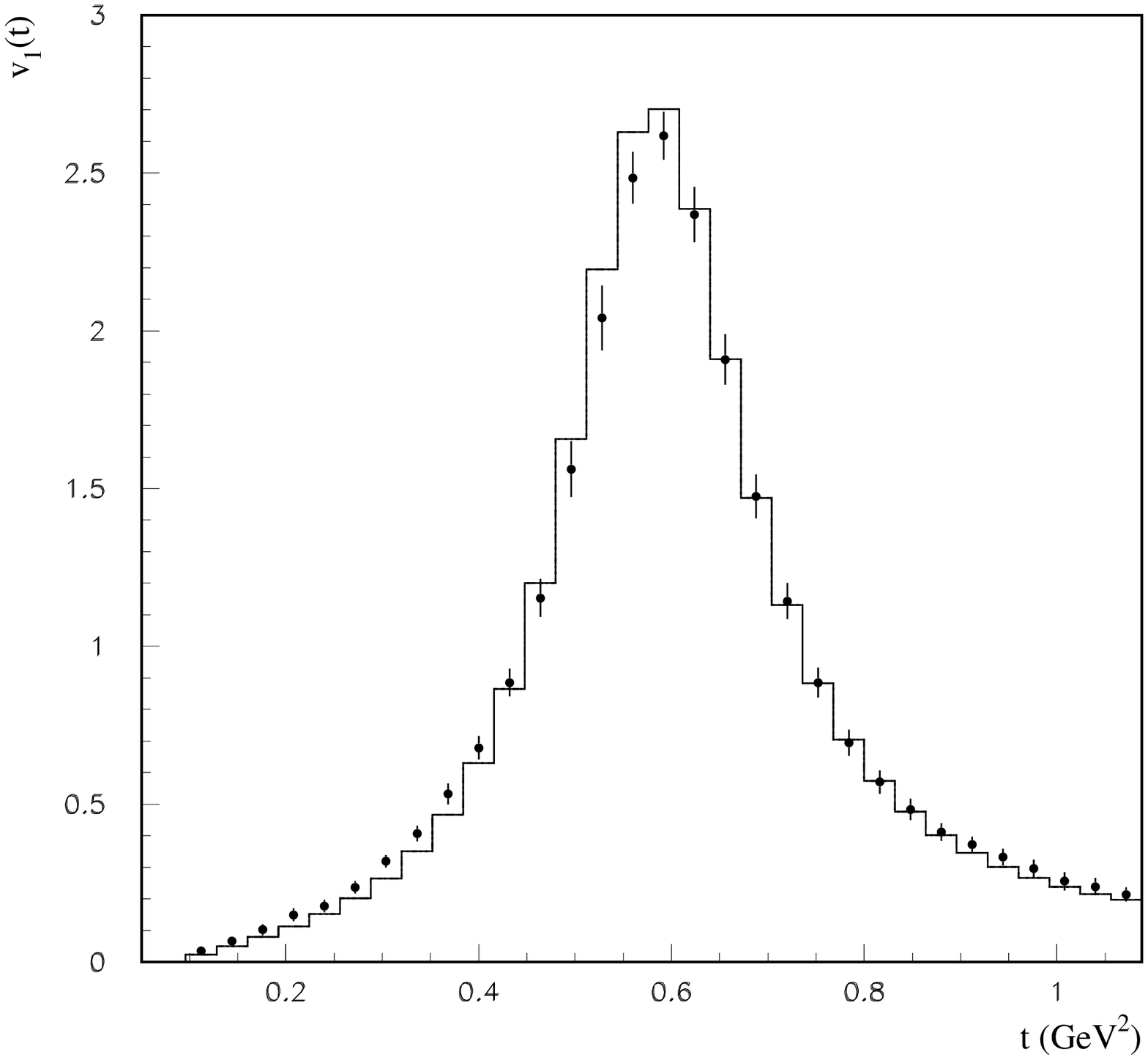}}\hfil}}
\setbox6=\vbox{\hsize 10truecm\captiontype\figurasc{Figure 7.2.2 }{Plot 
of the fits to $v_1(t)$ (histograms), and data from $\tau$ decay (black dots).\hb
Left: Aleph data. Right: Opal data.}
The theoretical values (histograms) are results of the {\sl same} 
calculation, with the same parameters, so the differences between the two merely 
reflect the  variations between the two experiments. 
We do not include the result of the fit to the third existing set of data
 (Anderson et al.,~2000),  which is much like the ones depicted here.} 
\bigskip
\centerline{\box0}
\centerline{\box6}
\bigskip}
\box9
}\endinsert 

However, for our purposes here it is more interesting to consider two other possibilities. 
We may use only  $\pi^+\pi^-$ data (possibility A) or we may use both $\pi^+\pi^-$
 and $\pi^+\pi^0$  data, {\sl neglecting isospin breaking effects}, in particular 
with the same $\rho$ 
parameters (possibility B); this would then represent a 
kind of isospin averaged result. 
The departure of A from B will be a measure of isospin breaking effects.

For the parameters $B_i$, $c_i$ we get, in case A, 
and without including systematic experimental errors,\fnote{The experimental numbers are from Barkov et al.~(1985), 
Akhmetsin et al.~(1999), Amendolia et al.~(1986), Anderson et. al,~2000,
 Barate et al.~(1997) and 
Ackerstaff et al.~(1999). There appears to be 
an inconsistency between the old an new versions 
of Akhmetsin et al.~(1999), related to whether the radiative corrections
 have been fully incorporated. 
The fit given in the present paper uses the old set of data; we 
have checked that replacing them by the new one leaves the values
 of the parameters $B_i$, $c_i$, $M_\rho$ 
essentially unchanged.}
$$\eqalign{
c_1=0.19\pm0.04,\quad c _2=-0.15\pm0.10,\cr
B_0=1.070\pm0.006,\quad B_1=0.28\pm0.06\cr
}
\equn{(7.2.3a)}$$
and, in case B,
$$\eqalign{
c_1=0.23\pm0.01,\;c _2=-0.16\pm0.03,\cr
B_0=1.060\pm0.005,\;B_1=0.24\pm0.04.\cr
}
\equn{(7.2.3b)}$$
If we included systematic errors we would obtain the values already reported in 
\subsect~6.3.1, Eqs.~(6.3.5).

Another question that has to be taken into account is the relative normalization of the 
various experiments.
This is  particularly important for the 
P wave $\pi\pi$ parameters,  and for    
 the slope (quadratic charge radius, $\langle r^2_\pi\rangle$) and second derivative 
($c_\pi$) of the
electromagnetic pion form factor, with these last two defined by 
$$F^2_\pi(t)\simeqsub_{t\to0}1+\tfrac{1}{6}\langle r^2_\pi\rangle t+c_\pi t^2.
\equn{(7.2.4)}$$
What happens is that, as is clear from \fig~7.2.2, there is
 a small but systematic difference between Opal and Aleph data
and, as shown in de~Troc\'oniz and Yndur\'ain~(2002), 
the spacelike data on $F_\pi(t)$ do not agree well with the 
theoretical curve unless one takes into account systematic normalization effects.

In view of this, we present two sets of values for each quantity. 
In the first, the various experiments are fitted including only 
{\sl statistical} errors. In the second set we repeat the fit, including 
{\sl systematic} normalization effects. 

It is in principle unclear which of the two sets of results is to be preferred; 
however, we will see later (\sect~7.5.3) that analyticity, in 
the form of the Froissart--Gribov representation, favours
 evaluations with systematic normalization errors
taken into account. 
 The results of all three procedures are 
presented in the following tables:

\medskip
\setbox0=\vbox{
\setbox1=\vbox{\petit \offinterlineskip\hrule
\halign{
&\vrule#&\strut\hfil\quad#\quad\hfil&\vrule#&\strut\hfil\quad#\quad\hfil&
\vrule#&\strut\hfil\quad#\quad\hfil&\vrule#&\strut\hfil\quad#\quad\hfil\cr
 height2mm&\omit&&\omit&&\omit&&\omit&\cr 
&\hfil $\pi^+\pi^-$ only\hfil&&\hfil Only stat. errors\hfil&
&\hfil With normalization errors\hfil& \cr
 height1mm&\omit&&\omit&&\omit&&\omit&\cr
\noalign{\hrule}height1mm&\omit&&\omit&&\omit&&\omit&\cr
&\vphantom{\Big|}$10^3\times a_1$ &&$(42\pm2)\;M_{\pi}^{-3}$&
&\hfil $(39\pm2)\;M_{\pi}^{-3}$ \hfil& \cr
\noalign{\hrule}height1mm&\omit&&\omit&&\omit&&\omit&\cr
&\vphantom{\Big|}$10^3\times b_1$ &&$(4.5\pm0.4)\;M_{\pi}^{-5}$&
&\hfil $(4.5\pm0.3)\;M_{\pi}^{-5}$ \hfil& \cr
\noalign{\hrule}height1mm&\omit&&\omit&&\omit&&\omit&\cr
&\vphantom{\Big|}$\langle r^2_\pi\rangle$ (fm$^2$)&&$0.433\pm0.002$&
&\hfil $0.426\pm0.003$ \hfil& \cr
\noalign{\hrule}height1mm&\omit&&\omit&&\omit&&\omit&\cr
&\vphantom{\Big|}$c_\pi\;({\gev}^{-4})$ &&\hfil $3.58\pm0.04$\hfil&
&\hfil $3.49\pm0.06$& \cr
\noalign{\hrule}}}
\centerline{\box1}
\medskip
\centerline{{\sc Table} I{\sc a}}
}\centerline{\box0}

\medskip
\setbox0=\vbox{
\setbox1=\vbox{\petit \offinterlineskip\hrule
\halign{
&\vrule#&\strut\hfil\quad#\quad\hfil&\vrule#&\strut\hfil\quad#\quad\hfil&
\vrule#&\strut\hfil\quad#\quad\hfil&\vrule#&\strut\hfil\quad#\quad\hfil\cr
 height2mm&\omit&&\omit&&\omit&&\omit&\cr 
&\hfil $\pi^+\pi^-$ \& $\pi^+\pi^0$\hfil&&\hfil Only stat. errors\hfil&
&\hfil With normalization errors\hfil& \cr
 height1mm&\omit&&\omit&&\omit&&\omit&\cr
\noalign{\hrule}height1mm&\omit&&\omit&&\omit&&\omit&\cr
&\vphantom{\Big|}$10^3\times a_1$ &&$(40.6\pm1.4)\;M_{\pi}^{-3}$&
&\hfil $(38.6\pm1.2)\;M_{\pi}^{-3}$ \hfil& \cr
\noalign{\hrule}height1mm&\omit&&\omit&&\omit&&\omit&\cr
&\vphantom{\Big|}$10^3\times b_1$ &&$(4.18\pm0.43)\;M_{\pi}^{-5}$&
&\hfil $(4.47\pm0.29)\;M_{\pi}^{-5}$ \hfil& \cr
\noalign{\hrule}height1mm&\omit&&\omit&&\omit&&\omit&\cr
&\vphantom{\Big|}$\langle r^2_\pi\rangle$ (fm$^2$)&&$0.438\pm0.003$&
&\hfil $0.435\pm0.003$ \hfil& \cr
\noalign{\hrule}height1mm&\omit&&\omit&&\omit&&\omit&\cr
&\vphantom{\Big|}$c_\pi\;({\gev}^{-4})$ &&\hfil $3.64\pm0.05$\hfil&
&\hfil $3.56\pm0.04$& \cr
\noalign{\hrule}}}
\centerline{\box1}
\medskip
\centerline{{\sc Table} I{\sc b}}
}\centerline{\box0}
\medskip

The lack of dependence of $a_1$ on the procedure used to obtain it is a bit
 fictitious as the fits were obtained including the 
constraint $a_1=(38\pm3)\times10^{-3}\,M_{\pi}^{-3}$. 
If we had not included it, the fits would have yielded values as high as 
$(43\pm3)\times10^{-3}\,M_{\pi}^{-3}$, 
whose central value is difficult to reconcile with $\pi\pi$ scattering data.

The parameters given above are the ones that, in particular, 
 produce the excellent {\sl prediction} of 
P wave phase shifts shown in \fig~6.3.2, as well as the precise values of 
some of  the chiral 
parameters $\bar{l}_i$ that we will give in \sect~9.4.

\booksubsection{7.2.2. The scalar form factor and radius of the pion}

\noindent
An important quantity in chiral perturbation theory calculations is the quad\-ratic 
scalar radius of the pion. 
To define it, we start with the pion 
scalar form factor, $F_S$, given by
$$\langle\pi(p_1)|m_u\bar{u}(0)u(0)+m_d\bar{d}(0)d(0)|\pi(p_2)\rangle=F_S(t),\quad
t=(p_1-p_2)^2.
\equn{(7.2.5)}$$
As $t$ goes to zero we write
$$F_S(t)\simeqsub_{t\to0}F_S(0)\Big\{1+\tfrac{1}{6}t\langle r^2_{{\rm S},\pi\rangle}\Big\}.
\equn{(7.2.6)}$$
One can obtain $\langle r^2_S\rangle$ in two ways. 
Theoretically, we may relate it to meson masses and decay constants, using chiral perturbation theory:
$$\eqalign{
\langle r^2_{{\rm S},\pi}\rangle=&\,\dfrac{6}{m^2_K-M^2_\pi}
\left(\dfrac{f_K}{f_\pi}-1\right)+\delta_3;\cr
\delta_3=&\,-\dfrac{1}{64\pi^2f^2_\pi}\,\dfrac{1}{m_K^2-M^2_\pi}\Bigg\{
6(2m_K^2-M^2_\pi)\log\dfrac{m_K^2}{M^2_\pi}+
9m^2_\eta\log\dfrac{m^2_\eta}{M^2_\pi}\cr
-&\,
2(m_K^2-M^2_\pi)\left(10+\tfrac{1}{3}\dfrac{M^2_\pi}{m^2_\eta}\right)\Bigg\}.
\cr}
\equn{(7.2.7a)}$$ 
The details may be found in Gasser and Leutwyler~(1985b) who using this obtain
$$\langle r^2_{{\rm S},\pi}\rangle=0.55\pm0.15\;{\rm fm}^2\quad\hbox{(GL)},
\equn{(7.2.7b)}$$
and the  error measures the dependence of the result on the 
estimated higher order effects.

Alternatively, Donoghue, Gasser and Leutwyler~(1990) 
calculate $\langle r^2_{{\rm S},\pi}\rangle$ from experiment, using a dispersive method 
based on the Omn\`es--Muskhelishvili procedure, to give what is 
presented as an accurate number:
$$\langle r^2_{{\rm S},\pi}\rangle=0.61\pm0.04\;{\rm fm}^2\quad\hbox{(DGL)}.
\equn{(7.2.7b)}$$
This calculation, however, neglects inelastic (in particular, $4\pi$) states 
and gives an overoptimistic treatment of quasielastic $\bar{K}K$ contributions, 
which are much worse known that what these authors think. 
The high energy contributions they take are also suspect: 
the central value may be biased, and the error in (7.2.7b) 
is certainly underestimated.  

We will here give a brief account of a calculation with the 
 Omn\`es--Muskhelishvili method; since its application it is very similar to that 
for the electromagnetic form factor, we may skip details. 
First of all, we remark that unitarity implies that,  for 
$4M_{\pi}^2\leq t\leq s_0$, the phase of $F_S(t)$ equals the phase of the S0 
wave in $\pi\pi$ scattering,
$\delta_0^{(0)}(t)$. Here $s_0$ is the energy squared where inelastic contributions 
begin to be nonnegligible; in our case, 
this happens at $\bar{K}K$ threshold, so $s_0=4m^2_K$. 
We will assume a behaviour of the scalar form factor similar to that 
of the electromagnetic one, as follows from the counting rules, \equn{(2.2.20)}.
From it, it follows that, if we denote by $\delta(t)$ to the phase of $F_S(t)$, 
one must have
$$\delta(t)\simeqsub_{t\to\infty}
\pi\left\{1+\nu\dfrac{\log\log t/\hat{t}}{\log t/\hat{t}}\right\}.
\equn{(7.2.8)}$$

Given this condition,  $\delta$ determines uniquely $F_S$: one has 
$$F_S(t)=F_S(0)\exp\left\{\dfrac{t}{\pi}
\int_{4M^2_\pi}^\infty\dd s\,\dfrac{\delta(s)}{s(s-t)}\right\}.
\equn{(7.2.9)}$$
From this we get a simple sum rule for the square radius 
$\langle r_S^2\rangle$ corresponding to $F_S(t)$:
$$\langle r_S^2\rangle=\dfrac{6}{\pi}\int_{4M^2_\pi}^\infty\dd s\,\dfrac{\delta(s)}{s^2}.
\equn{(7.2.10)}$$
We will split $\langle r^2_{{\rm S},\pi}\rangle$ as follows:
$$\langle r^2_{{\rm S},\pi}\rangle=Q_J(s_0)+Q_\phiv(s_0)+Q_G(s_0).
\equn{(7.2.11)}$$
Here $Q_J$ is the piece in (7.2.10) coming from the region 
where we know $\delta$,
$$Q_J(s_0)\equiv\dfrac{6}{\pi}\int_{4M^2_\pi}^{s_0}\dd s\,\dfrac{\delta(s)}{s^2};
\quad \delta(s)=\delta_0^{(0)}(s).
\equn{(7.2.12a)}$$
$Q_\phiv$ is obtained defining, as in (7.1.10), an effective phase 
that interpolates linearly (in $t^{-1}$) between the values 
of $\delta(t)$ at $s_0$ and $\infty$: we write
$$\delta_{\rm eff}(t)\equiv\pi+\Big[\delta_0^{(0)}(s_0)-\pi\Big]\dfrac{s_0}{t},
\equn{(7.2.13b)}$$
and then set
$$Q_\phiv(s_0)\equiv\dfrac{6}{\pi}\int_{s_0}^{\infty}\dd s\,\dfrac{\delta_{\rm eff}(s)}{s^2}.
\equn{(7.2.13c)}$$
Finally, $Q_G$ corrects for the difference between $\delta$ and $\delta_{\rm eff}$:
$$Q_G(s_0)\equiv\dfrac{6}{\pi}\int_{s_0}^{\infty}\dd s\,\dfrac{\delta(s)-\delta_{\rm eff}(s)}{s^2}.
\equn{(7.2.13d)}$$
$Q_J$, $Q_\phiv$ are known; $Q_G$ has to be fitted or estimated. 
The decomposition (7.2.11) is equivalent to decomposing $F_S$ as a product,
as we did for $F_\pi$:
$$\eqalign{
F_S(t)=&\,F_S(0)J_S(t)\phiv_S(t) G_S(t);\cr
J_S(t)=&\,\exp\left\{\dfrac{t}{\pi}\int_{4M^2_\pi}^{s_0} \dd s\;
\dfrac{\delta_0^{(0)}(s)}{s(s-t)}\right\},\cr
\phiv_S(t)=&\,\ee^{1-\delta_0^{(0)}(t_0)/\pi}
\left(1-\dfrac{t}{t_0}\right)^{[1-\delta_0^{(0)}(t_0)/\pi]t_0/t}
\left(1-\dfrac{t}{t_0}\right)^{-1}\cr
}
\equn{(7.2.14a)}$$
(we have integrated explicitly $\delta_{\rm eff}$), and $G_S(t)$ is defined by
$$G_S(t)=\exp\left\{\dfrac{t}{\pi}\int_{s_0}^\infty \dd s\;
\dfrac{\delta(s)-\delta_{\rm eff}(s)}{s(s-t)}\right\}.
\equn{(7.2.14b)}$$  
What we know about $G(t)$ is that $G(0)=1$, and that it 
 is analytic except for a cut $s_0\leq t<\infty$. 
Unlike for the case of the electromagnetic form factor, however, now we do not 
have experimental information on $F_S$ to which 
we could fit $G_S$, so one has to rely on models or approximations for 
it.

$Q_J$ is easily evaluated with the 
parametrizations of \sect~6.5; likewise, one can get $Q_\phiv$ using  
the value $\delta_0^{(0)}(4m^2_K)=3.14\pm0.52$, which comprises all experimental 
determinations. 
We find,
$$Q_J=0.465\pm0.05\;{\rm fm}^2,\;Q_\phiv=0.237\pm0.02\;{\rm fm}^2;\quad 
Q_J+Q_\phiv=0.70\pm0.06\;{\rm fm}^2.
\equn{(7.2.15)}$$

This equation should be interpreted a {\sl lower bound}
 on $\langle r_{{\rm S},\pi}\rangle$; it assumes that
the  phase of $F_S(s)$ does not increase 
for $s$ beyond $\bar{K}K$ threshold, while from (7.2.8) 
we expect $\delta(s)$ to increase somewhat before decreasing to its asymptotic value, 
$\delta(\infty)=\pi$. 
We have therefore found the result,
$$\langle r^2_{{\rm S},\pi}\rangle\geq 0.70\pm0.06\;{\rm fm}^2.
\equn{(7.2.16)}$$
To get a value for $\langle r^2_{{\rm S},\pi}\rangle$
 we need an estimate for $G_S$ or, alternatively, for 
the phase $\delta(t)$ between $\bar{K}K$ threshold and the asymptotic region, say 
$s\simeq 2\;\gev^2$. 
We will not present here the details, that the reader may find in Yndur\'ain~(2003). 
One gets,
$$\langle r^2_{{\rm S},\pi}\rangle= 0.75\pm0.07\;{\rm fm}^2.
\equn{(7.2.17)}$$

\booksubsection{7.2.3. The mixed $K\pi$ scalar form factor}

\noindent
The mixed $K\pi$ scalar form factor and quadratic radius are  defined by
$$\eqalign{\langle\pi(p)|(m_s-m_q)\bar{q}s(0)|K(p')\rangle=&\,
(2\pi)^{-3}f_{K\pi}(t),\;q=u,\,d;\cr
f_{K\pi}(t)\simeqsub_{t\to0}&\,f_{K\pi}(0)
\Big\{1+\tfrac{1}{6}\langle r^2_{{\rm S},K\pi}\rangle\,t\Big\}.\cr
}
\equn{(7.2.18a)}$$
To lowest order in chiralperturbation theory (Chapter~9),
$$f_{K^0\pi^+}(0)=M^2_K-M^2_\pi,\quad 
f_{K^+\pi^0}(0)=\sqrt{2}(M^2_K-M^2_\pi).
\equn{(7.2.18b)}$$
The mixed quadratic scalar
 radius $\langle r^2_{{\rm S},K\pi}\rangle$ can be evaluated in terms of its phase, 
$\delta(t)$, which, when we can neglect inelasticity, equals the phase shift 
for  the S wave $K\pi$ scattering with isospin $\tfrac{1}{2}$, 
$\delta_0^{(1/2)}(t)$:
$$\langle r^2_{{\rm S},K\pi}\rangle=
\dfrac{6}{\pi}\int_{(M_\pi+m_K)^2}^\infty\dd t\,\dfrac{\delta(t)}t^2,\qquad
\delta(t)=\delta_0^{(1/2)}(t)\;{\rm for}\;t\leq s_0;
\equn{(7.2.19)}$$
cf.~(7.2.10).
Experimentally,
$$\langle r^2_{{\rm S},K\pi}\rangle=0.31\pm0.06\;{\rm fm}^2.
\equn{(7.2.20)}$$
We will write, as for the pion radius,
$$\langle r^2_{{\rm S},K\pi}\rangle=Q_J+Q_\phiv+Q_G.
\equn{(7.2.21)}$$

For the low energy piece we consider two possibilities. 
First, 
we assume the phase $\delta(t)$ to be given, for $t^{1/2}\leq 1.5\,\gev$, 
by the resonance $K^*(1430)$, whose properties we take from the Particle Data Tables. 
Its mass is $M_*=1412\pm6 \mev$, and its width $\gammav_*=294\pm23\,\mev$; 
we neglect its small inelasticity ($\sim7\%$). 
We write a Breit--Wigner formula for the phase:

$$\cot\delta_0^{(1/2)}(t)=\dfrac{t^{1/2}}{2q}(1-s/M^2_*)B_0,\quad
 q=\dfrac{\sqrt{[s-(M_K-M_\pi)^2][s-(M_K+M_\pi)^2]}}{2s^{1/2}}
$$
and $B_0=2q(M^2_*)/\gammav_*=4.15\pm0.35.$
We take $s_0^{1/2}=1.5\,\gev$, and then we
 have
$$
Q_J=0.050\pm0.025,\quad Q_\phiv=0.087\pm0.001;\qquad
Q_J+Q_\phiv=0.137\pm0.03.
\equn{(7.2.22a)}$$
This means that $Q_G$ is large; in fact, 
on comparing with the experimental value, \equn{(7.2.20)}, 
we find
$$Q_G=0.175\pm0.03.
\equn{(7.2.22b)}$$

The sum of $Q_J$ and $Q_\phiv$ substantially {\sl underestimates} 
the value of the mixed scalar square radius: 
the true phase $\delta(t)$ of the form factor would have to
 go on growing 
a lot before  
setting to the asymptotic regime (7.2.8). 
 The size of the phase necessary 
to produce the large $Q_G$ required appears excessive. 

An alternate possibility is the existence of a lower  energy resonance (or enhancement; 
we denote it by $\kappa$),  
{\sl below} the $K^*(1430)$, which some analyses suggest,\fnote{From 
a theoretic analysis, Oller, Oset and 
Pel\'aez (1999) give $M_\kappa\simeq1.01\,\gev$; 
the experimental analysis of Aitala et al.~(2002) gives 
$M_\kappa=0.80\,\gev$, $\gammav_\kappa=400\pm100\,\mev$. 
Note that (7.2.23) should be interpreted as an 
{\sl effective} parametrization; the 
experimental phase does not cross 90\degrees\ at $t=M^2_\kappa$.}
with $M_\kappa\sim1\,\gev\,\mev$ and $\gammav_\kappa=400\pm100\,\mev$. 
In this case, we approximate the low energy phase, $s\leq s_0=1\,\gev^2$, by writing
$$\cot\delta_0^{(1/2)}(t)=\dfrac{t^{1/2}}{2q}(1-s/M^2_\kappa)B_\kappa,\quad
B_\kappa=1.8\pm0.5
\equn{(7.2.23)}$$
and find
$$
Q_J=0.07\pm0.03,\quad Q_\phiv=0.18\pm0.006;\qquad
Q_J+Q_\phiv=0.25\pm0.04,
\equn{(7.2.24a)}$$
which reproduces well the experimental number with a small $Q_G$, compatible with zero: 
$$Q_G\sim 0.06\pm0.07.
\equn{(7.2.24c)}$$
It would thus seem that the experimental data on 
$\langle r^2_{{\rm S},K\pi}\rangle$ supports the existence of this $\kappa$  
enhancement.

\booksection{7.3. Dispersion relations and Roy equations}

\noindent A possible way to improve the quality of the analysis of experimental
 data is to use what are known
as {\sl dispersion relations}, either at fixed $t$ or in the form 
of the so-called Roy equations. We start with the first. 

\booksubsection{7.3.1. Fixed $t$ dispersion relations}

\noindent The analyticity properties of $F(s,t)$, as discussed in \sect~2.1,
 imply that we can write a Cauchy
representation for it, fixing $t$ and allowing $s$ to be complex. 
Starting with $s\to s+\ii\epsilon$, $s$ positive and $\epsilon>0,\,\epsilon\to0$, we have
$$F(s+\ii\epsilon,t)=\dfrac{1}{\pi}\int_{4\mu^2}^\infty\dd s'\,
\dfrac{A_s(s',t)}{s'-(s+\ii\epsilon)}+\dfrac{1}{\pi}\int_{4\mu^2}^\infty\dd s'\,\dfrac{A_u(s',t)}{s'-u}.$$
Here $A_s(s',t)=(1/2\ii)\{F(s'+\ii\epsilon,t)-F(s'-\ii\epsilon,t)\}$ 
is the so-called {\sl absorptive} part of the scattering amplitude across the 
right hand cut, which actually equals $\imag F(s',t)$. 
$A_u$ is the corresponding quantity connected  with the left hand cut. 
Taking $\epsilon=0$ above we find a relation between the 
{\sl dispersive} part of $F$, $D(s,t)$, which coincides with its real part, 
and the $A_{s,u}$. For $s$ physical this reads
$$\real F(s,t)=D(s,t)=\dfrac{1}{\pi}\pepe\int_{4\mu^2}^\infty\dd s'\,
\dfrac{A_s(s',t)}{s'-s}+\dfrac{1}{\pi}\int_{4\mu^2}^\infty\dd s'\,\dfrac{A_u(s',t)}{s'-u}
\equn{(7.3.1)}$$
($\pepe$ denotes Cauchy's principal part of the integral).
This is the fixed $t$ dispersion relation.

Actually, and because, in many cases, the $A(s,t)$ grow 
 with $s$, (7.3.1) is divergent. 
This is repaired by {\sl subtractions}; that is to say,
 writing the Cauchy representation not for $F$ itself,
but for 
$F(s,t)/(s-s_1)$ where  $s_1$ is a convenient subtraction point, usually taken
 to coincide with a  threshold. 
This introduces a constant in the equations (the value of $F(s,t)$ at $s=s_1$). 
Rewriting our equations with the appropriate subtraction incorporated
 is a technical problem, that we leave for the
reader to take into account; 
for the important case of forward dispersion relations we will 
perform explicitly the subtractions in next subsection.

Let us rewrite the dispersion relation 
in a form such that we separate out 
the  high energy contribution. 
We have
$$D(s,t)=\dfrac{1}{\pi}\pepe\int_{4\mu^2}^{s_0}\dd s'\,
\dfrac{A_s(s',t)}{s'-s}+\dfrac{1}{\pi}\int_{4\mu^2}^{s_0}\dd s'\,\dfrac{A_u(s',t)}{s'-u}+
V(s,t;s_0)
\equn{(7.3.2a)}$$
and
$$V(s,t;s_0)=\dfrac{1}{\pi}\int_{s_0}^\infty\dd s'\,
\dfrac{A_s(s',t)}{s'-s}+\int^{\infty}_{s_0}\dd s'\,\dfrac{A_u(s',t)}{s'-u};
\equn{(7.3.2b)}$$
we are assuming $s<s_0$.  
Both $D$ and the $A$ may be written in terms of the {\sl same} set of phase shifts
by expanding them in partial waves:\fnote{We are actually simplifying a little; 
(7.3.3) should take into account the isospin structure of $s$ and $u$ channels, 
which the reader may find in e.g. the text of Martin, Morgan and Shaw (1976),
or one can consider that we are studying $\pi^0\pi^+$ or  $\pi^0\pi^0$
 scattering, for which 
$s$ and $u$ channels are identical. 
Also, we are assuming that there is no appreciable inelasticity 
below $s_0$.}
$$A(s,t)=\dfrac{2s^{1/2}}{\pi k}\sum_{l=0}^\infty(2l+1)P_l(\cos\theta)\sin^2\delta_l(s),
\equn{(7.3.3a)}$$ 
$$D(s,t)=\dfrac{2s^{1/2}}{\pi k}\sum_{l=0}^\infty(2l+1)P_l(\cos\theta)\cos\delta_l(s)\sin\delta_l(s).
\equn{(7.3.3b)}$$

These equations provide {\sl constraints} for the phase shifts 
provided one knows (or has a reliable model) for the 
high energy term, $V(s,t;s_0)$. 
They enforce analyticity and $s\leftrightarrow u$ crossing symmetry, but not 
 $s\leftrightarrow t$ or  $t\leftrightarrow u$ crossing. 
This is very difficult to implement completely, as it would require analytical continuation, 
but 
a partial verification is possible through the Froissart--Gribov representation 
that we will discuss in \sect~7.5.

\booksubsection{7.3.2. Forward dispersion 
relations}

\noindent By far the more 
important case of dispersion relations is that in which we take $t=0$ ({\sl forward dispersion 
relations}), which we discuss now in some detail.

Let us denote by $F_{\rm o}(s,t)$ to a scattering amplitude which is odd 
under the exchange of $s\leftrightarrow u$, and by $F_{\rm e}(s,t)$ 
to an even one. 
An example of the first is the amplitude corresponding to isospin $I_t=1$ in the $t$ channel,
$$F^{(I_t=1)}=\tfrac{1}{3}F^{(I_s=0)}+ \tfrac{1}{2}F^{(I_s=1)}- \tfrac{5}{6}F^{(I_s=2)}.
\equn{(7.3.4)}$$
Examples of even amplitudes are those for $\pi^0\pi^0\to\pi^0\pi^0$,  
 $\pi^0\pi^+\to\pi^0\pi^+$:
$$\eqalign{
F_{0+}\equiv&\,
F(\pi^0\pi^+\to\pi^0\pi^+)=\tfrac{1}{2}F^{(I_s=1)}+\tfrac{1}{2}F^{(I_s=2)},
\cr
F_{00}\equiv&\,
F(\pi^0\pi^0\to\pi^0\pi^0)=\tfrac{1}{3}F^{(I_s=0)}+\tfrac{2}{3}F^{(I_s=2)}.\cr
}
\equn{(7.3.5a)}$$
In terms of isospin in the $t$ channel we have
$$F_{0+}=\tfrac{1}{3}F^{(I_t=0)}-\tfrac{1}{3}F^{(I_t=2)},
\quad F_{00}=\tfrac{1}{3}F^{(I_t=0)}+\tfrac{2}{3}F^{(I_t=2)}.
\equn{(7.3.5b)}$$
Because there are three isospin states for pions, the three amplitudes 
$F^{(I_t=1)}$, $F_{\pi^0\pi^+}$ and $F_{\pi^0\pi^0}$  form a complete set.
 
For odd amplitudes we may profit from the antisymmetry to write a Cauchy representation for  
$F_{\rm o}(s,0)$ and obtain
$$\real F_{\rm o}(s,0)=\dfrac{2s-4\mu^2}{\pi}\,\pepe\int_{4\mu^2}^\infty\dd s'\,
\dfrac{\imag F_{\rm o}(s',0)}{(s'-s)(s'+s-4\mu^2)}.
\equn{(7.3.6)}$$
The integral is convergent. 
As discussed in \sect~2.4,  Regge theory implies, for 
$F_{\rm o}(s,t)=F^{(I_t=1)}(s,t)$, the behaviour 
$$F_{\rm o}(s,t)\simeqsub_{s\to\infty}C s^{\alpha_{\rho}(0)+\alpha'_{\rho} t},
\qquad\alpha_{\rho}(0)\simeq0.52,\quad \alpha'_{\rho}\simeq 1\,{\gev}^{-2}.$$

For even amplitudes we have to {\sl subtract}, i.e., 
consider combinations $[F_{\rm e}(s,t)-F_{\rm e}(\hat{s},t)]/(s-\hat{s})$, 
where $\hat{s}$ is a convenient energy squared, usually taken in the range $0<\hat{s}\leq4\mu^2$. 
Two popular choices are the $s\leftrightarrow u$ symmetric point, $\hat{s}=2\mu^2$, and 
threshold, 
$\hat{s}=4\mu^2$. 
We then get the equations, respectively,
$$\eqalign{
\real F_{\rm e}(s,0)=&\,F_{\rm e}(2\mu^2,0)\cr
+&\,\dfrac{(s-2\mu^2)^2}{\pi}
\pepe\int_{4\mu^2}^\infty\dd s'\,
\dfrac{\imag F_{\rm e}(s',0)}{(s'-2\mu^2)(s'-s)(s'+s-4\mu^2)}\cr
}
\equn{(7.3.7a)}$$
and
$$\eqalign{
\real F_{\rm e}(s,0)=&\,F_{\rm e}(4\mu^2,0)\cr
+&\,\dfrac{s(s-4\mu^2)}{\pi}
\pepe\int_{4\mu^2}^\infty\dd s'\,
\dfrac{(2s'-4\mu^2)\imag F_{\rm e}(s',0)}{s'(s'-4\mu^2)(s'-s)(s'+s-4\mu^2)}.\cr
}
\equn{(7.3.7b)}$$
These integrals are  convergent; the behaviour expected from Regge theory is now
$$F_{\rm e}(s,t)\simeqsub_{s\to\infty}
C s^{1+\alpha'_P t},\quad \alpha'_P\simeq 0.11\,\; {\gev}^{-2}.$$
Actually, the convergence of (7.3.7) 
may be proved to follow in a general local field theory.

For a variety of other types of forward dispersion relations, 
see the article of Morgan and Pi\u{s}ut~(1970) or the text of Martin, Morgan and Shaw~(1976).

\booksubsection{7.3.3. The Roy equations}

\noindent
Eqs.~(7.3.2), (7.3.3) look rather cumbersome. Roy~(1971) remarked that they 
appear simpler if we project (7.3.2)  into partial waves: 
one finds the {\sl Roy equations}
$$\cos\delta_l(s)\sin\delta_l(s)=\sum_{l'=0}^\infty\int_{4\mu^2}^{s_0}\dd s'\,
K_{ll'}(s,s')\sin^2\delta_{l'}(s)+V_l(s;s_0).
\equn{(7.3.8)}$$
Here the kernels $K_{ll'}$ are known and the $V_l$ are the (still unknown) 
projections of $V$.

Eq.~(7.3.8) is valid in 
the simplified case we are considering here, i.e., without subtractions. 
If we had subtractions, the fixed $t$ dispersion 
relations would acquire an extra term, a function $g(t)$. 
This may be eliminated, using crossing symmetry, 
in favour of the S wave 
scattering lengths. 
\equn{(7.3.8)} would be modified accordingly.

It should be clear that there is no physics ingredient entering the Roy equations 
that is not present in the fixed $t$ dispersion relation, plus 
partial wave expansions; (7.3.8) is strictly equivalent to the pair (7.3.2b) and (7.3.3). 
(In fact, there is some loss of information when using the Roy equations:
 in (7.3.2b) we can also require 
agreement between the integral and the real part, at high energies, using also Regge 
theory to evaluate the last).

Roy equations became fashionable in the early seventies, but were soon abandoned; 
not only high energy physicists found other,  more interesting, 
fish to fry, but it soon became obvious 
that they produced no better results than 
dispersion relations and a straightforward phase shift analysis 
in which one parametrizes the $\delta_l$ in a way compatible with 
analyticity. 
 There are several reasons why this is so. 
First of all (7.3.8) (say) are highly nonlinear integral and matrix equations, 
and it is not clear that a solution to them exists {\sl for a general} set of  $V_l$. 
Solutions are known to exist in some favorable cases; but this constitutes the second 
problem: there are too many of them.\fnote{As a simple example, 
consider the toy model in Chapter~4. 
One can add to  the interaction with the rho 
a new term: either an interaction with a scalar field, 
$g_\sigma\rvec{\pi}\,\rvec{\pi}\sigma$, or a quartic interaction, 
$\lambda_1(\rvec{\pi}\,\rvec{\pi})^2+\lambda_2(\rvec{\pi}\times\rvec{\pi})^2$.
Both are renormalizable field theoretical models, 
therefore they will satisfy unitarity, Roy's equations,
 crossing sum rules and the whole
kit-and-caboodle as accurately as one may wish by going to high enough orders in the coupling. 
Moreover, they fit reasonably well the P wave and, by 
 tuning the parameters $g_\rho$ and $g_\sigma$, $\lambda_i$ 
one can get any desired values for  $a_0^{(0)}$,  $a_0^{(2)}$; 
yet the two models  give very different scattering amplitudes: 
one does, and the other does not have 
a scalar resonance. 
The statement, found at times in the literature, that the Roy equations plus the 
S wave scattering lengths fix the low energy scattering amplitude is plain nonsense.
Fit to experiment is essential!} 
In fact, Atkinson~(1968) proved a long time ago that, for any arbitrary $V(s,t;s_0)$ 
such that it is sufficiently smooth and decreasing at infinity, 
one can obtain by iteration a solution not 
only of the Roy equations, but of the full Mandelstam representation and 
compatible with inelastic unitarity for {\sl all} $s$ as well. 
Therefore, the solutions to the  equations (7.3.8) are ambiguous in an 
unknown function; only the fact that the phase shifts fit experiment 
really constrains the solution. 
Indeed, it was found in the middle seventies that solutions of the Roy equations
 with suspiciously small 
errors simply reflected the prejudice  
as to what is a {\sl reasonable} $V(s,t;s_0)$ and about which sets of experimental 
phase shifts one ought to fit. 
Recently, the Roy equations have been resuscitated thanks to 
the appearance of new experimental data that allow 
more meaningful constraints.

From a practical point of view, the Roy equations present two further drawbacks (with respect to 
the method of parametrizations based on the effective range formalism, plus  
straight dispersion relations). First, 
 they mix various waves and, hence, transmit uncertainties of (say) the S-waves to 
other ones, and they require information on the medium and high energy regions 
($s\gsim1\,\gev^2$) where the 
mixing of $\pi\pi$ with channels such as $\bar{K}K$ is essential. 
Second, in the integrals in the r.h. side in (7.3.8) 
we have to project over partial waves, 
hence integrate with Legendre polynomials 
which, for $l$ larger than 1, 
oscillate and thus create unstabilities, which are difficult to control,
 for the D and higher waves.

Note, however, that this should not be taken as criticism of 
the use of Roy's equations, 
that provide a  useful tool to analyze $\pi\pi$ scattering. 
In the present review, however, we prefer to concentrate on other 
methods:
 we  leave the implementation of the Roy 
equations (and of fixed $t$ dispersion relations, except 
in a few simple cases) outside the scope of 
these notes. The interested reader may consult the classic papers
 of Basdevant, Froggatt and Petersen~(1972,~1974), Pennington~(1975) or, 
more recently, the very comprehensive articles of 
 Colangelo, Gasser  and Leutwyler (2001), Ananthanarayan et al.~(2001) and 
Descotes et al.~(2002).

\booksection{7.4. Evaluation of the forward dispersion relation for $\pi\pi$ 
 scattering}
\noindent 
As  examples of application of forward dispersion relations 
we will evaluate here (7.3.7a) for the scattering $\pi^0\pi^+\to\pi^0\pi^+$ 
and  $\pi^0\pi^0\to\pi^0\pi^0$, subtracted at   
 $s=4M_{\pi}^2$, and the Olsson sum rule, connected with the $F^{(I_t=1)}$ amplitude.

\booksubsection{7.4.1. The Olsson sum rule}

\noindent
We will first consider the forward dispersion relation for the 
odd amplitude under $s\leftrightarrow u$, $F^{(I_t=1)}$, 
given in (7.3.6). 
At $s=4M_{\pi}^2$, 
we have $F^{(I_t=1)}=(8M_{\pi}/\pi)(\tfrac{1}{3}a_0^{(0)}-\tfrac{5}{6}a_0^{(2)})$ 
and we find the so-called Olsson~(1967) sum rule,
$$2a_0^{(0)}-5a_0^{(2)}=D_{\rm Ol.},\quad
D_{\rm Ol.}=3M_{\pi}\int_{4M_{\pi}^2}^\infty \dd s\,
\dfrac{\imag F^{(I_t=1)}(s,0)}{s(s-4M_{\pi}^2)}.
\equn{(7.4.1)}$$
If we take $2a_0^{(0)}-5a_0^{(2)}$ from (6.4.11) and (6.4.1c) we get 
$$2a_0^{(0)}-5a_0^{(2)}=(0.691\pm 0.042)\,\times M_{\pi}^{-1}.
\equn{(7.4.2)}$$ 
On the other hand, evaluating the dispersive integral also with the parametrizations 
 (6.4.11) and (6.4.1), the remaining waves as given in \sect~6, 
and the high energy contribution ($s^{1/2}\geq1.42\,\gev$) 
with the Regge formulas of \subsect~7.3.4, we find  $D_{\rm Ol.}=(0.659\pm0.020)\,\times M_{\pi}^{-1}$. 
Although the results are compatible within errors, the central values are certainly displaced. 
One may argue that this displacement is due to a bias of $a_0^{(2)}$, the only important quantity 
that was obtained fitting data with one pion off mass shell. 
If we accordingly repeat the fit to the S2 wave, {\sl including} 
fulfillment of the Olsson sum rule into the fit, we find the parameters for this wave reported in 
(6.4.2). 
With them we have (7.4.2) replaced by
$$2a_0^{(0)}-5a_0^{(2)}=(0.671\pm0.023)\,M_{\pi}^{-1}.
\equn{(7.4.3a)}$$
For the dispersive integral we then find
(we now present the results in detail)
\smallskip
$$\matrix{
{\rm PY},\;{\rm dispersive}&\cr
0.431\pm0.016&\quad\hbox{[PY: S, P, $\;s^{1/2}\leq0.82\;\gev$]}\cr
             0.148\pm0.004&\quad\hbox{[Rest, $s^{1/2}\leq 1.42\,\gev$
 (incl., D, F below 0.82 \gev)]}\cr
            0.073\pm0.010&\quad\hbox{[Regge, $\rho\; \;s^{1/2}\geq1.42\;\gev$]}\cr
             0.010\pm0.003&\quad\hbox{[Regge, Bk;$\;s^{1/2}\geq1.42\;\gev$]}\cr
\phantom{\Bigg{|}}             0.664\pm0.018&\quad\hbox{[Total, disp.]}\cr
}
\equn{(7.4.3b)}$$
We call ``Rest"  to the contributions  of the 
D, F waves below $1.42\,\gev$, plus the S, P waves between 
0.82 and 1.42 \gev. Of this ``Rest",  the largest 
contribution comes from the 
D0 and P waves. The piece labeled ``Regge, Bk" is a background to the 
rho Regge pole; see Pel\'aez and Yndur\'ain~(2003).

For future reference, we give the results we would have obtained using, 
for the S, P waves at energies below 0.82 \gev, 
the phase shifts of Colangelo, Gasser and Leutwyler~(2001), that we denote by CGL, 
or those of Descotes et al.~(2002). 
For the first, we have
$$2a_0^{(0)}-5a_0^{(2)}=(0.663\pm0.007)\,\times M_{\pi}^{-1}\quad\hbox{[CGL]},
\equn{(7.4.4a)}$$
 $$D_{\rm Ol.}=(0.632\pm0.014)\,\times M_{\pi}^{-1}.
\equn{(7.4.4b)}$$
We will later discuss the reasons for this mismatch. 
For the ones of Descotes et al.~(2002),
$$2a_0^{(0)}-5a_0^{(2)}=(0.646\pm0.031)\,\times M_{\pi}^{-1}\quad\hbox{[Descotes et al.]},
\equn{(7.4.5a)}$$
 $$D_{\rm Ol.}=(0.666\pm0.010)\,\times M_{\pi}^{-1},
\equn{(7.4.5b)}$$
and the error in the second only takes into account the error in the Regge contribution.
The phase shifts of Descotes et al. are perfectly compatible with the Olsson sum 
rule and standard Reggeistics.

\booksubsection{7.4.2. $\pi^0\pi^0$}

\noindent
Next, we consider the forward dispersion 
relation for  $\pi^0\pi^0$ scattering, subtracted at the symmetric point $2M^2_\pi$.
 We have, with $F_{00}(s)$ the forward $\pi^0\pi^0$ amplitude,
$$F_{00}(4M_{\pi}^2)=F_{00}(2M_{\pi}^2)+D_{00},\quad D_{00}=\dfrac{4M_{\pi}^4}{\pi}\int_{4M_{\pi}^2}^\infty\dd s\,
\dfrac{\imag F_{00}(s)}{s(s-2M_{\pi}^2)(s-4M_{\pi}^2)}.
\equn{(7.4.6)}$$
In a first approximation we neglect the dispersive integral,
and then get the approximate sum rule
$$\dfrac{8M_{\pi}}{3\pi}\left(a_0^{(0)}+2a_0^{(2)}\right)=F_{00}(4M_{\pi}^2)\simeq F(2M_{\pi}^2)
\simeq\tfrac{2}{3}f_0^{(0)}(2M_{\pi}^2).
$$
The 
$\pi^0\pi^0$ amplitude contains, in the 
S wave, an $I=2$ component. This we will fix 
as given by (6.4.2). Likewise, we fix the D waves as given by the 
parametrizations of \sect~6.4. Finally, 
 for the S wave with $I=0$, we  take  the parameters of (6.4.11).
Then $F_{00}(4\mu^2)=0.093$ and $F_{00}(2\mu^2)=0.072$, reasonably close.

A more precise test requires that we evaluate $D_{00}$. 
At high energy, i.e., for $s^{1/2}\geq1420\,\mev$, 
we use the Regge expression for $\imag F_{00}$. 
The amplitude for exchange of isospin 2 is evaluated as in 
Pel\'aez and Yndur\'ain~(2003). 
The bulk of the contribution to $D_{00}$ is that of
 the S wave with $I=0$:

$$\matrix{
&D_{00}\vphantom{\Bigg|}&\hbox{Contribution}\cr
&            37.84\times 10^{-3}&\quad\hbox{S0}\cr
&            3.12\times 10^{-3}&\quad\hbox{S2}\cr
&            0.65\times 10^{-3} &\quad  \hbox{D0}\cr
&            0.06\times 10^{-3} &\quad  \hbox{D2}\cr
&            0.046\times 10^{-3}&\quad\hbox{[Regge, Pomeron $\; \;s^{1/2}\geq1.42\;\gev$]}\cr
&            0.006\times 10^{-3}&\quad\hbox{[Regge, $P'$;$\;s^{1/2}\geq1.42\;\gev$]}\cr
&            0.005\times 10^{-3}&\quad\hbox{[Regge, $I=2$;$\;s^{1/2}\geq1.42\;\gev$]}\cr
&\phantom{\Bigg{|}}            41.73\times 10^{-3}&\quad\hbox{[Total]}\cr
}
\equn{(7.4.7a)}$$
This is to be compared with
$$\eqalign{
F_{00}(4M^2_\pi,0)-F_{00}(2M^2_\pi,0)=&\,[(123.6\pm8.5)-(79.1\pm7.7)]\times 10^{-3}\cr
=&\,
(44.5\pm11.4)\times 10^{-3}:\cr
}
\equn{(7.4.7b)}$$
we find full overlap.
 
\booksubsection{7.4.3. $\pi^0\pi^+$}

\noindent 
We write $F_{0+}(s)\equiv F_{\pi^0\pi^+}(s,0)$ and so we have
$$F_{0+}(4M_{\pi}^2)=\dfrac{4M_{\pi}}{\pi}a_0^{(2)}=F_{0+}(2M_{\pi}^2)+D_{0+},
\equn{(7.4.8a)}$$
where the dispersive integral is
$$D_{0+}=\dfrac{4M_{\pi}^4}{\pi}
\int_{4M_{\pi}^2}^\infty\dd s\,\dfrac{\imag F_{0+}(s)}{s(s-2M_{\pi}^2)(s-4M_{\pi}^2)}.
\equn{(7.4.8b)}$$
Before making a detailed evaluation we will make a quantitative one. 
Because the scattering lengths both for S2 and P waves are very small, and the 
{\sl imaginary} parts of the amplitudes are (at low energy) proportional to the scattering 
lengths squared, we can, in a first approximation, 
neglect $D$ altogether. Moreover, for $F_{0+}(2M_{\pi}^2)$, 
the S2 wave is very near its zero. If we therefore  
neglect it we have the approximate sum rule
$$\dfrac{4M_{\pi}}{\pi}a_0^{(2)}\simeq3f_1(2M_{\pi}^2).
\equn{(7.4.9)}$$
Using the parametrization for the P wave (\subsect~6.3.1) which, it 
will be remembered, converges down to the left hand cut, $s=0$, we find
$3f_1(2M_{\pi}^2)=-0.0742$ and thus the scattering length $a_0^{(2)}=-0.058\;M_{\pi}^{-1}$, 
a very reasonable number, agreeing with what we deduced from $\pi\pi$ scattering data and, 
to a 20\%, with what is expected in 
chiral perturbation theory.

We next proceed to a more accurate evaluation, for which we use the same
 input as in the previous subsection.
The calculation is now more precise because $D_{0+}$ is dominated by the P wave, very well
known.  We find, for the dispersive evaluation,
$$\matrix{
&D_{0+}\vphantom{\Bigg|}&\hbox{Contribution}\cr
&            2.339\times 10^{-3}&\quad\hbox{S2}\cr
&            7.928\times 10^{-3}&\quad\hbox{P}\cr
&            0.044\times 10^{-3} &\quad  \hbox{D2}\cr
&            0.007\times 10^{-3} &\quad  \hbox{F}\cr
&            0.046\times 10^{-3}&\quad\hbox{[Regge, Pomeron $\; \;s^{1/2}\geq1.42\;\gev$]}\cr
&            0.006\times 10^{-3}&\quad\hbox{[Regge, $P'$;$\;s^{1/2}\geq1.42\;\gev$]}\cr
&            -0.002\times 10^{-3}&\quad\hbox{[Regge, $I=2$;$\;s^{1/2}\geq1.42\;\gev$]}\cr
&\phantom{\Bigg{|}}            10.32\times 10^{-3}&\quad\hbox{[Total]}\cr
}
\equn{(7.4.10a)}$$
On the other hand, using directly the explicit parametrizations 
for the partial wave amplitudes in \sect~2, which are valid at and below threshold 
(provided $s>0$) one has
$$\eqalign{
F_{0+}(4M^2_\pi,0)-F_{0+}(2M^2_\pi,0)=&\,[(-53.7\pm2.8)+(67.9\pm2.6)]\times10^{-3}\cr
=&\,(14.2\pm4.0)\times 10^{-3}.\cr
}
\equn{(7.4.10b)}$$
This is  within less than $1\sigma$ from (7.4.10a).

The fulfillment of the dispersion relations with the values of the 
parameters we found in this and the previous Subsections is then,  for $\pi^0\pi^+$,
$\pi^0\pi^0$, and the Olsson sum rule very satisfactory; but  
perhaps the more impressive feature of the calculations is how little 
imposing the fulfillment of the dispersion relation
 affects the values of the parameters.\fnote{Note, however, that the fulfillment of 
the dispersion relations is slightly less good than what (7.4.7), (7.4.10) 
seem to imply. If correlations 
are taken into account, the sum rules are 
stisfied only at the level of $0.9\,\sigma$ and $1.2\,\sigma$ respectively.}
Those obtained from the fits to data, respecting the appropriate unitarity and 
analyticity requirements, wave by wave, are  
essentially compatible with the dispersion relations.

\booksection{7.5. The Froissart--Gribov representation and low energy\hb
 P, D, F wave parameters}
\vskip-0.5truecm
\booksubsection{7.5.1. Generalities}

\noindent
A reliable method to obtain the P and, especially, D and higher
  scattering lengths and effective range parameters,
 which incorporates simultaneously $s,\,u$ and $t$ crossing
symmetry, is the Froissart (1961)--Gribov (1962) representation, to which we now turn.
This method of analysis was developed long
 ago by Palou and Yndur\'ain~(1974) where, in particular, a rigorous proof of the 
validity of the equations (7.5.3,4) below may be found and, especially, 
by Palou, S\'anchez-G\'omez and Yndur\'ain~(1975), where a complete calculation of
 higher waves and 
effective range parameters was given. Also in the last  
reference  
the method is extended to evaluate the scattering
 lengths for the processes $\bar{K}K\to\pi\pi$.
The interest of the representation is that it ties together $s$, $u$ and $t$ channel 
quantities, without need of singular extrapolations.

Consider a $\pi\pi$ scattering amplitude, $F(s,t)$, symmetric 
or antisymmetric under the exchange $s\leftrightarrow u$, 
such as $\pi^0\pi^0$ or $\pi^0\pi^+$ (symmetric), or the amplitude with isospin
 1 in the $t$ channel
(antisymmetric).
  We may project $F(s,t)$  into the $l$th partial wave in the $t-$channel, 
which is justified for 
$t\leq4\mu^2$. We have,
$$f_l(t)=\tfrac{1}{2}\int^{+1}_{-1}\dd \cos\theta_t\,P_l(\cos\theta_t) F(s,t).
\equn{(7.5.1a)}$$
Here $\cos\theta_t=1+2s/(t-4\mu^2)$ is the $t$ channel scattering angle.
We then write a dispersion relation, in the variable $s$:
$$F(s,t)=\dfrac{1}{\pi}\int_{4\mu^2}^\infty\dd s'\,
\dfrac{\imag F(s',t)}{s'-s}+u\;\hbox{channel}.
\equn{(7.5.1b)}$$
We have not written subtractions that, for $l=1$ and higher do not alter anything, and 
we also have not written explicitly the $u$-channel contribution; 
 it simply multiplies
 by 2 the $s$ channel
piece, because,  for $t=4\mu^2$,  
$u$ and $s$ channel contributions to the final result are identical.

After substituting (7.5.1b) into (7.5.1a), the integral on 
$\dd \cos\theta_t$ can be made with the help of the formula
$$Q_l(x)=\tfrac{1}{2}\int^{+1}_{-1}\dd y\,\dfrac{P_l(y)}{x-y},$$
with $Q_l$ the Legendre function of the second kind. 
This produces the Froissart--Gribov representation,
$$f_l(t)=\dfrac{1}{2k_t^2}\dfrac{1}{\pi}\int_{4\mu^2}^\infty\dd s\,
\imag F(s,t)Q_l\left(\dfrac{s}{2k_t^2}+1\right)+u\;\hbox{channel},\qquad k_t=\dfrac{\sqrt{t-4\mu^2}}{2}.
$$
Taking now the limit $t\to4\mu^2$ in both sides of (6.3.4) and using that 
$$Q_l(z)\simeqsub_{z\to\infty} 2^{-l-1}\sqrt{\pi}\,\dfrac{\gammav(l+1)}{\gammav(l+\tfrac{3}{2})}z^{-l-1}$$
we find, in general,
$$\dfrac{f_l(t)}{k_t^{2l}}\eqsub_{t\to4\mu^2}\dfrac{\gammav(l+1)}{\sqrt{\pi}\,\gammav(l+3/2)}
\int_{4\mu^2}^\infty\dd s\,\dfrac{\imag F(s,t)}{(s+2k_t^2)^{l+1}}+\cdots.
\equn{(7.5.2)}$$
We then use the formula, that can be easily verified for $l\geq1$ from 
the effective range expression,
$$\dfrac{f_l(t)}{k_t^{2l}}\simeqsub_{t\to4\mu^2}\dfrac{4\mu}{\pi}\left\{a_l+k_t^2b_l\right\},
\equn{(7.5.3)}$$
to find the integral representation for $a_l,\,b_l$,
$$\eqalign{a_l=&\,\dfrac{\sqrt{\pi}\,\gammav(l+1)}{4\mu\gammav(l+3/2)}
\int_{4\mu^2}^\infty\dd s\,\dfrac{\imag F(s,4\mu^2)}{s^{l+1}},\cr
b_l=&\,\dfrac{\sqrt{\pi}\,\gammav(l+1)}{2\mu\gammav(l+3/2)}
\int_{4\mu^2}^\infty\dd s\,\left\{\dfrac{4\imag F'_{\cos\theta}(s,4\mu^2)}{(s-4\mu^2)s^{l+1}}-
\dfrac{(l+1)\imag F(s,4\mu^2)}{s^{l+2}}\right\}.\cr
}
\equn{(7.5.4)}$$
Here $F'_{\cos\theta}(s,4\mu^2)=(\partial/\partial\cos\theta)F(s,t)|_{t=4\mu^2}$, 
and an extra factor of 2 should be added to the l.h. 
side for 
identical particles (as occurs if 
the $a_l$, $b_l$ refer to a state with well defined isospin).
The method holds, as it is, for waves with 
$l=1$ and higher. For the S wave, the corresponding integrals are divergent;
 one thus needs subtractions and
the  method becomes much less useful.
We remark that the formulas (7.5.4) are valid, when $l=$ even, {\sl only} 
for amplitudes $F$ symmetric under $s\leftrightarrow u$ crossing, and, 
for $l=$ odd, for amplitudes $F$ which are antisymmetric.

For actual calculations we will, as before, replace $\mu$ 
by the charged pion mass, $M_\pi$.

\booksubsection{7.5.2. D waves}

\noindent
For the D wave scattering lengths, the  ones that we will 
calculate now,
$$a_2=\dfrac{4}{15M_{\pi}}\int_{4M_{\pi}^2}^\infty\dd s\,\dfrac{\imag F(s,4M_{\pi}^2)}{s^{3}}.
\equn{(7.5.5)}$$
For these D waves, we will consider the combinations 
$$a_{00}=2\left[\tfrac{1}{3}a_2^{(0)}+\tfrac{2}{3}a_2^{(2)}\right];
\quad
a_{0+}=2\left[\tfrac{1}{3}a_2^{(0)}-\tfrac{1}{3}a_2^{(2)}\right].
\equn{(7.5.6)}$$
They correspond, respectively, to the processes 
$\pi^0\pi^0\to\pi^0\pi^0$ and $\pi^0\pi^0\to\pi^+\pi^-$. 
The factor of 2 is due to the identity of the particles; 
it is introduced so that 
the projected amplitudes are $F_{0+}$, $F_{00}$ 
as given in (7.3.5) so, for example,
$$a_{0+}=\dfrac{4}{15M_{\pi}}\int_{4M_{\pi}^2}^\infty\dd s\,
\dfrac{\imag F_{0+}(s,4M_{\pi}^2)}{s^{3}}.
$$

We will here  illustrate the method with a detailed evaluation of $a_{0+}$ 
and  $a_{00}$; we start with the first.
The contribution of the high energy ($s^{1/2}\geq1.42\,\gev$) is obtained integrating 
(7.5.4) in that region with the Regge formulas of \sect~2.4. 
For the low energy pieces we  use, for the D, F waves and for the S, P waves above $s^{1/2}=0.82\,\gev$, 
the fits given in Chapter~6. For the S, P waves below 0.82 \gev, we 
use either the phase shifts of Colangelo, Gasser and Leutwyler~(2001), 
that we denote by CGL, or ours here, that we denote by PY (since they correspond to 
the calculation made in Pel\'aez and Yndur\'ain,~2003). 
We find, in units of $10^{-4}\;M_{\pi}^{-5}$
\smallskip
$$\matrix{
{\rm CGL},\;{\rm Froissart-Gribov}&
{\rm PY},\;{\rm Froissart-Gribov}\cr
8.43\pm0.09\quad\hbox{[CGL: S, P, $\;s^{1/2}\leq0.82\;\gev$]}&
8.09\pm0.15\quad\hbox{[PY S, P, ]}\cr
1.84\pm0.05\quad\hbox{[Rest, $s^{1/2}\leq 1.42\,\gev$]}&\cr
0.68\pm0.07\quad\hbox{[Regge, $I_t=0$]}\cr
-0.06\pm0.02\quad\hbox{[Regge, $I_t=2$]}\cr
\phantom{\Bigg{|}}10.90\pm0.13
\quad\hbox{[CGL, F.--G.]}&10.51\pm0.15
\quad\hbox{[PY, F.--G.]}\cr
}
\equn{(7.5.7)}$$

The value found with this method with the CGL phase shifts
 disagrees by several standard deviations with the value given in the paper of these authors
(Colangelo, Gasser and Leutwyler,~2001),
$$a_{0+}=(10.53\pm0.10)\times10^{-4}\;M_{\pi}^{-5}\;({\rm CGL}). $$ 
We will discuss this mismatch later on.

This combination we have calculated is the one that may be evaluated with less ambiguity; 
the values of other low energy  parameters depend 
substantially on the S wave scattering length. 
Thus, from the $\pi^0\pi^0$ scattering amplitude we  calculate the combination 
$$a_{00}=2\left[\tfrac{1}{3}a_2^{(0)}+\tfrac{2}{3}a_2^{(2)}\right].$$
We find, again in units of  $10^{-4}\;M_{\pi}^{-5}$,
\smallskip
$$\matrix{
{\rm CGL},\;{\rm Froissart-Gribov}&{\rm PY},\;{\rm  F.-G.}\cr
11.73\pm0.32\quad\hbox{[CGL S, P, $\;s^{1/2}\leq0.82\;\gev$]}&
12.24\pm0.62\quad\hbox{[PY S, P]}\cr
1.91\pm0.04\quad\hbox{[Rest, $s^{1/2}\leq 1.42\,\gev$]}&&\cr
0.68\pm0.07\quad\hbox{[Regge, $I_t=0$]}&\cr
0.12\pm0.04\quad\hbox{[Regge, $I_t=2$]}&\cr
\phantom{\Bigg{|}}14.44\pm0.33&14.95\pm0.65
\quad\hbox{[Total];}\cr
}
\equn{(7.5.8)}$$
The value given for this quantity in Colangelo, Gasser and Leutwyler~(2001) is
$$a_{00}=(13.94\pm0.32)\times10^{-4}\;M_{\pi}^{-5}.$$

The effective range parameters
$$b_{00}=2\left[\tfrac{1}{3}b_2^{(0)}+\tfrac{2}{3}b_2^{(2)}\right];
\quad
b_{0+}=2\left[\tfrac{1}{3}b_2^{(0)}-\tfrac{1}{3}b_2^{(2)}\right]
$$
 may also be calculated; we only give the final results, 
in units of  $10^{-4}\;M_{\pi}^{-7}$ now:
\smallskip
$$\eqalign{\matrix{
b_{0+}:&\;
\;\;[{\rm CGL},\;{\rm direct}];&
\quad\hbox{[CGL, F.--G.];}&
\quad\hbox{[PY, F.--G.];}\cr
&-0.189\pm0.016&-0.233\pm0.036&-0.170\pm0.083;\cr
}\cr
\matrix{\vphantom{{L^L}^{L^L}}
b_{00}:&\;
\;\;[{\rm CGL},\;{\rm direct}];&
\quad\hbox{[CGL, F.--G.];}&
\quad\hbox{[PY, F.--G.].}\cr
&-6.72\pm0.22&-6.61\pm0.23&-6.85\pm0.47\cr
}\cr}
\equn{(7.5.9)}$$
Here the quantities labeled ``CGL, direct" are obtained from the values
 given in  Colangelo,
Gasser and Leutwyler~(2001) for scattering lengths and ranges.

\booksubsection{7.5.3.  P and F waves}

\noindent
We use now (7.5.4), with $\imag F\equiv\imag F^{(I_t=1)}$. 
For the P wave scattering length, the integral is slowly 
convergent; the integrand behaves like $\imag F^{(I_t=1)}/s^2\sim s^{-1.48}$, and 
we do not have the factor $s-4M_{\pi}^2$ in the denominator of the Olsson sum rule 
that favoured low energies. 
Because of this, the details of the energy region $1.42\leq s^{1/2}\leq 1.80\,\gev$ 
are not negligible. 
We have represented $\imag F^{(I_t=1)}(s,4M_{\pi}^2)$ there by  a Regge 
formula, given by the rho 
trajectory. 
To this we could add the 
resonances 
$\rho(1450)$, $\rho(1700)$ and $\rho_3(1690)$, 
whose contribution is easily evaluated 
in the 
narrow width approximation,
$$\imag f\simeq \dfrac{2s^{1/2}}{\pi k}\pi M\gammav_{2\pi}\delta(s-M^2),$$
with $M$, $\gammav_{2\pi}$ the mass and two-pion width of the resonance. 
Their contribution is small. 
Alternatively, we may supplement the rho Regge piece with a smooth 
background; see for example 
Pel\'aez and Yndur\'ain~(2003); this is the method we will use. 
The contribution of this background is also small. 

We find the results, in units of $10^{-3}\,\times M_{\pi}^{-3}$,
\smallskip
$$\matrix{
{\rm CGL},\;{\rm Froissart-Gribov}&{\rm PY},\;{\rm F.-G.}&
\hbox{TY(St.+Sys.)}\cr
18.5\pm0.2\;\hbox{[CGL S, P, $\;s^{1/2}\leq0.82\;\gev$]}&
19.4\pm0.3&&\cr
9.1\pm0.3\quad\hbox{[Rest, $s^{1/2}\leq 1.42\,\gev$]}&&\cr
8.3\pm1.1\quad\hbox{[Regge, $\rho$]}&&\cr
1.0\pm0.3\quad\hbox{[Regge, Bk]}&&\cr
\phantom{\Bigg{|}}37.0\pm1.3
\quad\hbox{[Total]}&38.1\pm1.4&38.6\pm1.2.\cr
}
\equn{(7.5.10)}$$
The number labeled ``TY~(St.+Sys.)"  refers to what we  obtained in \subsect~7.2.1
 from  
the fit to the pion form factor.
 
As we see, there is good agreement, within errors, among
 all determinations, and also with the 
value given by Colangelo, Gasser and Leutwyler~(2001), 
$$a_1=(37.9\pm0.5)\times10^{-3}\,\times M_{\pi}^{-3},\;{(\rm CGL)}.$$
The value coming from the pion form factor, if we had not taken into account 
systematic errors, would have been
$$a_1=(40.6\pm1.4)\times10^{-3}\,\times M_{\pi}^{-3};$$
this is a bit too high, which is one of the reasons why we prefer the 
fit {\sl including} systematic normalization errors. 

For the effective range parameter, we find, in units of $10^{-3}\,\times M_{\pi}^{-5}$ 
now,
\smallskip
$$\matrix{
{\rm CGL},\;{\rm Froissart-Gribov}&{\rm PY},\;{\rm F.-G.}
&\hbox{TY(St.+Sys.)}\cr
-.92\pm0.05\;\hbox{[CGL S, P, $s^{1/2}\leq.82\;\gev$]}&
-0.57\pm0.10&&\cr
1.02\pm0.04\quad\hbox{[Rest, $s^{1/2}\leq 1.42\,\gev$]}&&&\cr
4.82\pm0.86\quad\hbox{[Regge, $\rho$]}&&&\cr
0.50\pm0.16\quad\hbox{[Regge, Bk]}&&&\cr
\phantom{\Bigg{|}}5.50\pm0.82\quad\hbox{[Total.]}&5.15\pm0.90
&4.47\pm0.29.\cr
}
\equn{(7.5.11)}$$
Here the Regge contribution is particularly important because the lower energy pieces cancel 
almost completely; we use, as we did in \equn{(7.5.10)},
 the quadratic expression for the rho trajectory. The
value following from the pion form factor,  without taking into account systematic errors, would be
 $(4.18\pm0.43)\times10^{-3}\,\times M_{\pi}^{-5}$.

The value obtained with the Froissart--Gribov representation and 
the  phases of  Colangelo, Gasser and Leutwyler~(2001), 
or that given by these authors, $b_1=(5.67\pm0.13)\times10^{-3}\,\times M_{\pi}^{-5}$, 
is $4\,\sigma$ away from the value obtained  from  
the fit to the pion form factor, labeled ``TY~(St. +Sys.)", or from the same
without systematic errors, which gave $b_1=(4.18\pm0.43)\times10^{-3}\,M_{\pi}^{-5}$.

We conclude the present Subsection with the 
F wave scattering length. 
Here the high energy part is negligible; 
we give only those contributions that are sizable. 
We have, if using the phase shifts given here,
$$
a_3=(6.00\pm0.07)\times10^{-5}\;M_{\pi}^{-5}\quad\hbox{(PY)}.
\equn{(7.5.12)}$$
This is  displaced by about $2\,\sigma$ from the value given 
by Colangelo, Gasser and Leutwyler~(2001):
$$
a_3=(5.60\pm0.19)\times10^{-5}\;M_{\pi}^{-5}\quad\hbox{(CGL)}.
$$

\booksubsection{7.5.4.  G waves}

\noindent
The scattering lengths and effective range parameters may be calculated for the G 
waves; we only give the values of the first:
$$a_4^{(0)}=(8.0\pm0.2)\times10^{-6}\,M_{\pi}^{-9},\quad
a_4^{(2)}=(4.5\pm0.2)\times10^{-6}\,M_{\pi}^{-9}.
\equn{(7.5.13)}$$ 

\booksection{7.6. Summary and conclusions}

\noindent
In this Section we discuss two recent alternate sets of phase shifts 
due to the Bern group and to Descotes et al.~(2002), and compare them 
between themselves and with
what we have found.

\booksubsection{7.6.1. The S, P waves of  Colangelo, Gasser and Leutwyler}

\noindent
In a recent  paper, Ananthanarayan et al. (2001) have 
made a detailed calculation of $\pi\pi$ scattering using 
experimental information and the Roy equations (but {\sl not} full information on 
the pion form factor). 
This calculation has been refined by Colangelo, Gasser and Leutwyler~(2001) 
who use Roy equations and chiral perturbation theory,  including (estimated) terms of order
$p^6$.  In the last paper, that we will here denote by CGL, these authors 
pretend to obtain an extremely precise representation of low energy 
$\pi\pi$ parameters that, in some cases, reaches beyond the 1\% level.

\topinsert{
\setbox0=\vbox{{\epsfxsize 11.8truecm\epsfbox{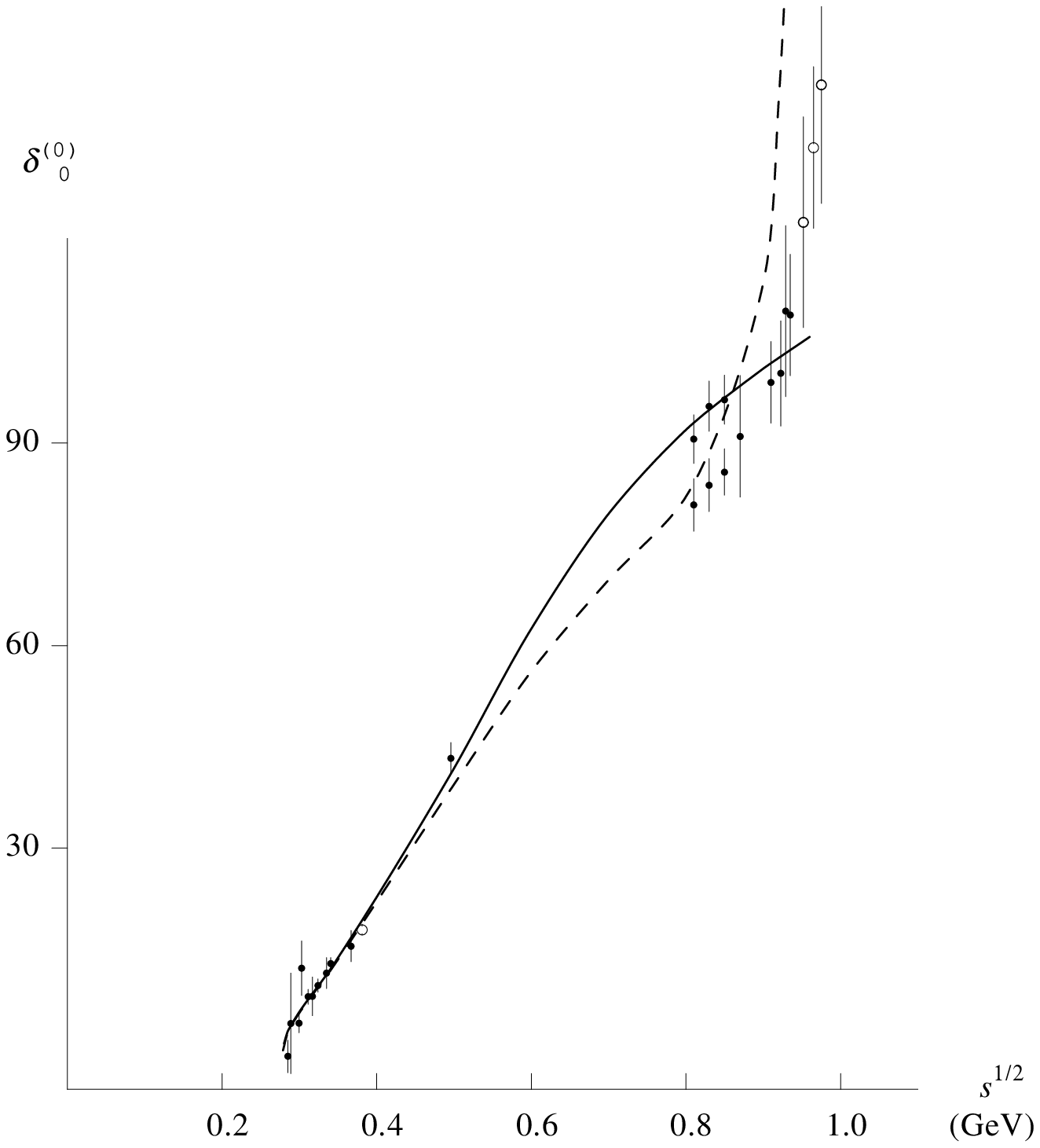}}} 
\setbox6=\vbox{\hsize 10truecm\captiontype\figurasc{Figure 7.6.1. }{
The  
$I=0$, $S$-wave phase shifts corresponding to Eq.~(6.4.11) (continuous line) 
and CGL, Eqs.~(7.6.1) (broken line). 
Note that CGL only fit up to 0.8 \gev, 
but we have represented the upper tail because its lashing out is very suggestive; see
text.}\hb} 
\centerline{\tightboxit{\box0}}
\bigskip
\centerline{\box6}
\medskip
}\endinsert

Unfortunately, and as made clear in particular in the articles of
 Pel\'aez and
Yndur\'ain~(2003), Yndur\'ain~(2003b) (and as we  will also see briefly here) 
the precision claimed by CGL is excessively optimistic. 
First of all, the high energy scattering amplitude that CGL 
(and also  Ananthanarayan et al.,~2001) use is excessively different from what 
standard Regge theory (or its QCD version) implies.\fnote{In section B.4 of their paper,
Ananthanarayan et al.~(2001)  
 explain that they take an asymptotic value  
for the total (Pomeron) cross 
section of $5\pm3$~mb. 
This is  a factor three smaller than the value implied 
by factorization, as discussed in our \sect~2.4 here. 
They also take a rho residue 50\% higher than the 
standard value and use  incorrect formulas for the slopes of the
Regge trajectories. 
The incompatibility of such Regge parameters with experimental data on $\pi\pi$ 
scattering makes all the results found by these authors very suspect.}  Secondly, it is not
clear that the accuracy of their chiral expansions  is what one can reasonably expect. 
Because of these reasons and, perhaps, also due to optimism of CGL when 
evaluating their errors, 
mismatches occur in the CGL phase shifts that reach the 2 to $4\,\sigma$ level. 
Some examples have already been seen; we come back to  them below.

CGL only present in their paper details for the S, P waves up to an energy of $0.82\,\gev$, 
in the form of the following parametrizations, whose form they have taken 
from  Schenk~(1991):
$$\eqalign{
\tan&\,\delta_l^{(I)}(s)=k^{2l}\sqrt{\dfrac{1-4M^2_\pi}{s}}\cr
\times&\,
\left\{A_l^I+B_l^Ik^2/M^2_\pi+C_l^Ik^4/M^4_\pi+D_l^Ik^6/M^6_\pi\right\}
\dfrac{4M^2_\pi-s_{lI}}{s-s_{lI}},\cr
}
\equn{(7.6.1a)}$$
$k=\sqrt{s/4-M^2_\pi}$, and the values of the parameters, as given by CGL, \equn{(17.2)},  are
$$\matrix{
A_0^0=0.220,&B_0^0=0.268,&C_0^0=-0.0139,&D_0^0=-1.39/10^3,\cr
A_0^2=-0.0444/10,&B_0^2=-0.0857,&C_0^2=-2.21/10^3,&D_0^2=-1.29/10^4,\cr
A_1=0.379/10,&B_1=0.140/10^4,&C_1=-6.73/10^5,&D_1=1.63/10^8,\cr
}
\equn{(7.6.1b)}$$
and
$$s_{00}=36.77\,M^2_\pi,\quad
s_{02}=-21.62\,M^2_\pi,\quad s_1=30.72\,M^2_\pi.
\equn{(7.6.1c)}$$
In  Appendix~B we will criticise this type of parametrization, which is very inefficient.

The values of the low energy parameters will be given below. 
The errors to \equn{(7.6.1)} may be found in an Appendix to the 
paper by Ananthanarayan et al. (2001).

As stated before, this CGL solution is slightly different from that given in 
Pel\'aez and Yndur\'ain~(2003), reproduced in the present notes: 
S0, S2 phase shifts are displaced with respect to the PY ones by  
about $1\, \sigma$ (\figs~7.6.1, 6.4.1), and the errors are about
 a half of the errors we have found 
with our direct fits.

An important fact to remark is that  Ananthanarayan et al. (2001) as well 
as CGL impose the (experimental) values of the 
phase shifts at the highest energy at 
which they cut off the Roy  equations, $s^{1/2}=0.8\,\gev$. 
The values they take for the S2, P waves are reasonable but, for the 
S0 wave they require$$\delta_0^{(0)}((0.8\,\gev)^2)=82.3\pm3.4\degrees.
$$
The error is a factor 2 or 3 times too small; cf. our discussion in 
\subsect~6.4.2. 
The fit is thus a forced fit, something that appears very clearly 
in \fig~7.6.1: the phase shoots  
up as soon as the energy is above $0.8 \gev$, where these authors constrain  
 it to go through the narrow corridor of a value of  
$82.3\pm3.4\degrees$.

\booksubsection{7.6.2. The S wave scattering lengths of Descotes et al., and 
Kami\'nski et al.}

\noindent
The Roy equations (and of course experimental data) have also 
been used in a recent paper by Descotes et al.~(2002) to 
find the low energy $\pi\pi$ scattering, especially the S0, S2 scattering lengths. 
The results these authors find are about two standard 
deviations different from those of  Colangelo, Gasser and Leutwyler~(2001), 
but more similar --both in central value and errors--
to the ones given in the present notes: with $M_\pi=1$, Descotes et al. give the values
$$a_0^{(0)}=0.228\pm0.012,\quad a_0^{(2)}=-0.0382\pm0.0038.
$$
The phase shifts of Descotes et al.~(2002) are also compatible 
with the Olsson sum rule evaluated with 
standard Regge behaviour, as we showed in \subsect~7.4.1.

A similar calculation, using also Roy equations, has been given by Kami\'nski, Le\'sniak and
Loiseau~(2003). These authors find 
$$a_0^{(0)}=0.224\pm0.013,\quad a_0^{(2)}=-0.0343\pm0.0036
$$
and the effective range parameters
$$b_0^{(0)}=0.252\pm0.011,\quad b_0^{(2)}=-0.075\pm0.015.
$$
These results are compatible with those of Descotes et al.~(2002), less so 
with our results here, and even less with the results of
  Colangelo, Gasser and Leutwyler~(2001). 
We will discuss more about this calculation below.

\booksubsection{7.6.3. Comparison of different calculations. 
Low energy parameters\hb for $\pi\pi$
scattering}

\noindent
In this Subsection we summarize and discuss the results 
obtained in the two last chapters. In them, 
we have found simple, explicit formulas that satisfy the general requirements of 
analyticity and unitarity and 
which fit well the experimental data; 
then we have verified that our solutions  
are compatible with a few crossing and 
analyticity constraints: forward dispersion relations at threshold and 
the Froissart--Gribov representation.

 It should be clear that we 
have not made an exhaustive analysis: 
nor was it intended. 
Thus, we have not tried to improve 
our parameters by fully imposing  
 consistency requirements. For example, it is easily verified that a change 
 in the parameters of the P and S0 waves to decrease (in the case of the first) 
the scattering length by of about \ffrac{1}{2}
$\sigma$ would improve agreement of the Froissart--Gribov and direct determinations of 
$a_1$. However, for the improvements to be more than cosmetic, we should   
 also consider dispersion relations in
a wider range of $s$ and 
$t$ values or, equivalently, 
the Roy equations.
This is the path followed by Ananthanarayan et al.~(2001), Colangelo, Gasser and
Leutwyler~(2001), and Descotes et al.~(2002),
  where the interested reader may find details. 
The results found in the first of these papers (which does not impose 
chiral perturbation theory) are compatible with ours, at the 1 to 1.5 $\sigma$ level. 
Also the errors are similar, with theirs generally smaller. 
The price they pay, however, is that all their numbers are correlated, 
whereas ours are not: in this sense, our results are more robust. 
The method of the Roy equations and ours here 
are complementary. 

\midinsert{
\bigskip
\setbox0=\vbox{
\setbox1=\vbox{\petit \offinterlineskip\hrule
\halign{
&\vrule#&\strut\hfil\ #\ \hfil&\vrule#&\strut\hfil\ #\ \hfil&
\vrule#&\strut\hfil\ #\ \hfil&
\vrule#&\strut\hfil\ #\ \hfil&\vrule#&\strut\hfil\ #\ \hfil\cr
 height2mm&\omit&&\omit&&\omit&&\omit&&\omit&\cr 
&\hfil \hfil&&\hfil DFGS \hfil&&ACGL&
&\hfil CGL\hfil&
&\hfil PY\hfil& \cr
 height1mm&\omit&&\omit&&\omit&&\omit&&\omit&\cr
\noalign{\hrule}
height1mm&\omit&&\omit&&\omit&&\omit&&\omit&\cr
&$a_0^{(0)}$&&\vphantom{\Big|}$0.228\pm0.012$&&$0.240\pm0.060$&
&\hfil$0.220\pm0.005$ \hfil&&
$0.230\pm0.010$& \cr 
\noalign{\hrule} 
height1mm&\omit&&\omit&&\omit&&\omit&&\omit&\cr
&\vphantom{\Big|}$a_0^{(2)}$ 
\phantom{\big|}&&\phantom{\Big|}$-0.0382\pm0.0038$&&$-0.036\pm0.013$&
&\hfil $-0.0444\pm0.0010$   \hfil&&\hfil 
$-0.0422\pm0.0022$& \cr 
height1mm&\omit&&\omit&&\omit&&\omit&&\omit&\cr
\noalign{\hrule} 
height1mm&\omit&&\omit&&\omit&&\omit&&\omit&\cr
&$b_0^{(0)}$&&\vphantom{\Big|}&&$0.276\pm0.006$&
&\hfil$0.280\pm0.001$ \hfil&&
$0.268\pm0.010$& \cr 
\noalign{\hrule} 
height1mm&\omit&&\omit&&\omit&&\omit&&\omit&\cr
&\vphantom{\Big|}$b_0^{(2)}$ 
\phantom{\big|}&&\phantom{\Big|}&&$-0.076\pm0.002$&
&\hfil $-0.080\pm0.001$   \hfil&&\hfil 
$-0.071\pm0.004$& \cr 
height1mm&\omit&&\omit&&\omit&&\omit&&\omit&\cr
\noalign{\hrule}}
\vskip.05cm}
\centerline{\box1}\medskip
\centerline{\sc Table~II}
\smallskip
\centerrule{5truecm}}
\centerline{\box0}
}\endinsert
\medskip

We will start the comparison with the S wave scattering lengths 
and effective rage parameters (Table~II). 
Here we compare the results of Descotes et al.~(2002), 
denoted by DFGS,  Ananthanarayan et al.~(2001) , denoted by ACGL, 
Colangelo, Gasser and Leutwyler~(2001), denoted by CGL, and 
the results in the present notes, denoted by PY.  
We use units with $M_\pi=1$.
As already remarked, the results of Descotes et al.~(2002) for the S waves
 are in disagreement, at the level of 
$1\,\sigma$ per wave, with those of CGL, 
but agree well with the results presented in the present work (PY). 
The effective range parameters, $b_0^{(I)}$, are also
 compared between ACGL, CGL and PY.

We next turn to higher waves. Here we have, in the 
case of the analysis of CGL, two results: those 
given in the paper of Colangelo, Gasser and Leutwyler~(2001), or those obtained 
using the Froissart--Gribov representation with the low energy S, P waves of CGL. 
 Both types of results are shown in Table~III.

\medskip
\smallskip
\setbox0=\vbox{
\setbox1=\vbox{\petit \offinterlineskip\hrule
\halign{
&\vrule#&\strut\hfil\ #\ \hfil&\vrule#&\strut\hfil\ #\ \hfil&
\vrule#&\strut\hfil\ #\ \hfil&
\vrule#&\strut\hfil\ #\ \hfil&\vrule#&\strut\hfil\ #\ \hfil\cr
 height2mm&\omit&&\omit&&\omit&&\omit&&\omit&\cr 
&\hfil \hfil&&\hfil Nagel \hfil&&PSGY&
&\hfil CGL\hfil&
&\hfil PY\hfil& \cr
 height1mm&\omit&&\omit&&\omit&&\omit&&\omit&\cr
\noalign{\hrule} 
height1mm&\omit&&\omit&&\omit&&\omit&&\omit&\cr
&$a_1$&&\vphantom{\Big|}$38\pm2$&&$38.5\pm0.6$&
&\hfil$37.0\pm0.13\; [37.9\pm0.5]^{\,\rm a}$ \hfil&&
$\displaystyle{38.1\pm1.4}\atop{\displaystyle[\,38.6\pm1.2]^{\,\rm b}\;\times\,10^{-3}}$& \cr 
\noalign{\hrule} 
height1mm&\omit&&\omit&&\omit&&\omit&&\omit&\cr
&\vphantom{\Big|}$b_1$ 
\phantom{\big|}&&\phantom{\Big|}&&$3.2\pm0.3$&
&\hfil ${\displaystyle 5.50\pm0.82}\atop{\displaystyle[5.67\pm0.13]^{\,\rm a}}$   \hfil&&\hfil 
${\displaystyle 5.15\pm0.90}\atop{\displaystyle [4.47\pm0.27]\;^{\rm b}\;\times\,10^{-3}}$& \cr
\noalign{\hrule} 
height1mm&\omit&&\omit&&\omit&&\omit&&\omit&\cr
&$a_2^{(0)}$&&\vphantom{\Big|}$17\pm3$&
&\hfil $18.5\pm0.6$  \hfil&&$17.5\pm0.3$&&$18.0\pm0.2$
$\;\times\,10^{-4}$& \cr
\noalign{\hrule} 
height1mm&\omit&&\omit&&\omit&&\omit&&\omit&\cr
&\vphantom{\Big|}$a_2^{(2)}$&&$1.3\pm3$&
&\hfil $1.9\pm0.6$ \hfil&&$1.70\pm0.13$&&$2.2\pm0.2$
$\;\times\,10^{-4}$& \cr
\noalign{\hrule}
height1mm&\omit&&\omit&&\omit&&\omit&&\omit&\cr
&\vphantom{\Big|}$a_{0+}$&&$10.5\pm3$&
&\hfil $11.07\pm0.52$ \hfil&&${\displaystyle10.90\pm0.13}\atop{\displaystyle[10.53\pm0.10]^{\,\rm a}}$&&
$10.51\pm0.15$
$\;\times\,10^{-4}$& \cr
\noalign{\hrule}
height1mm&\omit&&\omit&&\omit&&\omit&&\omit&\cr
&\vphantom{\Big|}$a_{00}$&&$13.1\pm5$&
&\hfil $14.9\pm0.8$ \hfil&&$ {\displaystyle 14.44\pm0.33}\atop{\displaystyle[13.94\pm0.32]^{\,\rm a}}$&&
$14.95\pm0.32$
$\;\times\,10^{-4}$& \cr
\noalign{\hrule}
height1mm&\omit&&\omit&&\omit&&\omit&&\omit&\cr
&\vphantom{\Big|}$b_{0+}$&&$ $&
&\hfil $-0.12\pm0.07$ \hfil&
&${\displaystyle-0.233\pm0.036}\atop{\displaystyle[-0.189\pm0.016]^{\,\rm a}}$&&$-0.170\pm0.083$
$\;\times\,10^{-4}$& \cr
\noalign{\hrule}
height1mm&\omit&&\omit&&\omit&&\omit&&\omit&\cr
&\vphantom{\Big|}$b_{00}$&&$ $&
&\hfil $-7.1\pm0.8$ \hfil&
&${\displaystyle-6.61\pm0.23}\atop{\displaystyle[-6.72\pm0.22]^{\,\rm a}}$&&$-6.85\pm0.47$
$\;\times\,10^{-4}$& \cr
\noalign{\hrule}
height1mm&\omit&&\omit&&\omit&&\omit&&\omit&\cr
&\vphantom{\Big|}$a_3$&&$6\pm2$&
&\hfil $6.2\pm0.9$ \hfil&&$5.60\pm0.19$&&$6.0\pm0.7$
$\;\times\,10^{-4}$& \cr
\noalign{\hrule}
height1mm&\omit&&\omit&&\omit&&\omit&&\omit&\cr
&\vphantom{\Big|} $b_3$&&\hfil $ $\hfil&&\hfil $-5.0\pm1.2$ \hfil&
&$-4.02\pm0.18$&&
\phantom{$-4.02\pm0.18$}$\times\,10^{-4}$& \cr
\noalign{\hrule}
height1mm&\omit&&\omit&&\omit&&\omit&&\omit&\cr
&\vphantom{\Big|} $a_4^{(0)}$&&\hfil $ $\hfil&&\hfil $0.9\pm0.1$ \hfil&
&$ $&&
$0.8\pm0.2\; \times\,10^{-5}$& \cr
\noalign{\hrule}
height1mm&\omit&&\omit&&\omit&&\omit&&\omit&\cr
&\vphantom{\Big|} $a_4^{(2)}$&&\hfil $ $\hfil&&\hfil $0.45\pm0.12$ \hfil&
&$ $&&
$0.45\pm0.2\; \times\,10^{-5}$& \cr
\noalign{\hrule}}
\vskip.05cm}
\centerline{\box1}
\medskip\noindent
Units of $M_\pi$. Nagel: Nagel et al.,~(1979); PSGY:
 Palou, S\'anchez-G\'omez and Yndur\'ain~(1975). 
(These two sets  are included for historical interest). CGL:
Colangelo,  Gasser and Leutwyler~(2001);  PY: 
the results  in Pel\'aez and Yndur\'ain~(2003), as given in the present review. 
The errors in PSGY only include the error due to 
variation of $a_0^{(0)}$ between $0.2$ and $0.3\,M_{\pi}^{-1}$; 
full errors could be a factor 2 larger.
\hb($^{\rm a}$) The numbers in braces are those given by 
CGL themselves; the numbers outside are from the Foissart--Gribov 
representation.\quad
($^{\rm b}$) Numbers in braces are  from the 
pion form factor (de~Troc\'oniz and Yndur\'ain,~2002). 
Other numbers in ``PY" are from the Froissart--Gribov projection.
\medskip
\centerline{\sc Table~III}
\smallskip
\centerrule{5truecm}}
\box0

We have also, not included in the Table~III,
 the results of 
 Kami\'nski, Le\'sniak and
Loiseau~(2003) for the P wave (the S waves were discussed already in \subsect~6.3.2). 
These authors give the numbers
$$a_1=(39.6\pm2.4)\times10^{-3}\,M_{\pi}^{-3},\quad
b_1=(2.63\pm0.67)\times10^{-3}\,M_{\pi}^{-3}.
$$
The central value for $a_1$ is too high, although it is compatible within its 
errors with other determinations (Table~III). 
The value of $b_1$, however, is almost $3\,\sigma$ below the lowest one 
in Table~III.  
The reason could be that
Kami\'nski, Le\'sniak and
Loiseau use some calculation techniques  (effective, separable potentials 
and Pad\'e approximants) which may bias their results.

We think the numbers in Table~III speak for themselves, but there are  
a few remarks that can be made.
First of all, we would also like to comment on the agreement of the 
old results of PSGY with 
the more modern determinations, both here and in CGL. 
This is noteworthy, and 
lends weight to the 
suitability of the 
Froissart--Gribov method to calculate low energy parameters for 
P, D and higher waves: compare for example PSGY with Nagel, in spite of the fact that the 
last is two year more modern (apparently, Nagel and collaborators were not aware of the 
PSGY evaluations). 
Secondly, we have not presented the results of  Ananthanarayan et al.~(2001). 
They are similar to those of CGL, but the errors are a factor $\sim2.5$ times larger.
Thirdly, we have the inconsistency of the results of CGL, both with 
themselves (different values for different determinations,
 as for $a_{0+}$, $a_{00}$) and with the 
TY value (for $b_1$), which reaches the $4\,\sigma$ level. 
This is discussed in detail in Pel\'aez and Yndur\'ain~(2003), Yndur\'ain~(2003b) and, together 
with the independent indications found by Descotes et al.,~(2002), indicates that the 
CGL solution is displaced, for some quantities,
 by about one to two sigma, and their error estimates are probably a factor 1.5 or 2 
too optimistic.

The reasons for this are, probably, a mixture of the following: 
first, use of an unrealistic set of Regge parameters 
and intermediate energy ($1.42\,\gev\leq s^{1/2}\leq 2\,\gev$) 
partial wave amplitudes. 
Secondly, the chiral perturbative expansions these authors use probably 
do not converge as fast as they assume. 
And thirdly, their experimental input is, at times, 
endowed with irrealistically small error estimates: 
see for example the discussion of this last in  Yndur\'ain~(2003b).

\bookendchapter

\brochureb{\smallsc chapter 8}{\smallsc qcd, pcac and chiral symmetry for pions and kaons}{93}
\bookchapter{8. QCD, PCAC and chiral symmetry for pions and kaons}
\vskip-0.5truecm
\booksection{8.1. The QCD Lagrangian. Global symmetries;
conserved currents}

\noindent
In the previous chapters we have discussed {\sl general} properties 
of pion interactions. 
In this  chapter\fnote{This chapter follows, to a large extent, 
the corresponding one in the text of the author, Yndur\'ain~(1999), where we 
send for more details on QCD.} we remember that pions are bound 
states of quark-antiquark, and that we have a theory of strong 
interactions, QCD. 
We will then study properties of 
pion physics which follow, in one way or another, from the QCD 
Lagrangian,
$${\cal L}=-\sum m_l\bar{q}_lq_l+\ii\sum_{l=1}^n\bar{q}_l\slash{D\!}q_l
-\tfrac{1}{4}(D\times B)^2+\hbox{gauge fixing + ghost terms},
\equn{(8.1.1)}$$
where $q_l$ is the quark operator for flavour $l$, $B$ is the gluon field operator, etc.;
sum over omitted colour indices is generally understood.
In (8.1.1) we assume that we have only {\sl light} 
quarks ($u,\,d$ and, eventually, $s$). The existence of 
heavy quarks has little influence in the physics of small momenta 
in which we are interested here.
 
In the present section we start with  the {\sl global} symmetries
of the QCD Lagrangian.  Since its form is unaltered by renormalization, we 
can neglect the distinction between bare and renormalized $\cal L$.

Clearly, $\cal L$ is invariant under Poincar\'e transformations, $x\rightarrow\Lambdav x+a$. 
The currents corresponding to (homogeneous) Lorentz 
transformations $\Lambdav$ are not of great interest for us here. Space-time 
translations generate the {\sl energy-momentum tensor}. Its form is fixed by Noether's 
theorem, which gives
$$\Thetav^{\mu\nu}=\sum_i \dfrac{\partial {\cal L}}{\partial(\partial_\mu \Phiv_i)}
\,\partial_\nu\Phiv_i-g^{\mu\nu}{\cal L},
\equn{(8.1.2a)}$$
and the sum over $i$ runs over all 
the fields in the QCD Lagrangian. These currents are conserved,
$$\partial_\mu\Thetav^{\mu\nu}=0,$$
and the corresponding ``charges" are the components of the four-momentum
$$P^\mu=\int\dd^3{\bf x}\,\Thetav^{0\mu}(x).$$
The explicit expression for $\Thetav^{\mu\nu}$ in QCD is
$$\eqalign{\Thetav^{\mu\nu}=&\,\ii\sum_q\bar{q}\gamma^\mu D^{\nu}q-
\ii g^{\mu\nu}\sum_q\bar{q}\slash{D\!}q+g^{\mu\nu}\sum_q m_q\bar{q}q\cr
&-g_{\alpha\beta}G^{\mu\alpha}G^{\nu\beta}+\tfrac{1}{4}g^{\mu\nu}G^2+\hbox{gauge fixing}+
\hbox{ghost terms}.\cr}\equn{(8.1.2b)}$$

In the 
quantum version, we understand that products are replaced by Wick ordered products. 
 $\Thetav$ is not unique and, as a matter of fact, direct application of (8.1.2a) does not 
yield a gauge invariant tensor. To obtain the gauge invariant expression (8.1.2b) one may proceed 
by replacing derivatives by covariant derivatives. A more rigorous procedure would be to 
reformulate (8.1.2a) in a way consistent with gauge 
invariance by performing gauge transformations simultaneously to the spacetime 
translation. For an infinitesimal one, $x^\mu\rightarrow x^\mu+\epsilon^\mu$, we then 
define
$$B_a^\mu\rightarrow B_a^\mu+(\epsilon_\alpha\partial^\alpha B_a^\mu\equiv 
D^\mu\epsilon_\alpha B_a^\alpha+\epsilon_\alpha G_a^{\alpha\mu}).$$
The term $D^\mu\epsilon_\alpha B_a^\alpha$ may be absorbed by a gauge transformation, so we may
 write the transformation as $B_a^\mu\rightarrow 
B_a^\mu+\epsilon_\alpha G_a^{\alpha\mu}$. For a discussion of the arbitrariness in the definition 
of the energy--momentum tensor, see Callan, Coleman and Jackiw (1970)
 and Collins, Duncan and Joglekar (1977).

Next, we have the currents and charges associated with
 colour rotations. We leave it to the reader to write them explicitly; 
they are particular cases of colour gauge transformations (with constant parameters). We 
now  pass 
over to a different set of currents {\sl not} associated with interactions 
of quarks and gluons among themselves.

If all the quark  masses vanished, we would have invariance of $\cal L$ under the 
transformations,
$$q_f\rightarrow\sum_{f'=1}^{n_f}W_{ff'}q_{f'},\quad 
q_f\rightarrow\sum_{f'=1}^{n_f}W^5_{ff'}\gamma_5q_{f'}\equn{(8.1.3)} $$
where $f,\,f'$ are flavour indices, and $W,\,W^5$ unitary matrices. This implies that the 
currents
$$\eqalign{V_{qq'}^{\mu}(x)=\bar{q}(x)\gamma^\mu q'(x),\cr
A_{qq'}^{\mu}(x)=\bar{q}(x)\gamma^\mu \gamma_5q'(x)\cr}\equn{(8.1.4)}$$
would be each separately conserved. When mass terms are taken into account, 
only the diagonal $V^{\mu}_{qq}$ are conserved; the others are what is 
called {\sl quasi-conserved} currents, i.e., their divergences are proportional to masses. 
These divergences are easily calculated: since the transformations in (8.1.3) commute 
with the interaction part of $\cal L$, we may evaluate them with free fields, 
in which case use of the free Dirac equation $\ii\slash{\partial}q=m_q q$ gives
$$\partial_\mu V^{\mu}_{qq'}=\ii(m_q-m_{q'})\bar{q}q',\quad
\partial_\mu A^{\mu}_{qq'}=\ii(m_q+m_{q'})\bar{q}\gamma_5q'.\equn{(8.1.5)}$$ 
In fact, there is a subtle point concerning the divergence of 
axial currents. \equ~(8.1.5) is correct as it stands for 
the nondiagonal currents, $q\neq q'$; for 
$q=q'$, however, one has instead
$$\partial_\mu A^{\mu}_{qq}(x)=\ii(m_q+m_q)\bar{q}(x)\gamma_5q(x)+
\dfrac{T_Fg^2}{16\pi^2}\epsilon^{\mu\nu\rho\sigma}G_{\mu\nu}(x)G_{\rho\sigma}(x),
\equn{(8.1.6)}$$
with $T_F=1/2$ a colour factor. This is the so-called Adler--Bell--Jackiw 
anomaly, that we will discuss later.

The {\sl equal time commutation relations} (ETC) of the $V,\,A$ with the fields are 
also easily calculated, for free fields. Using (8.1.4) 
and the ETC of quark fields, one finds,
$$\eqalign{\delta(x^0-y^0)[V^0_{qq'}(x),q''(y)]=&-\delta(x-y)\delta_{qq''}q'(x),\cr 
\delta(x^0-y^0)[A^0_{qq'}(x),q''(y)]=&-\delta(x-y)\delta_{qq''}\gamma_5q'(x),\;\hbox{etc.}\cr}
\equn{(8.1.7)}$$
The $V,\,A$ commute with gluon and ghost fields. The equal time commutation relations of the   
 $V,\,A$ among themselves (again for free fields) are best described for three 
flavours, $f=1,\,2,\,3=u,\,d,\,s$  
by introducing the Gell-Mann $\lambda^a$ matrices in 
flavour space (for two quarks $u,\,d$, replace the $\lambda^a$ by the $\tau^a$ of Pauli, 
and the $f_{abc}$ by $\epsilon_{abc}$ in (8.1.9) below).  So we
let
$$V^{\mu}_{a}(x)=\sum_{ff'}\bar{q}_f(x)\lambda^{a}_{ff'}\gamma^\mu q_{f'}(x),\quad
A^{\mu}_{a}(x)=\sum_{ff'}\bar{q}_f(x)\lambda^{a}_{ff'}\gamma^\mu \gamma_5q_{f'}(x),
\equn{(8.1.8)}$$
and we then obtain the commutation relations
$$\eqalign{\delta(x^0-y^0)[V^0_a(x),V^{\mu}_{b}(y)]=&
\,2\ii\delta(x-y)\sum f_{abc}V^{\mu}_{c}(x),\cr 
\delta(x^0-y^0)[V^0_a(x),A^{\mu}_{b}(y)]=&
\,2\ii\delta(x-y)\sum f_{abc}A^{\mu}_{c}(x),\cr 
\delta(x^0-y^0)[A^0_a(x),A^{\mu}_{b}(y)]=&
\,2\ii\delta(x-y)\sum f_{abc}V^{\mu}_{c}(x),\;\hbox{etc.}\cr} \equn{(8.1.9)}$$
Equations (8.1.7) and (8.1.9) have been derived only for free fields. However, 
they involve short distances; therefore in QCD, and because of asymptotic freedom, 
they will hold as they stand even 
in the presence of interactions.

Equal time commutation relations of conserved or quasi-conserved 
currents with the Hamiltonian may also be easily obtained. If 
$J^\mu$ is conserved, then the corresponding charge 
$$Q_J(t)=\int\dd^3{\bf x}\,J^0(t,{\bf x}),\quad t=x_0,$$
commutes with $\cal H$:
$$[Q_J(t),{\cal H}(t,{\bf y})]=0.$$
Here $\cal H$ is the {\sl Hamiltonian density}, ${\cal H}=\Thetav^{00}$. 
If $J$ is quasi-conserved, let ${\cal H}_m$ be the mass term in $\cal H$,
$${\cal H}_m=\sum_qm_q\bar{q}q.$$
Then,
$$[Q_J(t),{\cal H}_m(t,{\bf y})]=\ii\partial_\mu J^\mu(t,{\bf y}).\equn{(8.1.10)}$$
Of course, $Q_J$ still commutes with the rest of $\cal H$.

\booksection{8.2 Mass terms and invariances: chiral invariance}

\noindent In this section we will consider quarks with masses $m\ll\Lambdav$ 
(with $\lambdav$ the QCD mass parameter), to 
be referred to as {\sl light quarks}.\fnote{It 
is, of course, unclear whether the meaningful parameter in this respect is $\Lambdav$, 
connected to the strong interaction coupling by
$$\alpha_s(t)=\dfrac{12\pi}{(33-2n_f)\log t/\lambdav^2},$$
 or 
$\Lambdav_0$ defined by $\alpha_s(\Lambdav_0)\approx 1$. From considerations of chiral dynamics 
(see later), 
it would appear that the scale for smallness of quark masses is $4\pi f_\pi\sim1\,\gev$,
 where $f_\pi$ is the pion 
decay constant; but even if 
we accept this, it is not obvious at which 
scale $m$ has to be computed. 
 We will see that $m_u,\,m_d\sim4\,\hbox{to}\,10\,\mev$ so there is little doubt
 that $u,\,d$ quarks should be classed as ``light"  with any reasonable definition; but 
the situation is less definite for the $s$ quark with $m_s\sim 180\,\mev$.}  Because 
the only dimensional parameter intrinsic to QCD is, we believe, $\Lambdav$, we may expect 
that to some approximation we may neglect the masses of such quarks, which will yield only 
contributions  of order $m^2/\Lambdav^2$. 

To study this, we consider the QCD Lagrangian,
$${\cal L}=-\sum_{l=1}^n m_l\bar{q}_lq_l+\ii\sum_{l=1}^n\bar{q}_l\slash{D\!}q_l
-\tfrac{1}{4}(D\times B)^2+\hbox{gauge fixing + ghost terms}.
\equn{(8.2.1)}$$
The sum now runs only over {\sl light} quarks;
the presence of heavy quarks will have no practical effect in what follows 
and consequently we neglect them.
We may then split the quark fields into left-handed and right-handed components:
$$q_l=q_{L,l}+q_{R,l};\quad
q_{L,l}\equiv q_{-,l}=\dfrac{1-\gamma_5}{2}q_l,\quad
q_{R,l}\equiv q_{+,l}=\dfrac{1+\gamma_5}{2}q_l.$$
In terms of these, the quark part of the Lagrangian may be written as
$${\cal L}=-\sum_{l=1}^n m_l\left(\bar{q}_{R,l}q_{L,l}+\bar{q}_{L,l}q_{R,l}\right)
+\ii\sum_{l=1}^n\left(\bar{q}_{L,l}\slash{D\!}q_{L,l}+\bar{q}_{R,l}\slash{D\!}q_{R,l}\right)
+\cdots\,.$$

We then consider the set of transformations
 $W^\pm$  
(left-handed times right-handed) given by the independent 
transformations of the $q_{R,l}$, $q_{L,l}$: 
$$q_{R,l}\to\sum_{l'}W^{+}_{ll'}q_{R,l'},\quad
q_{L,l}\to\sum_{l'}W^{-}_{ll'}q_{L,l'};\quad
 W^\pm\;\hbox{unitary}.\equn{(8.2.2)}$$
Clearly, the only term in $\cal L$ that is not invariant under all the transformations 
(8.2.2) is the mass term,
$${\cal M}=\sum_{l=1}^n m_l\bar{q}_lq_l=
\sum_{l=1}^n m_l\left(\bar{q}_{R,l}q_{L,l}+\bar{q}_{L,l}q_{R,l}\right).\equn{(8.2.3)}$$

When written in this form, the mass term is invariant under the set of transformations $[U(1)]^n$,
$$q_l\to \ee^{\ii\theta_l}q_l,\equn{(8.2.4)}$$
but this would not have been the case if we had allowed for nondiagonal terms 
in the mass matrix. To resolve this question of which are the general invariance 
properties of a mass term, we will prove two theorems.\fnote{The theorems are valid for {\sl any}
 quark mass matrix, i.e., also including heavy flavours.}
\smallskip
\hangindent=0.5cm{\noindent\quad{\sc Theorem 1.}  {\sl Any general mass matrix 
can be written in the form (8.2.3) by appropriate redefinition 
of the quark fields. Moreover, we may assume that $m\geq0$. Thus, (8.2.3) is actually the 
most general mass term possible.}}
\smallskip\noindent
 For the proof we consider that 
the most general mass term compatible with hermiticity is
$${\cal M}'=\sum_{ll'}\left\{\bar{q}_{L,l}M_{ll'}q_{R,l'}+
\bar{q}_{R,l}M^*_{ll'}q_{L,l'}\right\}.\equn{(8.2.5)}$$
Let us temporarily denote matrices in flavour space by putting a tilde 
under them. If $\undertilde{M}$ is the matrix with components $M_{ll'}$, then the 
well-known polar decomposition, valid for any matrix, allows us to write
$$\undertilde{M}=\undertilde{m}\undertilde{U},$$
where $\undertilde{m}$ is positive-semidefinite, so all its eigenvalues are $\geq0$, 
and $\undertilde{U}$ is unitary. \equ~(8.2.5) may then be cast in the form 
$${\cal M}'=\sum_{ll'}\left\{\bar{q}_{L,l}m_{ll'}q'_{R,l'}+
\bar{q'}_{R,l}m_{ll'}q_{L,l'}\right\},\quad q'_{R,l}=\sum_{l'}U_{ll'}q_{R,l'},\equn{(8.2.6)}$$
and we have used the fact that $\undertilde{m}$ is Hermitian. Define 
$q'=q_L+q'_R$; because 
$\bar{q}_Rq_R=\bar{q}_Lq_L=0$, (8.2.6) becomes, in terms of $q'$,
$${\cal M}'=\sum\bar{q}'_lm_{ll'}q'_{l'}.$$
It then suffices to transform $q'$ by the matrix that diagonalizes $\undertilde{m}$ 
to obtain (8.2.3) with positive $m_l$. The term $\bar{q}\slash{D\!}q$ in 
the Lagrangian is left invariant by all these transformations, so the theorem is proved.
\smallskip
\hangindent=0.5cm{\noindent\quad{\sc Theorem 2.} {\sl If all the $m_l$ are nonzero 
and different, then the only invariance left is the $[U(1)]^n$ of (8.2.4).}}
\smallskip\noindent
Let us consider the $\undertilde{W}_\pm$ of (8.2.2), and assume that
 $\undertilde{W}_+=\undertilde{W}_-\equiv \undertilde{W}$; to show that this 
must actually be the case is left as an exercise. The condition of invariance of 
$\cal M$ yields the relation
$$\undertilde{W}^{\dag}\undertilde{m}\undertilde{W}=\undertilde{m},\quad {\rm i.e.},
\quad [\undertilde{m},\undertilde{W}]=0.
\equn{(8.2.7)}$$
It is known that any $n\times n$ diagonal matrix can be written as 
$\sum_{k=0}^{n-1}c_k\undertilde{m}^k$ if, as 
occurs in our case, all the eigenvalues of $\undertilde{m}$ are different and nonzero. 
Because of (8.2.7), it then follows that $\undertilde{W}$ commutes with all diagonal 
matrices, and hence it must itself be diagonal: because it is also unitary, it consists of 
diagonal phases, i.e., 
 it may be written as a product of transformations (8.2.4), as 
was to be proved. We leave it to the reader to check that the 
conserved quantity corresponding to the $U(1)$ that acts on 
flavour $q_f$ is the corresponding flavour number.

In the preceding theorems, we have not worried whether the masses $m$
 were bare, running or invariant masses. This is because, 
in the \msbar\ scheme, the mass matrix becomes renormalized as a whole:
$$\undertilde{M}=Z_m^{-1}\undertilde{M}_u,$$
where $Z_m$ is a {\sl number}. The proof of this last property is easy: all we have to do 
is to repeat the standard renormalization of the quark propagator, allowing for the 
matrix character of $M,\,Z_m$. We find, 
for the divergent part and in an arbitrary covariant gauge, 
with gauge parameter $\xi$,
$$\eqalign{\undertilde{S}_R^\xi(p)=&
\dfrac{\ii}{\slash{p}-\undertilde{M}}+
\dfrac{1}{\slash{p}-\undertilde{M}}\Bigg\{-[\undertilde{\Deltav}_F(\slash{p}-\undertilde{M})+
(\slash{p}-\undertilde{M}){\undertilde{\Deltav}}_F^{\dag}]-\delta\undertilde{M}\cr
&-(1-\xi)(\slash{p}-\undertilde{M})N_\epsilon C_F\dfrac{g^2}{16\pi^2}+
3N_\epsilon C_F\dfrac{g^2}{16\pi^2}\undertilde{M}\Bigg\}\dfrac{\ii}{\slash{p}-\undertilde{M}},\cr}$$
and we have defined
$$\undertilde{M}=\undertilde{M}_u+\delta\undertilde{M},\quad 
\undertilde{Z}_F=1-\undertilde{\Deltav}_F.$$
The renormalization conditions then yield
$$\eqalign{{\undertilde{\Deltav}_F}^{\dag}+\undertilde{\Deltav}_F=&\,
-(1-\xi)N_\epsilon C_F\dfrac{g^2}{16\pi^2}={\rm diagonal,}\cr
[\undertilde{\Delta}_F,\undertilde{M}]=&\,0,\quad [\undertilde{M},\delta\undertilde{M}]=0,\cr
\delta\undertilde{M}=&\,3N_\epsilon C_F\dfrac{g^2}{16\pi^2}\undertilde{M}.\cr}$$
Thus, the set of fermion fields and the mass matrix get renormalized as a whole:
$$\undertilde{Z}_F^{-1}=1+(1-\xi)N_\epsilon C_F\dfrac{g^2}{16\pi^2},\quad 
\undertilde{Z}_m=1-3N_\epsilon C_F\dfrac{g^2}{16\pi^2},\equn{(8.2.8a)}$$
i.e.,
$$\undertilde{Z}_F=Z_F\,1,\quad\undertilde{Z}_m=Z_m\,1.\equn{(8.2.8b)}$$
We have proved this to lowest order, but the renormalization group 
equations guarantee the result to leading order in $\alpha_s$.

This result can be understood in yet another way. The invariance of 
$\cal L$ under the transformations (8.2.4) implies that we may choose 
the counterterms to satisfy the same invariance, so the mass matrix will remain 
diagonal after renormalization. In fact, this proof shows that in mass independent 
renormalization schemes (such as the \msbar), \equs~(8.2.8b) actually hold to all orders.

The results we have derived show that, if all the $m_i$ are different and nonvanishing,\fnote{As 
seems to be the case in nature. As we will see, one finds ${m}_d/{m}_u\sim2$, 
 ${m}_s/{m}_d\sim20$, ${m}_u\sim 5\,\mev$.} the 
only global symmetries of the Lagrangian are those 
associated with flavour conservation, (8.2.4). As stated above, however, under 
certain conditions it may be a good approximation to neglect the $m_l$. In this case, all 
the transformations of \equ~(8.2.2) become symmetries of the Lagrangian. The measure of the accuracy of 
the symmetry is given by, for example, the divergences of the corresponding currents 
or, equivalently, the conservation of the charges. This has been 
discussed in \sect~8.1, and we now present some extra details.

Let us parametrize the $W$ as $\exp\{(\ii/2)\sum\theta_a\lambda^a\}$, where the $\lambda$ are the 
Gell-Mann matrices. (We consider the case $n=3$; for $n=2$, replace the $\lambda$ by the 
$\tau$ of Pauli). We may denote by $U_\pm(\theta)$  the operators that implement (8.2.2):
$$U_\pm(\theta)\dfrac{1\pm\gamma_5}{2}q_lU_\pm^{-1}(\theta)=
\sum_{l'}\left(\ee^{(\ii/2)\sum\theta_a\lambda^a}\right)_{ll'}\dfrac{1\pm\gamma_5}{2}q_{l'}.
\equn{(8.2.9)}$$
For infinitesimal $\theta$, we write
$$U_\pm(\theta)\simeq1-\dfrac{\ii}{2}\sum \theta_aL^a_\pm,\quad
(L^a_\pm)^{\dag}=L^a_\pm,$$
so that (8.2.9) yields
$$[L^a_\pm,q_{\pm,l}(x)]=-\sum_{l'}\lambda^a_{ll'}q_{\pm,l'}(x),\quad
q_{\pm,l}\equiv \dfrac{1\pm\gamma_5}{2}q_l.\equn{(8.2.10)}$$
Because the $U_\pm$ leave the interaction part of the Lagrangian invariant, and since QCD 
is a free field theory at zero distance, we may solve (8.2.10) using free-field 
commutation relations. The result is
$$L^a_\pm(t)=:\int\dd^3{\bf x}\sum_{ll'}\bar{q}_{\pm,l}(x)\gamma^0\lambda^a_{ll'}
q_{\pm,l'}(x):,
\quad t=x^0.\equn{(8.2.11)}$$
These will be recognized as the charges corresponding to the currents
$$J^{a\mu}_{\pm}(x)=
:\sum_{ll'}\bar{q}_l(x)\lambda^a_{ll'}\gamma^\mu\dfrac{1\pm\gamma_5}{2}q_{l'}(x):.
\equn{(8.2.12)}$$
If the symmetry is exact, $\partial_\mu J_{\pm}^{a\mu}=0$, and a standard calculation 
shows that the $L^a_\pm(t)$ are actually independent of $t$. Otherwise, we 
have to define {\sl equal time} transformations and modify (8.2.9, 10) writing, for 
example,
$$[L^a_\pm(t),q_{\pm,l}(x)]=-\sum_{l'}\lambda^a_{ll'}q_{\pm,l'}(x),\quad
t=x^0.\equn{(8.2.13)}$$
The set of transformations
$$U_\pm(\theta,t)=\exp\left\{-\dfrac{\ii}{2}\sum L^a_\pm(t)\theta_a\right\},$$
builds up the group of {\sl chiral transformations} generated 
by the currents (8.2.12). In our present case we find the 
chiral $SU_F^+(3)\times SU_F^-(3)$ group. Its generators may be rearranged
 in terms of the set of vector and axial currents $V^{\mu}_{ll'}(x)$, $A^{\mu}_{ll'}(x)$ 
introduced in \sect~8.2. (Actually, not
 all diagonal elements are in $SU_F^+(3)\times SU_F^-(3)$, but they are in the group 
 $U_F^+(3)\times U_F^-(3)$).  An important subgroup of 
 $SU_F^+(3)\times SU_F^-(3)$ is that generated by the vector currents, 
which is simply the flavour $SU(3)$ of Gell-Mann and 
Ne'eman.

The exactness of the symmetries is related to the time independence of the 
charges $L_\pm$, which in turn is linked to the divergence of the currents. 
These divergences are proportional to differences of
 masses, $m_l-m_{l'}$ for the vector, and sums $m_l+m_{l'}$ for the 
axial currents.
Thus, we conjecture that $SU_F(3)$ will be good to the extent that 
$|m_l-m_{l'}|^2\ll \Lambdav^2$ 
and chiral $SU_F^+(3)\times SU_F^-(3)$ to the extent that $m_l\ll \Lambdav$. In 
the real world, it appears that mass differences are of the same order as the 
masses themselves, so we expect chiral symmetries to be almost as good 
as flavour symmetries. This seems to be the case experimentally.

\booksection{8.3 Wigner--Weyl and Nambu--Goldstone realizations of symmetries}

\noindent The fact that flavour and chiral $SU(3)$ (or $SU(2)$) appear to 
be valid to similar orders of approximation does not mean that these symmetries are 
realized in the same manner. In fact, we will see that there are good 
theoretical and experimental reasons why they are very different.

Let us begin by introducing the charges with definite parity,
$$Q^a=L^a_++L^a_-,\quad Q^a_5=L^a_+-L^a_-.\equn{(8.3.1)}$$
Their {\sl equal time commutation relations} are
$$\eqalign{[Q^a(t),Q^b(t)]=&\,2\ii\sum f^{abc}Q^c(t),\cr 
[Q^a(t),Q^b_5(t)]=&\,2\ii\sum f^{abc}Q^c_5(t),\cr 
[Q^a_5(t),Q^b_5(t)]=&\,2\ii\sum f^{abc}Q^c(t).\cr }\equn{(8.3.2)}$$
The set $Q^a$ builds the group $SU_F(3)$. In the limit $m_l\to0$, all 
$Q$, $Q_5$ are $t$-independent and
$$[Q^a,{\cal L}]=[Q^a_5,{\cal L}]=0.\equn{(8.3.3)}$$
The difference between $Q^a$, $Q^a_5$, however, lies in the vacuum. 
In general, given a set of generators $L^j$ of symmetry
 transformations of $\cal L$, we have two possibilities:
$$L^j|0\rangle=0,\equn{(8.3.4)}$$ 
which is called a {\sl Wigner--Weyl} symmetry, or
$$L^j|0\rangle\neq0,\equn{(8.3.5)}$$
or {\sl Nambu--Goldstone} symmetry. Obviously,
 we will in general have a mixture of the two symmetries, 
with some $L^i$, $i=1,\dots,r$, verifying (8.3.4) and the 
rest, $L^k$, $k=r+1,\dots,n$, satisfying (8.3.5). Since the commutator
 of two operators that annihilate the 
vacuum also annihilates the 
vacuum, it follows that the subset of Wigner--Weyl symmetries forms a subgroup.  

Two theorems are especially relevant with respect to these questions.
 The first, due to Coleman (1966), asserts that ``the 
invariance of the vacuum is the invariance of the world", or, in more transparent terms, 
that the physical states (including bound states) 
are invariant under the transformations of a Wigner--Weyl group of symmetries. It follows that, 
if we assumed that chiral symmetry was all of it realized in the Wigner--Weyl mode, 
we could conclude that the masses of all mesons in a flavour multiplet would be degenerate,
 up to corrections of order $m_q^2/M_h^2$, with $M_h$ the (average) hadron mass. 
This is true of the $\omega$, $\rho$, $K^*$, $\phi$, but if we include parity doublets 
this is no longer the case. Thus, for example, there is no scalar 
meson with a mass anywhere near that of the pion, and the axial vector meson masses are 
more than half a \gev\ larger than the masses of $\omega$ or $\rho$. Thus 
it is strongly suggested that $SU_F(3)$ is a Wigner--Weyl symmetry, but chiral  
 $SU_F^+(3)\times SU_F^-(3)$ contains generators of the Goldstone--Nambu type. We assume, 
therefore,
$$Q^a(t)|0\rangle=0,\quad Q^a_5(t)|0\rangle\neq0.\equn{(8.3.6)}$$

The second relevant theorem is Goldstone's (1961). It states that, for 
each generator that fails to annihilate the vacuum, there must exist a massless boson with the 
quantum numbers of that generator. Therefore, we ``understand" the smallness of the 
masses of the pion or kaon\fnote{The particles with zero flavour quantum numbers 
present problems of their own (the so-called $U(1)$ problem) 
that will be discussed later.} because, in the limit $m_u,\,m_d,\,m_s\to0$, 
we would also have $\mu\to0$, $m_K\to0$. Indeed, we will later show that
$$\mu^2\sim m_u+m_d,\quad m^2_K\sim m_{u,d}+m_s.\equn{(8.3.7)}$$

We will not prove either theorem here, but we note that (8.3.7) affords a 
quantitative criterion for the validity of chiral  symmetries; they hold to 
corrections of order $\mu^2/m^2_\rho$ for $SU(2)$ and of $m^2_K/m^2_{K^*}$ for 
$SU(3)$.

We also note that a Nambu--Goldstone realization (Nambu, 1960; Nambu and 
Jona--Lasinio, 1961a,b) is never possible in perturbation theory.
 Since the symmetry generators are Wick-ordered products of 
field operators, it is clear that  
to all orders of perturbation theory  $Q^a_5(t)|0\rangle=0$. 
This means that the physical vacuum is different from the vacuum 
of perturbation theory in the 
limit $m\to0$. We emphasize this by writing $|0\rangle$ for 
the perturbation-theoretic vacuum and $|{\rm vac}\rangle$ 
for the physical one when there is danger of confusion.
 So we rewrite (8.3.6) as
$$Q^a(t)|{\rm vac}\rangle=0,\quad Q^a_5(t)|{\rm vac}\rangle\neq0.
\equn{(8.3.8)}$$
It is not difficult to see how this may come about in QCD. Let $a^{\dag}_P({\bf k})$ 
be the creation operator for a particle P with three-momentum $\bf k$. The states
$$a^{\dag}_P({\bf 0})\overbrace{\mathstrut\dots}^n a^{\dag}_P({\bf 0})|0\rangle=
|n\rangle$$
are all degenerate in the limit $m_P\to 0$. Therefore, the physical vacuum will be, 
in this limit, 
$$|{\rm vac}\rangle=\sum_nc_n|n\rangle,$$
i.e., it will contain zero-frequency massless particles (Bogoliubov's 
model). In QCD we have the gluons which 
are massless, and so will the light quarks be, to a good 
approximation, in the chiral limit.

\booksection{8.4. PCAC, $\pi^+$ decay, the pion propagator and light quark mass ratios}
\vskip-0.5truecm
\booksubsection{8.4.1.  The weak axial current and $\pi^+$ decay}
\noindent

\noindent We are now in a position to obtain quantitative results
 on the masses of the light quarks, 
relating them to the masses of pions and kaons.
 To do so, consider the current
$$A^{\mu}_{ud}(x)=\bar{u}\gamma^\mu\gamma_5d(x),$$
and its divergence
$$\partial_\mu A^{\mu}_{ud}(x)=\ii(m_u+m_d)\bar{u}\gamma_5d(x).$$
The latter has the quantum numbers of the $\pi^+$, so we can use 
it as a composite pion field operator. We thus write
$$\partial_\mu A^{\mu}_{ud}(x)=\sqrt{2}f_\pi \mu^2 \phi_\pi(x).\equn{(8.4.1a)}$$
The factors in (8.4.1a) are chosen for historical reasons (our convention is  
not universal, however). $\phi_\pi(x)$ is the pion field normalized to
$$\langle0|\phi_\pi(x)|\pi(p)\rangle=\dfrac{1}{(2\pi)^{3/2}}\ee^{-\ii p\cdot x},
\equn{(8.4.1b)}$$
with $|\pi(p)\rangle$ the state of a pion with momentum $p$. The constant $f_\pi$ 
may be obtained from experiment as follows. Consider the weak decay 
$\pi^+\to\mu^+\nu$. With the effective Fermi Lagrangian for weak interactions 
(see, e.g., Marshak, Riazzudin and Ryan, 1969)
$${\cal L}^{\rm Fermi}_{\rm int}=
\dfrac{G_F}{\sqrt{2}}\bar{\mu}\gamma_\lambda(1-\gamma_5)\nu_\mu
\bar{u}\gamma^\lambda(1-\gamma_5)d+\cdots,$$
we find the decay amplitude
$$F(\pi\to\mu\nu)=
\dfrac{2\pi G_F}{\sqrt{2}}\bar{u}_{(\nu)}(p_2)\gamma_\lambda(1-\gamma_5)v_{(\mu)}(p_1,\sigma)
\langle0|A^{\lambda}_{ud}(0)|\pi(p)\rangle.\equn{(8.4.2a)}$$
Now, on invariance grounds,
$$\langle0|A^{\lambda}_{ud}(0)|\pi(p)\rangle=\ii p^\lambda C_\pi;\equn{(8.4.2b)}$$
contracting with $p_\mu$ we find $C_\pi=f_\pi\sqrt{2}/(2\pi)^{3/2}$ and hence
$$\mu^2 C_\pi=\langle0|\partial_\lambda A^{\lambda}_{ud}(0)|\pi(p)\rangle=
\sqrt{2}f_\pi \mu^2\dfrac{1}{(2\pi)^{3/2}}.\equn{(8.4.2c)}$$
Therefore
$$\tau(\pi\to\mu\nu)=\dfrac{4\pi}{(1-m^2_\mu/\mu^2)^2G^2_Ff^2_\pi \mu m^2_\mu},$$
and we obtain $f_\pi$ from the decay rate. 
An accurate evaluation of $f_\pi$ requires taking into account 
the Cabibbo rotation and electromagnetic radiative corrections;
one gets $f_\pi\simeq 93\,\mev$. A 
remarkable fact is that, if we repeat the analysis for kaons,
$$\partial_\mu A^{\mu}_{us}(x)=\sqrt{2}f_Km^2_K\phi_K(x),\equn{(8.4.3)}$$
we find that, experimentally, $f_K\approx110\,\mev$: it agrees with 
$f_\pi$ to 20\%. Actually, this is to be expected because, in the limit 
$m_{u,d,s}\to0$, there is no difference 
between pions and kaons, and we would find strict equality. That $f_\pi$, $f_K$ are 
so similar in the real world is a good point in favour of $SU_F(3)$ chiral 
ideas.

The relations (8.4.1) and (8.4.3) are at times called 
PCAC\fnote{Partially conserved axial current. In fact, in the limit 
$\mu^2\to0$, the right hand side of 
(8.4.1a) vanishes.} but this is not very meaningful, for these 
equations are really {\sl identities}. One may use any pion field operator 
one wishes, in particular (8.4.1), provided that it has the right quantum numbers and its 
vacuum-one pion matrix element is not zero. The nontrivial part of PCAC 
will be described below.

\booksubsection{8.4.2.  The pion propagator; quark mass ratios}
\noindent

The next step is to consider the two-point function, or {\sl correlator} (we 
drop the $ud$ index from $A_{ud}$)
$$\piv^{\mu\nu}(q)=
\ii\int\dd^4x\,\ee^{\ii q\cdot x}\langle{\rm T}A^\mu(x)A^\nu(0)^{\dag}\rangle_{\rm vac},$$
and contract with $q_\mu,\,q_\nu$:
$$\eqalign{q_\nu q_\mu \piv^{\mu\nu}(q)=&
-q_\nu\int\dd^4x\,\ee^{\ii q\cdot x}\partial_\mu
\langle{\rm T}\,A^\mu(x)A^\nu(0)^{\dag}\rangle_{\rm vac}\cr
=&-q_\nu\int\dd^4x\,\ee^{\ii q\cdot x}\delta(x^0)
\langle[A^0(x),A^\nu(0)^{\dag}]\rangle_{\rm vac}\cr
&-q_\nu\int\dd^4x\,\ee^{\ii q\cdot x}
\langle{\rm T}\,\partial\cdot A(x)A^\nu(0)^{\dag}\rangle_{\rm vac}\cr
=&\,2\ii\int\dd^4x\,\ee^{\ii q\cdot x}\delta(x^0)
\langle[A^0(x),\partial\cdot A(0)^{\dag}]\rangle_{\rm vac}\cr
&+\ii\int\dd^4x\,\ee^{\ii q\cdot x}
\langle{\rm T}\,\partial\cdot A(x)\partial\cdot A(0)^{\dag}\rangle_{\rm vac}.\cr}$$
Using \equs~(8.4.1, 2) and evaluating the 
commutator, we find
$$\eqalign{q_\nu q_\mu \piv^{\mu\nu}(q)=&\,2(m_u+m_d)\int\dd^4x\,\ee^{\ii q\cdot x}
\delta(x)\langle\bar{u}(x)u(x)+\bar{d}(x)d(x)\rangle_{\rm vac}\cr
&+2\ii f^2_\pi \mu^4\int\dd^4x\,\ee^{\ii q\cdot x}
\langle{\rm T}\,\phi_\pi(x)\phi_\pi(0)^{\dag}\rangle_{\rm vac},\cr}$$
or, in the limit $q\to0$,
$$\eqalign{2(m_u+m_d)\langle:\bar{u}(0)u(0)+\bar{d}(0)d(0):\rangle_{\rm vac}\cr
=-2\ii f^2_\pi \mu^4\int\dd^4x\,\ee^{\ii q\cdot x}
\langle{\rm T}\,\phi_\pi(x)\phi_\pi(0)^{\dag}\rangle_{\rm vac}\big|_{q\to0},\cr}$$
and we have reinstated explicitly the colons of normal ordering.
The right hand side of this equality has  contributions from 
the pion pole and from the continuum; by writing 
a dispersion relation (Cauchy representation) for $\Piv(t)$, defined by
$$\Piv(q^2)=\ii\int\dd^4x\,\ee^{\ii q\cdot x}
\langle{\rm T}\,\phi_\pi(x)\phi_\pi(0)^{\dag}\rangle_{\rm vac},$$
they can be expressed  as\fnote{The equation 
below should have  been written with subtractions, to compensate for the growth 
of $\Piv(q^2)$ for large $q^2$; but these do not alter the conclusions.}
$$\eqalign{\ii\int\dd^4x\,\ee^{\ii q\cdot x}
\langle{\rm T}\,\phi_\pi(x)\phi_\pi(0)^{\dag}\rangle_{\rm vac}\big|_{q\to0}&
=\left\{\dfrac{1}{\mu^2-q^2}+
\dfrac{1}{\pi}\int\dd t\,\dfrac{\imag \Piv(t)}{t-q^2}\right\}_{q\to0}\cr
&=\dfrac{1}{\mu^2}+\dfrac{1}{\pi}\int\dd t\dfrac{\imag \Piv(t)}{t}.\cr
}$$
The order of the limits is essential; we first must take $q\to0$ and the chiral limit 
afterwards. In the limit $\mu^2\to0$, the first term on the right hand side 
above {\sl diverges}, and the second remains finite.\fnote{Properly speaking, 
this is the PCAC limit, for in this limit 
the axial current is conserved.} We then get
$$\eqalign{(m_u+m_d)\langle\bar{u}u+\bar{d}d\rangle=
-2f^2_\pi \mu^2\left\{1+O(\mu^2)\right\},\cr
\langle\bar{q}q\rangle\equiv\langle:\bar{q}(0)q(0):\rangle_{\rm vac},\quad q=u,d,s,\dots.\cr}
\equn{(8.4.4)}$$
This is a strong indication 
that $\langle\bar{q}q\rangle\neq 0$ because, in order to ensure that it vanishes, we would 
require $f_\pi=0$ (or very large $O(\mu^4)$ corrections). We also note that we have not 
distinguished in e.g. (8.4.4), between bare ($u$) or renormalized 
($R$) quark masses and operators; 
the distinction is not necessary because $m_q$ and 
 $\langle\bar{q}q\rangle$ acquire opposite renormalization, so that $m_u\langle\bar{q}q\rangle_u=
m_R \langle\bar{q}q\rangle_R$.

We may repeat the derivation of (8.4.4) for 
kaons. We find, to leading order in $m_K^2$,
$$\eqalign{(m_s+m_u)\langle\bar{s}s+\bar{u}u\rangle\simeq&-2f_{K^+}^2m_{K^+}^2,\cr
(m_s+m_d)\langle\bar{s}s+\bar{d}d\rangle\simeq&-2f_{K^0}^2m_{K^0}^2.\cr}\equn{(8.4.5)}$$
We may assume $f_{K^+}=f_{K^0}$ since, in the limit $m_{u,d}^2\ll \Lambdav^2$ they should 
be strictly equal. For the 
same reason, one can take it that the VEVs $\langle\bar{q}q\rangle$ are equal 
for all light quarks. Under these circumstances, we may eliminate the VEVs and 
obtain
$$\dfrac{m_s+m_u}{m_d+m_u}\simeq\dfrac{f_K^2m^2_{K^+}}{f^2_\pi \mu^2},\quad
\dfrac{m_d-m_u}{m_d+m_u}\simeq\dfrac{f_K^2(m^2_{K^0}-m^2_{K^+})}{f^2_\pi \mu^2}.$$
A more careful evaluation requires consideration of electromagnetic contributions 
to the observed $\pi$, $K$ masses (Bijnens, 1993; Donoghue, Holsten and Wyler, 1993) 
and higher order chiral corrections (Kaplan and Manohar, 1986;
 Bijnens, Prades and de~Rafael, 1995).\fnote{The method  originates
 in the work of Glashow and Weinberg (1968a,\/b) and Gell-Mann, Oakes and Renner (1968). In QCD, see 
Weinberg (1978a), Dom\'\i nguez (1978) and Zepeda (1978). Estimates of the 
quark masses essentially agreeing with (8.4.6, 7) below had been obtained even before QCD by
 e.g. Okubo (1969), but nobody knew what to do with them. The first 
evaluation in the context of QCD is due 
to Leutwyler (1974).} In this way we find
$$\dfrac{m_s}{m_d}=18\pm5,\quad \dfrac{m_d}{m_u}=2.0\pm0.4.\equn{(8.4.6)}$$
If we couple this with the phenomenological estimate (coming from meson and baryon 
spectroscopy) $m_s-m_d\approx 100$ to $200$ \mev, $m_d-m_u\approx 4\,\mev$, we obtain the 
masses (in \mev)
$$\bar{m}_u(Q^2\sim m^2_\rho)\approx 5,\quad
\bar{m}_d(Q^2\sim m^2_\rho)\approx 9,\quad
\bar{m}_s(Q^2\sim m^2_\rho)\approx 190,\equn{(8.4.7)}$$
where the symbol $\approx$ here means that a 50\% error 
would not be very surprising.

This method for obtaining light quark masses is admittedly very rough; 
in the next section we will describe more sophisticated ones.

To conclude this section we make a few comments concerning light quark condensates, 
$\langle\bar{q}q\rangle$. The fact that these do not vanish implies spontaneous breaking of 
chiral symmetry because, under $q\to\gamma_5q$, $\langle\bar{q}q\rangle\to-\langle\bar{q}q\rangle$. 
One may thus wonder whether chiral symmetry would not be restored in 
the limit $m_q\to0$, which would 
imply
$$\langle\bar{q}q\rangle\rightarrowsub_{m_q\to0}0.\equn{(8.4.8)}$$
This possibility is discussed for example by Gasser and Leutwyler (1982). 
The equation (8.4.8) 
is highly unlikely. If it held, one would expect in particular the ratios,
$$\langle\bar{s}s\rangle:\langle\bar{d}d\rangle:\langle\bar{u}u\rangle\sim
190:9:5,$$
which runs contrary to all evidence, from hadron spectroscopy 
to SVZ sum rules which suggest
$$\langle\bar{s}s\rangle\sim\langle\bar{d}d\rangle\sim\langle\bar{u}u\rangle$$
to a few percent. Thus we obtain an extra indication that chiral symmetry is indeed 
spontaneously broken in QCD.

\booksection{8.5. Bounds and estimates of light quark masses in terms of the pion and kaon masses}

\noindent In this section we describe a method for obtaining bounds and estimates 
of light quark masses. The method was first used (to get rough estimates)
 by Vainshtein et al. (1978) and 
further refined by Becchi, Narison, de Rafael and Yndur\'ain (1981),
 Gasser and Leutwyler (1982), etc. One
 starts with the correlator,
$$\eqalign{\Psiv^5_{ij}(q^2)=&\;
\ii\int\dd^4x\,\ee^{\ii q\cdot x}
\langle{\rm T}\partial\cdot A_{ij}(x)\partial\cdot A_{ij}(0)^{\dag}\rangle_{\rm vac}\cr
=&\;\ii(m_i+m_j)^2\int\dd^4x\,\ee^{\ii q\cdot x}
\langle{\rm T}J^5_{ij}(x)J^5_{ij}(0)^{\dag}\rangle_{\rm vac},\cr}\equn{(8.5.1)}$$
where $A_{ij}^\mu=\bar{q}_i\gamma^\mu\gamma_5q_j$, 
$J^5_{ij}=\bar{q}_i\gamma_5q_j$, $i,j=u,d,s$.

To all orders of perturbation theory, the function
$$F_{ij}(Q^2)=\dfrac{\partial^2}{\partial(q^2)^2}\Psiv^5_{ij}(q^2), \quad Q^2=-q^2,$$
vanishes as $Q^2\to\infty$. Hence, we may write a dispersion relation of the form
$$F_{ij}(Q^2)=
\dfrac{2}{\pi}\int_{m_P^2}^{\infty}\dd t\,\dfrac{\imag\Psiv^5_{ij}(t)}{(t+Q^2)^3},
\quad P=\pi,\,K. 
\equn{(8.5.2)}$$

For large values of $Q^2$, $t$ we may calculate $F_{ij}(Q^2)$, 
$\imag\Psiv^5_{ij}(t)$. The calculation has been improved along the 
years due to increasing precision of the QCD evaluations of
 these quantities.\fnote{Broadhurst (1981) and Chetyrkin et al. (1995) 
for subleading mass corrections; Becchi, Narison, de Rafael and Yndur\'ain (1981), 
 Generalis (1990), Sugurladze and Tkachov (1990), Chetyrkin, Gorishnii and Tkachov (1982), 
Groshny, Kataev, Larin and Sugurladze (1991) and Pascual and de~Rafael (1982) for 
radiative corrections to various terms.} Here, however, we will consider only
 leading effects and first order subleading corrections. We then have, 
$$\eqalign{F_{ij}(Q^2)=&
\,\dfrac{3}{8\pi^2}\,\dfrac{[\bar{m}_i(Q^2)+\bar{m}_j(Q^2)]^2}{Q^2}\cr
&\times\Bigg\{1+\tfrac{11}{3}\dfrac{\alpha_s(Q^2)}{\pi}+
\dfrac{m_i^2+m_j^2+(m_i-m_j)^2}{Q^2}+
\dfrac{2\pi}{3}\,\dfrac{\langle\alpha_sG^2\rangle}{Q^4}\cr
&-\dfrac{16\pi^2}{3Q^4}
\left[\left(m_j-\dfrac{m_i}{2}\right)\langle\bar{q}_iq_i\rangle+
\left(m_i-\dfrac{m_j}{2}\right)\langle\bar{q}_jq_j\rangle\right]\Bigg\}\cr}
\equn{(8.5.3a)}$$
and
$$\imag \Psiv^5_{ij}(t)
=\dfrac{3[\bar{m}_i(t)+\bar{m}_j(t)]^2}{8\pi}
\left\{\left[1+\tfrac{17}{3}\,\dfrac{\alpha_s(t)}{\pi}\right]t-
(m_i-m_j)^2\right\}.\equn{(8.5.3b)}$$
The contributions containing the condensates are easily evaluated 
taking into account the nonperturbative parts of the quark and gluon 
propagators. The quantities $m_i\langle\bar{q}_jq_j\rangle$ 
may be reexpressed in terms of experimentally known quantities, $f_{K,\pi}$, $m_{K,\pi}$ 
as in (8.4.4, 5). For the case 
$ij=ud$, which is the one we will consider in more detail, their 
contribution is negligible, as are the terms of order $m^2/Q^2$ in \equs~(8.5.3).
 We will henceforth 
neglect these quantities. 
Because one can write the imaginary part of the spectral function as
$$\imag\Psiv_{ij}^5(t)=\tfrac{1}{2}
\sum_{\Gammav}\left|\langle{\rm vac}|\partial^{\mu}A^{ij}_{\mu}(0)|\Gammav\rangle\right|^2
(2\pi)^4\delta_4(q-p_{\Gammav})$$
it follows that $\imag\Psiv_{ij}^5(t)\geq 0$: it is this
 positivity that will allow us to derive 
quite general bounds. To obtain tight ones it is important to use the information contained in 
both \equs~(8.5.3a,b); to this end, we define the function
$$\eqalign{\varphi_{ij}(Q^2)=&\;F_{ij}(Q^2)-\int_{Q^2}^{\infty}\dd t\,
\dfrac{1}{(t+Q^2)^3}\,\dfrac{2\imag\Psiv^5_{ij}(t)}{\pi}\cr
=&\int_{m_P^2}^{Q^2}\dd t\,\dfrac{1}{(t+Q^2)^3}\,\dfrac{2\imag\Psiv^5_{ij}(t)}{\pi}.\cr}
\equn{(8.5.4)}$$
For sufficiently large $Q^2$ we may use (8.5.3) and integrate the imaginary part
 to obtain, for $ij=ud$,
$$\eqalign{\varphi_{ud}(Q^2)=\dfrac{3}{8\pi^2}
\left\{\dfrac{[\bar{m}_u(Q^2)+\bar{m}_d(Q^2)]^2}{Q^2}
\left[\tfrac{1}{4}+\left(\tfrac{5}{12}+2\log 2\right)\dfrac{\alpha_s}{\pi}\right]\right.\cr
\left. +\,\dfrac{1}{3Q^6}\left[8\pi^2f^2_\pi \mu^2+
2\pi\langle\alpha_sG^2\rangle\right]\right\}.\cr}\equn{(8.5.5a)}$$
For the $ij=us,ds$ cases, and neglecting $m_{u,d}/m_s$,
$$\eqalign{\varphi_{us,ds}(Q^2)=&\,\dfrac{3}{8\pi^2}
\Bigg\{\dfrac{\bar{m}_s^2}{Q^2}
\left[\tfrac{1}{4}+\left(\tfrac{5}{12}+2\log 2\right)\dfrac{\alpha_s}{\pi}\right]\cr
-&\,\dfrac{2\bar{m}_s^4}{Q^4}\left[\tfrac{3}{4}+\left(6+4\log 2\right)
\dfrac{\alpha_s}{\pi}\right]
 +\dfrac{1}{3Q^6}\left[8\pi^2f_K^2m_K^2+
2\pi\langle\alpha_sG^2\rangle\right]\Bigg\}.\cr
}\equn{(8.5.5b)}$$
We can extract the pion (or kaon, as the case may be) pole explicitly from 
the low energy dispersive integral in (8.5.4) thus getting 
 for e.g., $\varphi_{ud}$
$$\varphi_{ud}(Q^2)=\dfrac{4f^2_\pi \mu^4}{(\mu^2+Q^2)^3}+
\int_{t_0}^{Q^2}\dd t\,\dfrac{1}{(t+Q^2)^3}\,\dfrac{2\imag\Psiv^5_{ij}(t)}{\pi};
\equn{(8.5.6)}$$
the continuum threshold $t_0$ is  $3\mu^2$ for
 $ij=ud$ or $(m_K+2\mu)^2$ for $ij=(u,d)s$. Because of the positivity of 
$\imag \Psiv$ this immediately gives bounds on $m_i(Q^2)+m_j(Q^2)$ as soon 
as $Q^2\geq Q_0^2$, where $Q_0^2$ is a 
momentum  large enough for the QCD estimates (8.5.5) to be valid: thus,
to leading order,
$$\bar{m}_u(Q_0^2)+\bar{m}_d(Q_0^2)\geq\left\{\dfrac{2^7\pi^2f^2_\pi \mu^4}{3}\,
\dfrac{Q_0^2}{(Q_0^2+\mu^2)^3}\right\}^{\frac{1}{2}};\equn{(8.5.7a)}$$
for the combination $us$,
$$\bar{m}_s(Q_0^2)\geq\left\{\dfrac{2^7\pi^2f^2_K m^4_K}{3}\,
\dfrac{Q_0^2}{(Q_0^2+m^2_K)^3}\right\}^{\frac{1}{2}}.\equn{(8.5.7b)}$$
The bound depends a lot on the value of $Q_0^2$. We find, for example, the bounds
$$\eqalign{\bar{m}_u(1\,\gev^2)+\bar{m}_d(1\,\gev^2)\geq&\, 13\,\mev,\quad Q^2_0=1.75\;\gev^2,\cr
\bar{m}_u(1\,\gev^2)+\bar{m}_d(1\,\gev^2)\geq&\, 7\;\mev,\quad Q^2_0=3.5\;\gev^2\cr}
\equn{(8.5.8a)}$$
and
$$\eqalign{\bar{m}_s(1\,\gev^2)\geq&\, 245\;\mev,\quad Q^2_0=1.75\;\gev^2,\cr
\bar{m}_s(1\,\gev^2)\geq&\, 150\;\mev,\quad Q^2_0=3.5\;\gev^2.\cr}
\equn{(8.5.8b)}$$
As is customary, we have translated the 
bounds (as we will also do for the estimates later on) 
to bounds on the running masses defined at 1 \gev.
To do so, we have used the three loop expression for the 
running quark masses,
$$\eqalign{m(t)=\hat{m}\left(\tfrac{1}{2}\log t/\Lambdav^2\right)^{-d_m}
\left[1-d_1\dfrac{\log\log t/\Lambdav^2}{\log t/\Lambdav^2}+
d_2\dfrac{1}{\log t/\Lambdav^2}\right];\cr
d_m=\dfrac{4}{\beta_0},\;
d_1=8\,\dfrac{51-\tfrac{19}{3}n_f}{\beta_0^3},\;d_2=\dfrac{8}{\beta_0^3}
\Big[\left(\tfrac{101}{12}-\tfrac{5}{18}n_f\right)\beta_0
-51+\tfrac{19}{3}n_f\Big],\cr
}$$
 and also the three loop running coupling constant,
$$\eqalign{
\alpha_s(Q^2)=&\,\dfrac{4\pi}{\beta_0L}\left\{1-\dfrac{\beta_1\log L}{\beta_0^2L}+
\dfrac{\beta_1^2\log^2L-\beta_1^2\log L+\beta_2\beta_0-\beta_1^2}{\beta_0^4L^2}\right\};\cr
 L=&\,\log\dfrac{Q^2}{\lambdav^2}.\cr
}
$$
Here 
$$\eqalign{\beta_0=&\,11-\tfrac{2}{3}n_f,\quad
\beta_1=102-\tfrac{38}{3}n_f,\cr
\beta_2=&\,\tfrac{2857}{2}-\tfrac{5033}{18}n_f+\tfrac{325}{54}n_f^2.\cr
}$$
  The value for the QCD
parameter that we use is 
$$\lambdav(3\;{\rm loop},\,n_f=3)=340\pm120\;\mev;$$
see for example Yndur\'ain~(1999) or the more recent Particle Data Tables values, that 
essentially agree with this.

The bounds can be stabilized somewhat by considering derivatives of $F^5_{ij}$, 
but (8.5.8) do not change much.

To get {\sl estimates} for the masses, a model is necessary for the 
low energy piece of the dispersive integral (8.5.6). At very low energy, one can calculate 
$\imag \Psiv^5$ using chiral perturbation theory (see for example Pagels and Zepeda, 1972; 
Gasser and Leutwyler, 1982); the contribution is minute. The important region is that where 
the quasi-two body channels are open, the $(\rho,\omega)-\pi$ channels for the 
$ud$ case. This is expected to be dominated by the $\pi'$ resonance, with a mass of 
$1.2\,\gev$. One can take the residue of the resonance as a free 
parameter, and fit the QCD expression (8.5.5). This is the procedure 
followed by Narison and de~Rafael (1981), Hubschmid and Mallik (1981), 
Gasser and Leutwyler (1982), Kataev, Krasnikov and 
Pivovarov (1983),  Dom\'\i nguez and de~Rafael (1987), 
Chetyrkin, Pirjol and Schilcher (1997), etc. The errors 
one finds in the literature are many times {\sl overoptimistic} because they do not 
take into account the important matter of the value $Q_0^2$ at which the 
perturbative QCD evaluation is supposed to be valid (Yndur\'ain, 1998).
 Now, as is clear from 
\equ~(8.5.5), the radiative corrections feature a large coefficient, so 
it is difficult to estimate reliably a  figure  for $Q_0^2$. Both bounds (as shown above) and 
estimates will depend on this. As reasonably safe estimates we may quote the values
$$\eqalign{\bar{m}_u(Q^2=1\,\gev^2)=&\,4.2\pm 2\;\mev,\cr
\bar{m}_d(Q^2=1\,\gev^2)=&\,8.9\pm 4.3\;\mev,\cr
\bar{m}_s(Q^2=1\,\gev^2)=&\,200\pm 50\;\mev,\cr}\equn{(8.5.9a)}$$
and we have, to reduce  the errors a bit, taken also into 
account the chiral theory estimates of the 
mass ratios given in the previous section, \equ~(8.4.6).

For the $s$ quark,  independent estimates (Chen et al., 2001, Narison, 1995) following from 
$$\tau\to\nu_\tau+\;\hbox{strange 
particles},\quad e^+e^-\to\;\hbox{strange 
particles}$$
give slightly smaller numbers. 
Taking them into account we may write
$$\bar{m}_s(Q^2=1\,\gev^2)=183\pm 30\;\mev.\equn{(8.5.9b)}$$

\booksection{8.6. The triangle anomaly; $\pi^0$ decay. The gluon anomaly. The $U(1)$ problem}
\vskip-0.5truecm
\booksubsection{8.6.3. The triangle anomaly and the $\pi^0$ decay}

\noindent
 Historically, one of the first motivations for the
 colour degree of freedom came from the study of the 
decay $\pi^0\to \gamma\gamma$, which we now consider in  some detail.

The amplitude for the process $\pi^0\to\gamma\gamma$ may be written, using the 
reduction formulas, as
$$\eqalign{\langle\gamma(k_1,\lambda_1),\gamma(k_2,\lambda_2)|S|\pi^0(q)\rangle
=\dfrac{-\ii e^2}{(2\pi)^{9/2}}\,
\epsilon^*_{\mu}(k_1,\lambda_1)\epsilon^*_{\nu}(k_2,\lambda_2)\cr
\times\int\dd^4x_1\,\dd^4x_2\,\dd^4z\,\ee^{\ii(x_1\cdot k_1+x_2\cdot k_2-z\cdot q)}
(\dal_z+\mu^2)\langle{\rm T}J^{\mu}_{\rm em}(x_1) J^{\nu}_{\rm em}(x_2)\phi_{\pi^0}(z)\rangle_0,\cr}
\equn{(8.6.1)}$$
and we have used the relation $\dal A^{\mu}_{\rm ph}(x)=J^{\mu}_{\rm em}(x)$,
 with $ A^{\mu}_{\rm ph}$ the photon field. We leave it as an exercise for the 
reader to check this, as well as  to verify that, in our particular case, one can replace
$$\dal_{x_1}\dal_{x_2}{\rm T}\{ A^{\mu}_{\rm ph}(x_1) A^{\nu}_{\rm ph}(x_2)\phi_{\pi^0}(z)\}
\to{\rm T}\{(\dal A^{\mu}_{\rm ph}(x_1))(\dal A^{\nu}_{\rm ph}(x_2))\phi_{\pi^0}(z)\},
$$
i.e., that potential delta function terms that appear 
when the derivatives in the d'Alembertians act on the 
theta functions $\theta(x_1-z),\dots$ implicit in the T-product make no 
contribution. Separating off the delta of four-momentum 
conservation, we then find
$$F\left(\pi^0\to\gamma(k_1,\lambda_1),\gamma(k_2,\lambda_2)\right)=
\dfrac{e^2(q^2-\mu^2)}{\sqrt{2\pi}}\epsilon^*_{\mu}(k_1,\lambda_1)\epsilon^*_{\nu}(k_2,\lambda_2)
F^{\mu\nu}(k_1,k_2),\equn{(8.6.2a)}$$
$ q=k_1+k_2,$
where we have defined the VEV
$$F^{\mu\nu}(k_1,k_2)=\int\dd^4x\,\dd^4y\,\ee^{\ii(x\cdot k_1+y\cdot k_2)}
\langle{\rm T}J^{\mu}(x)J^{\nu}(y)\phi(0)\rangle_0.\equn{(8.6.2b)}$$
We henceforth suppress the indices ``em" and ``$\pi^0$" in $J$ and $\phi$ respectively.

We next use the equation (8.4.3), generalized to include the $\pi^0$:
$$\eqalign{\partial_\mu A_0^\mu(x)=\sqrt{2}f_\pi \mu^2\phi(x),\quad
\phi\equiv\phi_{\pi^0},\cr    
A_0^{\mu}(x)=\dfrac{1}{\sqrt{2}}\left\{\bar{u}(x)\gamma^\mu\gamma_5u(x)-
\bar{d}(x)\gamma^\mu\gamma_5d(x)\right\}.\cr}\equn{(8.6.3a)}$$
It will prove convenient to use, instead of $A_0$, the current $A_3$, defined as
$$A_3^{\mu}(x)=\left\{\bar{u}(x)\gamma^\mu\gamma_5u(x)-
\bar{d}(x)\gamma^\mu\gamma_5d(x)\right\};\equn{(8.6.3b)}$$
with  it, we write
$$\eqalign{F^{\mu\nu}(k_1,k_2)=&\,\dfrac{1}{f_\pi \mu^2}T^{\mu\nu}(k_1,k_2),\cr  
T^{\mu\nu}(k_1,k_2)=&\,\tfrac{1}{2}
\int\dd^4x\,\dd^4y\,\ee^{\ii(x\cdot k_1+y\cdot k_2)}
\langle{\rm T}J^{\mu}(x)J^{\nu}(y)\partial\cdot A_3(0)\rangle_0.\cr}
\equn{(8.6.4)}$$

Up to this point, everything has been exact. The next step involves using the PCAC 
hypothesis in the following form: we assume that $F(\pi\to\gamma\gamma)$ can 
be approximated by its leading term in the limit $q\to0$. On purely 
kinematic grounds, this is seen to imply that also $k_1,k_2\to0$. One may write
$$T^{\mu\nu}(k_1,k_2)=\epsilon^{\mu\nu\alpha\beta}k_{1\alpha}k_{2\beta}\Phiv+O(k^3).
\equn{(8.6.5)}$$
The PCAC hypothesis means that we retain only the first
 term in \equ~(8.6.5). As will be seen presently, this 
will lead us to a contradiction, the resolution of which will involve introducing the so-called 
{\sl axial}, or {\sl triangle anomaly},
 and will allow us actually to calculate $T^{\mu\nu}$ exactly to all orders 
of perturbation theory (in the PCAC approximation).

The first step is to consider the quantity
$$R^{\lambda\mu\nu}(k_1,k_2)=
\ii\int\dd^4x\,\dd^4y\,\ee^{\ii(x\cdot k_1+y\cdot k_2)}
\langle{\rm T}J^\mu(x)J^\nu(y)A_3^\lambda(0)\rangle_0.\equn{(8.6.6)}$$
On invariance grounds, we may write the general decomposition,
$$R^{\lambda\mu\nu}(k_1,k_2)=\epsilon^{\mu\nu\lambda\alpha}k_{1\alpha}\Phiv_1+
\epsilon^{\mu\nu\lambda\alpha}k_{2\alpha}\Phiv_2+O(k^3),\equn{(8.6.7)}$$
where the $O(k^3)$ terms are of the form
$$\epsilon^{\mu\lambda\alpha\beta}k_{i\alpha}k_{j\beta}k_{l\lambda}\Phiv_{ijl}+
\hbox{ permutations of}\;i,j,l=1,2,3,$$
and, for quarks with nonzero mass, the $\Phiv$ are regular as $k_i\to0$. 

The conservation of the e.m. current, $\partial\cdot J=0$, yields two equations:
$$k_{1\mu}R^{\mu\nu\lambda}=k_{2\nu}R^{\mu\nu\lambda}=0.\equn{(8.6.8)}$$
The first implies
$$\Phiv_2=O(k^2);\equn{(8.6.9a)}$$
the second gives
$$\Phiv_1=O(k^2).\equn{(8.6.9b)}$$
Now we have, from (8.6.4) and (8.6.6),
$$q_\lambda R^{\lambda\mu\nu}(k_1,k_2)=T^{\mu\nu}(k_1,k_2),\quad {\rm i.e.},\;\Phiv=\Phiv_2-\Phiv_1,
\equn{(8.6.10)}$$
and hence we find the result of Veltman (1967) and Sutherland (1967),
$$\Phiv=O(k^2).\equn{(8.6.11)}$$
Because the scale for $k$ is $\mu$, this means that $\Phiv$ should be of order $\mu^2/M^2$, 
where $M$ is a typical hadronic mass. Thus, we expect that $\Phiv$ would be 
vanishing in the chiral limit, and hence very small in the real world. Now, this is in 
disagreement with experiment, as the decay $\pi^0\to2\gamma$
 is in no way suppressed; but worse still, 
(8.6.11) contradicts a direct calculation. In fact, we may use the equations of motion and write
$$\partial_\mu A_3^\mu(x)=
2\ii\left\{m_u\bar{u}(x)\gamma_5u(x)-m_d\bar{d}(x)\gamma_5d(x)\right\}.\equn{(8.6.12)}$$
We will calculate first neglecting strong interactions; (8.6.11) should certainly 
be valid in this approximation. This involves the diagrams of \fig~8.6.1 with a $\gamma_5$ 
vertex. The result, as first obtained by Steinberger (1949) is, in the limit $k_1,k_2\to0$,  
and defining $\delta_u=1$, $\delta_d=-1$,

\topinsert{
\setbox0=\vbox{\epsfxsize 10.6cm\epsfbox{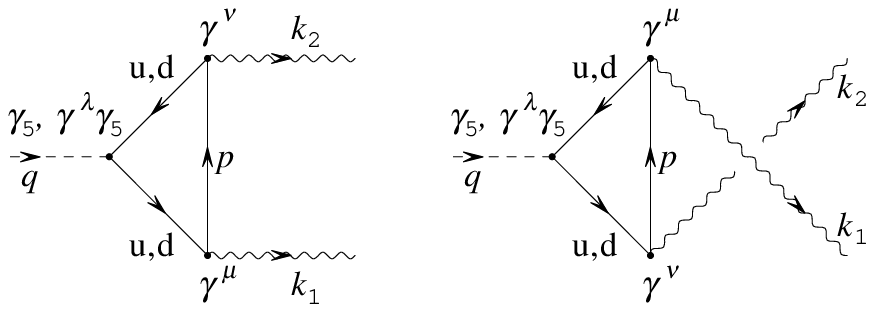}\hfil}
\centerline{{\box0}}
\setbox1=\vbox{\hsize 10truecm\captiontype\figurasc{Fig. 8.6.1. }
 {Diagrams connected with the anomaly ($\pi^0\to\gamma\gamma$ decay).}\hb
\vskip.1cm}
\centerline{\box1}
}\endinsert

$$\eqalign{T^{\mu\nu}(k_1,k_2)&=2N_c\sum_{f=u,d}\delta_fQ^2_fm_f\cr
\times&\int\dfrac{\dd^4p}{(2\pi)^4}\,
\dfrac{\trace\gamma_5(\slash{p}+\slash{k}_1+m_f)\gamma^\mu
(\slash{p}+m_f)\gamma^\nu(\slash{p}-\slash{k}_2+m_f)}{[(p+k_1)^2-m_f^2][(p-k_2)^2-m_f^2]
(p^2-m_f^2)}\cr
=&-\dfrac{1}{4\pi^2}
\epsilon^{\mu\nu\alpha\beta}k_{1\alpha}k_{2\beta}\left\{3(Q_u^2-Q_d^2)\right\}+O(k^4)\cr
=&-\dfrac{1}{4\pi^2}
\epsilon^{\mu\nu\alpha\beta}k_{1\alpha}k_{2\beta}+O(k^4).
\cr}$$
The factor $N_c=3$ comes from the sum over the three colours of the 
quarks and the factor 2
 from the two diagrams in \fig~8.6.1 (which in fact contribute equally to the amplitude). We 
thus find that
$$\Phiv=-\dfrac{1}{4\pi},\equn{(8.6.13)}$$
which contradicts (8.6.11). This is the {\sl triangle anomaly} (Bell and Jackiw, 1969; Adler, 1969).

What is wrong here? Clearly, we cannot maintain (8.6.12), which was 
obtained with free-field equations of motion, $\ii\slash{\partial}q=m_qq$; 
we must admit that in the presence of interactions with vector fields (the 
photon field in our case), \equ~(8.6.12) is no longer valid. To 
obtain agreement with (8.6.13) we have to write (Adler, 1969)
$$\eqalign{\partial_\mu A_3^\mu(x)=&\,
2\ii\left\{m_u\bar{u}(x)\gamma_5u(x)-m_d\bar{d}(x)\gamma_5d(x)\right\}\cr
&+N_c(Q_u^2-Q_d^2)\dfrac{e^2}{16\pi^2}F_{\mu\nu}(x)\widetilde{F}^{\mu\nu}(x),\cr}\equn{(8.6.14)}$$
where the {\sl dual} $\widetilde{F}$ has been defined as
$$\widetilde{F}^{\mu\nu}=\tfrac{1}{2}\epsilon^{\mu\nu\alpha\beta}F_{\alpha\beta},\quad 
F^{\mu\nu}=\partial^\mu A^{\nu}_{\rm ph}-\partial^\nu A^{\mu}_{\rm ph}.$$
More generally, for fermion fields interacting with vector fields with strength $h$, 
we find
$$\partial_\mu\bar{f}\gamma^\mu\gamma_5f=
2\ii m_f\bar{f}\gamma_5f+\dfrac{T_Fh^2}{8\pi^2}H^{\mu\nu}\widetilde{H}_{\mu\nu};\equn{(8.6.15)}$$
$H^{\mu\nu}$ is the vector field strength tensor.

Let us return to the decay $\pi^0\to 2\gamma$. From (8.6.13) we calculate
 the amplitude, in the PCAC limit 
$\mu\sim0$,
$$F(\pi^0\to2\gamma)=\dfrac{\alpha}{\pi}\,
\dfrac{\epsilon^{\mu\nu\alpha\beta}k_{1\alpha}k_{2\beta}\epsilon^*_\mu(k_1,\lambda_1)
\epsilon^*_\mu(k_2,\lambda_2)}{\sqrt{2\pi}},
\equn{(8.6.16)}$$
and the decay rate
$$\Gammav(\pi^0\to2\gamma)=\left(\dfrac{\alpha}{\pi}\right)^2\dfrac{m^3_\pi}{64\pi f^2_\pi}
=7.25\times10^{-6}\,\mev,$$
to be compared with the experimental figure,
$$\Gammav_{\rm exp}(\pi^0\to2\gamma)=7.95\times10^{-6}\,\mev.$$
Actually, the sign of the decay amplitude can also be measured (from the Primakoff effect) 
and it agrees with the theory. It is important to note that, if we had no colour,
our result would have decreased by a factor $1/N_c^2$, i.e., it would have been 
off experiment by a full order of magnitude.

One may wonder what credibility to attach to this calculation: after all,
 it was made to zero order in $\alpha_s$. In fact, the calculation is exact to all 
orders in QCD;\fnote{The proof is essentially contained in the original paper of 
Adler and Bardeen (1969). See also Wilson (1969), 
Crewther (1972) and Bardeen (1974).} the only approximation is the PCAC one 
$\mu\approx0$. To show this we will give an alternate
 derivation of the basic result, \equ~(8.6.13). We then return to (8.6.6). To zero 
order in $\alpha_s$,
$$\eqalign{R^{\mu\nu\lambda}=&\sum\delta_fQ^2_f\cr
\times&\int\dfrac{\dd^4p}{(2\pi)^4}\,
\dfrac{\trace\gamma^\lambda\gamma_5(\slash{p}+\slash{k}_1+m_f)\gamma^\mu
(\slash{p}+m_f)\gamma^\nu(\slash{p}-\slash{k}_2+m_f)}{[(p+k_1)^2-m_f^2][(p-k_2)^2-m_f^2]
(p^2-m_f^2)}\cr
+&\;\hbox{crossed term}\cr}$$
(\fig~8.6.1 with the $\gamma^\lambda\gamma_5$ vertices). More generally, we 
regulate the integral by working in dimension $D$, and consider 
an arbitrary axial triangle with
$$R^{\mu\nu\lambda}_{ijl}=2\int\dfrac{\dd^Dp}{(2\pi)^D}\,
\trace\gamma^\lambda\gamma_5\dfrac{1}{\slash{p}+\slash{k}_1-m_i}\gamma^\mu
\dfrac{1}{\slash{p}-m_j}\gamma^\nu\dfrac{1}{\slash{p}-\slash{k}_2-m_l}.\equn{(8.6.17)}$$
We would like to calculate $q_\lambda R^{\mu\nu\lambda}_{ijl}$. Writing identically
$$(\slash{k}_1+\slash{k}_2)\gamma_5=
-(\slash{p}-\slash{k}_2-m_l)\gamma_5+(\slash{p}+\slash{k}_1+m_i)\gamma_5-
(m_i+m_l)\gamma_5,$$
we have
$$\eqalign{q_\lambda R^{\mu\nu\lambda}_{ijl}=&\,-2(m_i+m_l)\cr
\times&\int\dfrac{\dd^Dp}{(2\pi)^D}\,
\dfrac{\trace\gamma_5(\slash{p}+\slash{k}_1+m_i)\gamma^\mu
(\slash{p}+m_j)\gamma^\nu(\slash{p}-\slash{k}_2+m_l)}{[(p+k_1)^2-m_i^2][(p-k_2)^2-m_l^2]
(p^2-m_j^2)}\cr
&+a^{\mu\nu\lambda}_{ijl},\cr}\equn{(8.6.18a)}$$
$$\eqalign{a^{\mu\nu\lambda}_{ijl}=-2&\int\dd \hat{p}\,
\trace\left\{(\slash{p}-\slash{k}_2-m_l)\gamma_5-(\slash{p}+\slash{k}_1+m_i)\gamma_5\right\}\cr
&\times\dfrac{1}{\slash{p}+\slash{k}_1-m_i}\gamma^\mu\,\dfrac{1}{\slash{p}-m_j}\gamma^\nu
\dfrac{1}{\slash{p}-\slash{k}_2-m_l}.\cr}\equn{(8.6.18b)}$$
The first term on the right hand side of (8.6.18a) is what 
we would have obtained by naive use of the equations of motion, 
$\partial_\mu\bar{q}_i\gamma^\mu\gamma_5q_l=\ii(m_i+m_l)\bar{q}_i\gamma_5q_l$; 
$a^{\mu\nu\lambda}_{ijl}$ is the anomaly. If we accepted the commutation
 relations $\{\gamma^\mu,\gamma_5\}=0$ also for dimension $D\neq4$, we could rewrite it as
$$\eqalign{a^{\mu\nu\lambda}_{ijl}=-2\int\dd\hat{p}\,\Bigg\{&\trace\gamma_5
\dfrac{1}{\slash{p}+\slash{k}_1-m_i}\gamma^\mu\dfrac{1}{\slash{p}-m_l}\gamma^\nu\cr
+&\trace\gamma_5\gamma^\mu\dfrac{1}{\slash{p}-m_j}\gamma^\mu
\dfrac{1}{\slash{p}-\slash{k}_2-m_l}\Bigg\}.\cr}\equn{(8.6.18c)}$$
Then we could conclude that $a^{\mu\nu\lambda}_{ijl}$ vanishes 
because each of the terms in (8.6.18c) consists of an
 antisymmetric tensor that depends on a single vector ($k_1$ for 
the first term, $k_2$ for the second) and this is zero. It is thus 
clear that the nonvanishing of $a^{\mu\nu\lambda}_{ijl}$ is due to the fact that 
it is given by an ultraviolet divergent integral: if it was convergent, 
one could take $D\to4$ and $a^{\mu\nu\lambda}_{ijl}$ would vanish. Incidentally, 
this shows that $a^{\mu\nu\lambda}_{ijl}$ is actually independent of the masses 
because $(\partial/\partial m)a^{\mu\nu\lambda}_{ijl}$ is convergent, 
and thus the former argument applies. We may therefore write $a^{\mu\nu\lambda}_{ijl}=a^{\mu\nu}$, 
where $a^{\mu\nu}$ is obtained by setting all masses to zero. A similar 
argument shows that $a^{\mu\nu}$ has to be of the form
$$a^{\mu\nu}(k_1,k_2)=a\epsilon^{\mu\nu\alpha\beta}k_{1\alpha}k_{2\beta},\quad a={\rm constant},
\equn{(8.6.19a)}$$
and thus we may obtain $a$ as
$$a\epsilon^{\mu\nu\alpha\beta}=
\dfrac{\partial^2}{\partial k_{1\alpha}\partial k_{2\beta}}a^{\mu\nu}(k_1,k_2)\Big|_{k_i=0}.
\equn{(8.6.19b)}$$ 
If we could write the formula (8.6.18c) for $a$, we would immediately conclude from 
(8.6.19b) that $a=0$, in contradiction with the Veltman--Sutherland theorem.
 But this is easily seen to be inconsistent: if we would have shifted 
variables in (8.6.18c), say $p\to p-\xi k_2$, we would have found a finite but 
nonzero value, actually $\xi$-dependent  for $a$, $a=-\xi/2\pi^2$. This shows
 that the commutation relations\fnote{These commutation relations are 
actually self-contradictory. For example, using only the commutation relations of 
the $\gamma_\mu$, $\mu=0,\dots,D-1$ for $D\neq4$, we have
$$\trace \gamma_5\gamma^\alpha\gamma^\mu\gamma^\nu\gamma^\rho\gamma_\alpha\gamma^\sigma=
(6-D)\trace\gamma_5\gamma^\mu\gamma^\nu\gamma^\rho\gamma^\sigma,$$
while, if we allow $\gamma_5$ anticommutation, we can obtain
$$\trace\gamma_5\gamma^\alpha\gamma^\mu\gamma^\nu\gamma^\rho\gamma_\alpha\gamma^\sigma=
-\trace\gamma_5\gamma^\mu\gamma^\nu\gamma^\rho\gamma_\alpha\gamma^\sigma\gamma^\alpha=(D-2)
\trace\gamma_5\gamma^\mu\gamma^\nu\gamma^\rho\gamma^\sigma,$$
which differs from the former by a term $O(D-4)$. These problems, however, only arise for 
arrays with an odd number of $\gamma_5$ and at least four other 
gammas.} 
$\{\gamma^\mu,\gamma_5\}=0$ cannot be accepted for $D\neq0$, for 
they lead to an undefined value for the anomaly. If, however,
 we start from (8.6.18b) and refrain from commuting $\gamma_5$ and $\gamma^\mu$s, 
$$a\epsilon^{\mu\nu\alpha\beta}=-2\int\dd\hat{p}\trace\gamma_5
\left\{\dfrac{1}{\slash{p}}\gamma^\alpha
\dfrac{1}{\slash{p}}\gamma^\mu
\dfrac{1}{\slash{p}}\gamma^\nu
\dfrac{1}{\slash{p}}\gamma^\beta-
\dfrac{1}{\slash{p}}\gamma^\mu
\dfrac{1}{\slash{p}}\gamma^\nu
\dfrac{1}{\slash{p}}\gamma^\beta
\dfrac{1}{\slash{p}}\gamma^\alpha\right\}.$$
Performing symmetric integration and using only the commutation rules 
 for $D\neq4$, we obtain an unambiguous result:
$$\eqalign{a\epsilon^{\mu\nu\alpha\beta}=&\,
\dfrac{8(D-1)(4-D)}{D(D+2)}\,\dfrac{\ii}{16\pi^2}\dfrac{2}{4-D}\,
\trace\gamma_5\gamma^\mu\gamma^\nu\gamma^\alpha\gamma^\beta+O(4-D)\cr
&\rightarrowsub_{D\to4}-\dfrac{1}{2\pi^2}\epsilon^{\mu\nu\alpha\beta}.\cr}$$
This is one of the peculiarities of the anomaly: a {\sl finite} Feynman 
integral whose value depends on the regularization prescription. Fortunately,
 we may eschew the problem by using the Veltman--Sutherland theorem to conclude 
that, at any rate, there is a {\sl unique} value of $a^{\mu\nu}$ 
compatible with gauge invariance for the e.m. current, viz.,
$$a^{\mu\nu}_{ijl}=a^{\mu\nu}=
-\dfrac{1}{2\pi^2}\epsilon^{\mu\nu\alpha\beta}k_{1\alpha}k_{2\beta}.\equn{(8.6.20)}$$
We have explicitly checked that our regularization leads to 
precisely this value; to verify that it also respects gauge invariance is left as as simple exercise.

Before continuing, a few words on the Veltman--Sutherland theorem for zero quark 
masses are necessary. In this case, the first term on the 
right hand side of (8.6.18a) is absent: it would appear that we could not maintain our result for the anomaly, 
\equ~(8.6.20), because this would imply
$$q_\lambda R^{\mu\nu\lambda}_{ijl} =
-\dfrac{1}{2\pi^2}\epsilon^{\mu\nu\alpha\beta}k_{1\alpha}k_{2\beta}\neq0,$$
thus contradicting the Veltman--Sutherland conclusion, $q_\lambda R^{\mu\nu\lambda}_{ijl}=0$. 
This is not so. The relation $q_\lambda R^{\mu\nu\lambda}_{ijl}=a^{\mu\nu}$ 
and the value of $a^{\mu\nu}$ {\sl are} correct. What occurs is that 
for vanishing masses the functions $\Phiv_i$ in (8.6.7) possess singularities of the type 
$1/k_1\cdot k_2$, singularities coming from the denominators in, 
for example, \equ~(8.6.17) 
when $m_i=0$. Therefore, the 
Veltman--Sutherland theorem is {\sl not} applicable. This is yet another 
peculiarity of the anomalous triangle: we have the relation 
$$\lim_{m\to0}q_\lambda R^{\mu\nu\lambda}_{ijl}=0$$
 but, 
if we begin with $m=0$,
$$q_\lambda R^{\mu\nu\lambda}_{m\equiv0}=a^{\mu\nu}\neq0.$$
\topinsert{
\setbox0=\vbox{{\epsfxsize11.truecm\epsfbox{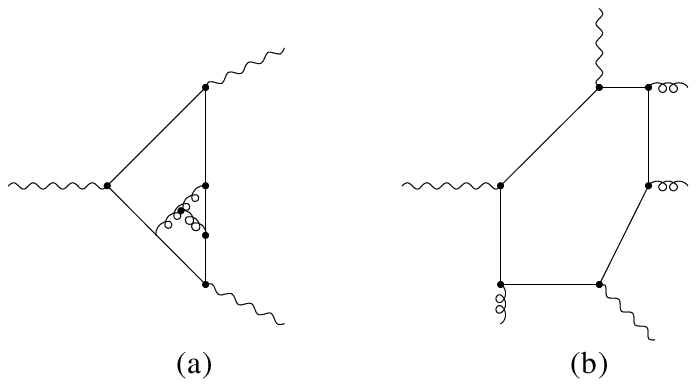}}\hfil}
\centerline{{\box0}}
\setbox1=\vbox{\hsize 10truecm\captiontype\figurasc{Fig. 8.6.2. }
 {(a) A nonanomalous diagram. (b) ``Opened" diagram\hb 
corresponding to (a).}\hb
\vskip.4cm}
\centerline{\box1}
\vskip0.4cm
}\endinsert

Let us return to our original discussion, in particular for $m\neq0$. 
The present method shows how one can prove that the result does not get renormalized. 
The Veltman--Sutherland theorem is exact; so we have actually shown that 
it is sufficient to prove that (8.6.20) is not altered by higher orders in $\alpha_s$.
 Now, consider a typical 
higher order contribution (\fig~8.6.2a). It may be written as an integral 
over the gluon momenta and an integral over the quark momenta. But for the latter, the 
triangle has become an hexagon (\fig~8.6.2b) for 
which the quark integral is convergent and here the limit $D\to4$ 
may be taken: it vanishes identically. In addition, the above  
arguments have shown that the anomaly is in fact related 
to the large momentum behaviour of the theory and thus we 
expect that the exactness of (8.6.13) will not be spoiled by nonperturbative effects.
We will not make the proof more precise, but refer to the literature.\fnote{For 
a detailed discussion, see the reviews of Adler (1971) and Ellis (1976). 
The triangle graph is the only one that has {\sl primitive} anomalies; it does, 
however, induce secondary anomalies 
in square and pentagon graphs. The triangle with three axial currents has
 an anomaly closely related to the one we have discussed, cf. the text of Taylor (1976).
 An elegant discussion 
of currents with anomalies for arbitrary interaction may be found in Wess and Zumino (1971).
The derivation of the anomaly in the context of the path integral formulation of field 
theory, where it is connected with the {\sl divergence} of the measure, 
may be found in  Fujikawa (1980,~1984,~1985).}

\booksubsection{8.6.2. The $U(1)$ problem and the gluon anomaly}

\noindent In the previous section, we discussed the triangle anomaly
 in connection with the decay $\pi^0\to\gamma\gamma$. 
As remarked there, the anomaly is not restricted to photons; in particular, we 
have a gluon anomaly. 
Although this lies outside the scope of the present review, we will say a few words on the subject.
 Defining the current
$$A_0^\mu=\sum_{f=1}^n\bar{q}_f\gamma^\mu\gamma_5q_f,\equn{(8.6.23)}$$
we find that it has an anomaly
$$\partial^\mu A_0^\mu=\ii\sum_{f=1}^n2m_f\bar{q}_f\gamma_5q_f+
\dfrac{ng^2}{16\pi^2}\widetilde{G}G,\equn{(8.6.24)}$$
where
$$\widetilde{G}_a^{\mu\nu}\equiv \tfrac{1}{2}\epsilon^{\mu\nu\alpha\beta}G_{a\alpha\beta},
\quad \widetilde{G}G=\sum_a\widetilde{G}_a^{\mu\nu}G_{a\mu\nu}.$$
The current (8.6.23) is the so-called $U(1)$ current (pure flavour singlet) and is 
atypical in more respects than one. In particular, it is associated with the 
$U(1)$ problem, to which we now turn.

Assume that we have $n$ light quarks; we only consider these and will  neglect 
(as irrelevant to the problem at hand) the existence of heavy flavours.
 We may take $n=2$ ($u,d$) and then we speak of 
``the $U(1)$ problem of $SU(2)$" or $n=3$ ($u,d,s$), which is the $SU(3)$ $U(1)$ problem. 
 Consider now the $n^2-1$ matrices in flavour space $\lambda_1,\dots,\lambda_{n^2-1}$;
for $SU(3)$ they coincide 
with the Gell-Mann matrices, and for $SU(2)$ with the Pauli matrices. 
 Define further $\lambda_0\equiv1$. Any 
$n\times n$ Hermitian matrix may be written as a linear combination of the $n^2$ matrices
 $\lambda_\alpha,\;\alpha=0,1,\dots,n^2-1$.
 Because of this completeness, it is sufficient to consider 
the currents
$$A_{\alpha}^{\mu}=\sum_{ff'}\bar{q}_f\gamma^\mu\gamma_5\lambda^{\alpha}_{ff'}q_{f'};
\quad \alpha=0,1,\dots,n^2-1.$$
Of course, only $A_0$ has an anomaly. 

Now let $N_1(x),\dots,N_k(x)$ 
denote local operators (simple or composite) and 
consider the quantity
$$\langle{\rm vac}|{\rm T}A_{\alpha}^{\mu}(x)\prod_jN_j(x_j)|{\rm vac}\rangle.
\equn{(8.6.25)}$$
For $\alpha=a\neq0$, the Goldstone theorem implies that the masses of the pseudoscalar particles 
$P_a$ with the quantum numbers of the $A_a$ vanish in the chiral limit; introducing a common 
parameter $\epsilon$ for all the quark masses by letting $m_f=\epsilon r_f$, 
$f=1,\dots,n$, where the 
$r_f$ remain fixed in the chiral limit, we have
$$m^2_{P_a}\approx \epsilon.\equn{(8.6.26)}$$
Therefore, 
in this limit, the quantity (8.6.25) develops a pole at $q^2=0$, for $\alpha=a\neq0$. 
To be precise, what this means is that in the chiral limit (zero quark masses),
$$\lim_{q\to0}\int\dd^4x\,\ee^{\ii q\cdot x}
\langle{\rm vac}|{\rm T}A_{\alpha}^{\mu}(x)\prod_jN_j(x_j)|{\rm vac}\rangle
\approx \hbox{(const.)}\times\,q^\mu \dfrac{1}{q^2}.$$
If we neglect anomalies, the derivation of (8.6.26) can be repeated for the case 
$\alpha=0$ and we would thus find that the $U(1)$ (flavour singlet)
 particle $P_0$ would also have vanishing mass in the chiral limit (Glashow, 1968). 
This statement 
was made more precise by Weinberg (1975) who 
proved the bound $m_{P_0}\leq\sqrt{n}\times({\rm average}\;m_{P_a})$.
 Now, this is a catastrophe since, for the $SU(2)$ case,
$m_\eta\gg\sqrt{2}\,\mu$ and, for 
$SU(3)$, the mass of the $\eta'$ particle also violates the bound. This is the 
$U(1)$ problem. In addition, 
Brandt and Preparata (1970) proved that under these conditions the decay 
$\eta\to3\pi$ is forbidden, which is also in contradiction with experiment. We are thus led to 
{\sl assume} that (8.6.25) remains regular as 
$\epsilon\to0$ for 
$\alpha=0$. If we could {\sl prove} that this is so, we would have solved the 
$U(1)$ problem. We will not discuss this matter 
any more here, sending to the standard references.\fnote{Adler (1969); Bardeen
(1974); Crewther (1979b), etc.}

\bookendchapter

\brochureb{\smallsc chapter 9}{\smallsc  chiral perturbation theory}{117}
\bookchapter{9. Chiral perturbation theory}
\vskip-0.5truecm
\booksection{9.1 Chiral Lagrangians}
\vskip-0.5cm
\booksubsection{9.1.1. The $\sigma$ Model}

\noindent In this and the following sections we will  describe a method
 that has been devised to explore 
{\sl systematically} the consequences of the chiral symmetries of QCD, in the limit of 
small momenta and neglecting the light quark masses (or to leading order in 
 these). 
The method consists in writing Lagrangians consistent with chiral symmetry 
for pion field operators. 
These Lagrangians are not unique but, on the mass shell and for momenta $p^2$ 
much smaller than $\Lambdav^2$, all produce the {\sl same} results 
(Coleman, Wess and Zumino, 1969; Weinberg, 1968a). 
The Lagrangians are not renormalizable, but 
this is not important as they are to be used only at tree level
 (actually, it turns out to be possible to go beyond tree level,  at the 
cost of introducing a number of phenomenological constants, as we will discuss later). 
One can then use these Lagrangians to calculate low energy quantities 
involving pions, if the symmetry we consider is chiral $SU(2)$, 
reproducing the results obtained in a more artisanal way with the help of 
current algebra and soft pion (PCAC) techniques. 
This general formulation of chiral dynamics was first proposed by 
Weinberg~(1979) and later developed  in much greater detail by 
Gasser and Leutwyler~(1984,~1985a,\/b). 
We will begin in this section with a few examples, to proceed in next section to 
contact with PCAC and present a first application; the general formulation of 
chiral perturbation theory will be left for \sect~9.3.

The starting point to formulate the effective chiral Lagrangian theories is to write the chiral 
transformation properties of pions,\fnote{We will consider here explicitly only
 chiral $SU(2)$; the extension to  chiral $SU(3)$, 
that is to say, to processes involving also kaons and the 
$\eta$, is straightforward.}  whose field we denote by $\vec{\varphi}$, with 
the vector representing an isospin index, 
and a fictitious, scalar particle that we will 
denote by $\sigma$. 
This is the so-called {\sl sigma model} for spontaneous  symmetry breaking, devised by 
Gell-Mann and L\'evy~(1960). For infinitesimal chiral (i.e., parity changing) transformations we write,
$$\eqalign{
\sigma\to& \sigma+\delta\sigma,\quad \delta\sigma=-\vec{\alpha}\vec{\varphi}\cr
\vec{\varphi}\to&\vec{\varphi}+\delta\vec{\varphi},\quad \delta\vec{\varphi}=\vec{\alpha}\sigma.\cr
}
\equn{(9.1.1a)}$$
The $\vec{\alpha}$  are the parameters of the chiral transformations in $SU(2)\times SU(2)$, 
which would correspond, in a quark formulation, to the transformations 
involving $\gamma_5$. 

For ordinary isospin transformations, with parameters $\vec{\theta}$, we have
$$\delta\sigma=0,\qquad\delta\vec{\varphi}=\vec{\theta}\times\vec{\varphi}.
\equn{(9.1.1b)}$$
Because we suppose invariance under the full  $SU(2)\times SU(2)$ transformations 
it follows that $\sigma$ and $\vec{\varphi}$ fields 
should have the same mass, that (in a first approximation) we take to be zero. 

We now assume that the interaction is such that the field $\sigma$ acquires a vacuum expectation value, 
$\langle\sigma\rangle=k\neq0$; this will provide a large (i.e., of order $\Lambdav$) mass 
for the sigma field, which will then disappear from the
 low energy effective theory.\fnote{Alternatively, 
we could interpret it as the enhancement experimentally observed in 
the isospin zero S-wave in 
pion-pion scattering at an energy around 750 \mev. 
The key point, of course, is that 
at low energies only the pions give appreciable contributions; 
those from other particles are suppressed by powers $p^2/M_{\sigma}^2$.} 
To formulate the last, we want to redefine fields which do no more mix 
under chiral transformations. It happens that this is not possible if using {\sl linear} transformations; 
but can be achieved if  
nonlinear ones are allowed ({\sl nonlinear sigma models}).
 A simple choice is to set $\sigma'=\sigma-k$ 
(so the VEV of $\sigma'$ vanishes) 
and then  define
$$\eqalign{R=&\sqrt{(\sigma'+k)^2+\vec{\varphi}^2}-k,\cr
\vec{\pi}=&\dfrac{k}{\sqrt{(\sigma'+k)^2+\vec{\varphi}^2}}\,\vec{\varphi}.\cr}
\equn{(9.1.2)}$$
For small energies we can expand the new fields in terms of the old 
(in effect, this is an expansion in powers of $k^{-1}$), 
$$R\simeq\sigma'+\cdots,\qquad\vec{\pi}\simeq \vec{\varphi}+\cdots$$
so that the $R$, $\vec{\pi}$ coincide, at leading order, with the old fields. 
However, the new fields do not mix under chiral transformations: 
we get
$$\delta
R=0,\qquad\delta\vec{\pi}=\vec{\alpha}\,\dfrac{k\sigma}{\sqrt{\sigma^2+\vec{\varphi}^2}}=
\vec{\alpha}\sqrt{k^2-\vec{\pi}^2}.
\equn{(9.1.3a)}$$
Under ordinary isospin we still have,
$$\delta R=0,\qquad\delta\vec{\pi}=\vec{\theta}\times\vec{\pi}.
\equn{(9.1.3b)}$$

Because of these properties we can write a Lagrangian, invariant under chiral 
transformations, using 
only the field $\vec{\pi}$: we have 
succeeded in decoupling the sigma field. 
The Lagrangian is not unique; a choice, suggested by 
Coleman, Wess and Zumino (1969), is to take
$${\cal L}=\tfrac{1}{2}\dfrac{1}{(1+a^2\vec{\Piv}^2)^2}\left(\partial\vec{\Piv}\right)^2,\quad 
(\partial\vec{\Piv})^2\equiv (\partial_\mu\vec{\Piv}) (\partial^\mu\vec{\Piv});\qquad
a=1/2k,
\equn{(9.1.4a)}$$
with $\vec{\Piv}$ a reparametrization of $\vec{\pi}$:
$$\vec{\Piv}=\dfrac{2\vec{\pi}}{1+\sqrt{1-\vec{\pi}^2/k^2}};$$
it transforms chirally as
$$\delta\vec{\Piv}=\dfrac{1}{a}
\left[\vec{\alpha}\left(1-a^2\vec{\Piv}^2\right)
+2a^2\vec{\Piv}(\vec{\alpha}-\vec{\Piv})\right].$$

We may expand $\cal{L}$ in powers of $a$ getting
$${\cal L}=\tfrac{1}{2}\left(\partial\vec{\Piv}\right)^2-
\dfrac{a^2}{2}\vec{\Piv}^2\left(\partial\vec{\Piv}\right)^2
+\dfrac{a^4}{2}\vec{\Piv}^4\left(\partial\vec{\Piv}\right)^2+\cdots
\equn{(9.1.5)}$$

\topinsert{
\setbox0=\vbox{\hsize8truecm{\epsfxsize 6.truecm\epsfbox{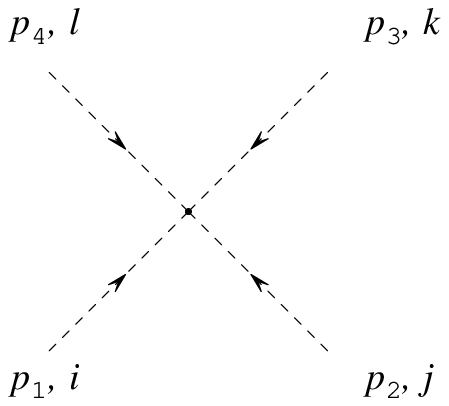}}} 
\setbox6=\vbox{\hsize 5truecm\captiontype\figurasc{Figure 9.1.1 }{\hb
The  four 
pion graph.\hb
\phantom{XX}}\hb
\vskip.1cm} 
\medskip
\line{
{\box0}\hfil\box6}
\medskip
}\endinsert

To show the usefulness  
of the effective Lagrangian formulation, we
 calculate $\pi\pi$ scattering to lowest order in $a$. 
Denote by $i,\,j,\,k,\,l$ to the isospin indices, 
varying from 1 to 3. 
The Feynman rule corresponding to (9.1.5) to lowest order 
in $a$ is, for a four-pion vertex with 
momenta $p_1,\,p_2,\,p_3,\,p_4$, all incoming (\fig~9.1.1),
$$\eqalign{\ii a^2g_{\mu\nu}\Big[
&\delta_{ij}\delta_{kl}\left(p_3^\mu p_4^\nu+p_1^\mu p_2^\nu\right)\cr
+&\delta_{ik}\delta_{jl}\left(p_2^\mu p_4^\nu+p_1^\mu p_3^\nu\right)\cr
+&\delta_{il}\delta_{jk}\left(p_2^\mu p_3^\nu+p_1^\mu p_4^\nu\right)
\Big].\cr}
\equn{(9.1.6)}$$
In terms of the Mandelstam variables
$$s=(p_1+p_2)^2,\quad t=(p_2+p_4)^2,\quad u=(p_2+p_3)^2,$$
 we can write the scattering amplitude that follows from (9.1.6) to lowest order as
$$F(i+j\to k+l)=
\dfrac{a^2}{(2\pi)^2}
\left\{\delta_{ij}\delta_{kl}s+\delta_{ik}\delta_{jl}t+\delta_{il}\delta_{jk}u\right\}.
\equn{(9.1.7)}$$
We will later identify $a$ with $1/f_\pi$, the inverse of the pion decay constant, so 
(9.1.7) gives the low energy ($s,\,t,\,u\ll \Lambdav^2$) pion-pion collision amplitude.
The simplicity of this evaluation contrasts with that based on ``old fashioned" 
PCAC, current algebra and soft pion techniques (Weinberg, 1966). 

\booksubsection{9.1.2. Exponential formulation}

\noindent A more elegant, but equivalent formulation uses a matrix representation of the pion field. 
Letting $\vec{\tau}$ be the Pauli matrices, for isospin space, 
we construct the $2\times2$ matrix
$$\pi=\vec{\tau}\vec{\varphi}
\equn{(9.1.8a)}$$
with $\vec{\varphi}$ the pion field. 
We then exponentiate $\pi$ and set the matrix 
$$\Sigmav=\exp2\ii\pi/F.
\equn{(9.1.8b)}$$ 
The chiral $SU(2)\times SU(2)$ transformations are defined in terms of the 
unitary matrices $W_L,\,W_R$:
$$\Sigmav\to\Sigmav'\equiv W_L\sigmav W_R^{\dag}.
\equn{(9.1.8c)}$$
The symmetry breaking condition is implemented by assuming a nonzero VEV for 
$\Sigmav$:
$$\langle \Sigmav\rangle=
\pmatrix{F&0\cr
0&F\cr}.$$
The advantage of the present method is that we only work with the pion field from the 
beginning. 

It is convenient to parametrize the $W_{R,L}$ as 
$$\eqalign{
W_L=&\ee^{\vec{\alpha}\vec{\tau}}\ee^{\vec{\theta}\vec{\tau}},\cr
W_R=&\ee^{-\vec{\alpha}\vec{\tau}}\ee^{\vec{\theta}\vec{\tau}}.\cr}
\equn{(9.1.9)}$$
For ordinary isospin transformations we simply set
 $\vec{\alpha}=0$ so that $W_L$ and $W_R$ coincide 
and (9.1.8) is equivalent to $\pi'=W(\vec{\theta}\/\/)\pi W^{-1}(\vec{\theta}\/\/)$. 
Then, 
for the pion field itself we have
$\vec{\pi}'=R(\vec{\theta}\/)\vec{\pi}$ with $R(\vec{\theta}\/\/)$ the three-dimensional
rotation  corresponding to the $SU(2)$ matrix $W(\vec{\theta}\/\/)$ given by the relation
$$W(\vec{\theta}\/)\tau_i W^{-1}(\vec{\theta}\/\/)=\sum_j
R^{-1}_{ij}(\vec{\theta}\/\/)\tau_j.$$  Under an infinitesimal chiral transformation, (9.1.8)
gives, after expanding,
$$\vec{\pi}'=\vec{\pi}+F\vec{\alpha}+\cdots.$$

Next we construct a Lagrangian invariant under 
(9.1.8). The one which contains {\sl less} derivatives is
$${\cal L}=\dfrac{F^2}{4}\trace \left(\partial_\mu\Sigmav^+\right)\partial^\mu\Sigmav,
\equn{(9.1.10)}$$
and the overall constant is chosen so that, after expanding, the kinetic 
term is $\tfrac{1}{2}(\partial_\mu\vec{\pi})\partial^\mu\vec{\pi}$.
This shows clearly the arbitrariness of the method: we can add 
extra terms with higher derivatives to (9.1.10). 
However, they will, on dimensional grounds, contribute to higher orders in the momenta. 
But it is important to realize that the effective Lagrangian methods are only 
valid to give the first orders in the expansion in powers of the momenta, $p^2/\Lambda^2$. 
The theory says nothing {\sl a priori} about higher corrections, which involve more and more 
arbitrary parameters. 

In this formalism we can introduce in a natural manner leading order 
symmetry breaking by considering that it is due to a quark mass 
matrix,
$$M=\pmatrix{m_u&0\cr0&m_d\cr}.$$
This is not invariant under chiral (or even ordinary isospin) 
transformations. 
We may couple $M$ and $\Sigmav$; the lowest dimensionality scalar that can be formed is the 
function
$$v^3\trace (\Sigmav^+M+M\Sigmav).$$
$v$ is a constant with dimensions of mass, that we will identify later.
Expanding in powers of $\pi$, we find that the first 
nonzero term is the quadratic one,
$$-\dfrac{4v^3}{F^2}\trace M\pi^2=-\dfrac{4v^3}{F^2}(m_u+m_d)\vec{\pi}^{\/2},
\equn{(9.1.11)}$$
and we have used that $(\vec{\lambda}\vec{\tau})^{\/2}=\vec{\lambda}^{\/2}$ for any
$\vec\lambda$.
\equn{(9.1.11)} provides the lowest order mass term for the pions; it has the nice feature 
that it reproduces (as it should) the result we had obtained with the help of PCAC and 
current algebra in (8.2.4). This allows us to 
realize that $v^3$ is proportional to the quark condensate.
Applications of this to calculate some hadronic corrections 
to low energy weak interactions may be found in the book of Georgi (1984).

An alternate to the exponential formulation presented here will be given in \sect~9.3.

\booksection{9.2. Connection with PCAC, and a first application}

\noindent Before starting to calculate with the chiral Lagrangians described in the 
previous section we have to interpret the constant ($F$ or $a$) that 
appears there. 
For this we have to introduce the {\sl axial current} in the present formalism, 
which we choose to do in the original Coleman--Wess--Zumino version;  
the derivation in the exponential version, somewhat messier, 
may be found in the text of Georgi~(1984).
  To do so we use a method which is a variant of
Noether's method, due to Adler (for more  details on it, see Georgi, 1984 or Adler, 1971). 
Let us consider a general Lagrangian ${\cal L}(\phi)$ depending on the field $\phi$, 
and make 
an infinitesimal transformation on the fields, characterized by the 
infinitesimal parameters
$\epsilon_i$:
$$\delta\phi=\sum_i\epsilon_i\xi_i(\phi).$$
The corresponding variation of the Lagrangian is then
$$\delta{\cal L}= K_i(\phi)\epsilon_i+L_i^\mu(\phi)\partial_\mu\epsilon_i+
M_i^{\mu\nu}(\phi)\partial^\mu\partial^\nu\epsilon_i+\hbox{higher derivatives}.$$
(sum over repeated indices understood). 
The variation of the action can then be written, after integrating by parts, as
$$\delta{\cal A}=\int\dd^4x\,\left\{K_i+\partial_\mu J_i^\mu\right\}\epsilon_i$$
and we have defined the current $J$ by
$$J_i^\mu=-L_i^\mu+\partial_\nu M_i^{\mu\nu}+\cdots.$$
For a symmetry of the system, the change must leave 
the action unchanged, hence $\partial_\mu J_i^\mu=-K_i$.
Moreover, choosing $\epsilon$ constant, $\cal L$ will be invariant only if 
$K_i=0$. 
In this case, $J_i^\mu$ is obtained simply as the coefficient of 
$\partial_\mu\epsilon_i$ in the variation of $\cal L$. 
It is interesting to note that, if $\cal L$ only contains 
first order derivatives of the field $\phi$, then all the terms $M,$ etc. above vanish 
so  $J_i^\mu$  coincides with $-L_i^\mu$. 

This can be immediately applied to the Lagrangian (9.1.4a). 
Working to lowest order in $\Piv$, we find immediately the 
axial current to be
$$\vec{A}_\mu=-\dfrac{1}{a}\partial_\mu\vec{\Piv}+\hbox{higher orders}=
-\dfrac{1}{a}\partial_\mu\vec{\varphi}+\hbox{higher orders}.$$
Taking derivatives of both sides and using the equations of motion this gives
$$\partial^\mu\vec{A}_\mu=\dfrac{1}{a}\mu^2\vec{\varphi}.$$
On comparing with the definitions in \sect~7.3, we can identify
$$\dfrac{1}{a}=f_\pi,$$
$f_\pi$ the pion decay constant, $f_\pi\simeq 93\;\mev$. 
(The factor $\sqrt{2}$ in the definitions of \sect~7.3 has disappeared because 
the physical pion states are related to the ones used now, $\vec{\pi}$, by 
$\pi^\pm=\mp2^{-1/2}(\pi_1\pm\ii \pi_2)$).    

With this identification we get the pion-pion scattering amplitude, 
given in  
\equn{(9.1.7)}, as
$$F(i+j\to k+l)=
\dfrac{1}{4\pi^2f^2_\pi}
\left\{\delta_{ij}\delta_{kl}s+\delta_{ik}\delta_{jl}t+\delta_{il}\delta_{jk}u\right\}.
\equn{(9.2.1)}$$
From this one can evaluate the low energy parameters for $\pi\pi$ scattering. 
For example, the isospin 1, P wave scattering length is 
calculated as follows. First, we identify the physical pion states in terms
 of the $i,\,j,\,\dots=1,\,2,\,3$ ones
 as 
$$|\pi^0\rangle=
|3\rangle,\qquad|\pi^{\pm}\rangle=\mp2^{-1/2}\left\{|1\rangle\pm\ii|2\rangle\right\};$$
the isospin 1 state will appear in particular in the combination $|\pi^0\pi^+\rangle$ as 
$$|\pi^0\pi^+\rangle=2^{-1/2}|I=1\rangle+2^{-1/2}|I=2\rangle.$$ 
Moreover, we have the partial wave expansion, for states with well defined isospin $I$, 
$$F^{(I)}=2\sum_l(2l+1)P_l(\cos\theta)f_l^{(I)};\qquad f^{(I)}_l=
\dfrac{2s^{1/2}}{\pi k}\sin \delta_l^{(I)}\ee^{\ii \delta^{(I)}_l},
\equn{(9.2.2)}$$
 with $\delta_l^{(I)}$ the phase shifts.\fnote{Recall
 that the factor 2 in the partial wave expansion is due to
the identity 
of the particles, in states with well-defined isospin.}

At small energy we write 
the {\sl partial wave amplitudes}, $f_l^{(I)}$, in terms of the 
{\sl scattering lengths}, $a_l^{(I)}$:
$$f_l^{(I)}(s)\simeqsub_{s\to 4M_{\pi}^2}\;\dfrac{4M_{\pi} k^{2l}}{\pi}a_l^{(I)}.$$
$k$ is the center of mass momentum; for massless pions, we can take $k^2=s/4$. 
Following the custom in modern chiral perturbative calculations, we express our numbers in terms of
$M_\pi=m_{\pi^+}$.  With all this we find, for the P wave
$$a_1=\dfrac{1}{24\pi f_\pi^2M_{\pi}}\simeq 0.031\, M_{\pi}^{-3}.
\equn{(9.2.3)}$$
Experimentally, and from the analysis of \sect~6.8, we know that 
$$a_1({\rm exp.})=(0.0386\pm0.0012)\,M_{\pi}^{-3}.$$ 
The agreement between theory and experiment improves if including pion mass corrections, 
and higher order 
chiral perturbative theory terms 
(to be discussed later).

The S-wave phase shifts are similarly calculated, and 
 we get,
$$\eqalign{
a_0^{(0)}=&\dfrac{7M_{\pi}}{32\pi f^2_\pi}\simeq0.157\; M_{\pi}^{-1};\cr
a_0^{(2)}=&-\dfrac{M_{\pi}}{16\pi f^2_\pi}\simeq-0.045\; M_{\pi}^{-1}.
\cr}$$
The agreement of these with experiment is less good than before.
 Including corrections, the
predicted  value for $a_0^{(0)}$ (for example) could go up 
to $0.22\,M_{\pi}^{-1}$, while  experiment gives 
values in the  range $0.215\, M_{\pi}^{-1}$ to $0.240\,M_{\pi}^{-1}$, 
as we saw in Chapter~6. Corrections will be discussed 
in more detail later on.

\booksection{9.3 Chiral perturbation theory: general formulation}
There is a large number of further applications of chiral perturbation theory  
(at times also denoted by the name of {\sl $\chi$PT}\/), to leading order, which the interested reader 
may find in the text of Georgi (1984). But one may ask if 
it is possible to go beyond. 
In fact, an enormous amount of work has been devoted to the matter in recent years,
 particularly following the basic 
papers of Gasser and Leutwyler (1984, 1985a,\/b).\fnote{We will not be able 
to give an amount of information comparable to that
 presented in these papers; we urge the reader to consult
them for a more detailed treatment and further applications. 
The subject has had an enormous growth in the 
last years; a recent review, with references, 
is that by Scherer~(2002). An introductory one is 
the text by Dobado et al.~(1997).} 
In the present section we will indeed describe  the general formalism of chiral
 perturbation theory, following, precisely, 
the excellent expos\'e of these authors. 
We will restrict ourselves to chiral isospin; the extension to chiral $SU(3)$ may be found in 
Gasser and Leutwyler (1985a).

The idea is the following: we will first extend the chiral symmetry in QCD 
to a gauge symmetry. Then we will construct the more general Lagrangians involving 
pions (for chiral $SU(2)$), first to 
leading order and then to higher orders, consistent with the 
PCAC definition $\partial\cdot \rvec{A}= f_\pi mu^2 \rvec{\phi_\pi}$ 
and verifying the gauge chiral symmetry. 
Because these Lagrangians share the symmetry with the QCD one, it will follow 
that the theory based on pions will satisfy identical Ward identities and commutation relations 
as QCD; 
therefore they will show the same low energy properties.

\booksubsection{9.3.1. Gauge extension of chiral invariance}
As stated, we start by extending the $SU(2)\times SU(2)$ symmetry to a gauge symmetry. 
We do so by introducing {\sl sources} in the QCD Lagrangian. 
We denote by ${\cal L}_{{\rm QCD}0}$ to the QCD Lagrangian for massless $u,\,d$ quarks,
$${\cal L}_{{\rm QCD}0}=\sum_{\alpha=u,d}\bar{q}_\alpha\ii \Slash{D}q_\alpha-\tfrac{1}{4}G^2.
\equn{(9.3.1a)}$$
Then we consider ${\cal L}(v_\mu,a_\mu,s,p)$ where 
$v_\mu,\,a_\mu,\,s,\,p$ are, respectively, vector, axial, scalar and pseudoscalar sources, 
and we define
$$\eqalign{
{\cal L}(v_\mu,a_\mu,s,p)=&\,{\cal L}_{{\rm QCD}0}\cr
+&\,\sum_{\alpha,\beta}
\bar{q}_\alpha\gamma_\mu \left(v^\mu_{\alpha\beta}+a^\mu_{\alpha\beta}\gamma_5\right)q_\beta
+\sum_{\alpha,\beta}
\bar{q}_\alpha \left(-s_{\alpha\beta}+\ii p_{\alpha\beta}\gamma_5\right)q_\beta.\cr
}
\equn{(9.3.1b)}$$
We include the mass matrix in $s_{\alpha\beta}$ so that 
$$s_{\alpha\beta}=m_\alpha\delta_{\alpha\beta}+\widetilde{s}_{\alpha\beta}.
\equn{(9.3.1c)}$$
$\alpha,\,\beta$ are flavour indices that run over the values $u,\,d$, in our case.

The Lagrangian (9.3.1b) is invariant under independent local gauge transformations of the left and 
right components of the $q$, {\sl provided} we at the same time transform the sources: 
$$\eqalign{q\to q'=&\,\left\{\tfrac{1}{2}(1+\gamma_5)W_R(x)+\tfrac{1}{2}(1-\gamma_5)W_L(x)\right\}q;\cr
v^\mu\pm a^\mu\to&\, v'^\mu\pm a'^\mu=W_{R,L}\left(v^\mu\pm a^\mu\right)W^{\dag}_{R,L}
+\ii W_{R,L}\partial^\mu W^{\dag}_{R,L},\cr
s+\ii p\to&\, s'+\ii p'=W_R(s+\ii p)W^{\dag}_L.\cr
}
\equn{(9.3.2)}$$
Here the $W_{R,L}$ are independent $SU(2)$ matrices. 
The symmetry may be extended to a $U(2)\times U(2)$ symmetry; however, the current associated 
with the diagonal piece presents an anomaly, as we know. 
We will not study this piece here, but refer to Gasser and Leutwyler (1985a). 
To avoid it we will restrict the $v^\mu$, $a^\mu$ to be traceless. 
This is automatic if we parametrize them in terms of the three-vectors $v_i^\mu$, $a_i^\mu$ writing
$$v^\mu=\tfrac{1}{2}\sum_iv_i^\mu \tau_i,\quad
a^\mu=\tfrac{1}{2}\sum_ia_i^\mu \tau_i
\equn{(9.3.3)}$$
and the $\tau_i$ are the Pauli matrices in flavour space. 
The $s,\,p$ may likewise be parametrized in terms of the ({\sl Euclidean}) four dimensional vectors 
$s_A$, $p_A$ with
$$s=\sum_As_A\tau_A,\quad
p=\sum_Ap_A\tau_A;\qquad \tau_0\equiv1.
\equn{(9.3.4)}$$

At low energy the only degrees of freedom are those associated with the pions; 
moreover, we have to take also into account that, in QCD, the 
scalar densities have a nonzero expectation value in the ground state (the physical vacuum). 
We will use the quantity $B$ defined as
$$B=-\dfrac{\langle\bar{q}q\rangle}{f^2}.
\equn{(9.3.5)}$$
We write $f$ for the pion decay constant in the chiral limit 
($m_{u,d}\to0$). 
In \subsect~9.3.3 we will see the connection with the 
physical decay constant, whose value we take to be $f_\pi\simeq93\,\mev.$
In the chiral limit, $B$ is independent of which $q$ ($u$ or $d$) we take. 
Comparing with (9.2.4) we have
$$B=\mu^2/(m_u+m_d).$$

\booksubsection{9.3.2. Effective Lagrangians in the chiral limit}

\noindent
We will start by working in the chiral limit, $m_{u,d}=0$. 
At low energies an effective Lagrangian should only include 
pion fields and, apart from the nonzero value of the 
condensate, should respect chiral gauge invariance.\fnote{This is, of course, a limitation 
of the chiral dynamics approach; 
it must fail at distances where the {\sl composite} character 
of the pions becomes relevant; thus certainly at energies 
of the order of the $\rho$ mass, as this particle is 
a quark-antiquark bound state, and decays into two pions.}
To construct this Lagrangian we proceed as for the 
nonlinear $\sigma$-model of \sect~9.1. 
We define a chiral four-dimensional vector $\varphi_A$, $A=0,\,1,\,2,\,3$ such that  
$\vec{\varphi}=\vec{\pi}$ (the pion field) 
and $\varphi_0=\sigma$ (the $\sigma$ field). 
We get rid of the last by imposing the invariant constraint
$$\sum_A\varphi_A\varphi_A=f^2.
\equn{(9.3.6a)}$$
We could include this into the Lagrangian, using a multiplier or, 
more simply, by admitting that 
$\varphi_0$ is not an independent field, but one has
$$\varphi_0=\sqrt{f^2-\vec{\varphi}^2}.
\equn{(9.3.6b)}$$

The transformation properties of $\varphi$ under  $SU(2)\times SU(2)$ 
imply the following values for the chiral covariant derivative, 
that we denote by $\nabla^\mu$:
$$\eqalign{
\nabla^\mu\varphi_0=&\,\partial^\mu\varphi_0+\vec{a}^\mu(x)\vec{\varphi},\cr
\nabla_\mu\vec{\varphi}=&\,\partial^\mu\vec{\varphi}+\vec{v}^\mu(x)\times\vec{\varphi}-
\vec{a}^\mu(x)\varphi_0.\cr
}
\equn{(9.3.7)}$$

We then construct the more general Lagrangians which are  
compatible with \equn{(9.3.2)}, and involve only $\varphi_A$. 
We start at lowest order in the momenta, $O(p^2)$. 
If we only allow two powers of the momenta at tree level, then only two derivatives can occur and the more
general form 
of this first order Lagrangian is, simply,
$${\cal L}_{{\rm ch.}1}=\tfrac{1}{2}\sum_A(\nabla_\mu\varphi_A)\nabla^\mu\varphi_A.
\equn{(9.3.8)}$$
The index ``ch." reminds us that this is valid in the chiral limit, 
and the factor $\ffrac{1}{2}$ is included so that the kinetic energy term 
agrees with that for three real, (pseudo-)scalar fields. 
One can evaluate the axial current from (9.3.8) and identify 
$f$ with the value of the pion decay constant, $f_\pi$, in the chiral limit. 
(In this case the identification of the axial current is simpler than before, as it is 
the current coupled to the
axial source, $\vec{a}^\mu$).

In particular, to lowest order and replacing $\varphi_0$ in terms of 
$\vec{\varphi}$, this gives
$${\cal L}_{{\rm ch.}1}\simeq\tfrac{1}{2}(\partial_\mu\vec{\varphi})\partial^\mu\vec{\varphi}
+\dfrac{1}{2f^2}\left(\vec{\varphi}\partial_\mu\vec{\varphi}\right)
\left(\vec{\varphi}\partial^\mu\vec{\varphi}\right)+\;\hbox{source terms}\;+\;\hbox{higher orders}.
\equn{(9.3.9)}$$
To order $p^2$, this is equivalent to (9.1.5).

Let us next consider $O(p^4)$. 
Simple power counting shows that the loop corrections generated by (9.3.8) 
are of relative order $p^2$ for each new loop; hence, one loop corrections induced by 
${\cal L}_{{\rm ch.}1}$ will be of order $p^4$. 
These corrections (which are necessary in order to respect unitarity of the 
effective theory) are, generally speaking, divergent. 
However, if we use a regularization that respects gauge invariance (such as 
dimensional regularization, in the absence of anomalies) these divergences will multiply 
chiral gauge invariant polynomials of degree $p^4$. 
They can thus be absorbed into suitable counterterms.

This leads us to construct all possible terms of order $p^4$ which will build the 
second order effective Lagrangian, ${\cal L}_{{\rm ch.}2}$. 
After use of the equations of motion it can be seen (Gasser and Leutwyler, 1984) 
that its most general form will be (sum over repeated indices $A,\,B,\,C$ understood)
$$\eqalign{
{\cal L}_{{\rm ch.}2}=&\,\dfrac{1}{f^4}\Big\{l_1\left(\nabla^\mu\varphi_A \nabla_\mu\varphi_A\right)^2
+l_2\left(\nabla^\mu\varphi_A\nabla^\mu\varphi_A\right)\left(\nabla_\mu\varphi_B\nabla_\mu\varphi_B\right)\cr
+&\,l_5\varphi_A F^{\mu\nu}_{AB}F_{BC,\mu\nu}+l_6\nabla_\mu\varphi_A F^{\mu\nu}_{AB}\nabla_\nu\varphi_B\cr
+&\,h_2\trace F_{\mu\nu}F^{\mu\nu}\Big\}.\cr
}
\equn{(9.3.10a)}$$
Here $F$ is defined by
$$\left(\nabla^\mu\nabla^\nu-\nabla^\nu\nabla^\mu\right)\varphi_A=F^{\mu\nu}_{AB}\varphi_B
\equn{(9.3.10b)}$$
and the reason for the
numbering of the constants $l_1,\,\dots,\,h_2$ (that agrees with the definitions of Gasser and
Leutwyler, 1984) will be seen below.

The constants $l_1,\,\dots,\,h_2$ will be divergent: 
their divergence is to be adjusted so that it cancels the one loop divergences generated by 
${\cal L}_{{\rm ch.}1}$. The theory will, therefore, {\sl predict} 
the coefficients of terms of type $p^4\log p^2/\nu^2$, with $\nu$ a renormalization scale  
(and, when we take into account leading symmetry breaking by the pion mass, also terms in  
$\mu^4$ and $p^2\mu^2$ multiplied by either 
$\log p^2/\nu^2$ or $\log \mu^2/\nu^2$).
However, the finite parts of the constants $l_1,\,\dots,\,h_2$ are {\sl not} 
 given by the theory. 
In fact, what one does is to {\sl fix} these constants  
by requiring agreement of the predictions using ${\cal L}_{{\rm ch.}1}$, 
${\cal L}_{{\rm ch.}2}$ with experiment. 
Chiral dynamics does {\sl not} allow an evaluation from first principles of 
corrections of order $p^4$. 
What it does is to {\sl correlate} these corrections to all
 processes in terms of a finite number of
constants, the  $l_1,\,\dots,\,h_2$.

In principle one can extend this procedure to higher orders and, indeed, the 
$O(p^6)$ corrections have been considered in the literature,\fnote{Akhoury and Alfakih (1991); 
 Fearing and Scherer (1996); Knecht et al.~(1995,~1996); Bijnens et al.~(1996);
 Bijnens, Colangelo and Eder
(2000).}  but we will not discuss this in any detail here.
  Not only the number of constants to be fitted to experiment
 grows out of hand, but it is practically impossible
to separate the 
$O(p^4)$ and $O(p^6)$ pieces of the   $l_1,\,\dots$, as we will see in 
some examples later. 
More interesting is to take into account the corrections due to 
the nonzero masses of the $u,\,d$ quarks (or, equivalently, of the pions) to which we now turn.

\booksubsection{9.3.3. Finite pion mass corrections}
Because the mass of the pion will appear in pion propagator denominators, $1/(p^2-\mu^2)$, 
a consistent way to treat the finiteness of the pion mass
 requires that we consider $p^2$ and $\mu^2$ to be
of the same order of magnitude, and calculate to all orders in their ratio; 
otherwise we would be replacing
$$\dfrac{1}{p^2-\mu^2}\quad\hbox{by}\quad\dfrac{-1}{\mu^2}\left\{1+\dfrac{p^2}{\mu^2}+
\dfrac{p^4}{\mu^4}+\cdots\right\},$$
 not  a very accurate procedure.

To leading order we have to find the lowest order  terms that can be added to 
${\cal L}_{{\rm ch.}1}$ and which contain $s_0$; we recall that $s_0$ included the quark masses. 
There is only one such term that also preserves parity,
${\rm Constant}\times (s_0\varphi_0+\vec{p}\vec{\varphi})$. 
The constant may be identified requiring that the new term reproduce the 
equality (9.2.4) for the pion propagator. 
We then have the full ${\cal L}_{1}$, correct to $O(p^2)$, $O(\mu^2)$,
$${\cal L}_{1}={\cal L}_{{\rm ch.}1}+2Bf\left(s_0\varphi_0+\vec{p}\vec{\varphi}\right),
\equn{(9.3.11a)}$$
which corresponds to the pion mass
$$\mu^2=(m_u+m_d)B.
\equn{(9.3.11b)}$$
To next order, 
$${\cal L}_{2}={\cal L}_{{\rm ch.}2}+\dfrac{1}{f^4}\Big\{l_3(\xi_A\varphi_A)^2
+l_4\nabla^\mu\xi_A\nabla_\mu\varphi_A+l_7(\eta_A\varphi_A)^2+h_1\xi_A\xi_A+h_3\eta_A\eta_A\Big\}.
\equn{(9.3.12a)}$$
We have defined
$$\xi_0=2Bs_0,\quad\vec{\xi}=2Bp;\qquad
\eta_0=2Bp_0,\quad\vec{\eta}=-2B\vec{s}
\equn{(9.3.12b)}$$
and ${\cal L}_{{\rm ch.}1}$, ${\cal L}_{{\rm ch.}2}$ are as given in (9.3.8), (9.3.10).

For reference, we note the correspondence between our definitions and those of 
Gasser and Leutwyler (1984):
$$U_A=\dfrac{1}{f}\,\varphi_A,\quad \chi_A=\dfrac{1}{f}\,\xi_A,\quad 
\widetilde{\chi}_A=\dfrac{1}{f}\,\eta_A.
\equn{(9.3.13)}$$

\booksubsection{9.3.4. Renormalized effective theory}
Renormalization for the one loop graphs generated by ${\cal L}_1$  
proceeds  in the usual manner. 
The divergences, as stated in the previous subsection, can be 
canceled by divergent pieces in the $l_i,\,h_j$. 
One finds (Gasser and Leutwyler, 1984, where the $c_i$ are denoted by $\gamma_i$ and the 
$d_j$ by $\delta_j$)
$$\eqalign{
l_i=&\,l_i^{\rm loop.}(\nu)=\dfrac{c_i}{32\pi^2}\left\{\dfrac{2}{D-4}+\log\nu^2-
(\log4\pi-\gammae+1)\right\},\cr
 h_j=&\,h_j^{\rm loop.}(\nu)=
\dfrac{d_j}{32\pi^2}\left\{\dfrac{2}{D-4}+\log\nu^2-(\log
4\pi-\gammae+1)\right\};\cr
}
\equn{(9.3.14a)}$$
$\nu$ is the renormalization point and
$$\eqalign{
c_1=&\,\tfrac{1}{3},\quad c_2=\tfrac{2}{3},\quad c_3=-\tfrac{1}{2},\quad c_4=2,\quad
c_5=-\tfrac{1}{6},\quad c_6=\tfrac{1}{3},\quad c_7=0;\cr
d_1=&\,2,\quad d_2=\tfrac{1}{12},\quad  d_3=0.\cr}
\equn{(9.3.14b)}$$
The renormalized constants $l_i^{\rm ren.}$ may be obtained 
by comparing with experimental quantities.\fnote{The $h_j$ 
depend on the renormalization scheme and, in fact, 
do not intervene in any physical observable. 
This is discussed in Gasser and Leutwyler, 1984.} 
They depend on the renormalization point, $\nu$. Alternatively, one may replace them by the 
quantities $\bar{l}_i$, defined as (proportional to) the $l_i^{\rm ren.}(\nu)$ 
with $\nu=\mu_{\rm ch.}$. (Here we denote by $\mu_{\rm ch.}$ to the pion mass in the leading order in
chiral symmetry breaking, 
that is to say, using (9.3.11b) but evaluating $B=-\langle\bar{q}q\rangle/f$ 
in the chiral limit).  Then, we have
$$l_i^{\rm ren.}(\nu)=\dfrac{c_i}{32\pi^2}\left\{\bar{l}_i+\log\dfrac{\mu_{\rm ch.}}{\nu^2}\right\}.
\equn{(9.3.14c)}$$
We remark that this implies that the 
$\bar{l}_i$ are divergent in the chiral limit, as we are renormalizing at 
$\nu=\mu_{\rm ch.}$ which vanishes in this limit:
$$\bar{l}_i\;\simeqsub_{m_{u,d}\to0}\;-\log \mu_{\rm ch.}.$$

We can now compare the results of calculations made 
with ${\cal L}_1$ and ${\cal L}_2$  with experimental quantities, and obtain the $\bar{l}_i$. 
As an example we consider $\pi\pi$ scattering. 
If we use the full ${\cal L}_1$ and ${\cal L}_2$ we obtain, after a
 straightforward but tedious calculation
(Gasser and Leutwyler, 1984)
$$F(i+j\to k+l)=
\dfrac{1}{4\pi^2}
\left\{\delta_{ij}\delta_{kl}A(s,t,u)+\delta_{ik}\delta_{jl}A(t,s,u)+\delta_{il}\delta_{jk}A(u,t,s)\right\}
\equn{(9.3.15a)}$$
where now
$$A(s,t,u)=\dfrac{s-\mu^2_{\rm ch.}}{f^2}+B(s,t,u)+C(s,t,u).
\equn{(9.3.15b)}$$
Here $B$, $C$ are, respectively, the logarithmic and polynomial fourth order corrections:
$$\eqalign{
B(s,t,u)=&\,\dfrac{1}{96\pi^2f^4_\pi}\Bigg\{3(s^2-\mu^2)I(s)\cr
+&\,
\left[t(t-u)-2\mu^2 t+4\mu^2 u-2\mu^4\right]I(t)\cr
+&\,\left[u(u-t)-2\mu^2 u+4\mu^2 t-2\mu^4\right]I(u)\Bigg\};\cr
I(s)=&\,\beta\log\dfrac{\beta-1}{\beta+1}+2,\quad 
\beta=\sqrt{1-4\mu^2_{\rm ch.}/s};
\cr}
\equn{(9.3.16a)}$$
$$\eqalign{
C(s,t,u)=&\,\dfrac{1}{96\pi^2f^4_\pi}\Bigg\{2\left(\bar{l}_1-\tfrac{4}{3}\right)(s-2\mu^2)^2\cr
+&\,\left(\bar{l}_2-\tfrac{5}{6}\right)\left[s^2+(t-u)^2\right]-12 \mu^2 s+15 \mu^4\Bigg\}.
\cr}
\equn{(9.3.16b)}$$
The expression for $B$ in the chiral limit ($\mu=0$) has been known for a long time 
(Lehmann, 1972).
 To leading order $B,\,C\to0$, $\mu\to0$, and (9.3.15) of course reproduces (9.2.1).

The extension to $SU(3)$ (i.e., including kaons and $\eta$) may be found, for 
chiral perturbation theory, in 
Gasser and Leutwyler~(1985a); 
for $\pi\pi$ scattering  
in Bernard, Kaiser and Meissner~(1991) for some cases and, in general, in 
the paper of G\'omez-Nicola and Pel\'aez~(2002).

\booksubsection{9.3.5. The parameters of chiral perturbation theory}

\noindent
To one loop, the $\pi\pi$ scattering amplitude depends
 on the two unknown constants $\bar{l}_1,\,\bar{l}_2$ 
(besides, of course, $f_\pi$ and $M_{\pi}$). 
A technical point to be cleared is that, in the amplitude $A$ in 
\equn{(9.3.15b)}, we have the quantities 
$f$ and $\mu_{\rm ch.}$ which we have to 
relate to the physical ones. 
The details may be found again in the paper of Gasser and Leutwyler (1984); 
we have, writing as usual\fnote{In 
the limit in which we forget isospin breaking interactions, one would have $\mu=M_\pi$. 
Gasser and Leutwyler, and, following them, most modern authors, take $M_\pi$ as the starting point 
from which to perturb with isospin breaking interactions, 
instead of using --as would appear more natural-- $\mu$ as the starting point. 
As stated several times, we follow this custom 
when giving numerical results, for ease of comparison with other calculations.}
our results in terms of $M_\pi$
$$M_{\pi}^2=\mu^2_{\rm ch.}\left\{1-\dfrac{M_{\pi}^2}{32\pi^2f^2_\pi}\bar{l}_3\right\},\quad
f_\pi=f\left\{1+\dfrac{M_{\pi}^2}{16\pi^2f^2_\pi}\bar{l}_4\right\}. 
\equn{(9.3.17)}$$
Thus, $F$ in (9.3.15) depends also indirectly on the constants $\bar{l}_3,\,\bar{l}_4$. 
We can, however, obtain directly  $\bar{l}_1,\,\bar{l}_2$ by selecting an 
observable that depends only on second order effects. 
Such observables are the D waves  at low energy:
$$f_2^{(I)}\simeq \dfrac{(s-4M_{\pi}^2)^2}{4\pi}M_{\pi} a_2^{(I)},$$
and $I=0,\,2$ is the isospin index. 
Because this vanishes (for $M_{\pi}=0$)
  as $s^2=p^4$, the contributions to them  start at second order 
and we find
$$\eqalign{
a_2^{(0)}=&\,\dfrac{M_{\pi}^{-1}}{1440\pi^3f^4_\pi}\left\{\bar{l}_1+4\bar{l}_2-\tfrac{53}{8}\right\},\cr
a_2^{(2)}=&\,\dfrac{M_{\pi}^{-1}}{1440\pi^3f^4_\pi}\left\{\bar{l}_1+\bar{l}_2-\tfrac{103}{40}\right\}.\cr
}
\equn{(9.3.18)}$$

We can also improve our previous determination
 of the scattering lengths; for example, for the S and  P waves, 
including pion mass and $O(p^4)$ terms gives
$$\eqalign{
a_0^{(0)}=&\,\dfrac{7M_{\pi}}{32\pi f^2_\pi}\left\{1+\dfrac{5M_{\pi}^2}{84\pi^2 f^2_\pi}\left[
\bar{l}_1+2\bar{l}_2-\tfrac{3}{8}\bar{l}_3+\tfrac{21}{10}\bar{l}_4+\tfrac{21}{8}\right]\right\},\cr
a_0^{(2)}=&\,-\dfrac{M_{\pi}}{16\pi f^2_\pi}\left\{1-\dfrac{M_{\pi}^2}{12\pi^2f^2_\pi}\left[
\bar{l}_1+2\bar{l}_2+\tfrac{3}{8}\right]+\dfrac{M_{\pi}^2}{32\pi^2f^2_\pi}\left[\bar{l}_3+4\bar{l}_4\right]
\right\};\cr}
\equn{(9.3.19a)}$$
$$a_1=\dfrac{M_{\pi}^{-1}}{24\pi f^2_\pi}
\left\{1-\dfrac{M_{\pi}^2}{12\pi^2f_\pi^2}\left[\bar{l}_1-\bar{l}_2+\tfrac{65}{48}\right]
+\dfrac{M_{\pi}^2}{8\pi^2f^2_\pi}\bar{l}_4\right\}.
\equn{(9.3.19b)}$$
Note that we 
here use the  definition (cf.~(3.1.7, (7.5.3))
$$\dfrac{\pi}{4M_\pi k^{2l}}\real f_l^{(I)}(s)=a_l^{(I)}+b_l^{(I)}k^2+\cdots\,.
$$ 
This may be compared with the standard effective range expansion: 
$$k^{2l+1}\cot\delta_l^I(s)\,\simeqsub_{k\to 0}\,\dfrac{1}{a_l^{(I)}}+\tfrac{1}{2}r_0k^2+O(k^{4}).
$$
The connection between the corresponding
 parameters $a^I_l|_{\rm G.\&L.},\,b_l^{I}|_{\rm G.\&L.}$
 of Gasser and Leutwyler (1984) and our $a_l^{(I)},\,b_l^{(I)}$, 
and also with $r_0$ and with
 the parameters $b_l^{I}|_{\rm P.S\!-\!G.Y.}$
 of Palou, S\'anchez-G\'omez and
Yndur\'ain~(1975) is
$$
a^I_l|_{\rm G.\&L.}=M_{\pi} a^{(I)}_l,\quad 4b_l^{I}|_{\rm P.S\!-\!G.Y.}=b_l^{I}|_{\rm G.\&L.}= 
b_l^{(I)}=a_l^{(I)}\dfrac{1-M_{\pi}^2 a^{(I)}_lr_0}{2M_{\pi}}.$$

For the parameters $b_l^{(I)}$ one loop chiral perturbation theory gives
$$\eqalign{
b_0^{(0)}=&\,\dfrac{M_{\pi}^{-1}}{4\pi f^2_\pi}
\left\{1+\dfrac{M_{\pi}^2}{12\pi^2f^2_\pi}\left[2\bar{l}_1+3\bar{l}_2-\tfrac{13}{16}\right]
+\dfrac{M_{\pi}^2}{8\pi^2f^2_\pi}\bar{l}_4\right\},\cr
b_0^{(2)}=&\,-\dfrac{M_{\pi}^{-1}}{8\pi f^2_\pi}
\left\{1-\dfrac{M_{\pi}^2}{12\pi^2f^2_\pi}\left[\bar{l}_1+3\bar{l}_2-\tfrac{5}{16}\right]
+\dfrac{M_{\pi}^2}{8\pi^2f^2_\pi}\bar{l}_4\right\};\cr
\cr}
\equn{(9.3.20a)}$$
$$b_1=\dfrac{M_{\pi}^{-1}}{288\pi^3f^4_\pi}\left\{-\bar{l}_1+\bar{l}_2+\tfrac{97}{120}\right\}.
\equn{(9.3.20b)}$$

The values of the $a_0,\,b_0$ given above imply that the S waves with isospin $I$ 
have a zero each, for
$s=z_I^2$ in the range
$0<s<4M_{\pi}^2$, located at
$$\eqalign{z_0^2=&\,4M_{\pi}^2-\dfrac{7M_{\pi}^2}{2}\Bigg\{1+\dfrac{5M_{\pi}^2}{84\pi^2f^2_\pi}
\left[\bar{l}_1+2\bar{l}_2-\tfrac{3}{8}\bar{l}_4+\tfrac{21}{8}\right]\cr
-&\,\dfrac{M_{\pi}^2}{12\pi^2f^2_\pi}
\left[2\bar{l}_1+3\bar{l}_2-\tfrac{13}{16}\right]+
\dfrac{M_{\pi}^2}{8\pi^2f^2_\pi}\bar{l}_4\Bigg\};\cr
z_2^2=&\,4M_{\pi}^2-2M_{\pi}^2\left\{1+\dfrac{M_{\pi}^2}{32\pi^2f^2_\pi}\bar{l}_3+
\dfrac{M_{\pi}^2}{12\pi^2f^2_\pi}\left[\bar{l}_2+\tfrac{1}{16}\right]\right\}.
\cr
}\equn{(9.3.21)}$$
These zeros are often called 
{\sl Adler zeros}, after the work of Adler~(1965)
 on zeros of scattering amplitudes implied by PCAC.
It should be clear, however, that while the location of $z_2$ 
is probably well described by (9.3.21), 
there is no reason why the same should be the case for $z_0$. Indeed, to get this last, 
we have used the expansion of $f_l^{(I)}$ for 
$s=z_0^2\simeq\tfrac{1}{2}M_{\pi}^2$ were, due to the vicinity of the 
left hand cut of $f_0^{(0)}(s)$, starting at $s=0$, we would 
expect it to give a poor approximation. 
Actually, while fits to data do confirm $z_2$ (\subsect~6.4.1), the 
situation is less clear for $z_0$.

\vfill\eject
\booksection{9.4. Comparison of chiral perturbation theory to one loop\hb 
with experiment}
\vskip-0.5truecm
\booksubsection{9.4.1. One loop coupling constants, 
and $\pi\pi$ scattering and the electromagnetic
form factor of the pion}

\noindent
Using the experimental values of the quantities evaluated in Chapters~6,\/7,
 and others as well, we can find
the  constants $\bar{l}_i$. In fact, there are many more observables than constants; 
for example, $a^{(2)}_2,\,a^{(0)}_2$ and $b_1$ depend only on the two $\bar{l}_1,\,\bar{l}_2$.
 So the agreement of 
various determinations among themselves is a nontrivial 
check of second order chiral perturbation theory.\fnote{The tests are less impressive 
than what they may look at first sight. 
In fact, chiral perturbation theory is just a (very convenient) way to 
summarize properties that hold in any local field theory: 
analyticity, crossing and unitarity, 
plus the dynamical properties embodied in the 
constants $f_\pi$, $\mu$ and 
the $\bar{l}_i$. 
Thus for example, by comparing the r.h. sides of the 
Olsson sum rule (7.4.8) and the 
Froissart-Gribov representation for $a_1$, (7.5.4), we discover that, 
in any local field theory we must have the 
equality
$2a_0^{(0)}-5a_0^{(2)}=18\mu^2a_1+O(\mu^4)$.}

We collect here some recently obtained values of the constants $\bar{l}_i$; the
reader interested in  the details of the calculations should
 consult the original papers. 
We have, from Bijnens, Colangelo and Talavera~(1998) and Colangelo, Gasser and Leutwyler~(2001),
$$\eqalign{
\bar{l}_1=&\,-0.4\pm0.6,\quad \bar{l}_2=6.0\pm1.3,\quad \bar{l}_3=2.9\pm2.4,\quad
\bar{l}_4=4.4\pm0.2\quad\hbox{(CGL)};\cr
 \bar{l}_5=&\,13.0\pm1.0,\quad \bar{l}_6=16\pm1\qquad\hbox{(BCT)}.\cr
}
\equn{(9.4.1)}$$
The value of $\bar{l}_7$ is not known with any accuracy; an estimate for it is 
$\bar{l}_7\sim5\times10^{-3}$ (Gasser and Leutwyler, 1984).

Actually, the determinations in (9.4.1) include estimates of 
two loop effects. 
One should however remember that, as discussed in \sect~7.6, 
the calculations of Colangelo, Gasser and Leutwyler~(2001) of $\pi\pi$ scattering 
--on which their estimates of several of the $\bar{l}_i$ are based-- 
are biased by up to 2 standard deviations, and their errors are underestimated
 by a factor between 1.5 and 2. 
Something similar happens for $\bar{l}_4$; see \subsect~9.4.2 below.
This should affect the numbers in (9.4.1).

Some of these constants can be calculated independently with greater accuracy 
(but  only at the one loop level)  using the 
results reported in \sect~7.6 (cf.~Table~III) here. So, 
from the combination  $a_{0+}=\tfrac{1}{3}[a_2^{(0)}-a_2^{(2)}]$,
that  we evaluated very precisely, there follows the value
$$\bar{l}_2=5.97\pm0.07.
\equn{(9.4.2a)}$$
Likewise, the constant $\bar{l}_1$ can be deduced from the value of 
 $a_{00}=\tfrac{1}{3}[a_2^{(0)}+2a_2^{(2)}]$ that follows from the
Froissart--Gribov  representation for $\pi^0\pi^0$,
 and the value of $\bar{l}_2$. 
 We find
$$\bar{l}_1=-1.47\pm0.24,
\equn{(9.4.2b)}$$
somewhat larger in magnitude than the value given in (9.4.1).

Use of the  quadratic charge radius of the pion as input 
(see \equn{(9.4.5)} below) allows us to get 
also an accurate evaluation of $\bar{l}_6$:
$$\bar{l}_6=16.35\pm0.14,
\equn{(9.4.2d)}$$
which, in turn, implies a slightly more precise 
value for $\bar{l}_5$:
$$\bar{l}_5=13.7\pm0.7,
\equn{(9.4.2e)}$$

Our improvements, however,  may be challenged 
because of the possible contributions of two loop corrections, 
that we have not taken into account, and of 
electromagnetic corrections, that may be 
important. For the 
second, see next section for a discussion of a few examples. 
As for higher order ch.p..t corrections, we will say a few words at the end of  
\subsect~9.4.3.

In what respects S and P waves scattering lengths 
and effective range parameters, substituting the $\bar{l}_i$ into (9.3.19)
 we find the one-loop chiral perturbation 
theory predictions: in units of $M_{\pi}$, 
$$\eqalign{
a_0^{(0)}=&\,(0.207\pm0.003),\quad
a_0^{(2)}=(-0.043\pm0.002),\quad
a_1=(38.0\pm0.4)\times10^{-3};\cr
b_0^{(0)}=&\,(0.255\pm0.003),\quad
b_0^{(0)}=(-0.076\pm0.001),\quad b_1=(4.69\pm0.14)\times10^{-3}. 
}
\equn{(9.4.3)}$$
In the calculation we have used the values of  the $\bar{l}_i$ in 
(9.4.2) and (9.4.12) below in preference to those of (9.4.1), when possible.

The predictions for the $a_0^{(2)}$, $b_0^{(I)}$, $b_1$
 are in agreement with the experimental
values we  found before (Table~II, \subsect~7.6.2); the value 
of $a_0^{(0)}$ in (9.4.3) indicates that the higher order 
 correction (two loop or otherwise) for this quantity must be relatively large.  
The value 
of the P wave scattering length  in (9.4.3) is also  compatible with the result of 
the direct fit, $a_1=(38.6\pm1.2)\times10^{-3}\,M_{\pi}^{-3}$. 
It is also compatible with the results of other authors:
$$
a_1=\cases{
(37.9\pm0.5)\times10^{-3}\;M_{\pi}^{-3}\quad\hbox{(Colangelo, Gasser and Leutwyler, 2001)}\cr
(37.2\pm2)\times10^{-3}\;M_{\pi}^{-3}\quad\hbox{(Ananthanarayan et al., 2001)}\cr
(38.0\pm2)\times10^{-3}\;M_{\pi}^{-3}\quad\hbox{(Amor\'os, Bijnens and Talavera, 2000).}\cr
}
$$

We also note, as  consistency tests, that the value of $\bar{l}_4$ 
that follows from  $a_1$, via \equn{(9.3.19b)}, is compatible, 
within the rather large error, with (9.4.1a), 
as one gets $\bar{l}_4=5.5\pm2.0$, and that $b_1$ would yield 
a number for  $\bar{l}_2-\bar{l}_1$ compatible 
(also within errors) with what we found before. 
In what respects to $b_1$, however, a better, 
 nontrivial consistency test is obtained by 
eliminating $\bar{l}_1$, $\bar{l}_2$ between (9.3.19a), (9.3.20b). 
We find the relation
$$b_1=\tfrac{5}{2}\left[3a_{0+}-a_{00}\right]+(\tfrac{97}{120}+\tfrac{1}{8})
\dfrac{1}{288\pi^3f^4_\pi M_\pi},
\equn{(9.4.6a)}$$
in which some of  the larger higher order corrections cancel and, 
moreover, the r.h.s. is dominated by $a_{0+}$, which is known accurately. 
This gives, using the PY values for the  $a_{00}$, $a_{0+}$
 described in Table~III (\subsect~7.6.3), 
$$b_1=(4.68\pm0.20)\times10^{-3}\,M_{\pi}^{-5},
\equn{(9.4.6b)}$$
in excellent agreement with the value deduced from the 
electromagnetic form factor of the pion (see again Table~III), 
$b_1=(4.47\pm0.27)\times10^{-3}\,M_{\pi}^{-5}$.

The very precise calculation of the pion form factor possible with the 
Omn\`es--Muskhelishvili techniques also allows a direct 
determination of a second order (two loop) 
parameter. 
According to Gasser and Meissner~(1991), Colangelo, Finkelmeir and Urech~(1996),  and 
Fearing and Scherer~(1966), 
one has
$$c_\pi=\dfrac{1}{16\pi^2f^2_\pi}\left\{\dfrac{1}{60M_{\pi}^2}+\dfrac{1}{16\pi^2f^2_\pi}\bar{f}_2\right\}.
\equn{(9.4.5a)}$$
With the value  $c_\pi=3.60\pm0.03\,\gev^{-4}$ 
given in de~Troc\'oniz and Yndur\'ain~(2002), this 
implies
$$\bar{f}_2=5.520\pm0.056.
\equn{(9.4.5b)}$$
Note, however, that this result is purely formal; 
indeed, the nominally leading term ($1/60M_{\pi}^2$) 
is in fact {\sl smaller} than the nominally second order one, $\bar{f}_2/(16\pi^2f^2_\pi)$.
This shows clearly the limitations of chiral perturbation theory.
 
Another example is the charge radius of the pion, for which one has, to 
second order (Gasser and Meissner,~1991, and Colangelo, Finkelmeir and Urech,~1996),
$$\langle
r^2_\pi\rangle=\dfrac{1}{16\pi^2f^2_\pi}\left[\bar{l}_6-1+
\dfrac{M_{\pi}^2}{16\pi^2f^2_\pi}\bar{f}_1\right].
\equn{(9.4.6)}$$
Here the two loop term is  smaller than the leading one, for reasonable values 
of $\bar{f}_1$, but perhaps not totally negligible, 
given the accuracy of the experimental result for $\langle r^2_\pi\rangle$.
 The value of $\bar{l}_6$ given above was obtained 
neglecting $\bar{f}_1$;  a value of this quantity of the order of 
  $\bar{f}_2$  would alter $\bar{l}_6$  by  1\%, a 
variation slightly larger than the nominal 
error in (9.4.2d).

It is also possible to give a prediction, based on chiral perturbation theory and 
the Froissart--Gribov representation, for 
scattering lengths for large $l$. 
We will give some details for the amplitude for isospin 1 in the $t$ channel,
$$F^{(I_t=1)}=\tfrac{1}{3}F^{(I_s=0)}+\tfrac{1}{2}F^{(I_s=1)}-\tfrac{5}{6}F^{(I_s=2)}.$$
The corresponding scattering lengths are given by \equn{(7.5.4)},
$$2a_l^{(1)}=\dfrac{\sqrt{\pi}\,\gammav(l+1)}{4M_{\pi}\gammav(l+3/2)}
\int_{4M_{\pi}^2}^\infty\dd s\,\dfrac{\imag F^{(I_t=1)}(s,4M_{\pi}^2)}{s^{l+1}}
$$
and the factor 2 in the l.h. side is due to the identity of the pions. 
As $l\to\infty$, only the behaviour of $\imag F^{(I_t=1)}(s,4M_{\pi}^2)$ 
near threshold matters; hence we can replace
$$\eqalign{
\imag F^{(I_t=1)}(s,4M_{\pi}^2)\simeq&\,\tfrac{1}{3}\imag F^{(I_s=0)}-\tfrac{5}{6} \imag F^{(I_s=2)}\cr
\simeq&\, 2\times\dfrac{2s^{1/2}}{\pi k}
\left\{\tfrac{1}{3}\sin^2\delta_0^{(0)}(s)-\tfrac{5}{6}\sin^2\delta_0^{(2)}(s)\right\}\cr
\simeq&\,2\times\dfrac{2s^{1/2}}{\pi k}
\left\{\tfrac{1}{3}\left[a_0^{(0)}\right]^2-\tfrac{5}{6}\left[a_0^{(2)}\right]^2\right\}.\cr
}
\equn{(9.4.7)}$$
The factor 2 in the r.h. side also comes from the identity of the pions. 
Replacing the $a_0^{(I)}$ by their values at leading order in chiral perturbation theory, 
\equn{(9.3.19)}, we find that we can approximate
$$\imag F^{(I_t=1)}(s,4M_{\pi}^2)\simeq 2\dfrac{13M_{\pi}^2s^{1/2}k}{512\pi^3 f^4_\pi}.
$$
Substituting into the Froissart--Gribov representation,
 and performing the integral we get the result 
$$a_l^{(1)}\,\simeqsub_{l\to\infty}\,\dfrac{13M_{\pi}^{3-2l}}{2\times4^{l+5}\pi^2f^4_\pi}
\dfrac{\gammav(l+1)\gammav(l-1)}{\gammav(l+3/2)\gammav(l+1/2)}
\,\simeqsub_{l\to\infty}\,\dfrac{13M_{\pi}^{3-2l}}{2\times4^{l+5}\pi^2l^2f^4_\pi}
\equn{(9.4.8a)}$$
and, in the last step, we have used the asymptotic properties of the gamma function 
and replaced $\mu$ by $M_\pi$.
The same method gives a prediction for even waves; for example, for $I=0$ we have,
$$a_l^{(0)}\simeqsub_{l\to\infty}\dfrac{23M_{\pi}^{3-2l}}{2\times4^{l+5}\pi^2l^2f^4_\pi}.
\equn{(9.4.8b)}$$

Gasser and Leutwyler~(1983) have produced a formula for all 
scattering lengths with $I=1$ to leading order in chiral 
perturbation theory that is exact (and not only 
 valid for $l\to\infty$), based on a direct calculation.
They give the expression, valid for $l\geq3$,
$$\eqalign{
a_l^{(1)}({\rm G.- L.})=&\,\dfrac{M_{\pi}^{3-2l}}{512\pi^3 f^4_\pi}\,\dfrac{l!(l-3)!}{[(2l+1)!!]^2}
\,\big(13l^2+5l-22\big).
}
\equn{(9.4.9)}$$
If, in  (9.4.9), 
we replace $(2l+1)!!=\gammav(2l+2)/2^l\gammav(l+1)$ and use 
the duplication formula of the gamma function we find
$$\eqalign{
a_l^{(1)}({\rm G.- L.})=&\,
\dfrac{13M_{\pi}^{3-2l}}{2\times4^{l+5}\pi^2f^4_\pi}\,
\dfrac{\gammav(l+1)\gammav(l-1)}{\gammav(l+3/2)\gammav(l+1/2)}\,
\dfrac{l^2}{(l-2)(l+1/2)}\cr
\times&\,\left(1+\dfrac{5}{13l}-\dfrac{22}{13l^2}\right)\cr
}$$
which, as $l\to\infty$, agrees with (9.4.7a). 

The calculation  using leading order chiral perturbation theory 
yields the figure $a_3=1.8\times10^{-5}\,M_{\pi}^{-7}$. 
From the Froissart--Gribov representation 
we found in  \sect~7.6 results ranging between 
5.4 and 6.7, in units of $10^{-5}\,M_{\pi}^{-7}$. 
 A large disagreement (a factor of 3 to 4) is thus found between the 
leading chiral perturbation result and the  results 
based on experiment. 
It is not clear to the present author which is the reason for this disagreement; 
in fact, as far as I know, 
the Gasser--Leutwyler result in \equn{(9.4.9)} has not been checked by an independent 
calculation.

For $l=4$, our expression (9.4.7b) gives $a_4^{(0)}=1.4\times10^{-5}\,M_{\pi}^{-9}$ 
while the ``experimental" value (from the 
Froissart--Gribov representation) is $(0.8\pm0.2)\times10^{-5}\,M_{\pi}^{-9}$.
The disagreement for $a_3$, and the difference between the two values for $a_4^{(0)}$, 
 show that, in some cases, 
the corrections  due to subleading effects in chiral perturbation 
theory may be very large: 
for the quantity $a_3$, two to three times the nominally leading term. 
This is not surprising; as is clear in our derivation, the value we obtain depends on 
the {\sl square} of the S wave scattering lengths, for which 
one loop corrections are at least of 25\%.

\booksubsection{9.4.2. The scalar form factor of the pion}

\noindent
For the scalar form factor of the pion, defined in (7.2.5), chiral 
perturbation theory gives\fnote{Two 
loop evaluations of the scalar radii may be found 
in Gasser and Mei\ss ner~(1991) and Frink, Kubis and 
Mei\ss ner~(2002).}
$$\eqalign{
F_S(0)=&\,M^2_\pi\left\{1-\dfrac{M^2_\pi}{32\pi^2f^2_\pi}(\bar{l}_3-1)\right\},\cr
\langle r^2_{{\rm S},\pi}\rangle=&\,\dfrac{3}{8\pi^2f^2_\pi}\left\{\bar{l}_4-\tfrac{13}{12}\right\}.\cr
}
\equn{(9.4.9)}$$
It is also possible to give a formula for the quadratic scalar radius in terms of 
observable quantities (Gasser and Leutwyler,~1985b):
$$\eqalign{
\langle r^2_{{\rm S},\pi}\rangle=&\,\dfrac{6}{m^2_K-M^2_\pi}
\left(\dfrac{f_K}{f_\pi}-1\right)+\delta_3;\cr
\delta_3=&\,-\dfrac{1}{64\pi^2f^2_\pi}\,\dfrac{1}{m_K^2-M^2_\pi}\Bigg\{
6(2m_K^2-M^2_\pi)\log\dfrac{m_K^2}{M^2_\pi}+
9m^2_\eta\log\dfrac{m^2_\eta}{M^2_\pi}\cr
-&\,2(m_K^2-M^2_\pi)\left(10+\tfrac{1}{3}\dfrac{M^2_\pi}{m^2_\eta}\right)\Bigg\}
\cr}
\equn{(9.4.10a)}$$
and $m_\eta=547\,\mev$. 
The same authors give also a formula for the mixed kaon-pion radius, 
defined by 
$$\langle\pi(p)|(m_s-m_u)\bar{u}s(0)|K(p')\rangle\simeqsub_{t\to0}f_{K\pi}(0)
\Big\{1+\tfrac{1}{6}\langle r^2_{{\rm S},K\pi}\rangle\,t\Big\}:
$$
$$\eqalign{
\langle r^2_{{\rm S},K\pi}\rangle=&\,\dfrac{6}{m^2_K-M^2_\pi}
\left(\dfrac{f_K}{f_\pi}-1\right)+\delta_2;\cr
\delta_2=&\,-\dfrac{1}{192\pi^2f^2_\pi}\left\{
15h_2(M^2_\pi/m^2_K)+\dfrac{19 m_K^2+3m^2_\eta}{m^2_K+m^2_\eta}
h_2(m^2_\eta/m_K^2)-18\right\},\cr
h_2(x)=&\,\tfrac{3}{2}\left(\dfrac{1+x}{1-x}\right)^2+
\dfrac{3x(1+x)}{(1-x)^3}\log x.
\cr}
\equn{(9.4.10b)}$$

From these formulas follow the values, respectively,
$$\langle r^2_{{\rm S},\pi}\rangle_{\rm GL}=0.55\pm0.15\;{\rm fm}^2.
\equn{(9.4.11a)}$$
$$\langle r^2_{{\rm S},K\pi}\rangle_{\rm GL}=0.20\pm0.05\;{\rm fm}^2,
\equn{(9.4.11b)}$$
 
The value for the first that we obtained from experiment in \equn{(7.2.17)},
$\langle r^2_{{\rm S},\pi}\rangle=0.75\pm0.07\,{\rm fm}^2$, and 
the one for the second following from $K_{l3}$ decays,\fnote{See e.g. Yndur\'ain~(2003) for
details.} 
$\langle r^2_{{\rm S},K\pi}\rangle=0.31\pm0.06\,{\rm fm}^2$, 
are therefore about $2\,\sigma$ above the chiral theory prediction to one loop, 
Eqs.~(9.4.11). 
This indicates that two loop corrections may be important for the relations (9.4.10). 
If we neglect them, however, we get a very precise value for 
the constant $\bar{l}_4$:
$$\bar{l}_4=5.4\pm0.5,
\equn{(9.4.12)}$$
a number substantially larger than that given in (9.4.1) which, unfortunately, 
is based partially on the evaluation of Donoghue, Gasser and Leutwyler~(1990) 
which, as discussed in Yndur\'ain~(2003a) is not quite reliable within its 
estimated errors.

\booksubsection{9.4.3. Summary of ch.p.t. predictions for $\pi\pi$ scattering}

\noindent
We will now present a summary of the two previous subsections. 
We will use the improved values of the $\bar{l}_i$ parameters we have obtained, {\sl to one
loop}\/:
$$\eqalign{
\bar{l}_1=&\,-1.47\pm0.24,\quad \bar{l}_2=5.97\pm0.07,\quad\bar{l}_3=2.9\pm2.4,\cr
\bar{l}_4=&\,5.4\pm0.5,\quad \bar{l}_5=13.7\pm0.7,\quad \bar{l}_6=16.35\pm0.14.\cr
}
\equn{(9.4.13a)}$$
(Actually, only the first four $\bar{l}_i$ enter for $\pi\pi$ scattering).
With this, we may evaluate the low energy $\pi\pi$ parameters, and we find  the values of
Table~IV:
\bigskip
\setbox0=\vbox{\petit
\medskip
\setbox1=\vbox{\offinterlineskip\hrule
\halign{
&\vrule#&\strut\hfil#\hfil&\quad\vrule\quad#&
\strut\quad#\quad&\quad\vrule#&\strut\quad#\cr
 height2mm&\omit&&\omit&&\omit&\cr 
&\hfil\/\/ Quantity \hfil&&\hfil Exp. value (PY)\hfil&
&\hfil 1 loop ch.p.t.\hfil& \cr
 height1mm&\omit&&\omit&&\omit&\cr
\noalign{\hrule} 
height1mm&\omit&&\omit&&\omit&\cr
&\phantom{\Big|}$a_0^{(0)}$&&\hfil $0.230\pm0.010$\hfil&&\hfil$0.207\pm0.003$
\quad[0.157] \hfil& \cr
\noalign{\hrule}
&\phantom{\Big|}$a_0^{(2)}$&&\hfil$-0.0422\pm0.0022$\hfil&&\hfil$-0.043\pm0.002
\quad[-0.045]$\hfil& \cr
\noalign{\hrule}
&\phantom{\Big|}$b_0^{(0)}$&&\hfil $0.268\pm0.010$\hfil&&\hfil $0.255\pm0.003
\quad[0.179]$ 
\phantom{l}\hfil& \cr
\noalign{\hrule}
&\phantom{\Big|}$b_0^{(2)}$&&\hfil $-0.071\pm0.004$\hfil&&\hfil$-0.076\pm0.001
\quad[-0.089]$\hfil&\cr
\noalign{\hrule}
&\phantom{\Big|}$a_1$&&\hfil $(38.3\pm0.8)\times10^{-3}$ \quad\hfil&&\hfil 
$(38.0\pm0.04)\times10^{-3}
\quad[33.6]\;$\hfil&\cr
\noalign{\hrule}
&\phantom{\Big|}$b_1$&&\hfil $(4.56\pm0.26)\times10^{-3}$\hfil&&\hfil 
$(4.69\pm0.14)\times10^{-3}$ \hfil&\cr
\noalign{\hrule}
&\phantom{\Big|}$a_2^{(0)}$&&\hfil $(18.0\pm0.2)\times10^{-4}$\hfil&&\hfil input
\quad \hfil& \cr
\noalign{\hrule}
&\phantom{\Big|}$a_2^{(2)}$&&\hfil $(2.2\pm0.2)\times10^{-4}$\hfil&&\hfil input
\quad\hfil& \cr
\noalign{\hrule}
&\phantom{\Big|}$\langle r^2_{{\rm S},\pi}\rangle$&
&\hfil $0.77\pm0.07\;{\rm fm}^2$\hfil&&\hfil input\hfil&
\cr
\noalign{\hrule}}
\vskip.05cm}
\centerline{\box1}
\bigskip
\noindent{\petit Comparison of evaluations of low energy parameters 
from experiment, and from one loop 
chiral perturbation theory. 
The experimental numbers for $\pi\pi$ scattering are taken from the Pel\'aez--Yndur\'ain 
$\pi\pi$ amplitudes in Tables~II,~III; $\langle r^2_{{\rm S},\pi}\rangle$ 
is as given in \equn{(7.2.17)}. 
The values of $a_1$, $b_1$ have been (very slightly) improved 
by compounding the independent determinations 
from the pion form factor, and from the Froissart--Gribov representation. 
In brackets: the quantities to  leading order. }
\medskip
\smallskip
\centerline{\sc Table IV}
\centerrule{6cm}}
\medskip
\box0
\medskip
Except for $a_0^{(0)}$, and a little for $b_0^{(0)}$, 
the agreement is perfect, so no there is
no need  of two loop corrections at the present level 
of accuracy for the remaining low energy parameters. 
For both  $a_0^{(0)}$ and $b_0^{(0)}$  already the one loop corrections are quite  
large,  so we expect also large two loop corrections, just by renormalization group arguments 
(Colangelo,~1995). 
In fact, a detailed estimate for $a_0^{(0)}$   (Bijnens et al.,~1996) gives 
$$\delta_{2\;{\rm loop}}\,a_0^{(0)}=0.017\pm0.002$$
which brings the ch.p.t. value of $a_0^{(0)}$ to 
$$a_0^{(0)}=0.224\pm0.003\quad\hbox{[incl. two loop]},$$ 
well inside the experimental error bars.

\booksection{9.5. Weak and electromagnetic interactions.\hb 
The accuracy of chiral perturbative calculations}

\noindent
Weak and electromagnetic interactions, at tree level, can be introduced by 
making the standard minimal replacement in the covariant derivatives; 
for e.g., electromagnetic interactions,  
$\nabla_\mu\to\nabla_\mu-eA_\mu$. In this way one can 
calculate chiral dynamics values of quantities like the 
pion electromagnetic form factor, or strong interactions corrections to 
weak decays. 
Another matter are virtual electromagnetic corrections. 
These break chiral invariance, and can be  large. 
For example, the $\pi^+-\pi^0$ mass difference is of order $(m_d-m_u)^2$ 
in chiral perturbation theory; the corresponding calculation yields a very small number,
$$m^2_{\pi^+}-m^2_{\pi^0}=(m_d-m_u)^2\dfrac{2B^2\bar{l}_7}{f^2_\pi},\quad
B=-\langle\bar{q}q\rangle/f^2$$
which would give $m_{\pi^+}-m_{\pi^0}\sim0.2\,\mev.$
However, the experimental value is $m_{\pi^+}-m_{\pi^0}=4.6 \mev$. 
In this case one can use current algebra techniques to estimate the electromagnetic 
contribution, which is indeed of the right order of magnitude (Das, Mathur and Okubo, 1967), 
but in general this is not possible; we expect (generally unknown) 
electromagnetic corrections of something up to this order of magnitude, 
$\sim3.4\%$, to chiral perturbation theory calculations.

A case in which the electromagnetic corrections to the constants $\bar{l}_i$ is 
known is that of  
$\bar{l}_6$. The value reported in (9.4.2d) above is actually an average of those 
obtained from the charge radii of the pion with $\pi^0\pi^+$ and $\pi^+\pi^-$. 
If we use only the last (associated with the $\rho^0$), 
hence the parameters of (6.3.5c), we find 
instead 
$$\bar{l}_6=16.07\pm0.18.
\equn{(9.5.1)}$$
The difference between the two, a 1.5\%, is the minimum
 extra error due to electromagnetic corrections 
that we should append to all the determinations of chiral perturbation theory parameters.

A place where isospin violation corrections are 
potentially large are the scattering lengths. 
If we repeat the fits of de~Troc\'oniz and Yndur\'ain~(2002) without 
imposing the constraint
$a_1=(38\pm3)\times10^{-3}\,M_{\pi}^{-3}$, 
and fit separately $\pi^+\pi^-$ and 
$\tau$ decay data we find the numbers,
$$\eqalign{
a_1(\pi^+\pi^-)=&\,(37\pm3)\times10^{-3}\,M_{\pi}^{-3},\cr
a_1(\pi^+\pi^0)=&\,(43\pm3)\times10^{-3}\,M_{\pi}^{-3}.\cr
}$$
The two values overlap, but only barely; a difference of the order of 
$3\times10^{-3}$ (in units of $M_{\pi}$) 
cannot be excluded.

Another question is the scale of higher corrections in chiral perturbation theory. 
For the {\sl logarithmic} corrections 
we know that this scale is $1/(4\pi f_\pi)^2$, so for energies of the order 
of $M_{\pi}$ we expect corrections $O(M_{\pi}^2/(4\pi f_\pi)^2)\simeq1.4\%$. 
However, this estimate forgets the {\sl constant} contributions to the 
$\bar{l}_i$. There is no reason why 
they should be suppressed by powers of $1/(4\pi f_\pi)^2$; all we can expect  
is a suppression of order $O(M_{\pi}^2/\lambdav^2_0)$, with 
$\lambdav^2_0$ proportional to the QCD parameter $\lambdav\sim400\,\mev$
 (for 2 or 3 flavours). 
In some cases the coefficients of these terms will be small; in other they may be large. 
This last situation occurs for example for the  S0 wave in $\pi\pi$ scattering, 
where the correction necessary to get agreement between the leading 
value obtained from chiral dynamics, $a_0^{(0)}=0.16\,M_{\pi}^{-1}$, with the experimental 
values which  vary between 
$a_0^{(0)}\simeq0.24\,M_{\pi}^{-1}$ and $a_0^{(0)}\simeq0.21\,M_{\pi}^{-1}$
 is at least a third of the leading one.\fnote{
This possibility is particularly relevant in view of the doubts expressed by other researches 
on some aspects of chiral perturbation theory; see, 
for example, Fuchs, Sazdjian and Stern (1991); Knecht et al.~(1996).} 

In this context, we would like to emphasize that the situation is 
even worse for the quantities $a_3$, $c_\pi$,  for which the leading order calculations miss
the experimental values by a factor $\sim3$.

In some cases the size of the corrections may be gleaned from external arguments. 
For example, for the isospin zero S-wave in $\pi\pi$ scattering, 
chiral dynamics implies that its imaginary part should be suppressed with respect to 
the real part, at energy squared $s$, by powers $s/\lambdav^2_0$. 
However, already at $s^{1/2}=500\,\mev$, i.e., only $200\,\mev$ above threshold, 
real and imaginary part are of the same order of 
magnitude; so, we would expect poor convergence in this case, 
as indeed happens.

\bookendchapter
\brochureb{\smallsc appendices}{\smallsc  appendices}{139}
\bookchapter{Appendices}

\booksection{Appendix A: Summary of low energy, $s^{1/2}\leq1.42\,\gev$ partial waves}
\vskip-0.5truecm
\booksubsection{A.1. The S wave with isospin zero below $960$ \mev}

\noindent
We  impose the Adler zero at $s=\tfrac{1}{2}{M_\pi}^2$ (no attempt is made to vary this),
 and a resonance with mass
$M_\sigma$, a free parameter. 
We write
$$\cot\delta_0^{(0)}(s)=\dfrac{s^{1/2}}{2k}\,
\dfrac{{M_\pi}^2}{s-\tfrac{1}{2}{M_\pi}^2}\,\dfrac{M^2_\sigma-s}{M^2_\sigma}\,\psi(s),
\equn{(A.1)}$$
and
$$\psi(s)=\sum_n B_n\,w(s)^n;\quad
w(s)=\dfrac{\sqrt{s}-\sqrt{s_0-s}}{\sqrt{s}+\sqrt{s_0-s}},\quad
s_0=4m^2_K;
\equn{(A.2)}$$
we have taken $m_K=0.496\,{\gev}$. 
We will fit the phases that follow from $K_{l4}$ decays, with the P wave as given below. 
We  include in the fit 
 the value  $\delta_0^{(0)}(M_K^2)=43.3\pm2.3\degrees$, 
as discussed in the text. 
Finally, in the region\fnote{This 
is the energy region in which most {\sl experimental} phase shifts 
agree one with the other, within errors.}
 where $s^{1/2}$ is between 0.81 \gev\ and 0.98 \gev,
 we also include  some phases of 
Protopopescu et al.~(1973), and of the  $s$-wave solutions 
of Estabrooks and Martin~(1974), as given in \subsect~6.4.2.

\smallskip
\noindent{\sl Solution B2.}\quad
If we take two parameters in (A.4a) we find what we call {\sl solution 2B},
$$\eqalign{
\cot\delta_0^{(0)}(s)=&\,\dfrac{s^{1/2}}{2k}\,\dfrac{M_{\pi}^2}{s-\tfrac{1}{2}M_{\pi}^2}\,
\dfrac{M^2_\sigma-s}{M^2_\sigma}\,
\left\{B_0+B_1\dfrac{\sqrt{s}-\sqrt{s_0-s}}{\sqrt{s}+\sqrt{s_0-s}}\right\};\cr
{B}_0=&\,21.04,\quad {B}_1=6.62,\quad
M_\sigma=782\pm24\,\mev;\quad\dfrac{\chi^2}{\rm d.o.f.}=\dfrac{15.7}{19-3};\cr
a_0^{(0)}=&\,(0.230\pm0.010)\times M_{\pi}^{-1};\quad\delta_0^{(0)}(M_K)=41.0\degrees\pm2.1\degrees;
\cr    }
\equn{(A.3)}$$
this fit we take to be valid for $s^{1/2}\leq0.96\,\gev$. 
Uncorrelated errors are obtained if 
replacing the $B_i$ by the 
parameters $x,\,y$ with
$$B_0=y-x;\quad B_1=6.62-2.59 x;\quad y=21.04\pm0.75,\quad x=0\pm 2.4.
\equn{(A.4a)}$$
The corresponding phase shift is shown in \fig~6.4.3 in the main text.

\smallskip
\noindent{\sl Solution B3.}\quad
With three parameters a new minimum ({\sl solution 3B}\/) appears:
$$\eqalign{
\cot\delta_0^{(0)}(s)=&\,\dfrac{s^{1/2}}{2k}\,\dfrac{M_{\pi}^2}{s-\tfrac{1}{2}M_{\pi}^2}\,
\dfrac{M^2_\sigma-s}{M^2_\sigma}\cr
\times&\,
\left\{B_0+B_1\dfrac{\sqrt{s}-\sqrt{s_0-s}}{\sqrt{s}+\sqrt{s_0-s}}+
B_2\left[\dfrac{\sqrt{s}-\sqrt{s_0-s}}{\sqrt{s}+\sqrt{s_0-s}}\right]^2\right\};\cr
s_0^{1/2}=2M_K;&\quad\chi^2/{\rm d.o.f.}=11.1/(19-4).\cr
 M_\sigma=806\pm21,&\,\; B_0=21.91\pm0.62,\; B_1=20.29\pm1.55, \; B_2=22.53\pm3.48;\cr
a_0^{(0)}=&\,(0.226\pm0.015)\;M_{\pi}^{-1}.\cr
\cr
}
\equn{(A.5)}$$
The central values in (A.5) are  something between (A.4) and the solution of Colangelo, Gasser
and Leutwyler~(2001), which it comprises.

\topinsert{
\setbox0=\vbox{\hsize12truecm{\epsfxsize 10.5truecm\epsfbox{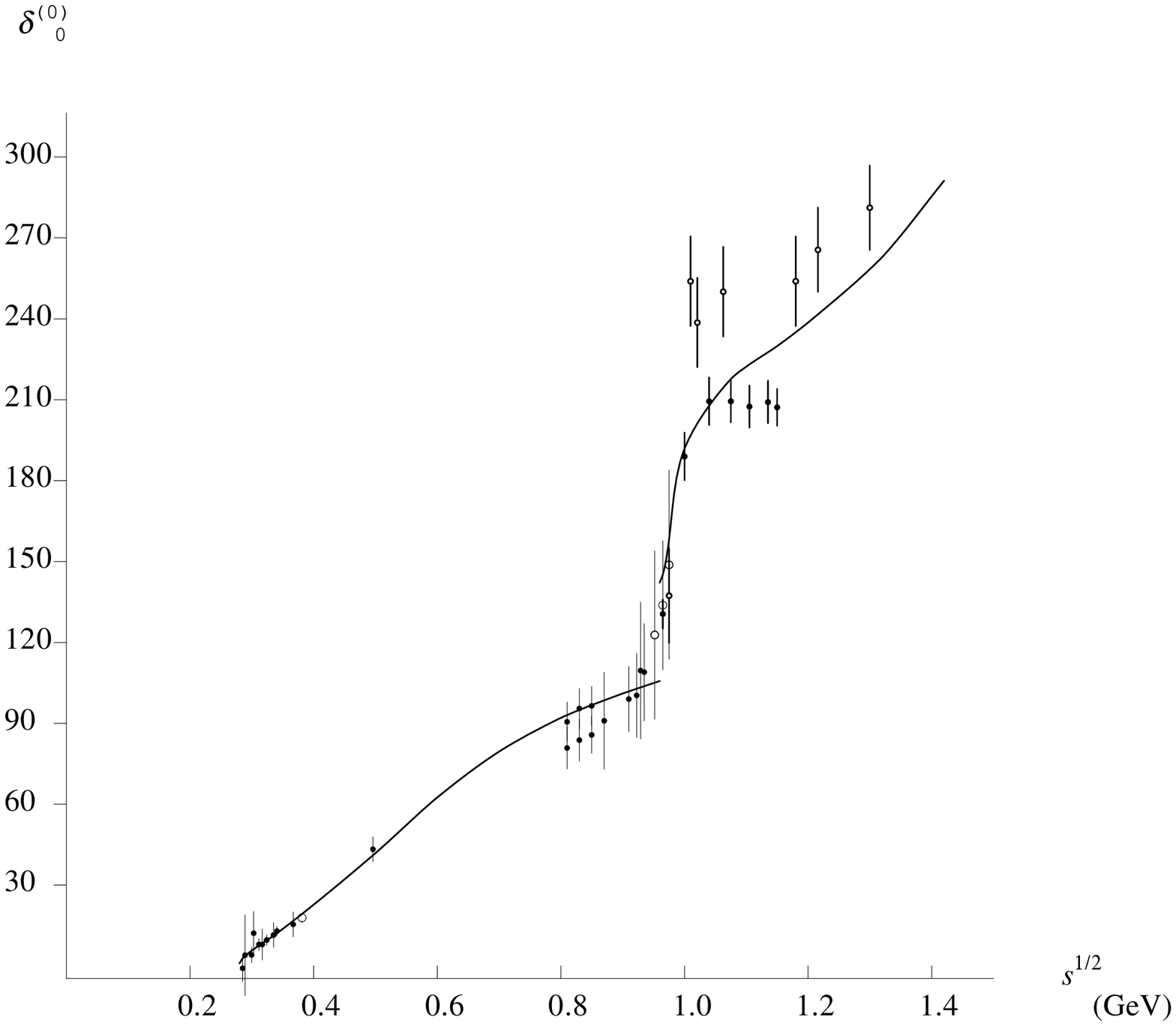}}} 
\setbox6=\vbox{\hsize 10truecm\captiontype\figurasc{Figure A.1. }{
The  
S0  phase shifts and inelasticities corresponding to Eqs.~(A.4), (A.6). 
Also shown are the experimental points   
 included in
the fits.}\hb} 
\centerline{\tightboxit{\box0}}
\bigskip
\centerline{\box6}
\medskip
}\endinsert

\booksubsection{A.2. The $I=0$ S wave between $960\,\mev$ and $1420\,\mev$}

\noindent
We here present a semi-phenomenological 
fit to $\delta_0^{(0)}$ and $\eta_0^{(0)}$, as discussed in the main text. 
We write
$$\cot\delta_0^{(0)}(s)=c_0\,\dfrac{(s-M^2_\sigma)(M^2_{f}-s)}{M^2_{f} s^{1/2}}\,
\dfrac{|k_2|}{k^2_2},\quad k_2=\dfrac{\sqrt{s-4m^2_K}}{2}
\equn{(A.6a)}$$
and
$$\eta_0^{(0)}=1-\left(c_1\dfrac{k_2}{s^{1/2}}+c_2\dfrac{k_2^2}{s}\right)
\,\dfrac{M'^2-s}{s}.
\equn{(A.6b)}$$
In the first, $c_0$ and $M_\sigma$ are free parameters and we fix $M_{f}=1320\,\mev$. 
In (A.6b),  the free parameters are  $c_1,\,c_2$ and we adjust $M'$ 
to  the inelasticity of Hyams et al. on the $f_0(1370)$. 
For the selection of data points, see the main text, subsect~6.4.3. 
We find,
$$\eqalign{
c_0=&\,1.36\pm0.05,\quad M_\sigma=802\pm11\,\mev,\; M'=1500\,\mev;
\quad \dfrac{\chi^2}{\rm d.o.f.}=\dfrac{36.2}{14-2};
\cr
c_1=&\,6.7\pm0.17,\quad c_2=-17.6\pm0.8;\quad\chi^2/{\rm d.o.f.}=7.7/(8-2).\cr
}
\equn{(A.6c)}$$
The errors for $c_0$, $M_\sigma$ correspond to {\sl three} standard deviations, 
since  we have a $\chidof\simeq3$.
The  value of $M_\sigma$ coincides, 
{\sl grosso modo}, with what we found below $\bar{K}K$ threshold.

The qualitative features of the fits to the S0 wave in the whole 
range $s^{1/2}\leq1.42\,\gev$ may be seen in \fig~A.1.

\booksubsection{A.3. Parametrization of the S wave for  $I=2$}

\noindent
As discussed in the main text, we consider three sets of experimental data.
 The first 
 corresponds to solution A in the paper by 
Hoogland et al.~(1977), who use the reaction $\pi^+ p\to\pi^+\pi^+n$; 
the set from the work of Losty
et al.~(1974), who analyze instead  $\pi^- p\to\pi^-\pi^-\Delta$; 
and the set of Cohen et al.~(1973), obtained fron pion-deuteron scattering 
(which, however, were {\sl not} included in the fit). 
We will not consider the so-called solution B in the paper of Hoogland et al.

For isospin 2, there is no low energy resonance, but $f_0^{(2)}(s)$
 presents the feature that a zero 
is expected (and, indeed, confirmed by the fits) 
in the region $0<s<4M^2_\pi$. 
This zero of $f_0^{(2)}(s)$ is related to the so-called Adler zeros 
and, to lowest order in chiral perturbation theory,  
occurs at $s=2z_2^2$ with $z_2={M_\pi}$. In view 
of this, 
we extract the zero  and write
$$\cot\delta_0^{(2)}(s)=\dfrac{s^{1/2}}{2k}\,\dfrac{{M_\pi}^2}{s-2z_2^2}\left\{{B}_0+{B}_1w(s)\right\}.
$$
The inelasticity of $\pi\pi$ scattering in this channel,
 say $\pi^{\pm,0}\pi^\pm$, is very small until 
one crosses the $\rho^{\pm,0}\rho^\pm$ threshold. In our calculations we will take 
$s_0^{1/2}=1.45\;\gev$ for the effective opening of the inelastic channels.

We can improve on the quality of the results by requiring, 
simultaneously with the fit to the data, 
fulfillment of the Olsson sum rule, 
within the errors produced by the remaining  waves. 
If moreover we fix $z_2=M_\pi$, and include all experimental data of 
Losty et al. and Hoogland et al., solution~A,  
(up to $s^{1/2}=1350\,\mev$) in the fit we find
\smallskip
$$\eqalign{
\cot\delta_0^{(2)}(s)=&\,\dfrac{s^{1/2}}{2k}\,\dfrac{M_{\pi}^2}{s-2z_2^2}\,
\left\{B_0+B_1\dfrac{\sqrt{s}-\sqrt{s_0-s}}{\sqrt{s}+\sqrt{s_0-s}}\right\};\cr
s_0^{1/2}=&\,1.45\;\gev;\quad\chi^2/{\rm d.o.f.}=17.2/(19-2).\cr
 B_0=&\,-118\pm2.5,\quad B_1=-105\pm2.5,\quad z_2=139.57\;\mev\;\hbox{[fixed]}.\cr
}
\equn{(A.7a)}$$
Then one has $a_0^{(2)}=-0.0422\pm0.0022$.

The inelasticity may be obtained fitting the data of 
Cohen et al.~(1973) and Losty et al.~(1974). One finds
$$\eta_0^{(2)}(s)=1-c\left(1-{M^2_{\rm eff}/s}\right)^{3/2};\quad 
M_{\rm eff}=0.96\;\gev,\quad 
c=0.28\pm0.12, 
\equn{(A.7b)}$$
which is valid for $s^{1/2}\geq0.96\;\gev$.
The plot of $\delta_0^{(2)}$ may be seen in \fig~7.6.2 in the main text. 
We will take (A.7) to be valid up to $1.42$ \gev. 

\goodbreak

\booksubsection{A.4. The P  wave below $1$ \gev}

\noindent
We will consider first the P wave for $\pi\pi$ scattering for energies below the 
region were the inelasticity reaches the 2\% level; say, below 
$s_0=1.1\,\gev^2$. 
We will neglect for the moment isospin invariance violations due 
to e.m. interactions or the  mass difference of the $u,\,d$ quarks.

The best values for our parameters are actually obtained from fits to the 
pion form factor.
 If we  take 
systematic normalization errors into account, but neglect isospin violation, 
we find
\smallskip
$$\eqalign{
\cot\delta_1(s)=&\,\dfrac{s^{1/2}}{2k^3}\,(M^2_\rho-s)\,
\left\{B_0+B_1\dfrac{\sqrt{s}-\sqrt{s_0-s}}{\sqrt{s}+\sqrt{s_0-s}}\right\};\cr
s_0^{1/2}=1.05\;\gev;&\quad\chi^2/{\rm d.o.f.}=227/(209-3).\cr
\quad M_\rho=773.5\pm0.85\;\mev,&\,\quad B_0=1.071\pm0.007,\quad B_1=0.18\pm0.05.\cr
[ s^{1/2}\leq&\, 1.0\;\gev]\cr
}
\equn{(A.8a)}$$
and 
$$\eqalign{
a_1=&\,(38.6\pm1.2)\times10^{-3}\;M_{\pi}^{-3},\quad b_1=(4.47\pm0.29)\times10^{-3}\;M_{\pi}^{-5},
\cr
 \gammav_{\rho}=&\,145.5\pm1.1\,\mev.\cr
}$$
Although the values of the experimental $\pi\pi$ 
phase shifts were {\sl not} included in the fit, the 
phase shifts that (A.8) implies are en very good agreement with them, 
as shown in \fig~6.3.2.

Eqs.~(A.8) above were evaluated with an average of information on $F_\pi(t)$ from the 
two channels that contain the $I=1$ P wave, that is, $\pi^+\pi^-$ (dominated by the 
$\rho^0$) and $\pi^0\pi^+$, dominated by the $\rho^+$.
Experimentally the first are obtained from 
processes $e^+e^-\to\pi^+\pi^-$, $\pi e\to\pi e$; 
the second from the decays $\tau\to\nu_\tau \pi^0\pi^+$. 
The values of the parameters for a pure $\rho^0$ ($\pi^+\pi^-$) are slightly different. 
Including systematic errors in the analysis we would find
$$\eqalign{B_0=&\,1.065\pm0.007,\quad B_1=0.17\pm0.05,\quad M_{\rho^0}=773.1\pm0.6,\cr 
\gammav_{\rho^0}=&147.4\pm1.0\,\mev;
}
\equn{(A.8b)}$$
$a_1,\,b_1$ do not change appreciably. 

\booksubsection{A.5. The P wave for $1\gev\leq s^{1/2}\leq 1.42\gev$}

\noindent
For the imaginary part of the P wave between 1 \gev\ and 1.42 \gev\ we use an empirical formula, 
obtained  adding a resonance (with mass 1.45 \gev) to a nonresonant background:
$$\eqalign{
\imag \hat{f}_1(s)=&\,\dfrac{1}{1+[\lambda+1.1k_2/s^{1/2}]^2}+
{\rm BR}\dfrac{M^2_{\rho'}\gammav^2
\left[k/k(M^2_{\rho'})\right]^6}{(s-M^2_{\rho'})^2+
M^2_{\rho'}\gammav^2\left[k/k(M^2_{\rho'})\right]^6};\cr
[1.0\leq s^{1/2}\leq1.42\;\gev]&\quad M_{\rho'}=1.45\,\gev,
\quad \gammav=0.31\,\gev,\quad \lambda=2.6\pm0.2.\cr }
\equn{(A.9)}$$
The parameters of the $\rho(1450)$  are poorly known.  
 We have taken ${\rm BR}=0.25\pm0.05$, but the error could 
well be twice as large.

\booksubsection{A.6. Parametrization of the D wave for  $I=0$}

\noindent
The D wave with isospin 0 in $\pi\pi$ scattering presents a resonance 
below $1.42\,\gev$: the $f_2(1270)$, 
that we will denote simply by $f_2$. 
Experimentally, 
$\gammav_{f_2}=185\pm4\,\gev$. 
The $f_2$  
couples mostly to $\pi\pi$; 
to a 15\% accuracy we may neglect inelasticity up to
$s_0^{1/2}=1.43\,\gev$.

We will first fit data on $\delta_2^{(0)}$ altogether neglecting inelasticity,
 which we will then add by hand.
The data are scanty, and of poor quality. The 
phase shifts of Protopopescu et al.~(1973)  cover 
only the range $810\leq s^{1/2}\leq 1150\,\mev$, and are incompatible with those of 
the Cern-Munich experiment, that we take as given in the $s$-channel solution of 
Estabrooks and Martin~(1974).\fnote{We take only the values of
 the phase shifts of these authors at the energies 0.63, 0.71, 
0.75, 0.79, 0.83, 0.87 and 0.91 \gev. Since they 
do not give errors for their numbers, we arbitrarily 
take a common error of 10\%.} 
 We  impose in the fit the scattering length,
 as obtained from the Froissart--Gribov 
representation, and the experimental width of the $f_2$:
$$a_2^{(0)}=(18.1\pm0.4)\times10^{-4}\,\times M_{\pi}^{-5},\quad
\gammav_{f_2}=185\pm4\,\gev.$$

We write
$$\cot\delta_2^{(0)}(s)=\dfrac{s^{1/2}}{2k^5}\,(M^2_{f_2}-s)\,{{M_\pi}^2}\,{\psi}(s),\quad
{\psi}(s)=B_0+B_1w(s)
\equn{(A.10a)}$$
and
$$w(s)=\dfrac{\sqrt{s}-\sqrt{s_0-s}}{\sqrt{s}+\sqrt{s_0-s}},\quad
 s_0=1430\,\mev;\quad
M_{f_2}=1275.4\,\mev.$$ 
We find,
$$\eqalign{
\dfrac{\chi^2}{\rm d.o.f.}=&\,74/(21-2),\quad B_0=22.4\pm0.1,\quad B_1=23.3\pm3.0;\cr
}
$$
The very poor \chidof\ is obviously due to the strong 
 bias of the data of Protopopescu et al.~(1973), clearly seen in \fig~6.4.2 in the main text.
Above values of the $B_i$ 
would give $\gammav_{f_2}=196\pm6\,\mev$. 
We will then take this solution up to $\bar{K}K$ threshold; on it, 
we  join 
the solution to a new one, for which we impose the $f_2$ width; we get 
$$B_0=22.5\pm0.1,\quad B_1=28.5\pm3.2.$$
Therefore, we have
$$B_0=\cases{22.4\pm0.1,\quad s< 4M_K^2,\cr22.5\pm0.1,\quad s>4M_K^2\cr};\qquad
B_1=\cases{23.3\pm3.0,\quad s< 4M_K^2,\cr28.5\pm3.2,\quad s>4M_K^2.\cr}
\equn{(A.10b)}$$ 
We take into account the inelasticity by writing
$$\eta_2^{(0)}(s)=\cases{1,\qquad  s< 4M_K^2;\cr
1-2\times\epsilon_f\,\dfrac{k_2(s)}{k_2(M^2_{f_2})},\;
\quad \epsilon_f=0.131\pm0.015,\quad s> 4M_K^2.\cr} 
\equn{(A.10c)}$$
$ k_2=\sqrt{s/4-M^2_K}.$
We have fixed the coefficient $\epsilon_f$ fitting the inelasticities of Protopopescu et al.,
and the experimental inelasticity of the $f_2$; the overall \chidof\ of this  
fit is $\sim1.8$. 
This parametrization is different from the corresponding formula used in 
Pel\'aez and Yndur\'ain~(2003) and,
 probably, more exact, 
although the influence of the change in the various sum rules is negligible. 
For example, the Olsson integral has only increased by $0.002$ by 
using the formulas given here.

The fit returns the values
$$\eqalign{
a_2^{(0)}=&\,(18.4\pm7.6)\times10^{-4}\, M_{\pi}^{-5},\quad
b_2^{(0)}=(-7.9^{+4.1}_{-11.0})\times10^{-4}\, M_{\pi}^{-7},\cr
\gammav_{f_2}=&\,185\pm5\,\mev.\cr
}
\equn{(A.10d)}$$

\booksubsection{A.7. Parametrization of the D wave for  $I=2$}

\noindent
For isospin equal 2, there are no resonances in the D wave. 
If we want a parametrization that 
applies down to threshold, we must incorporate the  
zero of the corresponding phase shift. So 
we write
$$\cot\delta_2^{(2)}(s)=
\dfrac{s^{1/2}}{2k^5}\,\left[B_0+B_1 w(s)\right]\,
\dfrac{{M_\pi}^4 s}{4({M_\pi}^2+\deltav^2)-s}
\equn{(A.11a)}$$
with $\deltav$ a free parameter and
$$w(s)=\dfrac{\sqrt{s}-\sqrt{s_0-s}}{\sqrt{s}+\sqrt{s_0-s}},\quad
 s_0=1450\,\mev.$$ 
 Moreover, we impose  the 
value for the scattering length 
that follows from the Froissart--Gribov representation.

\smallskip
\noindent{\sl Solution B2.}\quad
With two $b_i$s we   
 get a mediocre fit, $\chidof=72/(22-3)$, for $s^{1/2}$ below 
1 \gev, and the values of the parameters are
$$\eqalign{
B_0=&(2.30\pm0.17)\times10^3,\; B_1=-267\pm750,\;
\deltav=103\pm11\,\mev;\quad s^{1/2}\leq1.1\;\gev.\cr
}
\equn{(A.11b)}$$
Actually, (A.11b) is valid up to $s^{1/2}\sim1.2\,\gev$.
Above $1\,\gev$ we simply write a polynomial fit:
$$\eqalign{
\delta_2^{(2)}(s)=&\,(-0.051\pm0.004)+
a\left(\dfrac{s}{1\,{\gev}^2}-1\right)+b\left(\dfrac{s}{1\,{\gev}^2}-1\right)^2;\cr
a=&\,-0.081\pm0.033,\quad b=0.042\pm0.005;
\quad s\geq1.0\;\gev.\cr
}
\equn{(A.11c)}$$
The incompatibilities between the three sets of
 experimental data (obvious from a look 
at \fig~6.4.1 in the main text), 
probably related to those for the  S2 wave, 
preclude a better fit.
We can include inelasticity,
$$\eqalign{\eta_2^{(2)}(s)=&\,1-c\,(1-M^2_{\rm eff}/s)^{3/2},\quad M_{\rm eff}=0.96\;\gev,\quad
c=0.12\pm0.12;
\cr
s^{1/2}\geq&\,0.96\,\gev.\cr
}
\equn{(A.11d)}$$

  The fit returns a good value for the scattering length, and also for the effective range parameter,
$b_2^{(2)}$: 
$$a_2^{(2)}=(2.20\pm0.16)\times10^{-4}\,{M_\pi}^{-5};\quad
b_2^{(2)}=(-5.75\pm1.26)\times10^{-4}\,{M_\pi}^{-7},
\equn{(A.11e)}$$
to be compared with what we  found using the 
Froissart--Gribov representation (\sect~7.5), which gives very precise results,
$$a_2^{(2)}=(2.22\pm0.33)\times10^{-4}\,{M_\pi}^{-5};\quad
b_2^{(2)}=(-3.34\pm0.24)\times10^{-4}\,{M_\pi}^{-7}.
$$

\smallskip
\noindent{\sl Solution B4.}\quad
With {\sl four} parameters $B_i$, and  
including also the data of Cohen et al.~(1973) one gets,
$$\eqalign{
B_0=&\,(1.94\pm0.14)\times10^3,\quad B_1=(10.15\pm1.3)\times10^3,\quad
B_2=(18.68\pm2.4)\times10^3,\cr
B_3=&\,(-31.04\pm5.5)\times10^3;\qquad\deltav=218\pm22\,\mev.\cr
}
\equn{(A.12a)}$$
The errors here correspond to $3\,\sigma$. 
One has $\chidof=57/(25-5)$ and the fit returns the values 
of the low energy parameters
$$a_2^{(2)}=(2.04\pm0.5)\times10^{-4}\,M_{\pi}^{-5},\quad
b_2^{(2)}=(1.6\pm0.3)\times10^{-4}\,M_{\pi}^{-7}.
\equn{(A.12b)}$$
The large values of the parameters $B_i$, and the incompatibility of the three data sets, 
makes one suspect that the corresponding minimum is  spureous, 
but it represents resonably well the data.

\booksubsection{A.8. The F wave}

\noindent
For the imaginary part of the 
 F wave below $s^{1/2}=1.42\,\gev$ we write a background plus 
the tail of a Breit--Wigner formula for a resonance. 
The background is obtained fitting the low energy phase 
shifts of Protopopescu et al.~(1973), plus the scattering length as given 
by the Froissart--Gribov representation. The resonance is the $\rho_3$
 with its properties taken from the Particle Data Tables:
$$\eqalign{
\imag \hat{f}_3(s)=&\,\dfrac{1}{1+\cot^2\delta_3}+\left(\dfrac{k(s)}{k(M^2_{\rho_3})}\right)^{14}
{\rm
BR}\dfrac{M^2_{\rho_3}\gammav^2}{(s-M^2_{\rho_3})^2+M^2_{\rho_3}\gammav^2(k(s)/k(M^2_{f_4}))^{14}};\cr
\cot\delta_3(s)=&\,\dfrac{s^{1/2}}{2k^7}\,M^6_\pi\,
\left\{B_0+B_1\dfrac{\sqrt{s}-\sqrt{s_0-s}}{\sqrt{s}+\sqrt{s_0-s}}\right\};
\quad s_0^{1/2}=1.5\;\gev\cr
 M_{\rho_3}=&\,1.69\;\gev,
\quad \gammav=0.161\;\gev,\quad {\rm BR}=0.24;\cr
B_0=&\,(1.07\pm0.03)\times10^5,\quad B_1=(1.35\pm0.03)\times10^5.
\cr}
\equn{(A.13)}$$
This implies $a_3=(7.0\pm0.8)\times10^{-5}\,M^{-7}_\pi$.

The contribution of the F wave to all our sum rules is very small; the interest 
of calculating it lies in that it provides a test (by its very smallness)
 of the convergence of the partial wave expansions.

\booksubsection{A.9 The G waves}

\noindent 
For the G0 wave, we take its imaginary part to be given by the 
tail of the $f_4(2050)$ resonance, with its 
properties as given in the Particle Data Tables:
$$\eqalign{
\imag \hat{f}_4^{(0)}(s)=&\,\left(\dfrac{k(s)}{k(M^2_{f_4})}\right)^{18}
{\rm BR}\dfrac{M^2_{f_4}\gammav^2}{(s-M^2_{f_4})^2+M^2_{f_4}
\gammav^2(k(s)/k(M^2_{f_4}))^{18}};\cr
s^{1/2}\geq 1\;\gev.\qquad{\rm BR}=&\,17\pm2,\quad M_{f_4}=2025\pm8\;\mev,\gammav=194\pm13\;\mev.\cr}
\equn{(A.14)}$$

For the wave G2, we can write, neglecting its eventual inelasticity,
$$\cot\delta_4^{(2)}(s)=\dfrac{s^{1/2}M^8_\pi}{2k^9}\,B,\quad B=(-7.8\pm3.3)\times 10^6;\quad
s^{1/2}\geq 1\;\gev.
\equn{(A.15)}$$

It should be noted that this last, as well as the expression above for 
the G0 wave, are little more than order of magnitude estimates. 
Moreover, at low energies they certainly fail; 
an expression in terms of the scattering length 
approximation (cf.~\subsect~(6.5.4)) is more appropriate.
Thus, 
if, in a calculation, the value of either of the two G waves is 
important, it means that the calculation will have a large error.

\booksection{Appendix B: The conformal mapping method}

\noindent Let us consider a function, $f(z)$, 
analytic in a domain, $\cal{D}$; for example, this domain may be a plane with two cuts, as for 
the partial wave amplitudes; see e.g. Fig.~6.3.1 in the main text. 
According to general theorems (see, e.g., Ahlfors,~1953), 
it is always possible to map the interior of this domain into the interior 
of the  disk $\deltav(0,1)$, with center at the origin 
and unit radius. 
Let us call $w=w(z)$ to the corresponding variable. 
Then, in this variable, $f$ is analytic inside  $\deltav(0,1)$ 
and thus the ordinary Taylor expansion in terms of $w$ is absolutely and uniformly convergent 
in  $\deltav(0,1)$. 
Therefore, undoing the mapping, it follows that we can write
$$f(z)=\sum_{n=0}^\infty c_nw(z)^n,
\equn{(B.1)}$$
and this expansion is absolutely and uniformly convergent inside all of  $\cal{D}$.

It is important to realize that the representation (B.1) does not imply any 
supplementary assumption on $f(z)$ besides its analyticity properties; 
the convergence of (B.1) and the analyticity of $f$ in $\cal{D}$ are strictly 
equivalent statements.

We next say a few words about the specific situations we encountered in the 
main text.\fnote{ Further discussion
(with references) of the present method, and also of other 
similar ones (for example, mapping into an ellipse and expanding in 
a Legendre or Tchebycheff series there, the second in principle the more efficient 
procedure),  applied in particular to
$\pi\pi$ scattering, may be found in the reviews of Pi\u{s}ut~(1970) and Ciulli~(1973). 
The question of the stability of extrapolations is also discussed there.} 
In some cases we have a function $f$ analytic inside $\cal{D}$ 
except for a pole at $z_0$. 
Then the function
$$\varphi(z)=(z-z_0)f(z)$$
is analytic inside  $\cal{D}$ and it is $\varphi$ that can be expanded as in (B.1). 
In some other cases, we have a function $f(z)$ analytic inside  $\cal{D}$, with a zero at $z_0$. 
Of course, this zero does not spoil the analyticity, so we could expand $f$ itself. 
But, because the expansion of a function converges best if the function varies little, 
we have interest in extracting this zero and write
$$f(z)=(z-z_0)\psi(z),$$ 
expanding then $\psi$, which has the same analyticity properties as $f$.

The gain in convergence and stability obtained by expanding 
in the conformally transformed variable 
is enormous. 
The reader may verify this with
 the simple example of the function $\log(1+x)$. 
Here the region $\cal{D}$ is the complex plane cut from $-\infty$ to $-1$.
 If we expand in powers of
$x$,
$$\log(1+x)=-\sum_{n=1}^\infty\dfrac{(-x)^n}{n}
\equn{(B.2)}$$
 then for e.g. 
$x\sim1/2$ we need five terms for a 1\% percent accuracy.

In this case the expansion in the conformal variable can be made explicitly. 
The transformation that maps $\cal{D}$ into $\deltav(0,1)$ is 
$$w(x)= \dfrac{(1+x)^{1/2}-1}{(1+x)^{1/2}+1},$$
with inverse $x=2w/(1-w)$. 
Substituting this into $\log(1+x)$, we get the expansion
 in the conformally transformed variable 
$$\log(1+x)=4\sum_{n={\rm odd}}^\infty\dfrac{1}{n}\left[\dfrac{(1+x)^{1/2}-1}{(1+x)^{1/2}+1}\right]^n.
\equn{(B.3)}$$
If using this expansion, only two terms are necessary for an accuracy of a part in a thousand 
for $x\sim1/2$. 
Even for $x=25$, very far from the region of convergence of (B.2), the 
expansion (B.3) 
still represents the function closely: only three terms 
in (B.3) are necessary to get  a precision better than  2\%.\fnote{In this example 
we compare the virtues of (B.1), (B.3) as {\sl expansions}, for simplicity; in 
the main text, they are, however, used to {\sl fit}. 
Thus, we should really give ourselves the values of $\log(1+x)$ at a series of 
points, $x_1,\,x_2,\,\dots\,x_n$ and fit 
with (B.1), (B.3). 
The improvement is less spectacular than before, but 
it is still substantial.}

This economy is also apparent in our parametrizations of 
the partial waves or the Omn\`es auxiliary function $G(t)$, 
where only two, or in one case three terms, are necessary.
Indeed, 
the simplicity and economy of our parametrization contrasts 
with some of the complicated ones found in the literature. 
Thus, for example, Colangelo, Gasser and Leutwyler~(2001), who take 
it from Schenk~(1991), write
$$\cot\delta=\dfrac{2s^{1/2}}{k^{2l+1}}\dfrac{s-s_R}{4\mu^2-s_R}
\,\dfrac{1}{A+B k^2+C k^4+Dk^6}.
\equn{(B.4)}$$
For the P wave, they need these four parameters $A,\,B,\,C,\,D$ (apart from the 
squared mass of the resonance, $s_R$) when we only require two. 
Moreover, (B.3) only converges in the shaded disk in \fig~3.1.2 
(but it is used in the whole range, which is a recognized cause
 of unstability)
 and, in general, 
presents complex singularities, hence violating 
causality.

\topinsert{
\setbox0=\vbox{\hsize12truecm{\epsfxsize 11.3truecm\epsfbox{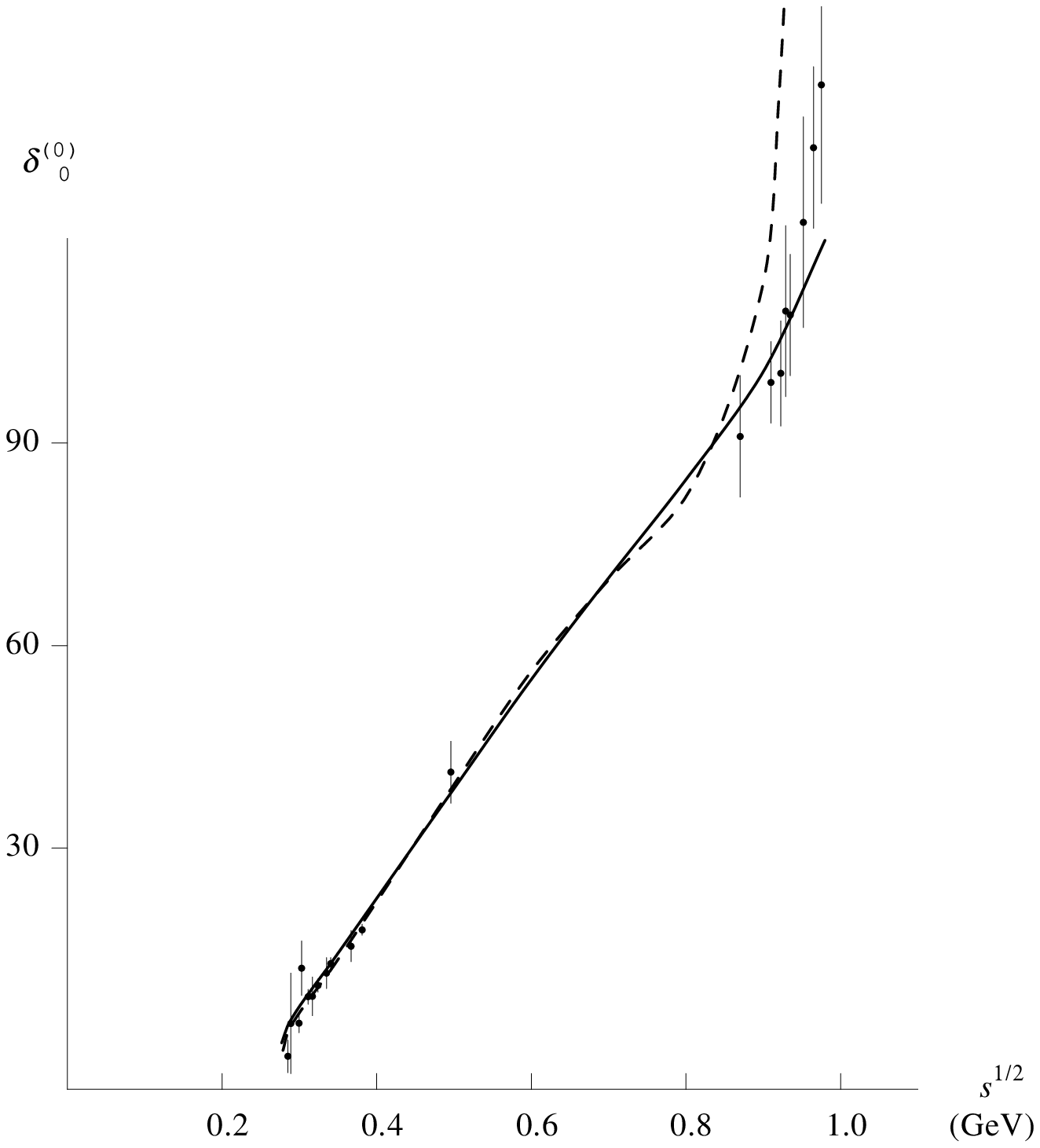}}} 
\setbox6=\vbox{\hsize 10truecm\captiontype\figurasc{Figure B.1.}{
The  
S0 wave phase shifts corresponding to (B.5) (continuous line) and 
 Colangelo, Gasser and Leutwyler~(2001), dashed line.}\hb} 
\centerline{\tightboxit{\box0}}
\bigskip
\centerline{\box6}
\vskip0.6truecm
}\endinsert

It s true that (B.4) is only used by Schenk and by Colangelo {\sl et alii} in the 
physical region; 
this, in fact, is one of its disadvantages: our parametrization can be used in all the 
cut complex plane and is therefore suited to discuss effects such as location of the 
poles associated with the resonances or the Adler zeros for the 
S waves. 
As a graphical example (\fig~B.1), consider the phase $\delta_0^{(0)}$ as given
 by Colangelo, Gasser and Leutwyler~(2001); for $s^{1/2}\leq0.8\,\gev$ 
it agrees, at the percent level, with the phase shift given by  
$$\eqalign{
\cot\delta_0^{(0)}(s)=&\,\dfrac{s^{1/2}}{2k}\,\dfrac{M_{\pi}^2}{s-\tfrac{1}{2}M_{\pi}^2}\,
\dfrac{M^2_\sigma-s}{M^2_\sigma}\cr
\times&\,
\left\{B_0+B_1\dfrac{\sqrt{s}-\sqrt{s_0-s}}{\sqrt{s}+\sqrt{s_0-s}}+
B_2\left[\dfrac{\sqrt{s}-\sqrt{s_0-s}}{\sqrt{s}+\sqrt{s_0-s}}\right]^2\right\};\cr
s_0^{1/2}=2M_K;&\quad\chi^2/{\rm d.o.f.}=11.1/(19-4).\cr
\quad M_\sigma=836,&\,\quad B_0=29,\quad B_1=39, \quad B_2=32.\cr
\cr
}
\equn{(B.5)}$$
[This is slightly displaced from the solution (6.4.12) in 
the main text].  
However, the CGL phase goes berserk above $0.9\,\gev$, while 
(B.5) continues to represent it fairly well up to the $\bar{K}K$ threshold: 
see the accompanying figure. 
It is also likely that at least some of the wiggles that the 
CGL solution presents (for example, the one below and around 0.8 \gev, 
clearly seen in the figure) are due to 
the unstability of the Schenk expansion.

It may be argued 
that, even if using Schenk's parametrization, 
one can get at $f_l(s)$ outside the physical region 
indirectly 
via Roy's equations. 
Using ours, however, you can get that {\sl both} 
directly 
and via Roy's 
equations, which 
provides useful consistency tests. 
As an example, we mention that the value we obtain for the 
Adler zero in \equn{(7.6.3)}, with a simple fit to data and only three 
parameters, namely $z_2=133\pm4.5\;\mev$, is consistent with (and the central 
value even slightly more 
accurate than) what Colangelo, Gasser and Leutwyler~(2001) get with 
the parametrization of Schenk, with five parameters,
 after imposing fulfillment of the Roy equations and a large number of 
crossing and analyticity sum rules: $z_2=136\,\mev$.

The fact that we manage with a smaller number of parameters is important 
not only as a matter of economy or consistency, but also in that 
we avoid  spureous minima which are liable to occur when 
large number of parameters are present.
 
\goodbreak
\booksection{Appendix C:  Sum rules and asymptotic behaviour}

\noindent
Long time ago it was remarked by Pennington~(1975) that one can use sum rules, based on
crossing symmetry, to relate low energy 
$\pi\pi$ physics to the high energy behaviour of the $\pi\pi$ scattering amplitudes. 
In Pennington's work, experimental phase shifts were used up to 
$s^{1/2}=2\,\gev$ and the conclusion was drawn that the Regge behaviour of 
$\pi\pi$ scattering was very different from what one
 could expect on the basis of factorization.

This conclusion could perhaps be maintained in 1974, when Pennington wrote his paper. 
First of all, the phase shifts used were those of the Cern--Munich experiment, 
which, as discussed in \sect~6.6 here,  bear little 
resemblance to reality --as we now know. Secondly, the QCD theory 
of strong interactions (in which factorization is automatic)
 was not established at the time,
when indeed it competed with other,  very different ones; 
string theories, for example. 
And finally (in both senses of the word), 
experimental data on various $\pi\pi$ cross sections 
between 1.2 and 6 \gev\ (Cohenet al.,~1973; 
Robertson, Walker and Davis,~1073; Hanlan et al.,~1976 and 
Abramowicz et al.,~1980) 
have fully confirmed the standard Regge picture, as described in \sect~2.4 in the main text.

Unfortunately, Ananthanarayan, B., et al. (2001) et al have, without looking carefully enough 
at the foundations of the paper of Pennigton~(1975), accepted its conclusions. 
And in this they have been followed by a number of modern authors, quoting the result
uncritically. 
In view of this we have considered useful to give
 a brief discussion of the matter in this 
Appendix, in which we will show that standard Regge theory, as reviewed here in \sect~2.4, 
 is perfectly compatible (within errors) with low energy $\pi\pi$ scattering 
provided we consider ``low energy" to mean $s^{1/2}$ less than $1.42\,\gev$, 
and we apply Regge formulas consistently above 
this energy.

First of all, we remark that the experimental cross sections, as deduced from the fits to
experiment carried in Chapter~6 here, produce cross sections that, at 
$s^{1/2}\sim 1.4\,\gev$, agree with what we get from the Regge formulas.
A complete set of figures may be seen in \sect~2.4 and in Pel\'aez and Yndur\'ain~(2003); 
here we only add that for isospin zero exchange, a fairly impressive one.

\topinsert{
\setbox3=\vbox{\hsize 11.5truecm\epsfxsize 10truecm\epsfbox{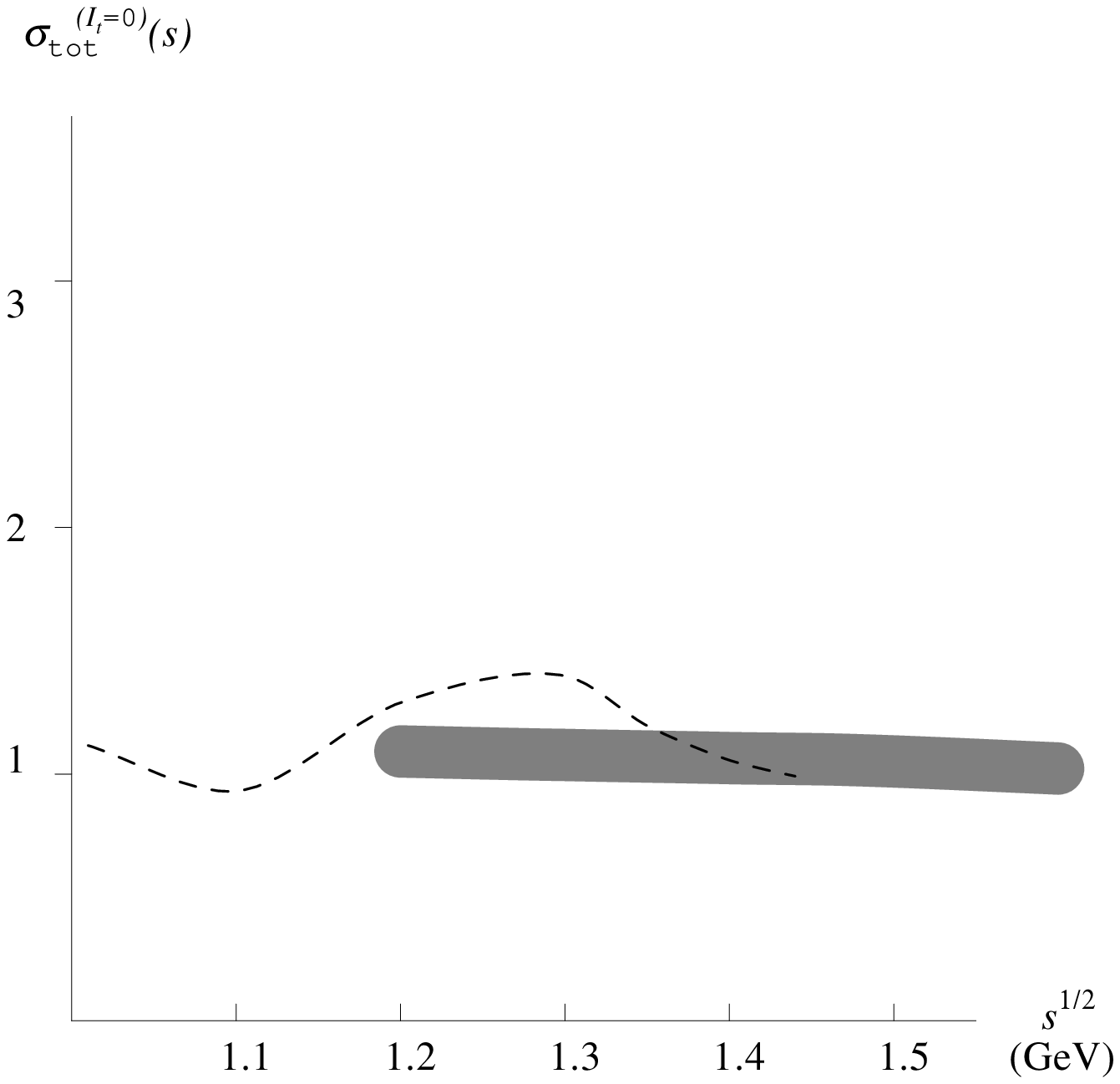}\hfil}
\setbox5=\vbox{\hsize 10truecm\captiontype\figurasc{Figure C.1. }
 {The average cross section $\tfrac{1}{3}[2\sigma_{\pi^0\pi^+}+\sigma_{\pi^0\pi^0}]$, 
which is pure $I_t=0$,  
 arbitrarily normalized. 
Broken line: experimental cross section. Note that the bump here 
is due to the coincidence of two resonances, $f_0(1270),\,f_2(1370)$, 
mostly elastic, around $s^{1/2}\sim 1.3\,\gev$. 
Thick gray line: Regge formula. The thickness of
 the line   covers the error in the theoretical value of the 
Regge residue.}}
\centerline{\box3}
\centerline{\box5}
\bigskip
}\endinsert

We next turn to the sum rules which, as explained,
 relate high ($s^{1/2}\geq1.42\,\gev$) and low energy, with the low energy 
given by the P, D, F waves in the region  $s^{1/2}\leq1.42\,\gev$
 and where the high energy 
is dominated  by, 
respectively, the rho and Pomeron Regge trajectories. 
We will explore explicitely only two of them; we remark that 
the equality of the determinations of the parameters $a_1$, $b_1$ from 
the pion form factor and from the Froissart--Gribov representations 
(\subsect~7.3.6) provide other, highly
nontrivial 
tests of the compatibility of Regge behaviour (in that case, of rho exchange) 
with $s$, $u$ and $t$ crossing, which all three enter into the 
Froissart--Gribov projection.

The first sum rule is obtained by profiting from the threshold behaviour to 
write an unsubtracted forward dispersion relation for the 
quantity $F^{(I_s=1)}(s,0)/(s-4M^2_\pi)$. 
One gets
$$\dfrac{6 M_\pi}{\pi}a_1=\dfrac{1}{\pi}\int_{M^2_\pi}^\infty\dd s\,
\dfrac{\imag F^{(I_s=1)}(s,0)}{(s-4M^2_\pi)^2}+\dfrac{1}{\pi}\sum_{I_s}C^{(su)}_{1I_s}
\int_{M^2_\pi}^\infty\dd s\,
\dfrac{\imag F^{(I_s)}(s,0)}{s^2},
\equn{(C.1)}$$
 which is known at times as the (second) Olsson sum rule; see e.g. Martin, Morgan and
Shaw~(1976).  $C^{(su)}_{1I_s}$ are the
$s-u$  crossing matrix elements.
Canceling $a_1$ with the Froissart--Gribov expression  for this quantity  
and substituting the $C^{(su)}_{1I_s}$ we find the result
$$\eqalign{
I\equiv I_1+I_2\equiv&\, \int_{M^2_\pi}^\infty\dd s\,
\dfrac{\imag F^{(I_t=1)}(s,4M^2_\pi)-\imag F^{(I_t=1)}(s,0)}{s^2}\cr
-&\,
 \int_{M^2_\pi}^\infty\dd s\,\dfrac{8M^2_\pi[s-2M^2_\pi]}{s^2(s-4M^2_\pi)^2}
\imag F^{(I_s=1)}(s,0)=0.\cr
}
\equn{(C.2)}$$
The second term, $I_2$, can also be expressed in terms of amplitudes 
with fixed isospin in the $t$ channel, writing
$$\imag F^{(I_s=1)}=\tfrac{1}{3}F^{(I_t=0)}+\tfrac{1}{3}F^{(I_t=1)}-\tfrac{5}{6}F^{(I_t=2)}.
\equn{(C.3)}$$

The contributions of the S waves cancel in (C.2), so only the P, D and F waves 
contribute (we systematically neglect waves G and higher). 
At high energy, $I_2$ contributes little since the corresponding integral converges rapidly: 
most of the high energy contribution comes from the first term, dominated by rho exchange.
We will use units so that $M_\pi=1$ and obtain the following results:
$$\eqalign{I(\hbox{low energy})=-3.5\times10^{-2},\cr
I(\hbox{high energy (Regge)}, \rho)=3.46\times10^{-2},\cr
I(\hbox{high energy (Regge)}, I=0)=-0.19\times10^{-2}.\cr
}
\equn{(C.4)}$$
By ``low energy" we understand, as usual, the contributions from energies below 
$1.42\,\gev$, where we use phase shifts and inelasticities to calculate the scattering 
amplitudes, and ``high energy" is 
above 1.42 \gev, where a Regge description is employed.
Our Regge parameters are those in \sect~2.4.

Adding the (small) contributions of the $I_t=1$ background and 
the (also small) $I_t=2$ piece we find
$$I=(0.02\pm0.4)\times10^{-2},
\equn{(C.5)}$$
that is to say, perfect consistency.

The second sum rule we discuss is that given in Eqs.~{(B.6), (B.7)} of 
Ananthanarayan et al. (2001), 
which these authors use to claim a Pomeron with a residue a third of its 
standard value. It reads,
$$\eqalign{
J\equiv\int_{4M^2_\pi}^\infty\dd s\,\Bigg\{&
\dfrac{4\imag F'^{(0)}(s,0)-10\imag F'^{(2)}(s,0)}{s^2(s-4M^2_\pi)^2}\cr
-&\,6(3s-4m^2_\pi)\,\dfrac{\imag F'^{(1)}(s,0)-\imag F^{(1)}(s,0)}{s^2(s-4M^2_\pi)^3}
\Bigg\}=0.\cr
}
\equn{(C.6)}
$$
Here $F'^{(I)}(s,t)=\partial F^{(I)}(s,t)/\partial\cos\theta$, 
and the  upper indices  refer to isospin in the $s$ channel.

We will separate $J$ into a low energy and a high energy piece:
$$J=J_{\rm l.e.}+J_{\rm h.e.};\quad
J_{\rm l.e.}=\int_{4M^2_\pi}^{s_h}\dd s\,\dots, \quad
J_{\rm h.e.}=\int_{s_h}^\infty\dd s\,\dots\,. 
\equn{(C.7)}$$ 
The low energy piece, $J_{\rm l.e.}$, only contains contributions of waves D 
and higher. We will show that, if we choose (as we are doing systematically)
 $s_h=1.42^2\,\gev^2$, 
then we find  cancellation, within errors.

For the low energy piece we use the parametrizations of 
Appendix~A and get, with $M_\pi=1$,
$$J_{\rm l.e.}=(1.15\pm0.05)\times10^{-4}. 
\equn{(C.8)}$$ 
For the high energy piece we 
first neglect the $P'$ and $I_t=2$ exchange pieces. Expanding in amplitudes with definite
isospin in the 
$t$ channel, and with the numbers in \sect~2.4 for the 
Pomeron and rho contributions, we then get  
$$J_{\rm h.e.}({\rm Pomeron})=-1.093\times10^{-4},
\quad J_{\rm h.e.}({\rm (Regge)}\;\rho)=0.034\times10^{-4}, 
$$
i.e., including errors, 
$$J_{\rm h.e.}=(-1.06\pm0.17)\times10^{-4}. 
\equn{(C.9)}$$
Thus, we have cancellation between (C.8) and (C.9),  
within errors:
 there is no reason  to justify departure off the expected Regge behaviour.

We next comment a little on the $P'$ and on the inclusion of 
the $I_t=2$ contribution. Because the high energy part of the sum rule  
is mostly given by the $t$ derivative of the even isospin amplitudes, 
a more precise evaluation than the one carried here requires 
 an accurate formula
for the $P'$. 
Unfortunately, the characteristics of this Regge pole are poorly known; see e.g. 
Rarita et al.~(1968). If we take for the 
 the $P'$ trajectory  a formula like that of the 
$\rho$, as discussed in \sect~2.4, then 
 (C.9) changes to
$$J_{\rm h.e.}(\hbox{With }\,P')=(-1.2\pm0.2)\times10^{-4}. $$
Including also the $I_t=2$ contribution, as given in Pel\'aez and 
Yndur\'ain~(2003), 
we would find 
$$J_{\rm h.e.}(\hbox{With }\,P',\,\hbox{and including PY}\,I_t=2) =
(-0.5\pm0.3)\times10^{-4}.
\equn{(C.10)}$$
This  only cancels the low energy piece, (C.8), at the $2\,\sigma$ 
level. This  discrepancy cannot be taken seriously,  because  
the $t$  slope for the $I_t=2$ exchange term of 
 Pel\'aez and 
Yndur\'ain~(2003) is little more than guesswork. 
In fact, one can reverse the argument and use (C.6) to get an idea of the 
parameters of isospin 2 exchange. 
Thus, if we take for $I_t=2$ exchange the parameters of   \sect~2.4, 
Eqs.~{(2.4.7)}, 
we get
$$J_{\rm h.e.}(\hbox{With }\,P',\,\hbox{and including (2.4.7)}\,I_t=2) =
(-0.93\pm0.24)\times10^{-4},
\equn{(C.11)}$$
i.e., $J_{\rm h.e.}+J_{\rm h.e.}=(0.22\pm0.24)\times10^{-5}$: 
 perfect cancellation, within errors.

\bookendchapter

\brochureb{\smallsc references}{\smallsc references}{153}
\booksection{Acknowledgments}

\noindent
I am grateful to J.~Gasser for information concerning, in particular, the 
best values for the $\bar{l}_i$ parameters, and to the same, to G.~Colangelo and 
to H.~Leutwyler for a number of comments
and  criticisms --constructive as well as destructive,  
but always of interest. 
A number of remarks by A.~Pich have been also very useful. 
 I am especially thankful to J.~Pel\'aez for very helpful discussions and 
suggestions.

Finally, I am grateful to the hospitality of CERN (Geneva) and NIKHEF (Amsterdam) 
where part of this work was done.

\booksection{References.}

{\petit
\noindent
The edition of the Particle Data Table we use is:\hb
Hagiwara, K.,  et al.  (2002). {\sl Phys. Rev.}, {\bf D66}, 010001.\hb
\medskip
\noindent
Abramowicz, H., et al. (1980). {\sl Nucl. Phys.}, 
{\bf B166}, 62.\hb
Adler, S. L. (1965). {\sl Phys. Rev.} {\bf 137}, B1022 and {\bf 139}, B1638.\hb
Adler, S. L. (1966). {\sl Phys. Rev.} {\bf 143}, 1144.\hb
Adler, S. L. (1969). {\sl Phys. Rev.} {\bf 177}, 2426.\hb
Adler, S. L. (1971). In {\sl Lectures in 
Elementary Particles and Quantum Field Theory} (Deser, Grisaru and Pendleton, eds.), 
MIT Press.\hb
Ackerstaff, K., et al. (1999). {\sl Eur.~Phys.~J.} {\bf C7}, {571}.\hb
Adler, S. L. and Bardeen, W. A. (1969). {\sl Phys. Rev.} {\bf 182}, 1517.\hb
Aguilar-Ben\'{\i}tez, M., et al. (1978). {\sl Nucl. Phys.} {\bf B140}, 73.\hb
Ahlfors, L. V. (1953). {\sl Complex Analysis}, McGraw Hill.\hb
Aitala, E.~M., et al., (2002). {\sl Phys. Rev. Lett.} {\bf 89}, 12801.\hb 
Akhmetsin, R. R.  et al. (1999). Budker INP 99-10 (hep-ex/9904027). 
New version: hep-ex/00112031.\hb 
Akhoury, R. and Alfakih, A. (1991). {\sl Ann. Phys.} (NY) {\bf 210}, 81.\hb
Aloisio, A., et al. (2002). {\sl Phys. Letters} {\bf B538}, 21.\hb
Altarelli, G., and Parisi, G. (1977). {\sl Nucl. Phys.} {\bf B126}, 298.\hb
Amendolia, S. R., et al. (1986). {\sl Nucl. Phys.} {\bf B277}, {168}.\hb
Amor\'os, G., Bijnens, J., and Talavera, P. (2000). 
{\sl Nucl. Phys.} {\bf B585}, 329 and (E) LU~TP~00-11.\hb
Ananthanarayan, B., et al. (2001). {\sl Phys. Rep.} {\bf 353}, 207.\hb
Anderson, S. et. al, (2000). {\sl Phys. Rev.} {\bf D61}, 112002\hb
Atkinson, D. (1968). {\sl Nucl. Phys.} {\bf B7}, 375.\hb
Atkinson, D. (1971). In {\sl Strong Interaction Physics} (H\"ohler, ed.), Springer, Berlin.\hb
Atkinson, D., Mahoux, G., and Yndur\'ain, F. J. (1973). {\sl Nucl. Phys.} {\bf B54}, 263; 
{\bf B98}, 521 (1975).\hb

\smallskip\noindent
Balitskii, Ya. Ya., and  Lipatov, L. N. (1978). {\sl  Sov. J. Nucl. Phys.} {\bf 28}, 822.\hb
Barate, R., et al. (1997). {\sl Z.~Phys.} {\bf C76}, {15}.\hb
Bardeen, W. A. (1974). {\sl Nucl. Phys.} {\bf B75}, 246.\hb
Barger, V. D., and Cline, D. B. (1969). {\sl Phenomenological Theories of High Energy 
Scattering}, Benjamin, New~York.\hb
Barkov, L.~M., et al. (1985). {\sl Nucl. Phys.} {\bf B256}, {365}.\hb
Barton, G. (1965). {\sl Dispersion Techniques in Field Theory},
 Benjamin, New~York, 1965.\hb
Basdevant, J. L., Froggatt, P. D., and Petersen, J. L. (1972). {\sl Phys. Letters} {\bf 41B},
173.\hb Basdevant, J. L., Froggatt, P. D., and Petersen, J. L. (1974). {\sl Nucl. Phys.}, {\bf
B72}, 413.\hb Baton, J. P., et al. (1970). {\sl Phys. Letters} {\bf 33B}, 528.\hb
Becchi, C., Narison, S., de Rafael, E., and Yndur\'ain, F. J. (1981). 
{\sl Z. Phys.} {\bf C8}, 335.\hb
Belavin, A. A., and Navodetsky, I. M. (1968). {\sl Phys. Letters} {\bf 26B}, 668.\hb
Bell, J. S., and Jackiw, R. (1969). {\sl Nuovo Cimento} {\bf 60A}, 47.\hb
Bernard, V., Kaiser, N., and Meissner, U. G. (1991). {\sl Phys. Rev.} {\bf D44}, 3698.\hb
Bijnens, J. (1993). {\sl Phys. Letters} {\bf B306}, 343.\hb
Bijnens, J.,  Prades, J., and  de Rafael, E. (1995). {\sl Phys. Letters} {\bf 348}, 226.\hb
Bijnens, J.,  et al. (1996). {\sl Phys. Letters} {\bf B374}, 210.\hb
Bijnens, J., Colangelo, G., and Eder, G. (2000). {\sl Ann. Phys.} (N.Y.) {\bf 280}, 100.\hb
Bijnens, J., Colangelo, G., and Talavera, P. (1998). {\sl JHEP} 9805: 014.\hb
Biswas, N. N., et al. (1967). {\sl Phys. Rev. Letters}, 
{\bf 18}, 273.\hb 
Bogoliubov, N.~N., Logunov, A.~A., and Todorov, I.~T. (1975).
{\sl Axiomatic Field Theory}, Benjamin, New~York.\hb
Brandt, R., and Preparata, G. (1970). {\sl Ann. Phys.} (N.Y.) {\bf 61}, 119.\hb
Broadhurst, D. J. (1981). {\sl Phys. Letters} {\bf B101}, 423.\hb

\smallskip\noindent
Chen, S., et al. (2001). {\sl Eur. J. Phys.} {\bf C22}, 3.\hb
Chetyrkin  K. G.,  Gorishnii, S. G., and  Tkachov, F. V. (1982). {\sl Phys. Lett.} {\bf B119}, 407.\hb
Chetyrkin, K. G.,  Pirjol, D., and  Schilcher, K. (1997). 
{\sl Phys. Lett.} {\bf B404}, 337.\hb
Chetyrkin  K. G.,  Groshny, S. G., and  Tkachov, F. V. (1982). {\sl Phys. Lett.} {\bf B119}, 407.\hb
Ciulli, S. (1973). In {\sl Strong Interactions}, 
Lecture Notes in Physics, Springer-Verlag, New~York.\hb
Callan, C. G., Coleman, S., and Jackiw, R. (1970). {\sl Ann. Phys.} (N.Y.) {\bf 59}, 42.\hb
Cirigliano, V., Donoghue, J. F., and Golowich, E. (2000). {\sl Eur. 
Phys. J.} {\bf C18}, 83.\hb
Cohen, D. et al.  (1973). {\sl Phys. Rev.} {\bf D7}, 661. \hb
Colangelo, G. (1995). {\sl Phys. Letters} {\bf B350}, 85 and 
(E) {\bf B361}, 234 (1995).\hb
Colangelo, G., Finkelmeir, M., and Urech, R. (1966). {\sl Phys. Rev.} {\bf D54}, 4403.\hb
Colangelo, G., Gasser, J.,  and Leutwyler, H. (2001).
 {\sl Nucl. Phys.} {\bf B603},  125.\hb
Coleman, S., Wess, J.,  and Zumino, B. (1969). {\sl Phys. Rev.} {\bf 177}, 2239 and 2247 
(with C. G. Callan).\hb
Collins, J. C., Duncan, A., and Joglekar, S. D. (1977). {\sl Phys. Rev.} {\bf D16}, 438.\hb
Crewther, R. J. (1972). {\sl Phys. Rev. Lett.} {\bf 28}, 1421.\hb
Crewther, R. J. (1979a). {\sl Riv. Nuovo Cimento} {\bf 2}, No. 7.\hb
Crewther, R. J. (1979b). In {\sl Field Theoretical Methods in Elementary 
Particle Physics,} Proc. Kaiserslautern School.\hb

\smallskip\noindent
Das, T., Mathur, V. S., and Okubo, S. (1967). {\sl Phys. Rev. Lett.} {\bf 18},761; {\bf 19},
859.\hb
Descotes, S., Fuchs, N. H.,  Girlanda, L., and   Stern, J., {Eur. Phys. J. C}, 
{\bf 24}, 469, (2002).\hb
 Dobado, A., Gomez-Nicola, A.,  Maroto, A., and   Pel\'aez, J.R. (1997).
{\sl Effective Lagrangians for the Standard Model} Springer, Berlin.\hb
Dokshitzer, Yu. L.  (1977). {\sl Sov. Phys. JETP} {\bf 46,} 641.\hb
Dom\'\i nguez, C. A. (1978). {\sl Phys. Rev. Lett.} {\bf 41}, 605.\hb
Dom\'\i nguez, C. A., and de Rafael, E. (1987). {\sl Ann. Phys.} (N.Y.) {\bf 174}, 372.\hb
Donnacie, S., Doch, G., Landshoff, P., and Nachtmann, O. (2002). 
{\sl Pomeron Physics and QCD}, Cambridge.\hb
Donoghue, J. F., Gasser, J., and Leutwyler, H. (1990). {\sl Nucl. Phys.} {\bf B343}, 431.\hb
Donoghue, J. F.,  Holstein, B. R., and  Wyler, D. (1993). {\sl Phys. Rev.} {\bf D47}, 2089.\hb

\smallskip\noindent
Ecker, G. (1995). {\sl Prog. Part. Nucl. Phys.} {\bf 35}, 71. \hb
Ecker, G., et al. (1989). {\sl Nucl. Phys.} {\bf B223}, 425 and {\bf B321}, 311.\hb
Eden, R. J., Landshoff, P. V., Olive, D. J., and Polkinghorne,~J.~C.~(1966). 
{\sl The Analytic $S$-Matrix}, Cambridge, 1966.\hb
Ellis, J. (1976). In 
{\sl Weak and Electromagnetic Interactions at High Energy}, North Holland, Amsterdam.\hb
Epstein, H.,  Glaser, V., and Martin, A. (1969). {\sl  Commun. Math. Phys.} 
{\bf 13}, 257.\hb
Estabrooks, P., and Martin, A. D. (1974). {\sl Nucl. Phys.} {\bf B79}, 301.\hb

\smallskip\noindent
Farrar, G., and Jackson, D. R. (1979). {\sl Phys. Rev. Letters}. {\bf 43}, 246.\hb
Fearing, H. W., and Scherer, S. (1996). {\sl Phys. Rev.} {\bf D53}, 315.\hb
Frink, M., Kubis, B., and Mei\ss ner,~U.~G. (2002). {\sl Eur. Phys. J.} {\bf C25}, 259.\hb
Froissart, M. (1961). Proc. la Jolla Conf. on Elementary Particles.\hb
Fujikawa, K. (1980). {\sl Phys. Rev.} {\bf D21}, 2848 and (E) {\bf D22}, 1499.\hb
Fujikawa, K. (1984). {\sl Phys. Rev.} {\bf D29}, 285.\hb
Fujikawa, K. (1985). {\sl Phys. Rev.} {\bf D31}, 341.\hb
Fuchs, N. H., Sazdjian, H., and Stern, J. (1991). {\sl Phys. Letters} {\bf 269}, 183.\hb

\smallskip\noindent
Galindo, A., and  and Pascual, P. (1978). {\sl Mec\'anica cu\'antica}, Alhambra, Madrid.\hb
\indent 
English translation: {\sl Quantum Mechanics}, Vols. I,~II. Springer, Berlin.\hb
Gasser, J., and Leutwyler, H. (1982). {\sl Phys. Rep.} {\bf C87}, 77.\hb
Gasser, J., and Leutwyler, H. (1983). {\sl Phys. Letters.} {\bf B125}, 321 and 325.\hb
Gasser, J., and Leutwyler, H. (1984). {\sl Ann. Phys.} (N.Y.) {\bf 158}, 142.\hb
Gasser, J., and Leutwyler, H. (1985a). {\sl Nucl. Phys.} {\bf B250}, 465.\hb
Gasser, J., and Leutwyler, H. (1985b). {\sl Nucl. Phys.} {\bf B250}, 517.\hb
Gasser, J., and Meissner, U.-G. (1991). {\sl Nucl. Phys.} {\bf B357}, 90.\hb
Gell-Mann, M. (1962). {\sl Phys. Rev. Lett.} {\bf 8}, 263.\hb
Gell-Mann, M., Oakes, R. L., and Renner, B. (1968). {\sl Phys. Rev.} {\bf 175}, 2195.\hb
Generalis,  S. C. (1990). {\sl J. Phys.} {\bf G16},  785.\hb
Georgi, H. D. (1984). {\sl Weak Interactions and 
Modern Particle Theory}. Benjamin, Menlo Park.\hb
Glashow, S. L., and Weinberg, S. (1968). {\sl Phys. Rev. Lett.} {\bf 20}, 224.\hb
Goldberger, M.~L. and Watson, K.~M. (1964). {\sl Collision Theory}, 
Wiley, New~York.\hb
G\'omez-Nicola, A., and Pel\'aez, J. (2002). {\sl Phys. Rev.} {\bf D65}, 054009.\hb
Gorishnii, S. G., Kataev, A. L., Larin, S. A., and Sugurladze, L. R. (1991).
 {\sl Phys. Rev.} {\bf D43}, 1633.\hb
Gounnaris, G. J., and Sakurai, J. J. (1968). {\sl Phys. Rev. Lett.} {\bf 21}, {244}.\hb
Grayer, G., et al., (1974). {\sl Nucl. Phys.}  {\bf B75}, 189.\hb
Gribov, V. N. (1962). Sov. Phys. JETP {\bf 14}, 1395.\hb
Gribov, V. N., and  Lipatov, L. N. (1972). {\sl Sov. J. Nucl. Phys.} {\bf 15}, 438 and 675.\hb
Gribov, V. N., and Pomeranchuk, I. Ya. (1962).  {\sl Phys. Rev. Lett.} {\bf 8}, 343.\hb
Guerrero, F., and Pich, A. (1997). {\sl Phys. Letters}, {\bf B412}, 382.\hb

\smallskip\noindent
Hanlon, J., et al. (1976).  {\sl Phys. Rev. Letters}, 
{\bf 37}, 967.\hb
Hoogland, W., et al. (1977). {\sl Nucl. Phys.} {\bf B126}, 109.\hb
Hubschmid, W., and Mallik, S. (1981). {\sl Nucl. Phys.} {\bf B193}, 368.\hb
Hyams, B., et al. (1973). {\sl Nucl. Phys.} {\bf B64}, 134.\hb

\smallskip\noindent
Kami\'nski, R., Le\'sniak, L., and Loiseau, B. (2003). {\sl Phys. Letters} {\bf B551}, 241.\hb
Kaplan, D. B., and Manohar,  A. V. (1986). {\sl Phys. Rev. Lett.} {\bf 56}, 2004.\hb
Kataev, A. L., Krasnikov, N. V., and Pivovarov, A. A. (1983). 
{\sl Phys. Lett.} {\bf 123B}, 93.\hb
Knecht, M., et al. (1995). {\sl Nucl. Phys.} {\bf 457}, 513.\hb
Knecht, M., et al. (1996). {\sl Nucl. Phys.} {\bf 471}, 445.\hb
Kokkedee, J. J. J. (1969). {\sl The Quark Model}, Benjamin, New York.\hb
Kuraev, E. A., Lipatov, L. N., and Fadin, V. S.
(1976). {\sl Sov. Phys. JETP} {\bf 44}, 443.\hb

\smallskip\noindent
Lehmann, H. (1972). {\sl Phys. Letters} {\bf B41}, 529.\hb
Leutwyler, H. (1974). {\sl Nucl. Phys.} {\bf B76}, 413.\hb
Levin, H. J., and Frankfurter, L. L. (1965). {\sl JETP Letters} {\bf 2}, 65.\hb
Losty, M.~J., et al. (1974). {\sl Nucl. Phys.} {\bf B69}, 185.\hb
Lovelace, C. (1968). {\sl Phys. Letters} {\bf B28}, 264.\hb

\smallskip\noindent
Marshak, R. E., Riazzuddin, and Ryan, C. P. (1969). 
{\sl Theory of Weak Interactions in Particle Physics}, Wiley, New~York.\hb
Martin, A. (1969). {\sl Scattering theory}, Lecture Notes in Physics, Springer, Berlin.\hb
Martin, B. R., Morgan, D., and and Shaw, G. (1976). {\sl 
Pion-Pion Interactions in Particle Physics}, Academic Press, New~York.\hb
Morgan, D., and Pi\u{s}ut, J. (1970). {\sl Low Energy Pion-Pion Scattering}, 
in ``Low Energy Hadron
Interactions", Springer, Berlin.\hb
Muskhelishvili, N.~I. (1958). {\sl Singular
 Integral Equations}, Nordhoof, Groningen.\hb

\smallskip\noindent
Nachtmann, O, and de Rafael, E. (1969). CERN preprint TH-1031 (unpublished).\hb
Nagel, M. M., et al. (1979). {\sl Nucl. Phys.} {\bf B147}, 189.\hb
Nambu, Y. (1960). {\sl Phys. Rev. Lett.} {\bf 4}, 380.\hb
Nambu, Y., and Jona-Lasinio, G. (1961a). {\sl Phys. Rev.} {\bf 122}, 345.\hb
Nambu, Y., and Jona-Lasinio, G. (1961b). {\sl Phys. Rev.} {\bf 124}, 246.\hb
Narison, S. (1995). {\sl Phys. Lett.} {\bf B358}, 112.\hb

\smallskip\noindent
Okubo, S. (1969). {\sl Phys. Rev.} {\bf 188}, 2295 and 2300.\hb
Oller, J.~A., Oset, E., and Pel\'aez, J. R.,  (1999). {\sl Phys. Rev.} {\bf D59}, 
074001.\hb
Olsson, M. G. (1967). {\sl Phys. Rev.} {\bf 162}, 1338.\hb
Omn\`es, R. (1958). {\sl Nuovo Cimento} {\bf 8}, 316.\hb
Omn\`es, R., and Froissart, R. (1963). {\sl Mandelstam Theory and 
Regge Poles}, Benjamin, New~York.\hb
Omn\`es, R., and De Witt, C. (Eds.) (1960). {\sl Dispersion Relations and 
Elementary Particles}, Les~Houches lectures, Wiley, New~York.\hb

\smallskip\noindent
Pagels, H. (1975). {\sl Phys. Rep.} {\bf C16}, 219.\hb
Palou, F. P., and Yndur\'ain, F. J. (1974). {\sl Nuovo Cimento} {\bf 19A}, 245.\hb
Palou, F. P., S\'anchez-G\'omez, J, L, and Yndur\'ain, F. J. (1975). {\sl Z. Phys.} {\bf A274},
161.\hb Pascual, P., and  de Rafael, E. (1982). {\sl Z. Phys.} {\bf C12}, 127.\hb 
Pascual, P., and Yndur\'ain, F. J. (1974). {\sl Nucl. Phys.} {\bf B83}, 362.\hb
Pel\'aez, J. R., and Yndur\'ain, F. J. (2003). {\sl Phys. Rev.} {\bf D}, in press 
(hep-ph/0304067).\hb 
Pennington, M. R., (1975). {\sl Ann. Phys.} (N.Y.), {\bf 92}, 164.\hb 
Pich, A. (1995). {\sl Rep. Progr. Phys}. {\bf 58}, 563.\hb
Pilkuhn, H. (1967). {\sl The Interaction of 
Hadrons}, North-Holland, Amsterdam.\hb
Pislak, S.,  et al. (2001). {\sl Phys. Rev. Lett.}, {\bf 87}, 221801.\hb
Pi\u{s}ut, J. (1970). {\sl Analytic Extrapolations and 
Determination of Pion-Pion Phase Shifts}, in ``Low Energy Hadron
Interactions", Springer, Berlin.\hb
Prokup, J. P., et al. (1991). Proc XVII London Conference on 
High Energy Physics.\hb
Protopopescu, S. D., et al. (1973). {\sl Phys Rev.} {\bf D7}, 1279.\hb

\smallskip\noindent
Rarita, W., et al. (1968). {\sl Phys. Rev.} {\bf 165}, 1615.\hb
Robertson, W. J., Walker, W. D., and Davis, J. L. (1973). {\sl Phys. Rev.} {\bf D7}, 2554.\hb
Rosselet, L., et al. (1977). {\sl Phys. Rev.} {\bf D15}, 574.\hb
Roy, S. M. (1971). {\sl Phys. Letters} {\bf 36B}, 353.\hb
Roy, S. M. (1990). {\sl Helv. Phys. Acta} {\bf 63}, 627.\hb

\smallskip\noindent
Scherer, S. (2002). MKPH-T-02-09 ({hep-ph/0210398}).\hb
Schenk, A. (1991). {\sl Nucl. Phys.} {\bf B363}, 97.\hb
Shapiro, J. A. (1969). {\sl Phys. Rev.} {\bf 179}, 1345.\hb
Sommer, G. (1970). {\sl  Fort. der Physik}, {\bf 18}, 577.\hb
Steinberger, J. (1949). {\sl Phys. Rev.} {\bf 76}, 1180.\hb
Sugurladze, L. R., and Tkachov, F. V. (1990). {\sl Nucl. Phys.} {\bf B331}, 35.\hb
Sutherland, D. G., (1967). {\sl Nucl. Phys.} {\bf B2}, 433.\hb

\smallskip\noindent
Taylor, J. C. (1976). {\sl Gauge Theories of Weak Interactions}, Cambridge.\hb
Titchmarsh, E. C. (1939). {\sl The Theory of Functions}, Oxford.\hb
de Troc\'oniz, J. F., and Yndur\'ain, F. J. (2002). {\sl Phys. Rev.}  {\bf D65}, 093001.\hb

\smallskip\noindent
Vainshtein, A. I., et al. (1978). {\sl Sov. J. Nucl. Phys.} {\bf 27}, 274.\hb
Veltman, M. (1967). {\sl Proc. Roy. Soc.} (London) {\bf A301}, 107.\hb
Veneziano, G. (1968). {\sl Nuovo Cimento} {\bf 57}, 190.\hb

\smallskip\noindent
Weinberg, S. (1966). {\sl Phys. Rev. Lett.} {\bf 17}, 616.\hb
Weinberg, S. (1968a). {\sl Phys. Rev.} {\bf 166}, 1568.\hb
Weinberg, S. (1968b). {\sl Phys. Rev.} {\bf 177}, 2247.\hb
Weinberg, S. (1975). {\sl Phys. Rev.} {\bf D11}, 3583.\hb
Weinberg, S. (1978a). In {\sl a Festschrift for I. I. Rabi},
 New York Academy of Sciences, New York.\hb
Weinberg, S. (1979). {\sl Physica A} {\bf 96}, 327.\hb
Wess, J., and Zumino, B. (1971). {\sl Phys. Letters} {\bf 37B}, 95.\hb

\smallskip\noindent
Yndur\'ain, F. J. (1972). {\sl Rev. Mod. Phys.} {\bf 44}, 645.\hb
Yndur\'ain, F. J. (1975). {\sl Nucl. Phys.} {\bf B88}, 318.\hb
Yndur\'ain, F. J. (1993). {\sl Relativistic Quantum Mechanics}, Springer, Berlin.\hb
Yndur\'ain, F. J. (1998). {\sl Nucl. Phys.} {\bf B517}, 324.\hb
Yndur\'ain, F. J. (1999). {\sl The Theory of Quark and Gluon Interactions},
Springer, Berlin.\hb
Yndur\'ain, F. J. (2003a). Preprint FTUAM 03-15 (hep-ph/03009039), to appear in Phys. Letters.\hb
Yndur\'ain, F. J. (2003b). Preprint FTUAM (hep-ph/0310206).\hb

\smallskip\noindent
Zepeda, A. (1978). {\sl Phys. Rev. Lett.} {\bf 41}, 139.\hb
}

\bookendchapter
\bye